\documentclass[a4paper,12pt,twoside]{book}
\usepackage{sebathesis}
\usepackage{amssymb}
\usepackage{amsmath}
\usepackage{lscape}
\usepackage{textcomp}
\usepackage{longtable}
\usepackage{ctable}
\usepackage{caption,subfig}
\usepackage{float}
\usepackage{appendix}

\headers                        

\begin{document}

\raggedbottom

\thispagestyle{empty}  

\begin{titlepage}
  \begin{center}

    \begin{LARGE} 
      \begin{bf}
        \begin{sc}
          UNIVERSIDADE DE LISBOA
        \end{sc}
      \end{bf}
    \end{LARGE}

\gap

    \begin{LARGE} 
      \begin{bf}
        \begin{sc}
          FACULDADE DE CI\^{E}NCIAS
        \end{sc}
      \end{bf}
    \end{LARGE}

\gap

    \begin{LARGE} 
      \begin{bf}
        \begin{sc}
          DEPARTAMENTO DE F\'ISICA
        \end{sc}
      \end{bf}
    \end{LARGE}

\gap

    \begin{figure}[ht]
      \centering\includegraphics[width=0.3\textwidth]{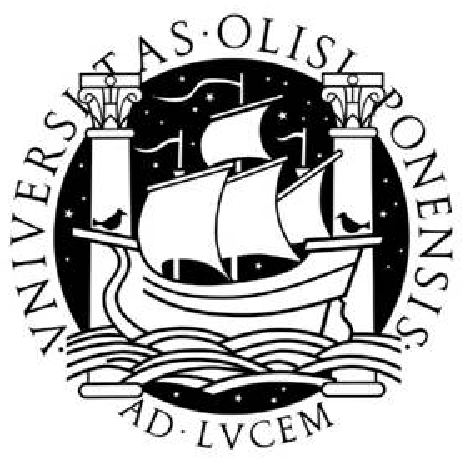}
    \end{figure}


    \begin{Huge}
      \begin{bf}
      \begin{sc}
        A multiwavelength study of\\ near- and mid-infrared\\ selected galaxies at high redshift:\\ ERGs, AGN-identification and\\ the contribution from dust\\
        \end{sc}
      \end{bf}
    \end{Huge}

\gap
    \gap

    \begin{large} 
      \begin{bf}
        \begin{sc}
        Hugo G. Messias\\[0.2cm]
        \end{sc}
      \end{bf}
      Doutoramento em Astronomia e Astrof\'isica\\
    \end{large}

    \vfill

    Ano 2011
  \end{center}

\blankpage
\clearpage
\thispagestyle{empty}
  \begin{center}

    \begin{LARGE} 
      \begin{bf}
        \begin{sc}
          UNIVERSIDADE DE LISBOA
        \end{sc}
      \end{bf}
    \end{LARGE}

\gap

    \begin{LARGE} 
      \begin{bf}
        \begin{sc}
          FACULDADE DE CI\^ENCIAS
        \end{sc}
      \end{bf}
    \end{LARGE}

\gap

    \begin{LARGE} 
      \begin{bf}
        \begin{sc}
          DEPARTAMENTO DE F\'ISICA
        \end{sc}
      \end{bf}
    \end{LARGE}

\gap

    \begin{figure}[ht]
      \centering\includegraphics[width=.3\textwidth]{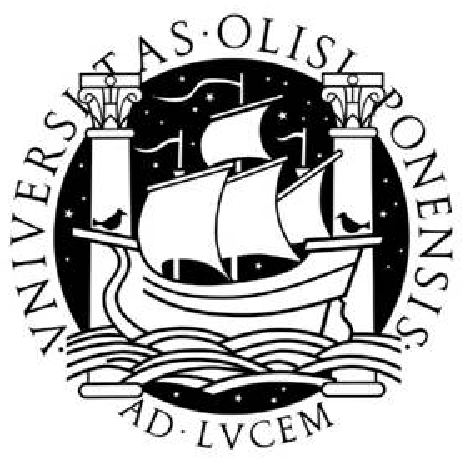}
    \end{figure}


    \begin{Huge}
      \begin{bf}
      \begin{sc}
        A multiwavelength study of\\ near- and mid-infrared\\ selected galaxies at high redshift:\\ ERGs, AGN-identification and\\ the contribution from dust\\
        \end{sc}
      \end{bf}
    \end{Huge}

\gap
    \gap

    \begin{large} 
      \begin{bf}
        \begin{sc}
        Hugo G. Messias\\[0.2cm]
        \end{sc}
      \end{bf}
      Doutoramento em Astronomia e Astrof\'isica\\
      Tese orientada pelo\\
      Prof. Doutor Jos\'e Manuel Afonso
    \end{large}

    \vfill

    Ano 2011
  \end{center}
\end{titlepage}

\blankpage
\dedicatoria
\blankpage
\newpage
\thispagestyle{empty}
\vspace*{.3\textheight}

\begin{flushright}

  {\it ``La science est faite d'erreurs, mais ce sont des erreurs utiles\\car elles m\`enent petit \`a petit \`a la v\'erit\'e.''\\``Science is made up of mistakes, but they are mistakes which it is useful to make,\\because they lead little by little to the truth.''}\\[1.5ex]
  Jules Verne, in \textit{Voyage au centre de la Terre\\A Journey to the Center of the Earth}

  \vspace{1cm}

  {\it ``La r\'ealit\'e fournit quelquefois des faits si romanesques que\\l'imagination elle-m\^eme ne pourrait rien y ajouter.''\\``Reality sometimes provides us with facts so romantic that\\imagination itself could add nothing to them.''}\\[1.5ex]
  Jules Verne

  \vspace{1cm}

  {\it ``Science is the belief in the ignorance of experts.'' ``If you thought that science was certain --- well, that is just an error on your part.'' ``We are trying to prove ourselves wrong as quickly as possible, because only in that way can we find progress.''}\\[1.5ex]
  Richard Feynman

  \vspace{1cm}

  {\it \textbf{``In the end, everything is a gag.''}}\\[1.5ex]
  Charles Chaplin

\end{flushright}
\singlespacing
\blankpage
\chapter*{Author's note}
\addcontentsline{toc}{chapter}{Author's note}
\thispagestyle{empty}
\pagenumbering{roman}
\setcounter{page}{1}

This thesis, by the title of ``A multiwavelength study of near- and mid-infrared selected galaxies at high redshift: ERGs, AGN-identification and the contribution from dust'', was supervised by Doctor Jos\'e Afonso and co-supervised by Doctor Bahram Mobasher. The work was developed at the Centre for Astronomy and Astrophysics of the University of Lisbon (CAAUL), research centre of the Faculty of Science of the University of Lisbon (FCUL).

On $21^{\rm st}$ July 2011, the work was presented and defended before a jury presided by Professor Doctor Maria Margarida Godinho and composed by Doctor Bahram Mobasher (Professor of University of California, Riverside), Doctor Duncan Farrah (Senior Research Fellow of University of Sussex), Doctor M\'ario Santos (Research Assistant of Technic University of Lisbon), Doctor Jo\~ao Yun (Associate Professor of FCUL), Professor Doctor Rui Agostinho (Assistant Professor of FCUL), Doctor Jos\'e Afonso (Assistant Professor of FCUL).


The current version which is now presented differs from the original one\footnote{The original version is available at: http://hdl.handle.net/10451/3824}. Corrections have been made mainly as a result from the PhD defence, the ongoing peer reviewal of a manuscript about the infrared selection of active galactic nuclei, and ongoing developments on the study of passive disc galaxies.
\blankpage
\chapter*{Acknowledgements/Agradecimentos}
\addcontentsline{toc}{chapter}{Acknowledgements/Agradecimentos}
\thispagestyle{empty}

Although the thesis is written in English, allow me to thank in Portuguese the people that have been part of my life for more than a thesis-time.

Obrigado M\~ae e Pai! Pelo esfor\c{c}o em fazer de n\'os o que somos. Por seguirem as vossas convic\c{c}\~oes metendo-nos sempre em primeiro lugar, independentemente das condi\c{c}\~oes em que o faziam. Por nos terem proporcionado todas as aventuras que vivemos, por nos terem levado por becos e ruelas, e caminhos de tr\^es estrelas, por termos dado a volta \`a Terra. Hei-de agradecer-vos com um forte abra\c{c}o sem mais dizer, mas queria que as primeiras linhas deste texto fossem vossas. Seja qual f\^or o resultado, Parab\'ens!

A voc\^es meus dois irm\~aos! \`As nossas tardes de improviso Ritchardeano, enquanto o Ritch\'e terminava o seu T\'ecnico, que agora muito tenho recordado e me t\^em posto a rir mesmo nas alturas em que isto corre pior. Obrigado aos dois pelo p\~ao-com-chouri\c{c}o \`a noite, pelos ``Salta puto!!'' e ``diverte-te puto!'', pelos treinos em que vamos os tr\^es deixar os pulm\~oes. Obrigado pelas sobrinhas que trouxeram ao Mundo. Parab\'ens Anuxas e BIpa!


A ti Juaninha! N\'os que ainda poucos passos demos, mas j\'a pass\'amos por muito. Obrigado \`a dan\c{c}a que nos uniu e que nos acompanhar\'a enquanto pudermos bater com a unha no bra\c{c}o cadeira. Os pr\'oximos tempos n\~ao parecem dos mais f\'aceis, mas h\~ao-de passar bem mais r\'apido que estes \'ultimos quatro anos. Tu vais voar! Vamos conseguir! Beijuhu! Obrigado L\'idia e Victor e Parab\'ens! 

Aos meus av\'os! Pelas bolachas Maria do ``pacote azul'', pelo assobio at\'e \`a m\'usica, pelas tardes de calor na Aldeia, pelas torradas na lareira. Por me terem dado os pais que deram!

Obrigado Teresa por tudo! Farei um esfor\c{c}o para te continuar a encher a caixa de correio. Boa sorte para o teu doc.

Ao grupo do Judo. Em especial, a voc\^es os dois, Moraes e Rui! Passaram bem para l\'a de simples treinadores. Cada gera\c{c}\~ao que vos passa nas m\~aos junta-se \`a pr\'oxima sem haver problemas de idade. Puxamos todos uns pelos outros, mesmo fora do tapete, e \'e a\'i que revelamos o grupo consistente que somos. Como o capit\~ao disse ``Levantem a cabe\c{c}a!'', porque para o ano regressa \`as nossas m\~aos. Matos, boa sorte para o doc! \`As nossas m\'usicas Hug\~ao! Beijo grande In\^es! Rosi, come\c{c}a a fazer a prancha.

Ao grupo do Tango. \`As minhas duas mentoras Dalila e Alexandra, \`a minha professora Miriam, ao meu curioso companheiro Francisco que ningu\'em consegue acompanhar a sua energia. A todos!

Um beijo enorme \`a Guida que tantas manias incutiu nesta fam\'ilia e outro \`a Maria Am\'elia. As duas aturaram um puto no seu auge de reguilice (mas sempre divertido, espero). Os meus parab\'ens! Queria ainda agradecer ao Col\'egio Moderno por me ter possibilitado seguir a op\c{c}\~ao Artes-F\'isica do 10$^{\rm o}$ ao 12$^{\rm o}$ ano, permitindo, deste modo, atrasar a derradeira decis\~ao at\'e ao \'ultimo ano de liceu.

This thesis would not be possible without the doctoral grant SFRH/BD/31338/2006 from Funda\c{c}\~ao para a Ci\^encia e Tecnologia. I am deeply thankful for the opportunity given by the institution. I wished I had produced more during the thesis itself, but all the work done is scientifically relevant and I will acknowledge your support in any of the outcomes resulting from my thesis work. Please, do continue to help students fulfilling their goals. As many as possible! Science is future and development, and we all need that right now. I also acknowledge support from FCT through the research grant PTDC/FIS/100170/2008, University of California Riverside for the support that enabled me to work with Bahram Mobasher, from the Space Telescope Science Institute and Anglo-Australian Observatory during a visit to the respective headquarters.

Obrigado Afonso! I am aware that in the beginning your patience was immense after all that knocking on your door. I hope I was a good first experience nonetheless, and I wish you all the luck, both professionally --- as the new CAAUL director --- and, more importantly, family-wise --- with the three kids to handle. Thank you for all the opportunities you gave me, allowing me to travel the world. You have taught me a lot, I just hope I make the most out of it. Hopefully, I will be back and help CAAUL grow even more. And thank you for the help throughout this last months of thesis writing.

``Merci'' Bahram, Azin, Armeen, and Tara for welcoming me to your home. I do appreciate that a lot! It was really a nice time with you and the Darc's, and I can't thank you enough. You have a friend in Lisbon waiting for a visit of yours. I hope we keep seeing each other, and never let contact go. Thank you Bahram for the opportunity to work with you (it was a great experience) and for the opportunities you have provided, allowing me to make contact with the people from the other side of the Atlantic. I do hope we keep collaborating along.

Thank you Andrew for always being the first to give comments to my ``hard'' drafts, the initial stacking code that forced me to understand the wonders of scripting, and welcoming me at AAO. Thank you so much Mara for the support, I owe the ESO presentation to you. Thank you for the galaxy templates and for guiding me throughout the Keck run. Thank you Dave and T\^ania for the help on the ERG work. Thank you Harry, Norman, and Tomas for the opportunity to work with you.

To both of my friends Fernando Buitrago and Antonio Cava, a big Obrigado for the help on the Passive Disk Galaxies project. I would not have made it this far if it wasn't for you two. I hope we keep collaborating from now on. Abrazo!

Thank you Jo\~ao Yun for bringing me to the OAL in the first place and introducing me to Afonso, Rui Agostinho for guiding me through the amazing Observat\'orio Astron\'omico de Lisboa and to show me how fun Astronomy is, Jo\~ao Retr\^e for the inhuman effort to bring together the team that now brings OAL to life at night, and Cristina Fernandes for the latex thesis template that helped a lot the thesis writing.

Thank you all my friends back in the USA. You made my life great there. It was really nice meeting you all! And for those who went with me on the New Zealand trip.... no words! Well.. a few, next time.. lets stop! Thank you for this great trip!

I would like to thank FIREWORKS, Norris \& Afonso et al., GOODS, MUSIC, COSMOS, Luo et al., SWIRE, SDSS, UKIDSS, 2MASS, Seymour et al., teams for producing the public catalogues and images on which this thesis is based. I have to thank COSMOS and SERVS teams for the opportunity to observe during their telescope times, respectively, at Keck\,II and Telescopio Nazionale Galileo.

I thank the following people for insightful conversations and comments that allowed this thesis to improve: Tommy Wiklind, Du\'{i}lia de Mello, Leonidas Moustakas, Tomas Dahlen, Fr\'ed\'eric Bournaud, Duncan Farrah, Harry Ferguson, Norman Grogin, Andrea Comastri, Jennifer Donley, Vernesa Smol\v{c}i\'{c}, Jennifer Lotz (also for providing the morphology code), Jessica Krick, and Pablo P\'erez-Gonz\'alez. And, although they made my life hard (and one still is), I have to thank the anonymous referees that pushed this work to a higher level.

I acknowledge the use of C language with which I wrote most of my codes, Virtual Observatory Tools (Topcat, VODesk, Aladin, VOconv, but specially Topcat!), Supermongo, Miriad, Karma, IRAF, NoMachine, IDL and above all, Ubuntu system. Thank you all the developers.

I acknowledge the frequent use of SAO/NASA ADS and B-ON online libraries, ESO data archive, Multimission Archive at STScI, NASA/IPAC Extragalactic Database, and Vizier.

This thesis is based on observations at: European Southern Observatory (ESO) Very Large Telescope (VLT) under the programs 074.A-0709, 168.A-0485, 170.A-0788, 171.A-3045, 175.A-0839, 275.A-5060, and the ESO Science Archive under programs 64.O-0643, 66.A-0572, 68.A-0544, 164.O-0561, 163.N-0210, and 60.A-9120; NASA/ESA \textit{Hubble Space Telescope} with ACS/NICMOS/WFC3 instruments, obtained at the Space Telescope Science Institute, operated by AURA Inc., under NASA contract NAS 5-26555;  \textit{Spitzer Space Telescope}, which is operated by the Jet Propulsion Laboratory, California Institute of Technology, under NASA contract 1407; Subaru Telescope, operated by the National Astronomical Observatory of Japan; Kitt Peak National Observatory, Cerro Tololo Inter-American Observatory, and the National Optical Astronomy Observatory, which are operated by the Association of Universities for Research in Astronomy, Inc. (AURA) under cooperative agreement with the National Science Foundation; Canada-France-Hawaii Telescope (CFHT) with MegaPrime/MegaCam operated as a joint project by the CFHT Corporation, CEA/DAPNIA, the NRC and CADC of Canada, the CNRS of France, TERAPIX, and the University of Hawaii; \textit{Chandra} X-ray Observatory Centre, which is operated by the Smithsonian Astrophysical Observatory; XMM-Newton, an ESA science mission with instruments and contributions directly funded by ESA Member States and NASA; Australia Telescope Compact Array which is part of the Australia Telescope, which is funded by the Commonwealth of Australia for operation as a National Facility managed by CSIRO; VLA which is a facility of the National Radio Astronomy Observatory, which is operated by Associated Universities, Inc., under a cooperative agreement with the National Science Foundation.

I apologize in case I have missed any important references throughout the thesis. It was not my intention. I only have few years of Astronomy experience and unfortunately, could not absorb all the information so far.

\'E com grande honra que acabo o doutoramento no ano de celebra\c{c}ao dos 150 anos do Observat\'orio Astron\'omico de Lisboa. Espero poder ajudar nos pr\'oximos.
\chapter*{Resumo}
\addcontentsline{toc}{chapter}{Resumo}
\thispagestyle{empty}

Com a primeira gera\c{c}\~ao de c\^ameras de infra-vermelho (IV) nos anos 70 e 80, como um melhoramento aos primeiros detectores de IR, possibilitou coberturas sistem\'aticas de grande \'area nesta regi\~ao do espectro. Esta nova janela que ent\~ao se abria mostrou \`a comunidade cient\'ifica qu\~ao limitada era a nossa vis\~ao do Universo quando restringida aos telesc\'opios de \'optico, mais desenvolvidos nessa altura. Hoje em dia sabemos que a maior parte da ac\c{c}\~ao acontece fora do regime do \'optico. Os raios-$\gamma$ e X mostram-nos os eventos mais energ\'eticos do Universo \citep[o mais distante remonta \`a \'epoca em que o Universo tinha somente 600 milh\~oes de anos,][]{Tanvir09}, o IV (1--1000\,$\mu$m) que revela quantidades enormes de poeira a reemitir luz absorvida do ultra-violeta/\'optica, e o r\'adio que at\'e aos meados dos anos 90 foi o recordista das fontes mais distantes observadas no Universo \citep{Stern99}. Esta tese est\'a focada no regime do IV, ao mesmo tempo que considera as restantes janelas espectrais de maneira a maximizar a caracteriza\c{c}\~ao das amostras de gal\'axias consideradas neste estudo.

Uma an\'alise multi-comprimento-de-onda (MCO, dos raios-X \`as frequ\^encias de r\'adio) das propriedades de popula\c{c}\~oes de gal\'axias extremamente vermelhas (GEVs) \'e apresentada de in\'icio. Um conjunto de dados entre os mais profundos alguma vez obtidos s\~ao tidos em conta neste trabalho. A regi\~ao do c\'eu \'e das mais intensamente observadas: o \textit{Great Observatories Origins Deep Survey}\footnote{Cobertura Profunda no Sul das Origens pelos Grandes Observat\'orios} / \textit{Chandra Deep Field South}\footnote{Campo Profundo no Sul do \textit{Chandra}. O telesc\'opio espacial \textit{Chandra} opera nos raios-X (0.5--8\,keV) e deve o seu nome ao astrof\'isico Subrahmanyan Chandrasekhar, http://chandra.harvard.edu/}. Ao adoptar uma metodologia puramente estat\'istica, considera-se toda a informa\c{c}\~ao fotom\'etrica e espectrosc\'opica dispon\'ivel em amostras numerosas de objectos extremamente vermelhos (OEVs, 553 fontes), IRAC\footnote{\textit{Infra-red array camera} (IRAC, c\^amera em grelha de IV) do telesc\'opio espacial $Spitzer$, http://irsa.ipac.caltech.edu/data/SPITZER/docs/irac/} OEVs (IOEVs, 259 fontes), e gal\'axias vermelhas distantes (GVDs, 289 fontes) de maneira a obter distribui\c{c}\~oes em dist\^ancia, identificar gal\'axias que alberguem um n\'ucleo gal\'actico activo (NGA) ou zonas de forma\c{c}\~ao estelar, e, utilizando observa\c{c}\~oes r\'adio neste campo, estimar densidades de taxa de forma\c{c}\~ao estelar ($\dot{\rho}_{\ast}$) robustas e independentes da exist\^encia de poeira nestas popula\c{c}\~oes de gal\'axias. As propriedades de sub-popula\c{c}\~oes de gal\'axias ``puras'' (aquelas que pertencem somente a um dos grupos referidos) e ``comuns'' (aquelas que s\~ao comuns aos tr\^es) s\~ao tamb\'em investigadas.

Em geral, um grande n\'umero de NGAs s\~ao identificados (at\'e 25\%, baseado em crit\'erios de raios-X e IV), sendo na sua maioria objectos de tipo-2 (obscurecidos). A emiss\~ao r\'adio oriunda de actividade NGA n\~ao \'e tipicamente forte, implicando um acr\'escimo de 10 a 25\% nas m\'edias/medianas das luminosidades r\'adio ao incluir-se GEVs que albergam AGN. Por\'em, os NGAs s\~ao frequentemente encontrados em GEVs, e a sua n\~ao identifica\c{c}\~ao poder\'a aumentar significativamente (em 200\% em alguns casos) as estimativas de $\dot{\rho}_{\ast}$ das GEVs. Este resultado pode ser interpretado de duas maneiras: ou a popula\c{c}\~ao GEV que alberga um NGA tem efectivamente uma grande componente de forma\c{c}\~ao estelar ou a emiss\~ao NGA est\'a a enviesar fortemente os resultados. Deste modo, apesar da contribui\c{c}\~ao da forma\c{c}\~ao estelar para a luminosidade r\'adio permane\c{c}a inconclusiva em gal\'axias que alberguem um NGA num estudo de r\'adio, pode-se ainda assim estimar limites superiores e inferiores de $\dot{\rho}_{\ast}$ em popula\c{c}\~oes GEV. S\~ao assim identificadas sub-popula\c{c}\~oes que cobrem uma larga escala de taxas de forma\c{c}\~ao estelar (TFE) m\'edias, desde menos de 10 massas solares (M$_\odot$) por ano (M$_\odot$\,ano$^{-1}$) at\'e 150 M$_\odot$\,ano$^{-1}$. Ao separar em intervalos de dist\^ancia ($1\leq{z}<2$ and $2\leq{z}\leq3$\footnote{O $redshift~(z)$ \'e uma unidade de dist\^ancia em astronomia que n\~ao e' linear com a dist\^ancia medida em metros, mas tem em conta a expans\~ao do Universo.}) obt\'em-se uma evolu\c{c}\~ao significante em $\dot{\rho}_{\ast}$. Enquanto OEVs e GVDs seguem a evolu\c{c}\~ao geral da popula\c{c}\~ao de gal\'axias observada no Universo, IOEVs aparentam uma evolu\c{c}\~ao constante. Contudo, os IOEVs s\~ao os maiores contribuidores para a $\dot{\rho}_{\ast}$ total a $1\leq{z}<2$ (at\'e um n\'ivel de 25\%), enquanto os OEVs poder\~ao contribuir at\'e 40\% a $2\leq{z}\leq3$.

A compara\c{c}\~ao de estimativas de TFEs no r\'adio com as de ultra-violeta confirma a natureza ``poeirenta'' das popula\c{c}\~oes comuns (com um obscurecimento m\'edio de E($B-V$)=0.5--0.6 e m\'aximos de E($B-V$)$\sim$1), e tamb\'em que a compara\c{c}\~ao directa destes dois regimes do espectro \'e v\'alida para obter uma estimativa de obscurecimento nas gal\'axias. GEVs s\~ao tamb\'em conhecidas por serem gal\'axias massivas a grande dist\^ancia, e, neste trabalho, obtemos fun\c{c}\~oes e densidades de massa estelar, mostrando que 60\% da massa estelar existente no Universo a $1\leq{z}\leq3$ est\'a em GEVs e que esta frac\c{c}\~ao aumenta em popula\c{c}\~oes de gal\'axias gradualmente mais massivas. \'E tamb\'em efectuado um estudo morfol\'ogico para uma caracteriza\c{c}\~ao mais completa de GEVs, que revela uma popula\c{c}\~ao de GVDs, que cont\'em uma mistura de popula\c{c}\~oes estelares jovem e adulta assim como actividade obscurecida NGA.

Estes resultados no c\^omputo geral poder\~ao apontar para o facto de OEVs, IOEVs, e GDVs serem de facto parte da mesma popula\c{c}\~ao, por\'em vista em fases diferentes de evolu\c{c}\~ao gal\'actica. Isto est\'a de acordo com o cen\'ario j\'a proposto por alguns autores que defendem as fases de gal\'axia de sub-mil\'imetro, gal\'axia obscurecida por poeira, GDV, e OEV como uma sequ\^encia de evolu\c{c}\~ao gal\'actica.

A segunda parte desta tese \'e dedicada a um trabalho que come\c{c}ou inicialmente como uma necessidade para a demografia de NGAs em GEVs, revelando-se como um dos grandes resultados desta tese, com grande relev\^ancia para o telesc\'opio espacial \textit{James Webb}\footnote{http://www.jwst.nasa.gov/} ($TEJW$), a ser lan\c{c}do no final desta d\'ecada (2018). \'E sabido que o IV possibilita a selec\c{c}\~ao de gal\'axias com actividade nuclear, que poder\'a nem ser detectada nas coberturas de raios-X mais profundas devido a extremo obscurecimento. Muitos crit\'erios de IV foram explorados para cumprir este objectivo e intensamente testados. A grande conclus\~ao \'e que a grandes dist\^ancias ($z\gtrsim2.5$) a contamina\c{c}\~ao por gal\'axias n\~ao activas \'e abundante. Isto n\~ao \'e de todo vi\'avel para estudos do Universo mais jovem, que \'e o grande objectivo de muitos estudos em curso hoje em dia e de futuras coberturas profundas. Ao utilizar modelos de distribui\c{c}\~ao espectral de energia que cobrem uma variedade de propriedades gal\'acticas, novas vers\~oes de crit\'erios de IV mais eficientes na selec\c{c}\~ao de NGAs a grandes dist\^ancias (at\'e $z\sim7$) s\~ao apresentadas. Com particular \^enfase nos comprimentos-de-onda cobertos pelo $TEJW$ (1--25\,$\mu$m), criou-se um crit\'erio IV (que usa bandas K e IRAC, KI) como alternativa aos crit\'erios existentes a $z<2.5$. \'E tamb\'em criado um crit\'erio IV que selecciona NGAs com grande fiabilidade desde dist\^ancias locais at\'e ao final da \'epoca de reioniza\c{c}\~ao ($z\sim7$). Tanto KI como KIM requerem filtros j\'a existentes, sendo poss\'ivel a sua aplica\c{c}\~ao no imediato. Amostras de controlo com cobertura MCO (desde os raios-X \`as frequ\^encias r\'adio) s\~ao tamb\'em utilizadas para estimar a fiabilidade destes novos crit\'erios em compara\c{c}\~ao aos j\'a existentes. Conclui-se que os modelos utilizados e amostras de controlo indicam um melhoramento significante do KI em compara\c{c}\~ao com outros crit\'erios de selec\c{c}\~ao NGA baseados somente em filtros IRAC, e que o KIM \'e fi\'avel mesmo a dist\^ancias maiores que $z\sim2.5$.

O \'ultimo cap\'itulo tem por objectivo alertar que a poeira existe e n\~ao deve ser subestimada. Este \'e e ser\'a ser sempre um facto que um astr\'onomo deveria manter-se ciente. Ao utilizar dados UKIRT/CFHT/\textit{Spitzer} no \textit{Cosmological Survey} (COSMOS), regimes de altas temperaturas de poeira (800--1500\,K) s\~ao investigados, ao inv\'es do regime mais frio normalmente referido na literatura ($<$100\,K). Fun\c{c}\~oes de luminosidade de IV (FLI) s\~ao obtidas (comprimentos-de-onda de repouso 1.6, 3.3, and 6.2\,$\mu$m) assim como \'e estimada a sua depend\^encia com a dist\^ancia e popula\c{c}\~oes de gal\'axias. A conhecida bimodalidade das FLI \'e observada. Frac\c{c}\~oes de poeira s\~ao extra\'idas por base num modelo de emiss\~ao puramente estelar, e as primeiras fun\c{c}\~oes de densidade de luminosidade de poeira quente alguma vez feitas s\~ao apresentadas. Ao separar em gal\'axias el\'ipticas, espirais, de forte forma\c{c}\~ao estelar e NGAs, mostra-se como a emiss\~ao NGA pode contribuir significativamente mesmo a 1.6\,$\mu$m, provocando um prov\'avel enviesamento (sistem\'atico e crescente) em qualquer estimativa de massa estelar baseada em luminosidades de IV. Este efeito, tal como a frac\c{c}\~ao de NGAs, aumenta com a dist\^ancia, sendo por isso de grande import\^ancia a adop\c{c}\~ao de um procedimento cuidado para a estimativa de massas estelares, mesmo numa an\'alise de ajuste \'a distribui\c{c}\~ao espectral de energia. Por fim, \'e apresentada a evolu\c{c}\~ao da densidade de luminosidade da poeira quente, revelando um decr\'escimo bem mais acentuado do que o da hist\'oria de forma\c{c}\~ao estelar no Universo. H\'a duas interpreta\c{c}\~oes v\'alidas para este resultado: ou a reduzida TFE no Universo local \'e incapaz de aquecer quantidades de poeira suficientes para esta dominar a 3.3\,$\mu$m ou h\'a efectivamente um decr\'escimo na quantidade de poeira existente nas gal\'axias no Universo local. Um estudo recente com o Observat\'orio Espacial \textit{Herschel} d\'a for\c{c}a ao \'ultimo cen\'ario.

Por \'ultimo, \'e apresentado um conjunto de projectos futuros que t\^em por objectivo tanto o melhoramento do trabalho aqui descrito, como a aplica\c{c}\~ao das t\'ecnicas desenvolvidas durante esta tese. Estas \'ultimas resultam em tr\^es projectos importantes: um estudo j\'a em curso de discos adultos a grandes dist\^ancias, sendo este um dos futuros campos de investiga\c{c}\~ao de grande relev\^ancia na altura em que o \textit{Atacama Large Millimeter Array} (ALMA) estiver completo; um censo dos NGA mais obscurecidos a grandes dist\^ancias; e uma compara\c{c}\~ao directa e consistente entre a emiss\~ao de poeira quente (800--1500\,K) e fria ($<100$\,K) dependendo n\~ao s\'o em luminosidade de IV como dist\^ancia.

\vspace{1cm}

\noindent PALAVRAS CHAVE: infra-vermelho; gal\'axias; evolu\c{c}\~ao; actividade nuclear; forma\c{c}\~ao estelar; poeira.

\chapter*{Abstract}
\addcontentsline{toc}{chapter}{Abstract}
\thispagestyle{empty}

The main focus of this thesis is the IR spectral regime, which since the 70's and 80's has revolutionised our understanding of the Universe.

A multi-wavelength analysis on Extremely Red Galaxy populations is first presented in one of the most intensively observed patch of the sky, the Chandra Deep Field South. By adopting a purely statistical methodology, we consider all the photometric and spectroscopic information available on large samples of Extremely Red Objects (EROs, 553 sources), IRAC EROs (IEROs, 259 sources), and Distant Red Galaxies (DRGs, 289 sources). We derive general properties: redshift distributions, AGN host fraction, star-formation rate densities, dust content, morphology, mass functions and mass densities. The results point to the fact that EROs, IEROs, and DRGs all belong to the same population, yet seen at different phases of galaxy evolution.

The second part of this thesis is dedicated to the AGN selection in the IR, with particular relevance to the \textit{James Webb Space Telescope}, to be launched in 2018. We develop an improved IR criterion (using $K$ and IRAC bands) as an alternative to existing IR AGN criteria for the $z\lesssim2.5$ regime, and develop another IR criterion which reliably selects AGN hosts at $0<z<7$ (using $K$, \textit{Spitzer}-IRAC, and \textit{Spitzer}-MIPS$_{24\,\mu\rm{m}}$ bands, KIM). The ability to track AGN activity since the end of reionization holds great advantages for the study of galaxy evolution.

The thesis then focus on the importance of dust. Based on deep IR data on the Cosmological Survey, we derive rest-frame 1.6, 3.3, and 6.2\,$\mu$m luminosity functions and their dependency on redshift. We estimate the dust contribution to those wavelengths and show that the hot dust luminosity density evolves since $z=1-2$ with a much steeper drop than the star-formation history of the Universe.

Future prospects are finally discussed in the last chapter.
\vspace{1cm}

\noindent KEY WORDS: infra-red; galaxies; evolution; active; starburst; dust.
\tableofcontents
\addcontentsline{toc}{chapter}{List of Figures}
\listoffigures
\addcontentsline{toc}{chapter}{List of Tables}
\listoftables


\chapter*{List of Abbreviations}
\addcontentsline{toc}{chapter}{List of Abbreviations}
\fancyhead[LE]{\textsc{\nouppercase{List of Abbreviations}}}
\fancyhead[RO]{\textsc{\nouppercase{List of Abbreviations}}}

\noindent FL --- Fun\c{c}\~ao de Luminosidade

\noindent FLI --- Fun\c{c}\~ao de Luminosidade de Infra-vermelho

\noindent GEV --- Gal\'axia Extremamente Vermelha

\noindent GVD --- Gal\'axia Vermelha Distante

\noindent IOEVs --- IRAC Objectos Extremamente Vermelhos

\noindent IV --- Infra-Vermelho

\noindent MCO --- Multi-Comprimento-de-Onda

\noindent NGA --- N\'ucleo Gal\'actico Activo

\noindent OEVs --- Objectos Extremamente Vermelhos

\vspace{1cm}

\noindent 2MASS --- Two Micron All Sky Survey

\noindent ACS --- Advanced Camera for Surveys

\noindent AGN --- Active Galactic Nuclei

\noindent ALMA --- Atacama Large Millimeter/submillimeter Array

\noindent ANN$z$ --- Artificial Neural Networks photometric redshift code

\noindent ASKAP --- Australian Square Kilometre Array Pathfinder

\noindent ATCA --- Australia Telescope Compact Array

\noindent Blazar --- Blazing Quasi-stellar Object

\noindent BLAGN --- Broad Line Active Galactic Nuclei

\noindent BQSO --- Bottom Quasi-stellar Object \citep[see][]{Polletta07}

\noindent BOOMERanG --- Balloon Observations of Millimetric Extragalactic Radiation and Geophysics

\noindent $\mathcal{C}$ --- Completeness

\noindent C\,IV --- Carbon IV ion

\noindent CANDELS --- Cosmic Assembly Near-infrared Deep Extragalactic Legacy Survey

\noindent CAS --- Concentration, Assymetry, Smoothness

\noindent CCD --- charge coupled divice

\noindent CDFs --- $Chandra$ Deep Field South

\noindent cERG --- Common Extremely Red Galaxy

\noindent CFHT --- Canada France Hawaii Telescope

\noindent CMB --- Cosmic Microwave Background

\noindent CO --- Carbon monoxide molecule

\noindent COBE --- Cosmic Bakground Explorer

\noindent COSMOS --- Cosmological Survey

\noindent CXB --- Cosmic X-ray Background

\noindent \textit{Chandra} --- $Chandra$ X-ray observatory

\noindent DM --- Dark Matter

\noindent DOG --- Dust Obscured Galaxy

\noindent DRG --- Distant Red Galaxy

\noindent ECDFs --- Extended $Chandra$ Deep Field South

\noindent ELAIS --- European Large Area ISO Survey

\noindent ERG --- Extremely Red Galaxy

\noindent ERO --- Extremely Red Object

\noindent ERS --- Early Release Science

\noindent FB --- X-ray Full-Band

\noindent FeLoBAL -Iron (Fe) Low-ionization Broad Absorption Line galaxy

\noindent FIR --- Far Infra-red

\noindent FIREWORKS --- catalogue assembling the data on the Faint InfraRed Extragalactic Survey (FIRES) fields $Hubble$ deep Field South and MS 1054–03

\noindent FORS2 --- FOcal Reducer and low dispersion Spectrograph 2

\noindent FWHM --- Full Width at Half Maximum

\noindent $f_{\rm obs}$ --- obscured fraction of AGN sources

\noindent G --- Gini coefficient

\noindent GATOR --- General Catalog Query Engine

\noindent GMRT --- Giant Metrewave Radio Telescope

\noindent GNS --- GOODS NICMOS Survey

\noindent GOODS --- Great Observatories Origins Deep Survey

\noindent HB --- X-ray hard-band

\noindent HDF --- Hubble deep field

\noindent HR --- Hardness Ratio

\noindent HR10 --- object number 10 of \citet{HuRidgway94}

\noindent HSO --- Herschel Space Observatory

\noindent HST --- $Hubble$ Space Telescope

\noindent H$z$RG --- High redshift ($z$) Radio Galaxy

\noindent IERO --- IRAC Extremely Red Object

\noindent IM --- IRAC+MIPS colour-colour space

\noindent IMF --- Initial Mass Function

\noindent IR --- Infra-Red

\noindent IRAC --- Infra-Red Array Camera

\noindent IRAS --- Infra-Red Astronomical Satellite

\noindent IRBG --- Infra-Red Bright Galaxy

\noindent IRS --- Infra-Red spectrograph

\noindent IRxs --- Infra-Red excess

\noindent ISAAC --- Infrared Spectrometer And Array Camera

\noindent ISM --- Inter Stellar Medium

\noindent ISO --- Infra-Red Space Observatory

\noindent JWST --- \textit{James Webb} Space Telescope

\noindent KI --- K+IRAC criterion

\noindent KIM --- K+IRAC+MIPS$_{\rm 24\mu{m}}$ criterion

\noindent L07 --- \citet{Lacy04,Lacy07} criterion

\noindent LDF --- Luminosity Density Function

\noindent LF --- Luminosity Function

\noindent LIRG --- Luminous Infra-Red Galaxy

\noindent M$_{20}$ --- the second-order moment value of the 20\% brigtest pixels

\noindent mag --- magnitude

\noindent MAST --- Multimission Archive at STScI

\noindent MC --- Monte Carlo

\noindent MCO --- Multi-Comprimento-de-Onda

\noindent MeerKAT --- Karoo Array Telescope

\noindent MF --- Mass Function

\noindent MIPS --- Multiband Imaging Photometer

\noindent MIR --- Mid-Infra-Red

\noindent mm --- millimeter


\noindent MUSIC --- MUltiwavelength Southern Infrared Catalogue

\noindent MWA --- MIT Haystack Observatory

\noindent $\rm{N_H}$ --- Hidrogen (H) column density

\noindent NICMOS --- Near Infrared Camera and Multi-Object Spectrometer

\noindent NIR --- Near-Infra-Red

\noindent NLAGN --- Narrow Line Active Galactic Nucleus

\noindent N\,V --- Nitrogen V ion

\noindent O\,III --- Oxigen III ion

\noindent $\mathcal{P}$ --- Probability

\noindent PAH --- Polycyclic Aromatic Hydrocarbon

\noindent PD --- Probability Distribution

\noindent PDG --- Passive Disc Galaxy

\noindent pDRG --- pure Distant Red Galaxy

\noindent pERO --- pure Extremely Red Object

\noindent PhD --- Latin Philosophi\ae Doctor

\noindent PIMMS --- Portable, Interactive Multi-Mission Simulator

\noindent PLE --- Pure Luminosity Evolution

\noindent QSO --- Quasi-Stellar Object

\noindent $\mathcal{R}$ --- Reliability

\noindent R$_{eff}$ --- effective radius

\noindent S05 --- \citet{Stern05} criterion

\noindent $\mathcal{S}_{12}$ --- type-1/type-2 relative sensitivity

\noindent $\mathcal{S}_{\rm HL}$ --- X-ray high/low-luminosity relative sensitivity

\noindent SB --- X-ray Soft-Band

\noindent SCUBA --- Submillimeter Common User Bolometer Array

\noindent SDSS --- Sloan Digital Sky Survey

\noindent SED --- Spectral Energy Distribution

\noindent SF --- Star Formation

\noindent SFH --- Star Formation History

\noindent SFR --- Star Formation Rates

\noindent Si$IV$ --- Silicon $IV$ ion

\noindent SKA --- Square Kilometre Array

\noindent SMBH --- Super Massive Black Hole

\noindent SMG --- Sub-Millimeter Galaxy

\noindent S/N --- Signal-to-Noise

\noindent \textit{Spitzer} --- \textit{Spitzer} Space telescope

\noindent SWIRE --- Spitzer Wide-area Infrared Extragalactic Survey

\noindent TEJW --- Telesc\'opio Espacial \textit{James Webb}.

\noindent TFE --- Taxa de Forma\c{c}\~ao Estelar

\noindent TP-AGB --- Thermally Pulsing - Asymptotic Giant Branch

\noindent TQSO --- Top Quasi-Stellar Object \citep[see][]{Polletta07}

\noindent type-1 --- unobscured AGN

\noindent type-2 --- obscured AGN

\noindent UDS --- Ultra Deep Survey

\noindent UKIDSS --- UKIRT Infrared Deep Sky Survey

\noindent UKIDSS-DXS --- UKIRT Infrared Deep Sky Survey - Deep Extragalactic Survey

\noindent UKIRT --- United Kingdom Infrared Telescope

\noindent ULIRG --- Ultra-Luminous Infra-Red Galaxy

\noindent UV --- Ultra-Violet

\noindent VIDEO --- VISTA Deep Extragalactic Observations

\noindent VIMOS --- VIsible MultiObject Spectrograph

\noindent VISTA --- Visible and Infrared Survey Telescope for Astronomy

\noindent VLBI --- Very Large Baseline Interferometer

\noindent VLA --- Very Large Array

\noindent VLT --- Very Large Telescope

\noindent VLT-UT --- Very Large Telescope --- Unit Telescope

\noindent WFC3 --- Wide Field Camera 3

\noindent WISE --- Wide-Field Infrared Survey Explorer

\noindent WMAP --- Wilkinson Microwave Anisotropy Probe

\noindent XMM-$Newton$ --- X-ray Multi-Mirror Mission --- $Newton$

\noindent $z$ --- redshift

\noindent $z$COSMOS --- spectroscopic catalogue of COSMOS

\noindent $\alpha$ --- spectral index

\noindent $\Gamma$ --- X-ray photon index

\noindent $\Lambda$CDM --- $\Lambda$ Cold Dark Matter

\noindent $\dot{\rho}_{\ast}$ --- star-formation rate density

\noindent ${\rho}_{\rm M}$ --- mass density

\chapter*{List of Unconventional Units}
\addcontentsline{toc}{chapter}{List of Unconventional Units}
\fancyhead[LE]{\textsc{\nouppercase{List of Unconventional Units}}}
\fancyhead[RO]{\textsc{\nouppercase{List of Unconventional Units}}}

Jansky --- 1\,Jy $\equiv$ 10$^{-23}$\,erg\,s$^{-1}$\,cm$^{-2}$\,Hz$^{-1}$

\noindent light-day --- 1\,light-day$\simeq2.59\times10^{13}$\,m

\noindent light-year --- 1\,light-year$\simeq9.461\times10^{15}$\,m

\noindent M$_\odot$ --- solar masses (1\,M$_\odot$ $\simeq$ $1.989\times10^{33}\,$g)

\noindent pc --- parsec (1\,pc $\simeq$ $3.086\times10^{18}\,$cm $\simeq$ 3.26\,light-years)

\noindent speed of light --- c$\simeq2.998\times10^8$\,m\,s$^{-1}$

\noindent yr --- year (1\,yr $\simeq$ 31557600\,s)

\chapter*{List of Conventions}
\addcontentsline{toc}{chapter}{List of Conventions}
\fancyhead[LE]{\textsc{\nouppercase{List of Conventions}}}
\fancyhead[RO]{\textsc{\nouppercase{List of Conventions}}}

AB magnitudes --- $\rm{m_{AB}}=2.5\times\log(f_\nu\rm{[erg\,s^{-1}\, cm^{-2}\,Hz^{-1}]})-48.6$

\noindent AB to Vega conversion --- (I, J, H, K, [3.6], [4.5], [5.8], [8.0])$_{AB}$ = (I, J, H, K, [3.6], [4.5], [5.8], [8.0])$_{Vega}$ + (0.403, 0.904, 1.373, 1.841, 2.79, 3.26, 3.73, 4.40)

\noindent \citep[][and http://spider.ipac.caltech.edu/staff/gillian/cal.html]{Roche03}

\vspace{1cm}

\noindent Power-law SED --- $f_\nu\propto\nu^{-\alpha}$

\noindent Luminosity --- $4\pi\rm{d_L}^2\times{f_\nu}\times{k}_{\rm corr}$

\noindent Radio $k_{\rm corr}$ --- $(1+z)^{\alpha-1}$

\noindent X-ray $k_{\rm corr}$ --- $(1+z)^{\Gamma-2}$

\noindent $\Gamma$ --- $\Gamma\equiv1-\alpha$, $\Gamma = 1.8$ \citep{Tozzi06}

\noindent Hardness Ratio --- $\frac{H-S}{H+S}$ (H $\equiv$ hard-band photon counts; S $\equiv$ soft-band photon counts)

\vspace{1cm}

\noindent Hubble constant --- H$_ 0=70$\,km\,s$^{-1}$\,Mpc$^{-1}$

\noindent Cosmological constant --- $\Omega_\Lambda=0.7$

\noindent Total matter density --- $\Omega_{\rm m}=0.3$

\blankpage
\doublespacing
\headers

\chapter{Introduction}
\label{ch:intro}
\thispagestyle{empty}
\pagenumbering{arabic}
\setcounter{page}{1}

\section{The $\Lambda$-Cold Dark Matter Universe}

\begin{figure}
  \begin{center}
    \includegraphics[width=0.9\columnwidth]{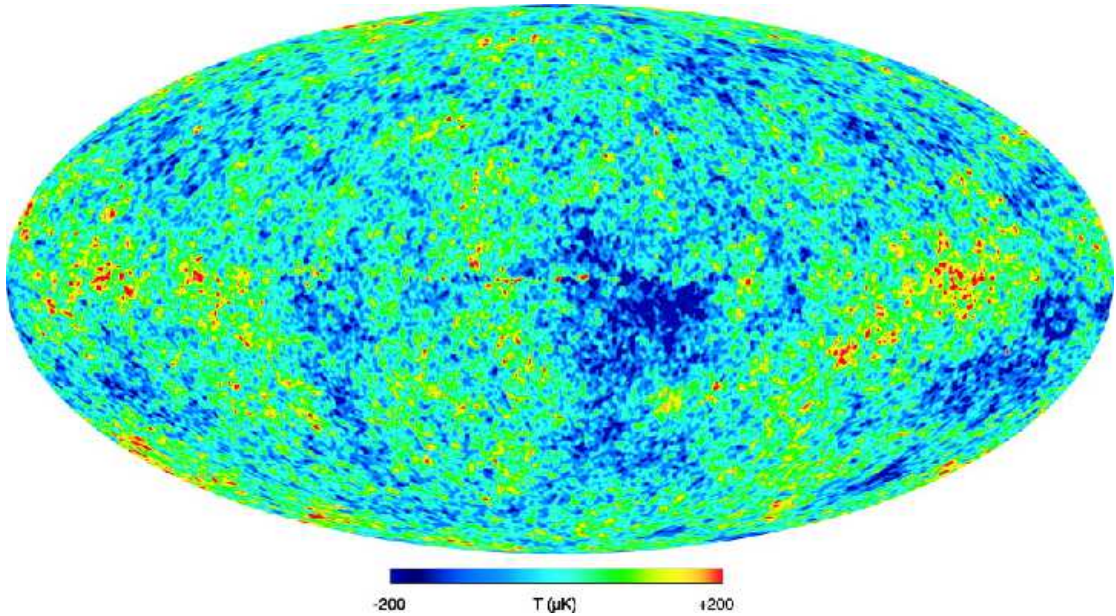}
  \end{center}
  \caption[The WMAP first data release]{The best picture yet of our hot young Universe. Note the colour scale varies between -200 and 200 $\mu$K. The hot over-densities are believed to be the predecessors of the galaxy clusters we see today. From \citet{Bennett03}.}
  \label{fig:wmap}
\end{figure}

The $\Lambda$ cold dark matter ($\Lambda$CDM) cosmology model is now widely accepted as the one that best explains our Universe, or at least what we know about it. The $\Lambda$ stands for a gravitationally repulsive effect (induced by the so called \textit{dark energy}) associated with the observed accelerating expansion of the Universe \citep{Riess98,Perlmutter99}. The cold dark matter is that needed to produce primordial deep gravitational potentials in which baryonic matter assembles, giving origin to the seeds of the very first galaxies. These potential wells are believed to be the cause for the fluctuations detected in what is known today as the cosmic microwave background (CMB). At present, due to the expansion of the Universe \citep{Tolman34}, this radiation peaks at longer wavelengths (1--2\,mm) equivalent to that emitted by a black body at a temperature of 2.725\,K \citep[][but see also \citealt{Regener31,Regener32,McKellar41}]{Penzias65,Dicke65,Mather99}. At the beginning of this millennium, and following the pioneering work of its predecessors (e.g., COBE\footnote{http://aether.lbl.gov/www/projects/cobe/} in 1992, BOOMERanG\footnote{http://cmb.phys.cwru.edu/boomerang/} in 2000), the Wilkinson Microwave Anisotropy Probe \citep[WMAP,][]{Bennett03} took an unprecedented detailed picture of the ``baby universe'', as the WMAP team likes to call it. Figure~\ref{fig:wmap} shows the best image we have so far of the (believed to be) Big Bang `afterglow', revealing temperature fluctuations on the order of a millionth of a degree. The patterns seen in the WMAP image are now believed to be density variations of matter, thus providing a means to track down the initial conditions of galaxy formation.

\begin{figure}[t]
  \begin{center}$
  \begin{array}{cc}
    \includegraphics[width=0.5\columnwidth]{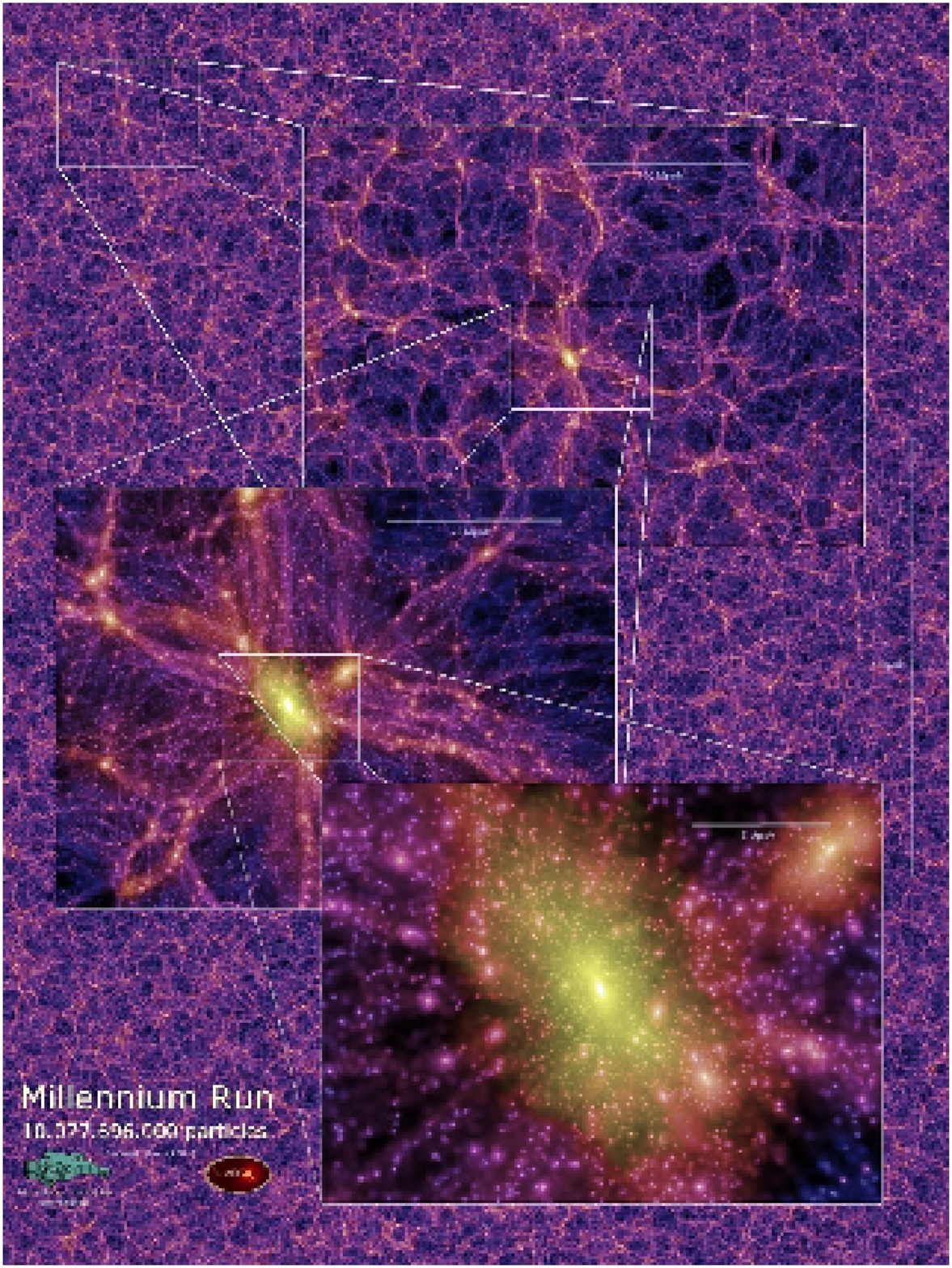} & \includegraphics[width=0.45\columnwidth]{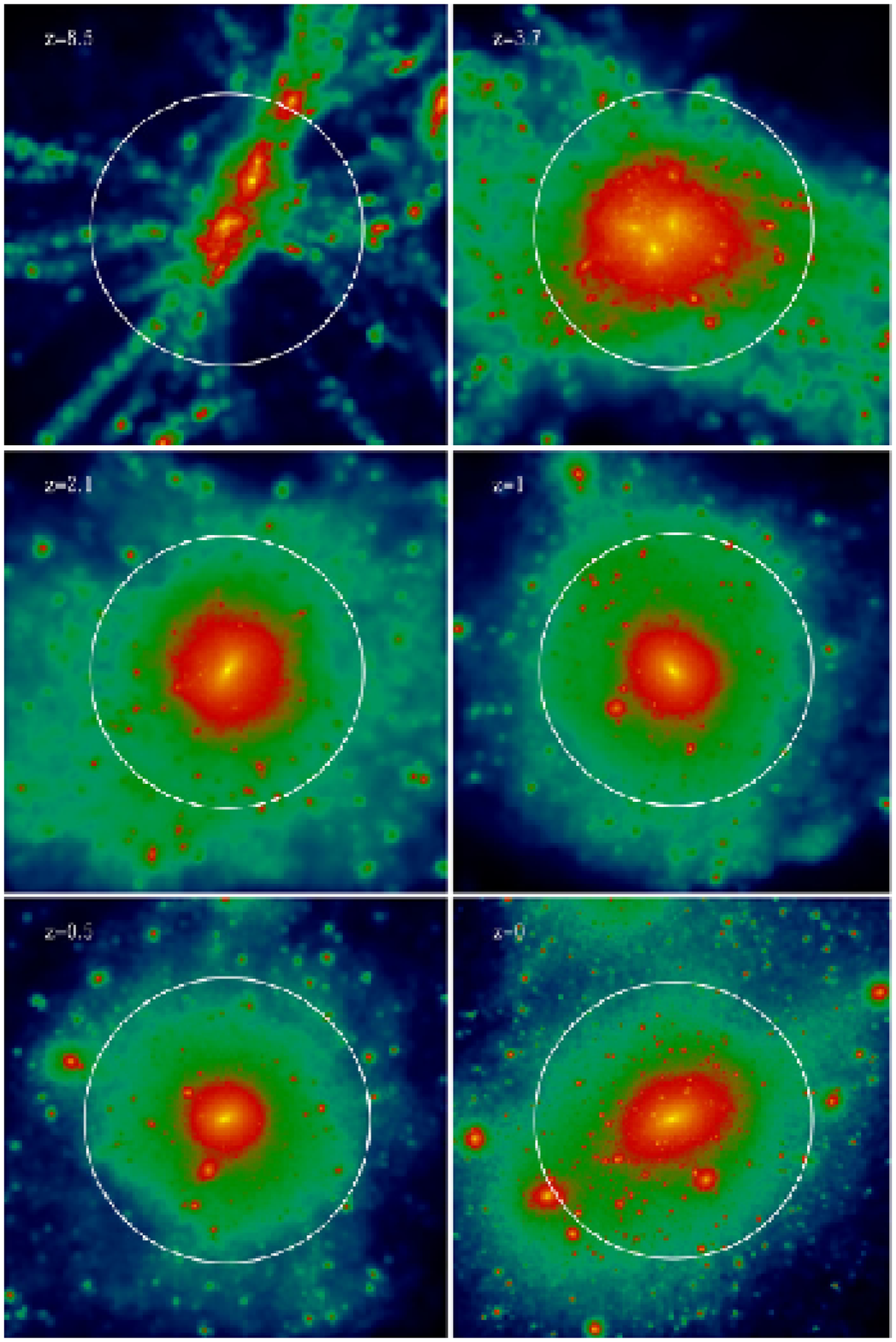}
  \end{array}$
  \end{center}
  \caption[N-body Simulations]{Zooming through a structured Universe into a dark matter halo at $z=0$ in the Millennium Simulation (left panel). Each zoom scales down to a factor of four (credit: Virgo Consortium). On the right panel, a high resolution N-body simulation showing different stages of the dark matter halo hierarchical merging (from $z=8.5$, top left, to $z=0$, bottom right). Redder regions indicate higher densities regions \citep[figure from][]{Baugh06}.}
  \label{fig:nbodysim}
\end{figure}

How these temperature fluctuations vary and how far apart they are in the sky (the so-called angular power spectrum) can be used, among other things, to derive cosmological parameters, implying a Universe composed by 75\% of `dark energy' ($\Omega_\Lambda$), and the remaining 25\% ($\Omega_{\rm M}$) in the form of either `dark matter' (DM, 21\%) or baryonic matter we see in galaxies and in the inter-galactic medium (only 4\%). The power-spectrum is also used as an input for any model attempting to trace back the origins of the Universe we see today. One recent work of reference is without any doubts the \textit{Millennium Simulation}\footnote{http://www.mpa-garching.mpg.de/galform/virgo/millennium/} \citep[Virgo Consortium,][but see also \citealt{Kang05,Croton06a,Croton06b,Bower06,deLucia07}]{Springel05b}. This numerical N-body simulation made use of enormous computer power at the Computing Centre of the Max-Planck Society (in Garching, Germany) to run a sizeable simulation (tracing $\sim10^{10}$ particles since redshift $z=127$) over the course of 28 days of continuous computation. This simulation assumed an hierarchical evolution of dark matter halos through dissipationless mechanisms of gravitational instability governed by the input power spectrum, the cosmology parameters, and the nature of the dark matter itself. This hierarchical dark matter halo assembly carries with it the gas which then cools and condenses to form galaxies (Figure~\ref{fig:nbodysim}). However, although now we (seem to) understand the evolution of DM, the baryonic evolution (hence, that of galaxies) is far more complex than a ``simple'' gravitationally induced evolution. The physics inherent to baryonic evolution comprise gas cooling, star-formation mechanisms resulting in the production of heavy elements (and consequently dust), feedback processes (such as super-nova winds and supermassive black hole activity), and mergers \citep[see][and references therein for a more detailed discussion on feedback models]{Kay02,Benson03,Benson10}.

\section{An unseen Universe}

\begin{figure}[t]
  \begin{center}
    \includegraphics[width=0.6\columnwidth]{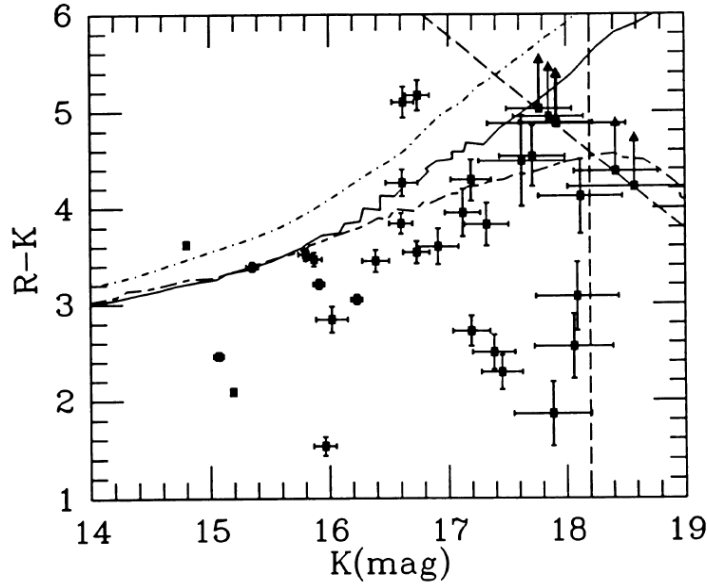}
  \end{center}
  \caption[The dawn of EROs]{The colour-magnitude diagram used to identify two high redshift candidates showing $K\sim16.7$ and $R-K\sim5$ above the model track for a central cluster elliptical (dotted-dashed line). Note the six optically undetected objects (with upward pointing arrows on the dashed line marking the $R$-band limit). The brighter source is at $z\sim0.3$ (L. L. Cowie \& S. J. Lilly 1988, private communication) and still shows a reasonable $R-K\sim3.6$ red colour. Credit: \citet{Elston88b}}
  \label{fig:elston88}
\end{figure}

Although the ideas which resulted in the development and belief of the $\Lambda$CDM model can be traced back to the 1970's \citep{Peebles80}, and even 1950's \citep{Hoyle51}, the first semi-analytical\footnote{The naming results from the trial-and-error strategy used in this models, making use of tunable physical parameters to fit the observations.} models to account for many of the ingredients of galaxy evolution appeared in the 1990's \citep{WhiteFrenk91,Cole91,LaceySilk91}, reporting successes (e.g., inter-galactic hot gas detectable in X-rays probably linked to the well-predicted star-formation rates in spirals) and acknowledging problems which still persist today (e.g., the overestimate of the faint-end of luminosity functions). Interestingly, this was close in time to the discovery (or recognition) of one of the biggest headaches hierarchical theorists have ever faced (and somehow still face). In the 1980's, the early stages of IR camera astronomy allowed the astronomers to better access the infra-red (IR, $\lambda>1\,\mu$m) spectral regime, which was about to reveal an unseen and unpredicted Universe. The PhD work of Elston in 1988 \citep{Elston88a}, making use of the first generation of IR cameras (instead of single-element detectors), revealed two objects with optical-to-near-IR colours \,($R-K$)\, redder than the massive central cluster galaxies seen locally, as well as a handful of objects undetected in $R$-band as candidates for $z>1$ passively evolved galaxies (Figure~\ref{fig:elston88}). At the time, the enthusiastic possibility for a detection of a primeval galaxy, with the Lyman limit redshifted between the $R$ and $K$ bands, took over the remainder interpretations of either passively evolved or dusty starburst galaxies at $z>1$. Soon afterwards, it was found that these galaxies were actually $z\sim0.8$ ``normal'' galaxies \citep{Elston89}. It should be mentioned that roughly ten years before the work of Elston et al., such red colours had been observed in luminous ultra-steep spectrum radio galaxies \citep{Rieke79}. Later, even more extreme colours ($R-K\sim6$--$7$) were found for distant radio galaxies \citep{Walsh85,Lilly85}. What was special about the Elston et al. work \citep[using a similar colour-magnitude plot as][]{Lilly85} was that, in just a 10\,arcmin$^2$ survey and to a limit of $K\sim17$ (Vega magnitudes), a numerous population of red galaxies was found. In case the sources happened to be $z>1$ central cluster galaxies, their number density was not expected even by the upper limits set by \citet[][50 cluster per square degree at $z=1$]{Gunn86}. The members of this red galaxy population are currently known as extremely red objects\footnote{The ERO nomenclature \citep[instead of extremely red galaxies,][]{HuRidgway94} owes its origin to the difficulty in disentangling red galaxies from cool galactic stars while using the $R-K$ colour alone. Current multi-wavelength surveys allow for a better, yet never perfect, separation.} \citep[EROs, probably introduced by][]{Dey99}. The name is broadly used in the literature to refer to many types of extreme red colour criteria using extremely red optical-to-IR colours ($R-K>5$, $R-K>6$, $I-K>4$, $I-H>3$, etc...). Later on, other criteria were proposed for the selection of ERGs (e.g., J-K>2.3, \citealt{Franx03}; $f_{3.6}/f_{z850}>20$, \citealt{Yan04}), resulting in a variety of acronyms (ERO; VRO; IERO; Hyper-ERO, \citealt{Im02}; DRG; VRG; FROG, \citealt{Moustakas98}; etc...). In this work we use the term ERG to refer the general extremely red galaxy population.

The $z>6$ dream of Elston et al. was made possible by the works of \citet{SteidelHamilton93}, \citet{Madau95} and \citet{Steidel96}, who showed that the Lyman continuum break was indeed an effective way to select high redshift sources, but the starting point was the $z>2$--$3$ Universe. This technique, selecting the so-called Lyman Break Galaxies (LBGs), together with the \textit{Hubble Space Telescope} (\textit{HST}), allowed the selection of $z\sim4$ galaxies still during the 1990's \citep{Madau96,Steidel99}, and, more recently, of $z\sim6-8$ galaxy candidates with the incorporation of the Wide Field Camera 3 (WFC3) on board \textit{HST}\footnote{The reason why we had to wait for WFC3 is due to the high thermal atmospheric IR background affecting ground-based telescopes \citep{Mountain09}, preventing even the 8--10\,m class telescopes, which have the increasing disadvantage of their strong telescope warm emission, to detect these faint high-redshift galaxies.} \citep[e.g.,][]{Oesch10,Bouwens10,McLure10}. However, these rest-frame ultra-violet/optical selected galaxies show a rather dust-free biased view of the Universe, and in that sense the work of Elston and others addressing optically faint galaxies (e.g., radio galaxies, see references above, and the luminous IR galaxies, \citealt{SandersMirabel96}) was truly pioneering. Since then, the IR (1--1000\,$\mu$m) spectral regime was acknowledged as one of the most relevant for the study of galaxy evolution, unveiling a significant population of both massive evolved systems, comprising the bulk of the stellar mass at such high redshifts \citep{Fontana04,Georgakakis06,vanDokkum06,Marchesini07}, and dusty starbursts, largely contributing to the star formation history of the Universe \citep[a contribution frequently larger than that from ultra-violet/optical selected galaxies,][]{Blain99,CharyElbaz01,Smail02,Chapman03,leFloch05}.

Understanding and modelling the IR Universe, however, has been everything but an easy task, and there are still missing pieces to the puzzle. This difficulty to understand what we actually observe originates in the basic concept of hierarchical models: smaller systems merge together to form larger ones. This implies that the last galaxies to form are the most massive ones and these are hence younger. However, in recent years, it has been shown that not only massive galaxies ($10^{10-11}$\,M$_\odot$) are already present at $z>2$ \citep[e.g.,][]{Lilly88,Mobasher05,Papovich06,vanDokkum06,Wiklind08,Wuyts09b,Marchesini09}, but the most massive ones ($>10^{11}$\,M$_\odot$) are mostly ellipticals \citep{Conselice06}, which show old stellar populations, and seem to be (fully) assembled by $z\sim1$. Furthermore, these apparently show (practically) no mass build-up activity since that epoch \citep[either by in-situ star-formation or even merger assembly, e.g.,][]{Cimatti06,Conselice08}. Smaller systems, on the contrary, continue to show significant specific star-formation \citep[the star-formation per unit mass,][]{Gavazzi96,Guzman97,Brinchmann00,Juneau05,Bauer05,Bundy06}. This is now called the ``downsizing'' scenario \citep{Cowie96}. However, there is still a lack of agreement in defining and characterizing downsizing. The actual evolution of massive galaxies since $z\sim1-2$ is unsettled. Some do defend there is no significant evolution for the most massive galaxies, implying a characteristic luminosity/mass above which the systems are fully assembled \citep{McCarthy04,Drory05,Damen09}. Others estimate a slight evolution resulting from residual star-formation\footnote{This was observed in the 1980's in radio galaxies whose IR colours revealed no evolution up to $z\sim1$, as opposed to their optical-IR colours showing a significant evolution indicative of a reminiscent younger stellar population \citep{Lilly84}.} \citep[e.g.,][]{Lilly84,Schweizer92,Barger96,Hopkins09} and minor-merger activity \citep{Naab07,Naab09,Bezanson09,vanDokkum10}. Others even support the ``dry'' merger scenario, where two (or more) galaxies, already deprived from gas supply, merge to form a larger system with no enhanced star formation \citep{Bell04,vanDokkum05,Bell06,deLucia07,Faber07}. To increase the clutter even more, many groups oppose to the downsizing concept. They find that, in reality, all galaxies present an equal decrement on star-formation rate (SFR) from high redshifts to the local Universe \citep[e.g.,][]{Zheng07,Damen09,Dunne09,Fontanot09,Karim11}. What they support is the scenario where the most massive systems (likely in the most massive dark matter halos) start their star-formation (and hence assembly) earlier than less massive ones \citep{Baugh99,Tanaka05,deLucia06,Neistein06}, explaining why, at each epoch, more massive galaxies present smaller specific SFRs than less massive galaxies. Still, both populations will present an equal decay of star-formation activity with time.

The wide variety of results and opinions may be related to a plethora of reasons, either technical or related to selection effects \citep{Conselice08b,vanderWel09,Hopkins10}. Most likely, it seems to be a cosmic variance problem. The under-cited work of \citet{Matsuoka10}, considering $\sim$60000 massive ($>10^{11}$\,M$_\odot$) galaxies found in 55.2\,deg$^2$ of UKIDSS and SDSS data, shows that the most massive galaxies ($>10^{11.5}$\,M$_\odot$) evolve in number rather faster when compared to less massive galaxies. This is in complete agreement with the results from hierarchical galaxy evolution models. Preliminary results from the Baryon Oscillation Spectroscopic Survey (BOSS), considering a 110\,deg$^2$ data set in Stripe 82 region, confirm the \citet{Matsuoka10} conclusions\footnote{Watch the presentation at the 2011 Hubble Fellows symposium by Kevin Bundy:\\https://webcast.stsci.edu/webcast/detail.xhtml;jsessionid=\\7009607225EFCBAF3A657CC31B49C119?talkid=2501\&parent=1}. Other technical factors may be evoked. Large uncertainties inherent to mass estimates \citep[highly dependent on template library, e.g.,][]{Marchesini09} may induce large --- and systematic --- variations in each data set. Selection of massive passively evolved galaxies is not homogeneous in the literature. Some groups use morphology to select spheroids (missing those with a recent merger history), others use rest-frame colours or even a spectral energy distribution (SED) fitting procedure \citep[missing those galaxies with reminiscent star-formation, which induces an UV excess, see discussion in][]{Conselice08b}. It should be stressed, however, that all agree on the existence of (extremely) massive (relatively old) galaxies at high redshifts, even at $z>3$ \citep[e.g.,][and references therein, but see \citealt{Lilly88} for one of the first examples at such high redshifts]{Marchesini09}.

Modelists, on the other hand, have to face a bigger problem: create a model able to match observations in the full observed redshift range, explaining along the process the disparity between models and observations and, if possible, that between conflicting observational results. Interpreting observations implies a proper prediction, for instance, of redshift and colour distributions, number densities, luminosity and mass functions (Section~\ref{intsec:lmfs}), for both massive and normal galaxies, both cluster and field samples. When considering EROs for the first time, hierarchical models did fail largely to predict number densities, redshift distributions, and morphologies of EROs \citep{Firth02,Roche02,Smith02}. This lead people to re-evoke monolithic collapse \citep{Eggen62,Tinsley72,Larson75,vanAlbada82} as the mechanism necessary to produce the properties of such massive galaxies at high-$z$ \citep[e.g., see the work by the K20 team,][and companion papers]{Cimatti02b,Pozzetti03}. Pure luminosity evolution (PLE) models (as in `monolithic models') did follow the basic requirements to form such exotic population (number densities and redshift distributions). However, PLE models fail to match the general picture of galaxy evolution \citep[][for a review on galaxy formation theory]{Benson10}. More recently, with the improvement of hierarchical models and the implied prescriptions (e.g., accounting for feedback processes and environment, Section~\ref{intsec:lmfs}), many authors have claimed success predicting red galaxy properties without the need of PLE. However, most results are either valid under limited conditions (either at specific magnitude limits or considering only a sub-set of galaxy type) or succeed only to predict specific properties \citep[number densities or redshift distribution; e.g., see \citealt{Gonzalez09} on][see also \citealt{Gabor10}]{Nagamine05,Kong06,KitzbichlerWhite07}.

Overall, the difficulty in explaining the red galaxy population, among other reasons, points to the need of understanding the IR as one of the best means to constrain any state-of-the-art model of galaxy evolution.

\section{The power of luminosity and mass functions} \label{intsec:lmfs}

\begin{figure}
  \begin{center}$
  \begin{array}{cc}
    \includegraphics[width=0.4\columnwidth]{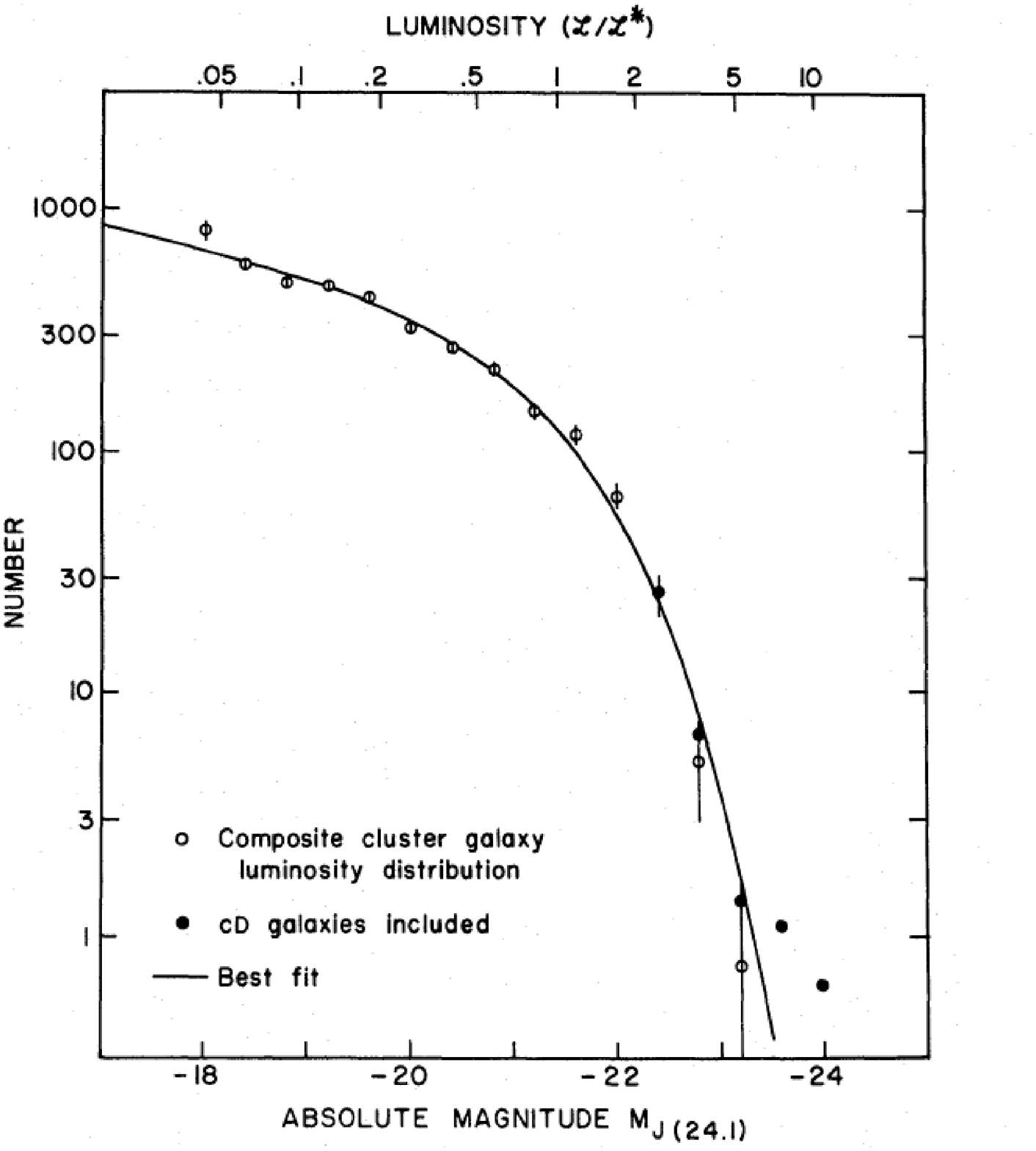} & \includegraphics[width=0.55\columnwidth]{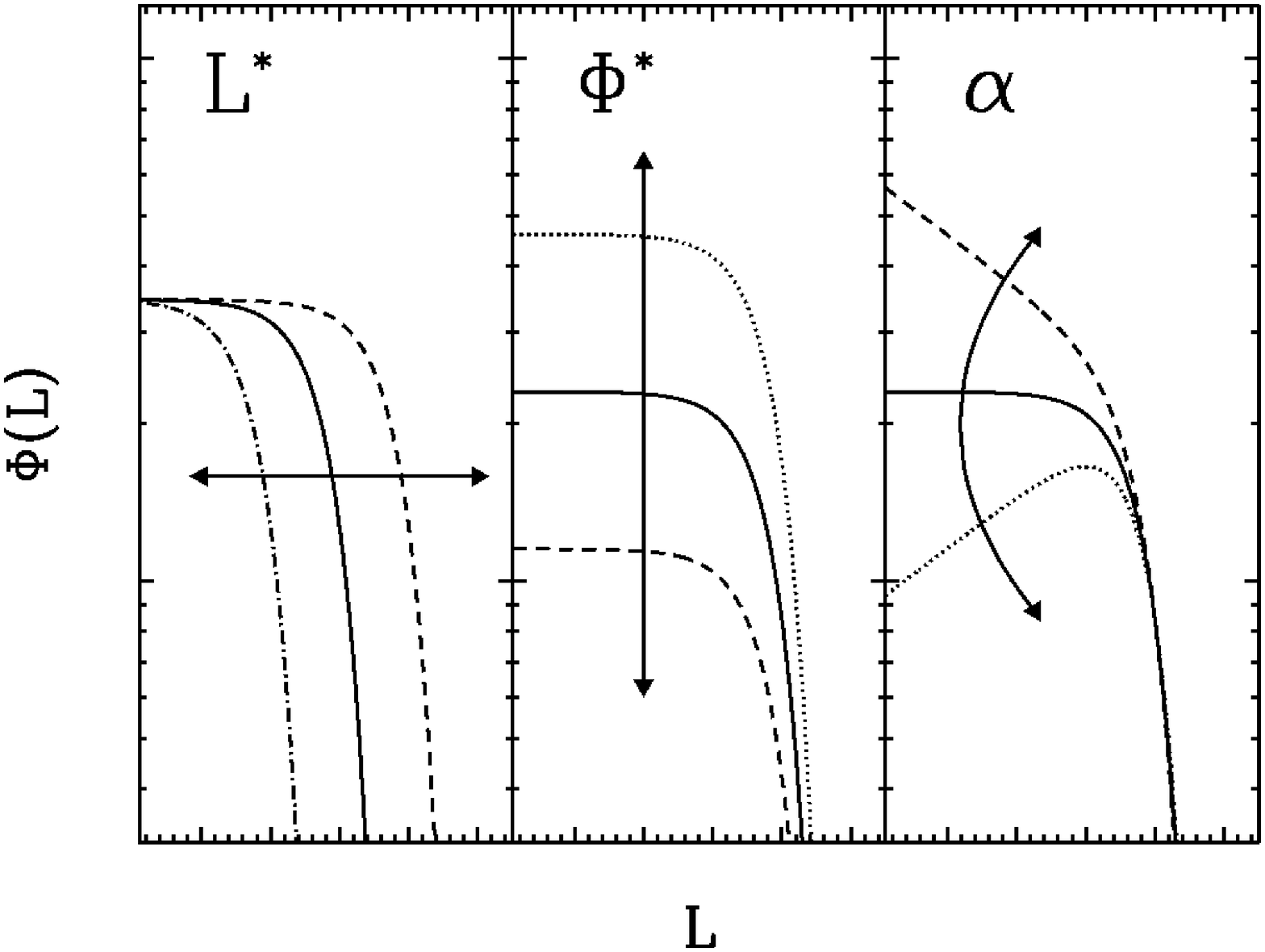}
  \end{array}$
  \end{center}
  \caption[Characterizing the LF]{On the left hand side, Figure~2 from \citet{Schechter76} is shown as an example for a LF \citep[in this case, at 5000\,\AA~ with observations with J(24.1) filter,][]{Oemler74}. On the right hand side, a simple sketch showing how changing the Schechter function parameters affects the shape of the LF: L$^*$ fixes the horizontal shift (left panel), $\Phi^*$ fixes the vertical shift (middle panel), while $\alpha$ determines the LF faint-end slope.}
  \label{c1fig:schechter76}
\end{figure}

One of the long-standing problems is, without any doubt, the ability to predict the galaxy luminosity function (and ultimately the mass function) from the highest redshift to the local Universe. Luminosity and mass functions are among the best tools for the study of galaxy evolution. They show how galaxies are distributed (or organised) in luminosity and mass. By providing the relative numbers between bright and faint or massive and light galaxies, they enable the determination of the evolution mechanism of galaxies. Sometimes, they may even allow an attempt to establish initial conditions of formation \citep{Binggeli88,Benson03}, and draw implications to the initial baryonic power spectrum \citep[e.g.,][]{Benson03}, which is directly correlated with the dark matter power spectrum \citep[see the discussion, for instance, by][on the correlation between halo and galaxy masses]{Drory09}. In the 1970's, Schechter \citep{Schechter75,Schechter76} proposed an analytical equation to describe the general shape of a LF:
\[\Phi({\rm L})=0.4~ln(10)~\Phi^*\times10^{\rm (L-L^*)(1+\alpha)}\times\rm{exp(-10^{(L-L^*)})}\]
\noindent where L is the luminosity (in logarithmic units) at which one wishes to estimate the galaxy number density $\Phi$, $\alpha$ is the slope of the faint-end of the LF, L$^*$ denotes the characteristic luminosity at which the LF exhibits a rapid change in the slope, and $\Phi^*$ is the normalization (Figure~\ref{c1fig:schechter76}). Although in specific occasions, multiple Schechter functions are necessary to fit the observations \citep[induced by the dependency on galaxy nature, e.g.,][]{Drory09}, one is generally enough, and is quite useful for further comparison between results of different research teams.

One of the first examples of the LF usefulness was its application to the Coma cluster, already since the beginning of the mid-20$^{\rm th}$ century \citep{HubbleHumason31,Zwicky51,Abell59}. The shape of Coma's LF faint-end slope has ``changed'' over the years as instrumentation improved and enabled the detection of fainter dwarfs \citep[e.g.,][]{Mobasher98}, forcing the theory to follow each new finding and peculiarity of Coma cluster \citep[see the review by][and references therein]{Biviano98}.

Current observational facilities have now reached incredible depth levels, enabling the characterisation of the LF faint-end to higher precision levels, and providing estimates of the galaxy LF (and MF) as far back as the first Gyr of universe time \citep[e.g.,][]{Bouwens07,Ouchi09,Oesch10}. Current large deep fields allow for a proper statistical study on the evolution of the galaxy population, enabling the community to probe well into the first half of Universe life \citep[e.g.,][]{Steidel99,Marchesini09,Cirasuolo10,Ilbert10}, and to assess evolution dependencies on environment, galaxy nature (early, late, starburst, AGN types) and stellar mass \citep[e.g.,][]{Zucca09,Bolzonella10,Peng10,Fu10,Strazzullo10,Ikeda11}. The reader can now realise the rather complex recipe needed to establish a good match between modelling and observations. Any state-of-the-art model today has to take into account the many physical mechanisms \citep[e.g.,][]{Kay02} and scenarios \citep[e.g.,][on dwarf galaxy disruption]{Henriques08}, each accounting for a specific feature of a given galaxy LF.

\begin{figure}
  \begin{center}
    \includegraphics[width=0.4\columnwidth]{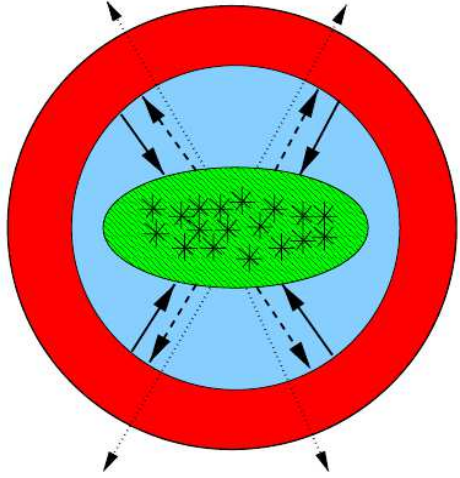}
  \end{center}
  \caption[Cooling and ejection of gas]{A sketch from \citet{Baugh06}, showing the cooling of gas from the outer hot halo (solid arrows). As the gas cools and settles into a disc (green region) to form stars, the hottest ones soon explode as super-nov\ae, reheating part of the cooled gas which then returns to the hot halo (dashed arrows) or is even completely expelled (dotted arrows). The blue region refers to the dark matter halo.}
  \label{c1fig:snfeedb}
\end{figure}

The two luminosity ends of the LF have always been (and still are) hard to predict by even the most elaborated models. Since the beginning of modelling era, the faint-end slope has frequently been overestimated (predicted slopes are too steep, i.e., small $\alpha$). Regarding the bright-end, we have set the scene already, a pure hierarchical model under-predicts luminous galaxies at high redshifts, while over-predicting them in the local Universe. In order to explain both extremes, alternative routes were taken, which led to the apparently crucial \textit{feedback} effects \citep[][and references therein]{Kay02,Benson03}. These are physically motivated and evidences for their existence have been observed \citep[][for a review]{Veilleux05}. On the one hand, the over-predicted faint-end can be explained if a star-forming bursting galaxy had its gas-supply ejected from its gravitational potential through super-nova winds blowing the gas to the outer regions of the halo \citep[][and references therein]{Kay02}. Two modes can then be identified, one of them (weaker) allowing the recapture of the ejected gas in a later stage of evolution (Figure~\ref{c1fig:snfeedb}) or during a merger, while the other completely expels the gas out of the gravitational potential. Both are used to efficiently explain different properties of the galaxy population. Dwarf galaxy disruption can also account for a flatter faint-end slope. In case dwarf galaxies happen to be falling into the core of a cluster, tidal interactions or ram pressure gas-stripping may occur, preventing more stars to be formed \citep[e.g.,][]{Boselli08,Henriques08}.

On the other hand, the quenching of star-formation in the most luminous and massive galaxies can not be explained based on stellar winds feedback. It is just too weak to expel the gas out of the deepest gravitational potentials. One possible mechanism is gas \textit{conduction} \citep{Loewenstein90,Narayan01,Gruzinov02,Voigt02,Benson03,Soker04}. Conduction in the ionized gas drives energy into the inner regions of the halo, producing a heating effect, consequently increasing the gas cooling time. Depending on the halo temperature, conduction may become relevant. For instance, in massive hot halos, likely hosting a more significant baryonic mass, massive galaxies assemble through mergers instead, as conduction may be too strong for the gas to condense through cooling flows. In fact, \citet{Benson03} reach a better matching to the observed LF (in both luminosity ends) when comparing with other kinds of feedback, with the caveat that it seems to require extremely high conductivity values. As an alternative, a \textit{super-wind} may be evoked \citep[see the seminal modelling work by][]{Granato01,Granato04}. The source of such magnitude is now believe to reside at the centre of each galaxy, in the form of a super massive black hole (SMBH). These are strong enough to deplete a galaxy halo of gas-supply (without recapture), quenching the star-formation activity, preventing a galaxy to grow larger, and hence producing the sharp edge of the LF bright-end. However, there is growing evidence for multiple active galactic nucleus (AGN) accretion modes, and each is applied differently depending on galaxy nature and cosmic time. For instance, the ``radio mode'' (the low accretion rate version) becomes gradually more significant toward lower redshifts, while the merger induced short blast-wave-like AGN feedback (the ``quasar mode'') peaks at high redshifts \citep[$2\lesssim{z}\lesssim4$,][]{Croton06a,Croton06b,Somerville08,Fontanot11}. The improvement is notable \citep[see][for probably the best matching result achieved by a model accounting for AGN feedback]{Bower06}, yet still not perfect at the LF faint-end \citep{Cirasuolo10}, thus still needing some fine-tuning. Hence, observations are fundamental to constrain the models. It is of the utmost importance to quantify and characterise the AGN population throughout Universe time, to pinpoint critical stages of evolution and to study how the interplay between host and AGN determines the evolution of both. As we shall see in the following section, and Nature would not do it in any other way, this is anything but straightforward.

\section{Finding AGN}

\begin{figure}
  \begin{center}$
  \begin{array}{cc}
    \includegraphics[width=0.4\columnwidth]{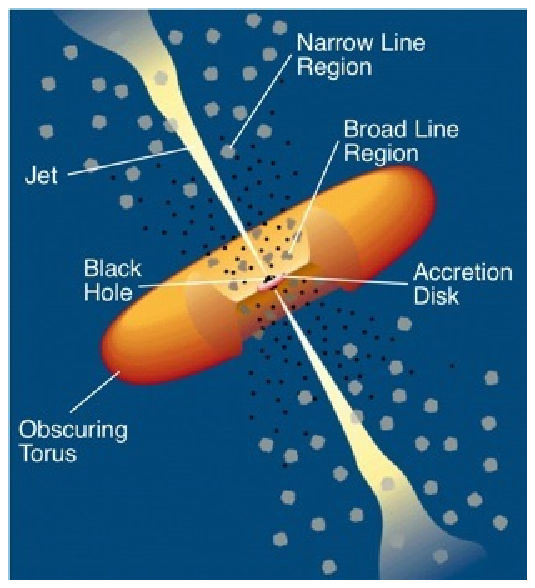} &
    \includegraphics[width=0.55\columnwidth]{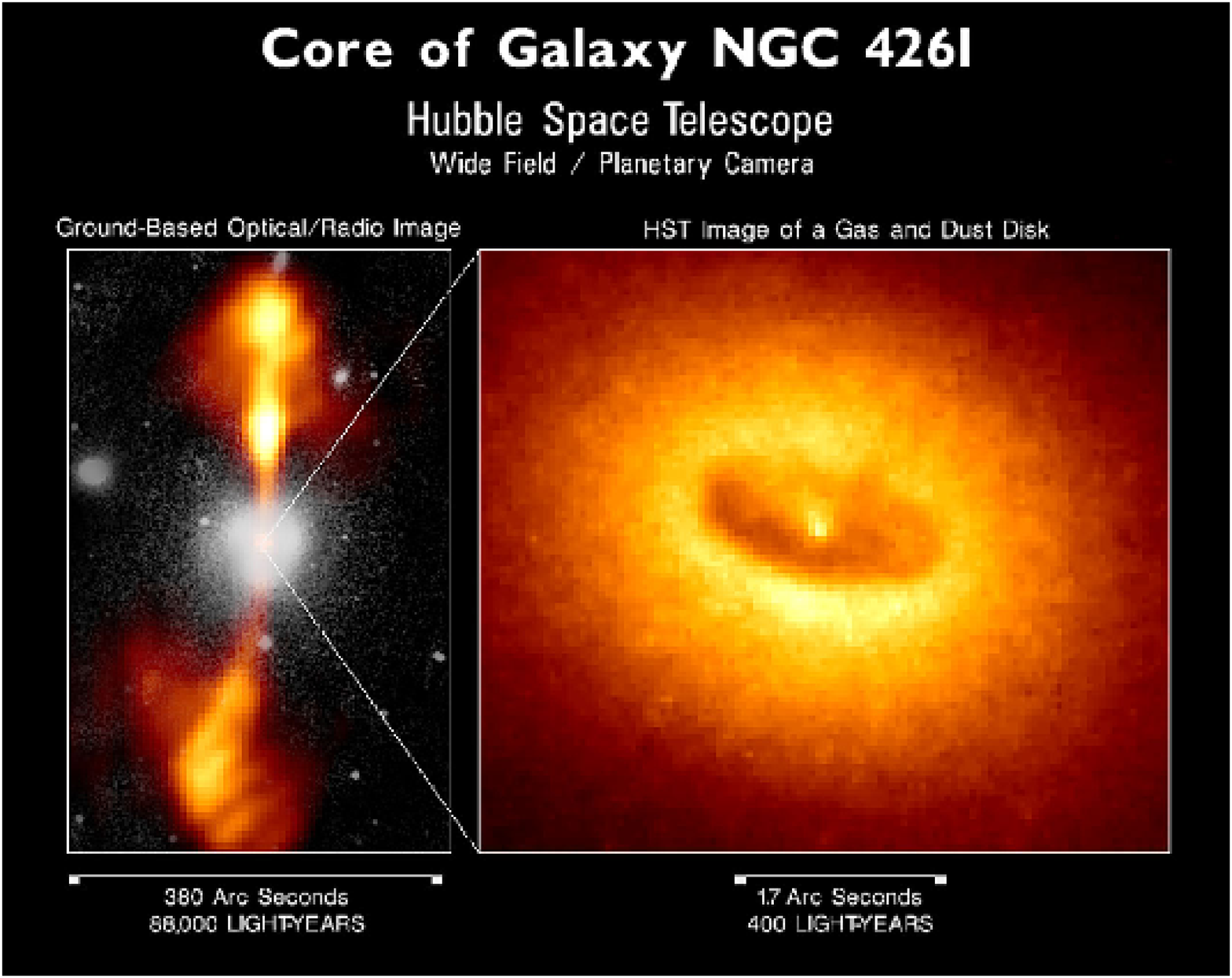}
  \end{array}$
  \end{center}
  \caption[The AGN unified model]{A representation (not to scale) of the unified AGN model (left hand side). The black hole and the accretion disc are indicated at the centre. Broad line features originate from clouds close to the nucleus ($\sim$100\,light-days) or from an accretion disc \citep{ShakuraSunyaev73}, but may be obscured by the dust torus ($\sim$100\,light-years in diameter) depending on the viewing angle. Narrow line regions are farther ($\sim$1000\,light-years) from the nuclear source. Also shown is a radio jet coming from the central engine. On the right hand side, the detection with $HST$ of a dusty thin disc surrounding the nuclear source of NGC 4261. Credit: \citet[][left hand side]{UrryPadovani95}, and \citet{Jaffe96} and \citet[][right hand side]{Ferrarese96}.}
  \label{c1fig:agnunif}
\end{figure}

AGN are an intriguing force of Nature. It is widely believed that accretion onto a nuclear SMBH is the key for AGN activity \citep{Rees84}. The study of AGN populations started back in the 1960's with the identification of the first quasi stellar object \citep[the radio source 3C-48,][]{Greenstein63,Matthews63}, or more accurately with Carl Seyfert 20\,years earlier \citep[][although the AGN nature was only acknowledged in the mid-1970's]{Seyfert43}. Now known as Seyfert galaxies, these systems were classified depending on the properties of their spectra: Seyfert 1's (showing broad and narrow emission lines) and Seyfert 2's (showing only narrow emission lines). Intermediate classifications were than needed owing to an apparent continuum of properties between these two classes \citep[e.g.,][]{OsterbrockKoski76}. A new paradigm was about to come to light after the study of optical spectra of polarized light from Seyfert 2 galaxies \citep{Miller83,Antonucci85}. \citet{Antonucci93} described it as the unified AGN model (see also \citealt{Antonucci85} and \citealt{UrryPadovani95}). In this scenario, the central engine, a SMBH, is common to all AGN, and the observed differences are assigned to different viewing angles of the central SMBH (Figure~\ref{c1fig:agnunif}). Nowadays, other AGN types (e.g., radio AGN and Blazars, X-ray type-1 and type-2 AGN) have also been linked to the unification model, and our understanding of it has improved by the consideration of clumpy dust torus models \citep{Nenkova08,HonigKishimoto10}, instead of the regular homogeneous dust torus assumption. However, more recent works, supported by observations \citep{Maccacaro82,Elvis04,Gaskell07,Risaliti07}, also point to the possibility that cold gas inflows, and not a dust torus, are the cause for the bulk of absorption of the X-ray emission from the accretion disc.

Most of the work done until the 1990's was based in ultra-violet (UV) or optical, while radio would only provide rare extreme objects. Today we know that light originated in such type of activity is seen throughout the complete electromagnetic spectrum and detectable up to the highest known redshifts \citep[e.g.,][and references therein]{Jiang06,Seymour07,Nenkova08,Schneider10,Ricci11}.

The X-rays regime is currently the most preferred one to study AGN evolution. This relates to the fact that, such high spectral energies can only hold for powerful mechanisms, which normal stellar populations are unable to achieve. Hence there is a stronger dominance in the X-rays from AGN emission over that of host galaxy, when compared to what happens at UV/optical wavelengths. Adding to that, obscuration will affect less the X-rays emission as opposed to UV and optical. This improves with increasing redshifts as more energetic rest-frame energies (hence less affected by dust) will be observed. Nonetheless, X-rays still surfer significant obscuration. One of the biggest evidence is the observed cosmic X-ray background (CXB), which has been resolved approximately 10\,years ago at a 70--90\% level (the upper limit being related to the softer part of the spectrum) by $Chandra$ and XMM space telescopes \citep{Mushotzky00,Giacconi01,Bauer04,Worsley04,Worsley05}. However, it was soon found out a decrease of that fraction with increasing spectral energies \citep[down to 50\% at $>8\,$keV,][and references therein]{Worsley04,Worsley05}. This is due to a high fraction of unobscured sources easily detected at softer energies, and high intrinsic obscuration column densities ($\rm{N_H}$) in the sources comprising the hard CXB. Seyfert\,2s are four times more numerous than Seyfert\,1s in the local Universe \citep{Maiolino95}, being half of the Seyfert\,2s classified as compton thick \citep[$\log(\rm{N_H[cm^{-2}]})>24$,][]{Maiolino98,Risaliti99}. At higher redshifts, obscured:unobscured source ratios of 3:1 to 4:1 are predicted based on the CXB and synthesis models \citep{Comastri01,Ueda03,Gilli04,Treister05,Tozzi06}. In deep fields, however, the ratio seems to be smaller (2:1, due to incompleteness toward obscured objects), but can get as high as 6:1 when considering specific galaxy populations \citep[SCUBA galaxies,][and this work, Chapter~\ref{ch:erg}]{Alexander05}. Hence, although reliable, X-rays AGN studies may be significantly affected by enhanced obscuration, specially at high redshifts.

The IR, on the other hand, shows a somewhat complementary view. Long known since the 1970's \citep[with ground-based telescopes,][and references therein]{KleinmannLow70,Rieke78} and 1980's \citep[with the start of IR space-based observations,][]{deGrijp85,Miley85,Neugebauer86,Sanders89}, active galaxies are prone to show intense emission at IR wavelengths. This is mostly due to the already referred dust obscuration hiding AGN signatures at optical and even X-ray wavelengths. The absorbed energy is subsequently reprocessed by the enshrouding dust and emitted at IR wavelengths, producing a power-law shaped emission excess beyond 1.6\,${\mu}$m\footnote{Blueward of this wavelength, the contribution of AGN emission through this reprocessed light mechanism diminishes significantly due to dust sublimation. Only scattered light and the tail of the Wien's thermal emission from the hottest dust grains are expected.}. This is a powerful tool as it allows the selection of AGN sources not revealed at other wavelengths. A major accomplishment in recent years has been the development of purely photometric techniques, in the 3--8~$\mu$m range, for the efficient selection of sources with enhanced IR emission redward of the 1.6\,${\mu}$m stellar peak, characteristic of an active galactic nucleus \citep[AGN; e.g.,][ and this work]{Lacy04,Stern05,Polletta06,Donley07,Fiore08}. These techniques effectively allow for the detection of a significant fraction of AGN sources missed even by the deepest X-ray-to-optical surveys.

It should be mentioned that none of the spectral regimes should be discarded in detriment to any of the remainder for the purpose of AGN selection (unless the science case implies such assumption). Each one of them is sensitive to specific (and sometimes distinct) AGN populations and/or phases and/or regions of emission (e.g., Figure~\ref{c1fig:agnunif}). Although some overlap between these AGN populations is expected, they should all be considered in ensemble (radio included), if the ultimate goal is the complete selection of AGN host galaxies. Geared with such tools and with the recent developments on clumpy dust torus models \citep[e.g.,][]{Nenkova08,HonigKishimoto10}, it is now possible to address with unprecedented detail the evolution of AGN host galaxies up to high redshifts.

\section{Dust everywhere}

As it was frequently highlighted in the previous sections, dust exists and its effects cannot be underestimated. There are clear evidences that dust is common in the Universe \citep{CharyElbaz01,HauserDwek01,leFloch05,Dole06,Franceschini08} and is observed even at high redshift \citep[e.g., in the radio-quiet QSO at $z=4.69$ announced by][]{Omont96}. Hence, not only a significant number of galaxies is missed even in the deepest UV/optical surveys due to extreme dust obscuration, but corrections have also to be applied to the light reaching us from the remainder galaxy population, bringing an undesired model-dependency to the process \citep[e.g.,][]{Buat05,Bouwens09}. This forced the community to turn its efforts to other wavelength regimes. Although optical telescopes have been those to provide the deepest and sharpest views of the sky, facilities at other wavelengths will soon catch up, such as IR (with \textit{James Webb Space Telescope} launch in 2018), millimetre (ALMA currently coming online), and radio observatories (the very large baseline arrays and coming facilities such as ASKAP, MeerKAT, and MWA as precursors of the long-waited SKA).

Hence, it is of great importance to determine, for example, how much dust is present in galaxies, how much does dust affects the light reaching us, and how it has evolved through cosmic time \citep{Dunne11}. For this purpose, the IR and millimetre (mm) spectral regimes have been the best unveiling the properties of dust in galaxies \citep[for a review, see][]{Hunt10}, as it mostly emits at these wavelengths. The X-ray-to-optical light absorbed by dust, is reprocessed and re-emitted at IR wavelengths. Hence, the IR is a viable tool to evaluate the dust content in galaxies. And in its turn, dust is believed to be produced either by supernov\ae\,\citep{Rho08,Kotak09,Barlow10,Meikle11} and/or low/intermediate mass asymptotic giant branch stars \citep{Gehrz89,FerrarottiGail06,Sargent10}. Dust itself may then be an indicator of the current and past star-formation history of a galaxy. However, much of the work done to this regard at IR and mm wavelengths \citep[e.g.,][]{Saunders90a,Saunders90b,Blain99,leFloch05,Jacobs11}, has relied on shallow data or in small number statistics when compared to optical-based studies. This is related to the yet unsolved lack of multiplexing spectral power and/or sensitivity of mid-IR ($>8\,\mu$m) facilities (space- and ground-based\footnote{Ground-based facilities have in addition to account for the strong atmospheric thermal background, preventing a proper study of the faintest galaxies.}), and the sensitivity of mm facilities. This implies that all but the brightest sources in the sky will be possible to study. Consequently, the conclusions arising from those studies can not be, by any means, generalised to the overall galaxy population. One way to solve the problem is through the application of stacking techniques \citep[e.g.,][]{Zheng06,Martin07,Martinez09,Lee10,Rodighiero10,Greve10,Bourne11}, allowing the estimate of the general properties of a given population, yet limiting any study relying on luminosity functions.

\section{Thesis outline}

This thesis is mostly focused on galaxy populations selected at IR wavelengths. As described above, recent years have assigned them a crucial roll on unveiling the mysteries of galaxy evolution from the early Universe to what we see locally.

\subsection{Extremely red galaxies}

Chapter~\ref{ch:erg} presents a multi-wavelength analysis of the properties of Extremely Red Galaxy (ERG) populations, selected in the GOODS-South/\textit{Chandra} Deep Field South field. In the literature, there are many criteria with which to select sources presenting extreme red colours ($(R-K)_{\rm vega}>5$, $(R-K)_{\rm vega}>6$, $(I-K)_{\rm vega}>4$, $(I-H)_{\rm vega}>3$, $(J-K)_{\rm vega}>2.3$, $f_{3.6}/f_{z850}>20$, etc...). This resulted in a plethora of acronyms (ERO, VRO, IERO, Hyper-ERO, DRG, VRG, FROG). Together with the fact that the AB photometry system and/or similar (yet different) filters are also often considered, it was difficult for the community to have a complete and uniform understanding of the ERG population (e.g., while comparing model and observational results). Here, we study and compare three of the most frequently referred criteria (Extremely Red Objects, $i_{775}-K_s>2.5$; IRAC Extremely Red Objects, $f_{3.6}/f_{z850}>20$; and Distant Red Galaxies, $J-K_s>1.35$) to assess their unique and common properties. A different statistical analysis from the one in \citet{Messias10} is adopted, where uncertainties related to low S/N photometry and limitations in modelling SEDs are accounted for. By using all the photometric and spectroscopic information available on large deep samples of Extremely Red Objects (EROs, 553 sources), IRAC EROs (IEROs, 259 sources), and Distant Red Galaxies (DRGs, 289 sources), we derive redshift distributions, identify AGN powered and star-formation powered galaxies, and, using the radio observations of this field, estimate robust dust-unbiased star formation rate densities ($\dot{\rho}_{\ast}$) for these populations. We also investigate the properties of ``pure'' (galaxies that conform exclusively to only one of the three ERG criteria considered) and ``combined'' (galaxies that verify simultaneously all three criteria) sub-populations. Overall, a large number of AGN are identified (up to $\sim25\%$, based on X-ray and mid-IR criteria), the majority of which are type-2 (obscured) objects. Among ERGs with no evidence for AGN activity, we identify sub-populations covering a wide range of average star-formation rates, from below 10\,M$_{\odot}$\,yr$^{-1}$ to as high as 140 M$_{\odot}$ yr$^{-1}$. Applying a redshift separation ($1\leq{z}<2$ and $2\leq{z}\leq3$) we find significant evolution in $\dot{\rho}_{\ast}$. While EROs and DRGs follow the general evolutionary trend of the galaxy population, no evolution is observed for IEROs. However, IEROs are the largest contributors (up to a 25\% level) to the global $\dot{\rho}_{\ast}$ at $1\leq{z}<2$, while EROs may contribute up to 40\% at $2\leq{z}\leq3$. The radio emission from AGN activity is typically not strong in the ERG population, with AGN increasing the average/median radio luminosity of ERG sub-populations by, nominally, between $\sim$10 and 25\%. However, AGN are common, and, if no discrimination is attempted, this could significantly increase the ERG $\dot{\rho}_{\ast}$ estimate (by 200\% in some cases). This can be understood in two ways: either the AGN host population is indeed actively forming stars or AGN emission can strongly bias such studies. Hence, although the contribution to the radio luminosity of star-forming processes in AGN host galaxies remains uncertain, one can still estimate lower and upper limits of $\dot{\rho}_{\ast}$ in ERG populations from the radio alone. A comparison between the radio estimates and the UV spectral regime confirms the dusty nature of the combined populations.

ERGs are known to be massive systems at high redshift, and, in this work, mass functions are produced and stellar mass densities estimated, showing that at $1\leq{z}\leq3$, 60\% of the mass of the universe resides in ERGs. A morphology study is pursued for a better characterization of this ERG sample, revealing an interesting population of DRGs, which show a mixture of young and old stellar populations together with obscured AGN activity. These results all together may point to the fact that EROs, IEROs, and DRGs are all the same population, yet seen in different phases of evolution.

\subsection{The IR selection of AGN}

Chapter~\ref{ch:agn} is focused on the AGN selection at IR wavelengths. It is widely accepted that the mid-IR (MIR) enables the selection of galaxies with nuclear activity, which may not be revealed even in the deepest X-ray surveys. Many MIR criteria have been explored to accomplish this goal and tested thoroughly in the literature. The main conclusion is that at high redshifts ($z\gtrsim2.5$) the contamination of these AGN selection criteria by non-active galaxies is abundant. This is not at all appropriate for the study of the early Universe, the main goal of many of the current and future deep surveys. Using state-of-the-art galaxy templates covering a variety of galaxy properties, we develop improved near- to mid-IR criteria for the selection of active galactic nuclei (AGN) out to very high redshifts. With a particular emphasis on the \textit{James Webb Space Telescope} (\textit{JWST}) wavelength range (1--25\,$\mu$m)\footnote{We note that even though the launch of $JWST$ is still unsettled (2018 is now the year of launch), we remember that there is still a wealth of data to be explored from IR space telescopes such as \textit{Spitzer}, \textit{Akari}, and WISE.}, we develop an improved IR criterion (using $K$ and IRAC bands, KI) as an alternative to existing MIR AGN criteria for the $z\lesssim2.5$ regime. We also develop a new MIR criterion which reliably selects AGN hosts from local distances to as far as the end of re-ionization ($0<z<7$, using $K$, IRAC, and MIPS-24\,$\mu$m bands, KIM). Both KI and KIM are based in existing filters and are suitable for immediate use with current galaxy observations. Control samples with deep multi-wavelength coverage (ranging from the X-rays to radio frequencies) are also utilized in order to assess the quality of the new criteria compared to existing ones. We conclude that the considered galaxy templates and control samples indicate a significant improvement for KI over previous IRAC-based AGN diagnostics, and that KIM is reliable even beyond $z\sim2.5$.

\subsection{The contribution of dust to the IR}

Chapter~\ref{ch:lfs} explores the extension of the current FIR/mm studies, on the cold (T$\lesssim100\,$K) dust re-emission dominating at those wavelengths, to the hot (T$\gtrsim1000\,$K) extremes of dust re-emission ($<8\,\mu$m) using observations on the Cosmic Evolution Survey \citep[COSMOS,][]{Scoville07}. The study is mostly based on data from the IR array camera (IRAC) on board the \textit{Spitzer Space Telescope} (\textit{Spitzer}), facility which, in less than a decade, has contributed so much to the field of galaxy evolution \citep[for a review, see][]{Soifer08}. The goal is to estimate the dust contribution to the SED of the galaxy population at shorter IR wavelengths, regime which has never been explored for such purpose. The sample is divided into redshift ranges where specific polycyclic aromatic hydrocarbons (PAHs) features (3.3, 6.2, and 7.7\,$\mu$m) are expected to be observed by \textit{Spitzer}-IRAC filters, and to which hot dust is known to contribute significantly. Although PAHs are not actual dust particles, they comprise a significant fraction of the Carbon existing in the universe, they are believed to be closely related to star-formation activity, and to reprocess a substantial fraction of UV-light into the IR wave-bands, hence being a major source of obscuration \citep{Tielens11}. The IR continuum comes from dust heated by energetic radiation fields. Vigorous obscured star-formation can account for such emission as well as AGN activity. However, the overall stellar population also emits at these wavelengths, even frequently dominating at $<3\,\mu$m and peaking at 1.6\,$\mu$m, due to the H$^-$ opacity minimum in stellar atmospheres. In this chapter, we describe how this is taken into account to derive the final dust luminosity density functions. Dependencies on both redshift and galaxy nature are estimated. We report a concerning AGN-induced source of significant bias to any mass estimate procedure relying on IR luminosities, specially at high redshifts. Valid counter arguments to other possible mechanisms giving origin to such effect are also discussed. Finally, evidences for the connection of the AGN population to the known bimodality of the IR LF \citep[][and references therein]{Drory09} are presented at both bright and faint ends.

\subsection{Future work}

A thesis work is never complete and there is always room for improvement. The work presented here is no exception.

In Chapter~\ref{ch:fut} we detail the many galaxy properties left to be explored in the ERG population, the questions still left to be answered on $K_s$-selected galaxy samples, and we describe the work currently being pursued for the development of a stacking algorithm for the application of stacking analysis on ASKAP data (one of the precursors of SKA).

The IR AGN selection may still require some fine tuning, as, for instance, it has never been tested against the emission from TP-AGB stars. Knowing that it peaks at $\sim2\,\mu$m, we believe the contamination will happen at higher redshifts (when the 8.0$\,\mu$m band probes the $2\,\mu$m rest-frame wavelength), unless even higher wavelength bands are used (e.g., $24\,\mu$m). Also, it is in the high-redshift regime where a larger incidence of systems with enhanced TP-AGB stellar emission is known to reside \citep{Maraston05,Henriques11}. If such effect in the IR regime significantly affects IR AGN selection, than we are forced to use only the most restrictive AGN criteria \citep[like the bright IR excess sources, e.g.,][]{Polletta06,Dey08} or to rely solely on the remainder spectral regimes, which sometimes is not the ideal scenario. On the other hand, if the criterion is confirmed to be efficient even when TP-AGB stellar emission is present, we will be able to track AGN activity from the earliest stages of cosmic time. This will provide AGN host populations with an enough number of sources to constrain any kind of model considering AGN activity, and in a large redshift range $0<z<7$. Also, as soon as $JWST$ becomes online, the filter set proposed in this work for the IR selection of AGN up to $z<7$, should be tested.

Taking advantage of the large source numbers we have studied in COSMOS field, we also describe in this chapter how we plan to use stacking analysis to directly compare the hot-dust regime (3--8\,$\mu$m) with the cold one emitting at FIR/mm wavelengths.

Finally, we describe future prospects as a result from this thesis. Among them, an on going project on passive disc galaxies at high redshifts ($1<z<3$). We propose an IR selection criterion, while providing evidences for its efficiency using the latest data from WFC3 on board \textit{HST}. Possible explanations are given and the implications for such a population to exist at these redshifts are discussed. These galaxies are one of the ultimate goals of ALMA science, thus being one of the most significant outcomes of this thesis.

Throughout this thesis we use the AB magnitude system\footnote{When necessary the following relations are used:\\(K, H, J, I)$_{AB}$ = (K, H, J, I)$_{Vega}$ + (1.841, 1.373, 0.904, 0.403) from \citet{Roche03};\\IRAC: ([3.6], [4.5], [5.8], [8.0])$_{AB}$ = ([3.6], [4.5], [5.8], [8.0])$_{Vega}$ + (2.79, 3.26, 3.73, 4.40) from \emph{http://spider.ipac.caltech.edu/staff/gillian/cal.html}}, we consider a $\Lambda$CDM cosmology is assumed with H$_{0} = 70$ km s$^{-1}$ Mpc$^{-1}$, $\Omega_{M} = 0.3$, $\Omega_{\Lambda} = 0.7$, and we adopt a Salpeter \citep{Salpeter55} initial mass function (IMF).
\chapter[A multi-wavelength approach to ERGs]{A multi-wavelength approach to Extremely Red Galaxies}
\label{ch:erg}
\thispagestyle{empty}

\section{Introduction}

In an attempt to constrain hierarchical models of galaxy formation, the last few years have seen optical-to-infrared or infrared-to-infrared colour criteria being used to find high-redshift galaxies hosting evolved stellar populations. Extremely Red Objects \citep[EROs,][]{Roche03}, IRAC-selected EROs \citep[IEROs, also known as IR EROs,][]{Yan04} and Distant Red Galaxies \citep[DRGs,][]{Franx03} were thought to identify old passively evolving galaxies at increasing redshifts (from $z>1$ for the EROs/IEROs to $z>2$ for the DRGs), for which a prominent 4000\,\AA\ break (as a result of absorption from ionized metals, e.g., the Ca\,II H and K breaks) would fall between the observed bands. These techniques, however, are also sensitive to active (star-forming, AGN or both) high-redshift dust-obscured galaxies \citep[$0.6<\rm{E}(B-V)_{\rm star}<1.1$,][]{Cimatti02a}, with intrinsically red spectral energy distributions \citep{Smail02,Alexander02,Afonso03,Papovich06}. These active members of the \textit{Extremely Red Galaxy} population (ERGs, as we will collectively call EROs, IEROs, and DRGs) are also important targets for further study given that they constitute a dusty population of galaxies easily missed at optical wavelengths \citep[e.g.][]{Afonso03}. A challenge in studying the nature of the red galaxy population is the difficulty in disentangling the effects due to redshift, dust-obscuration, and old stellar populations.

Identifying the so-called ``Passively Evolving'' and ``Dusty'' ERGs is a fundamental and particularly difficult task, where optical spectroscopic observations are of limited use. The identification and study of Active Galactic Nuclei (AGN) or star-formation (SF) activity in these galaxies, for example, requires multi-wavelength data from X-ray to radio frequencies. Long wavelength radio observations are of particular interest here, given the possibility to reveal the activity in these obscured systems and, for star-forming dominated galaxies, allowing for a dust-free estimate of their star-formation rates (SFR).

In this chapter we present a comparative study of different ERG populations. Using the broad and deep wavelength coverage in the Great Observatories Origins Deep Survey South (GOODSs) / $Chandra$ Deep Field South (CDFs), we select samples of EROs, IEROs, and DRGs and estimate their common and unique properties considering a new statistical approach. With the extensive photometric data available we explore the redshift distribution, SFR and mass densities, AGN activity, and dust content in these galaxies. The radio regime and a stacking procedure are considered to estimate dust-free SFRs and the contributions of the overall red galaxy populations and that with no detected AGN activity to the global star formation rate density ($\dot{\rho}_{\ast}$).

The structure of this chapter is as follows. Sample selection is described in Section~\ref{c2sec:sampsel}. Section~\ref{c2sec:agnid} addresses the AGN identification technique. In Section~\ref{c2sec:ergprop} the ERG sample is characterised, leading to the estimate of the dust-free contribution to $\dot{\rho}_{\ast}$, dust content, mass functions (MFs) and mass densities ($\rho_{\rm M}$), and morphology parameters. Finally, the conclusions are presented in Section~\ref{c2sec:conc}.

\section{Sample Selection} \label{c2sec:sampsel}

The GOODS was designed to assemble deep multi-wavelength data in two widely separated fields: the Hubble Deep Field North (HDFn) and the CDFs. Specifically the southern field includes X-ray observations with \textit{Chandra} X-ray Observatory ($Chandra$) and XMM-\textit{Newton}; optical ($BVIz$) high resolution imaging with the ACS on-board the \textit{HST}; NIR and mid-infrared (MIR) coverage with the \textit{Very Large Telescope} (\textit{VLT}) and the \textit{Spitzer Space Telescope} ($Spitzer$), respectively; and radio imaging with the \textit{ATCA}, \textit{VLA}, and \textit{GMRT}. These data are among the deepest ever obtained. Large programs aiming at comprehensive spectroscopic coverage of this field are also being performed. The quality and depth of such data make these fields ideal to perform comprehensive studies of distant galaxies and, in particular, of the ERG population.

\subsection{Methodology} \label{c2sec:method}

\begin{figure}
  \begin{center}
    \includegraphics[width=0.75\columnwidth]{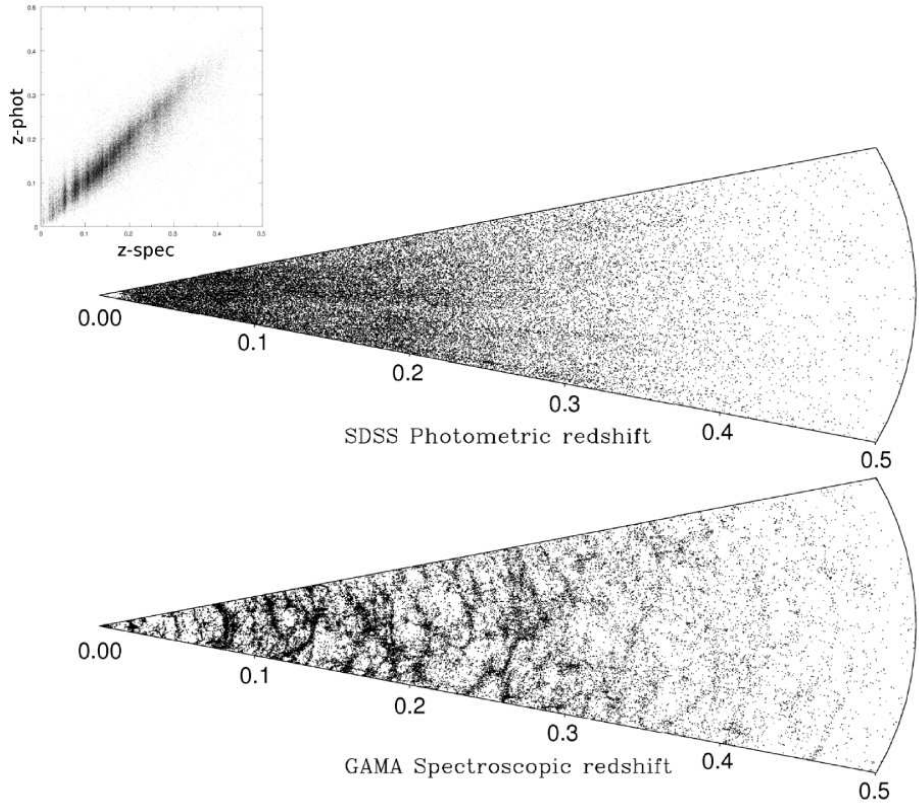}
  \end{center}
  \caption[The difference between $z_{\rm spec}$ and $z_{\rm phot}$]{Two light cones show how spectroscopic redshift estimates recover well the large scale structure of the Universe, while photometric ones do not, even when achieving acceptable results (small inset on the top left showing a close to one-to-one relation). The spectroscopic redshifts come from the GAMA survey, while the photometric redshifts estimated for the same sources were computed by Hannah Parkinson using ANNz. Credit: Simon Driver and the GAMA team.}
  \label{c2fig:gamasp}
\end{figure}

ERGs are in general faint galaxies. At $z=2.5$, for instance, Arp220 (M82) would present a colour $I-K\sim3.5\,(3)$ and a $K$-band flux of $f_K\sim1\,(2)\,\mu$Jy. Hence, although we consider a data-set amongst the deepest available, many ERGs will be found at a low signal-to-noise level. This implies larger errors in the photometry (from X-rays to radio frequencies) and any other estimate based on it. Take the photometric redshift for example. This technique is based on the broad-band spectral coverage available for a given galaxy (and, most times, in the adopted simple stellar population models), hence, highly dependent on the photometry precision. As one can expect, a photometric redshift estimate is less precise than a spectroscopic one. This is evident from Figure~\ref{c2fig:gamasp}. There, two light cones are presented, where the one seen at the bottom considers spectroscopic data from Galaxy and Mass Assembly \citep[GAMA,][]{Driver09}, while the light-cone at the top shows the estimated $z_{\rm phot}$ \citep[computed by Hannah Parkinson using ANNz,][and calibrated with the corresponding GAMA $z_{\rm spec}$]{Firth03} for the same Sloan Digital Sky Survey \citep[SDSS,][]{York00} sources. Although the ANNz algorithm produces predominantly a one-to-one $z_{\rm spec}-z_{\rm phot}$ match (small inset on the top left in the figure), the photometric procedure does not recover the large scale structure clearly evident in the spectroscopic light-cone.

When a photometric redshift is assigned to a galaxy, in reality, what is implied is a redshift probability distribution (PD) with a characteristic value $z_{\rm phot}$ (the value at which the integration of the PD reaches 0.5, i.e., 50\% probability) and lower/upper limits \citep[set by the values at which the integration from the edges of the PD reaches, for example, 0.317/2, i.e., $\sim16\%$, equivalent to $1\sigma$ confidence limits,][]{Wuyts08}. The accuracy and precision of the redshift estimates is important for the study of galaxy evolution with redshift, specially for samples mostly relying on photometric redshifts. Consider the following basic example, if one wants to know how many galaxies there are at $2<z<3$ in a sample of three galaxies with $z_{phot}={1.9,2.5,3.2}$, normally, the answer would be one galaxy. However, if we would estimate the probabilities ($\mathcal{P}$) of each one of these galaxies to be at $2<z<3$ and get $\mathcal{P}={45\%,100\%,45\%}$, the answer would have to be (approximately) two effective galaxies. Not only that, but if in a subsequent step one wishes to estimate average properties of the sample (luminosities, masses, etc...), the final value would not be based in just one object, yet in three measurements weighted by their probability, implying a much more reliable statistical value.

In this work we adopt the following methodology: whenever applying constraints to the sample (those being magnitude or colour cuts, redshift ranges, etc...), we assign to each source its probability to fulfil the imposed constraints. From that moment on, whenever a given source is taken into account while estimating general properties of a sample (effective source counts, redshift distributions, luminosities, masses, etc...), its contribution is weighted by the probability to fulfil such constraints.

Obviously, every source will have a non-zero probability to conform any considered constraint in this work. Hence, we adopt a limit below which a source is not considered. A source is only taken into account if its probability to fulfil a given sample constraint is $\mathcal{P}>0.317/2$ ($\sim$16\%), meaning that every considered source in this work is at most $1\,\sigma$ away from the established constraint.

Henceforth, every number presented in this chapter refers, unless the nominal value is stated, to the effective (or expected) number --- approximated to unit --- estimated after considering all the weights of the sources in a sample.

\subsection{The FIREWORKS catalogue}

We use the FIREWORKS $K_s$-band selected catalogue from GOODSs \citep{Wuyts08}. This provides reliable photometry from ultra-violet (UV) to IR wavelengths (0.2-24${\mu}$m) for each source detected in the $K_{s}$ ISAAC/\textit{VLT} maps \citep[ISAAC GOODS/ADP v1.5 Release,][]{Vandame02}, thus covering an area broadly overlapping the \textit{HST} ACS observations. The widely different resolutions between optical and IR bands are properly handled to allow consistent colour measurements. This is performed by adjusting the optical \textit{HST} and NIR \textit{VLT} images to a common resolution and by fitting the IRAC, and MIPS images, using the prior knowledge about position and extent of sources from the $K_{s}$-band image and considering the larger mid-IR point spread functions \citep[for a detailed description of the procedure, see][]{Wuyts07,Wuyts08}.

Resdshift estimates are also provided in the FIREWORKS catalogue. Recently, the VIMOS team \citep{Popesso09,Balestra10} has also released a set of spectroscopy data which is also considered in this work, as well as those referred by \citet{Silverman10} on $Chandra$ and VLA detected sources. Spectroscopic observations are used essentially for redshift information with which to derive the intrinsic luminosities of ERGs. Only good spectroscopic redshift determinations\footnote{Quality flag equal or greater than 0.5 in \citet{Wuyts08}, flag `A' in VIMOS catalogue, and flag `2' in \citet{Silverman10}.} were considered, comprising 22\% of the ERGs (Table~\ref{c2tab:tabctp}). For the remaining sources, photometric redshift estimates from the FIREWORKS catalogue and from \citet{Luo10} were considered. The redshift distributions will be discussed in Section~\ref{c2sec:reddis}.

\ctable[
   cap     = ERG number statistics,
   caption = {ERG number statistics: sample overlap, counterparts, and classification.},
   label   = {c2tab:tabctp},
   sideways,
   nosuper
]{cr|rrr|r|rrrrrr|rrr|r|r}{
  \tnote[Note.]{---The numbers displayed are effective counting and approximated to unit.}
  \tnote[$^a$]{Total number of sources in each (sub)sample and, in parenthesis, those which have good photometry in all bands involved in the ERG criteria: $i_{775}$, $z_{850}$, $J$, $K_s$, and $3.6{\mu}m$.}
  \tnote[$^b$]{Total number of X-ray identifications.}
  \tnote[$^c$]{Number of sources classified as type-1 or type-2 AGN (A1 or A2, respectively), type-1 or type-2 QSO (Q1 or Q2, respectively), and AGN with undetermined type (no $\rm{N_H}$ determination, column $\rm{nN_H}$) according to the \citet{Szokoly04} criterion.}
  \tnote[$^d$]{Number of sources selected as AGN by the KI criterion (final corrected number).}
  \tnote[$^e$]{Sources whose KI AGN probability has been corrected based on [8.0]-[24] colours.}
  \tnote[$^f$]{Total number of sources classified as AGN, considering all AGN identification criteria, along with the equivalent fraction in the total (sub)population.}
}{ \FL
POP & $\rm{N_{TOT}}$\tmark[a] & \tiny{ERO} & \tiny{IERO} & \tiny{DRG} & $\rm{N_{spec}}$ & \multicolumn{6}{c}{X-Ray} & KI\tmark[d] & MIPS & KIcr\tmark[e] & Radio & N\tiny{$_{\rm AGN}$}\tmark[f] \NN
&  &  &  &  &  & \tiny{XR}\tmark[b] & \tiny{A1}\tmark[c] & \tiny{A2}\tmark[c] & \tiny{Q1}\tmark[c] & \tiny{Q2}\tmark[c] & \tiny{$\rm{nN_H}$}\tmark[c] &  & $24{\mu}$m &  & \tiny{$1.4\,{\rm GHz}$} & \ML 
\bf{ERG} &  628 (607) & 553 & 259 & 289 & 140 & 72 & 14 & 26 & 3 & 17 & 7 & 111(27) & 338 & 16 & 24 & 154 (25\%) \NN
\bf{ERO} &  553 (541) & 553 & 249 & 212 & 130 & 67 & 13 & 25 & 3 & 15 & 7 & 85(24) & 308 & 12 & 23 & 127 (23\%) \NN
\bf{IERO} &  259 (258) & 249 & 259 & 163 & 30 & 39 & 9 & 11 & 2 & 9 & 5 & 64(17) & 167 & 10 & 16 & 85 (33\%) \NN
\bf{DRG} &  289 (280) & 212 & 163 & 289 & 33 & 40 & 10 & 11 & 2 & 10 & 4 & 94(19) & 175 & 16 & 14 & 114 (39\%) \NN
\bf{cERG} &  156 (156) & 156 & 156 & 156 & 14 & 29 & 8 & 9 & 1 & 6 & 4 & 51(13) & 109 & 9 & 10 & 66 (42\%) \NN
\bf{pERO} &  234 (234) & 234 & 0 & 0 & 90 & 21 & 2 & 12 & 0 & 4 & 2 & 8(4) & 110 & 0 & 6 & 24 (10\%) \NN
\bf{pDRG} &  61 (61) & 0 & 0 & 61 & 9 & 4 & 1 & 1 & 0 & 2 & 0 & 22(2) & 24 & 2 & 0 & 23 (38\%) \LL\NN
}

The FIREWORKS catalogue contains (nominally) 6308 $K_{s}$-selected sources. To allow for robust selection of our ERG populations the following requirements are considered: ($i$) a magnitude completeness limit of $K_{s,TOT}=23.8$\,AB, ($ii$) a flux error less than a third of the flux, ($iii$) no strong neighbouring sources affecting the flux estimates, and, ($iv$) following the prescription for robust photometric samples from \citet{Wuyts08}, adopt a pixel weight limit of $K_{s}$w$>0.3$, which takes into account local $rms$ and relative integration time per pixel and allows the rejection for bad/hot pixels and pixels with other kind of artefacts\footnote{See Section~3.4 of \citet{Wuyts08} for a description of the concept of pixel weight.}. This results in a $K_{s}$-selected sample of 4274 sources at $K_{s,TOT}<23.8$.

\subsection{Red Galaxy Samples} \label{c2sec:selerg}

Three categories of ERGs are considered:

\begin{itemize}
\item EROs: $i_{775}-K_{s}>2.5$ \citep{Roche03};
\item IEROs: $z_{850}-[3.6]$\,${\mu}m>3.25$ \citep{Yan04};
\item DRGs: $J-K_{s}>1.35$ \citep{Franx03}.
\end{itemize}

\begin{figure}
  \begin{center}
    \includegraphics[width=0.5\columnwidth]{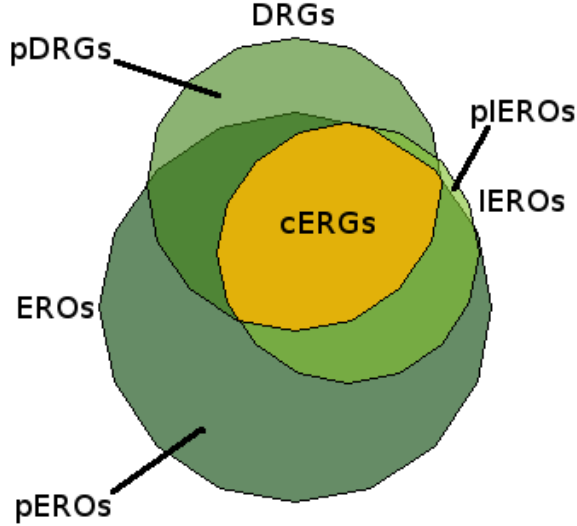}
  \end{center}
  \caption[The overlap between ERG populations]{The adopted ERG sub-sample nomenclature is ilustrated here through a Venn diagram. The overlap between the three ERG classes -- EROs, IEROs, and DRGs -- is significant (the common ERG population, labelled as cERGs). The outer non-overlapping regions represent the pure populations.}
  \label{c2fig:venncerg}
\end{figure}

Whenever a source is not detected in one of the bands ($i_{775}$, $z_{850}$, $J$ or 3.6\,$\mu$m), a limit to its magnitude is assumed (we adopt the $3\sigma$ flux level based on the local \textit{rms} provided in the catalogue). In the case of unreliable photometry (e.g., $K_s$w$\leq0.3$), the corresponding source is not considered further. Thus, only ERGs with robust photometry in both of the bands used for their identification are considered. The resulting ERG sample, with robust photometry, contains 628 objects: 553 EROs, 259 IEROs, and 289 DRGs, down to the adopted magnitude limit of $K_{s,TOT}=23.8$\,AB. These classifications are not exclusive, with individual objects potentially included in more than one classification, as illustrated in Figure \ref{c2fig:venncerg}.

It should be noted that FIREWORKS is a $K_{s}$-selected catalogue. As such, EROs and DRGs are selected according to the traditional definition, but IEROs selected from the FIREWORKS catalogue are in fact $K_{s}$-detected IEROs. This sample will only be representative of the true IERO population in the absence of a significant number of very red $K_s-[3.6]$ IEROs, which will be undetected in the $K_{s}$ image. One should also note that the $z_{850}$ detection limit (the current $K_s$ selected sample includes sources with up to $z_{850}\sim$27\,mag) imposes a [3.6]-band magnitude limit of $\sim$23.75\,mag for the IERO sample. 

\begin{figure}[t]
  \begin{center}
    \includegraphics[width=0.5\columnwidth]{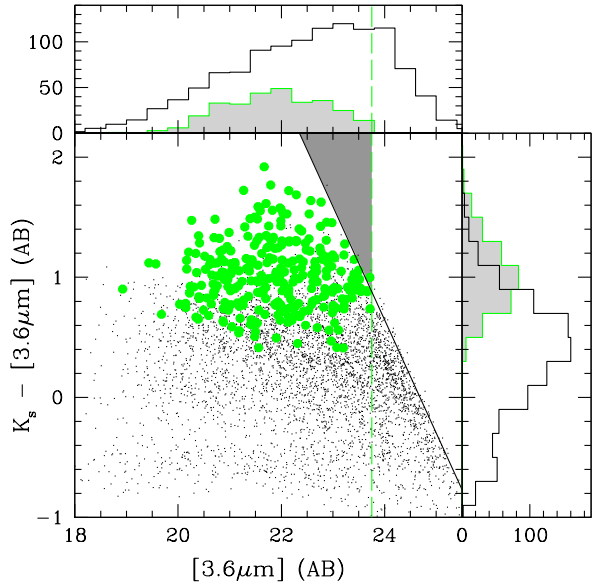}
  \end{center}
  \caption[$K_s$-selected IEROs]{$K_s-[3.6]$ colour-magnitude plot for sources in the FIREWORKS catalogue. Points and open histograms (scaled down by a factor of 4) represent the general $K_s$ population included in the catalogue, while filled circles and shaded histograms represent the selected IERO sample. The diagonal line corresponds to a $K_s$-band value of 23.8\,mag, the adopted magnitude limit of our sample. The dashed vertical line represents the practical [3.6]-band IERO magnitude limit (as imposed by the $z_{850}$ 3$\sigma$ magnitude limit and the IERO definition). The current $K_s$-detected IERO sample would differ significantly from the general IERO population if a large number of sources exists above the diagonal and to the left of the vertical line (shaded region). The $K_s-[3.6]$ colours present in the well sampled region of the diagram (below the diagonal line and to the left of the vertical one) argue against this scenario, implying that the current sample of IEROs is representative of the overall IERO population.}
  \label{c2fig:mciero}
\end{figure}

Figure~\ref{c2fig:mciero} shows the colour-magnitude distribution for sources in the FIREWORKS catalogue and for the $K_{s}$-detected IERO sample, displaying our adopted $K_{s}$-band magnitude limit (diagonal line) and the practical [3.6]-band magnitude limit (vertical line). The sampled region at $K_{s,TOT}-[3.6]<-[3.6]+23.8$ and $[3.6]<23.26$ (below the diagonal line and to the left of the vertical one) does not indicate a significant incompleteness for the critical region (above the diagonal line and to the left of the vertical one). For example, allowing all FIREWORKS sources to be considered (up to a $K_{s}$-band magnitude of 24.3\,mag), would only increase the IERO sample by 7\% (18 more objects). Consequently, we consider our sample of ($K_{s}$-detected) IEROs representative of the true IERO population and find it unnecessary to assemble a separate sample of IEROs from a 3.6\,$\mu$m selected catalogue, thus maintaining the photometric homogeneity within the ERG sample.

\subsection{Sub-classes of ERGs} \label{c2sec:comerg}

We refer to those sources that appear exclusively in one of these classes (ERO, IERO or DRG) as ``pure'' populations, while those that are simultaneously included in three ERG categories are referred to as the ``common'' population. In this work, the latter will be referred to as ``common'' ERGs, or cERGs. When addressing both the ``pure'' and ``common'' populations we will restrict ourselves to those sources which have sufficient information for a classification in each of the three red galaxy criteria (either good photometry or robust upper limits in \emph{all bands} used for classification). With such requirements, we find 607 ERGs: 541 EROs, 258 IEROs, and 280 DRGs. Practically all IEROs (249 out of 259\footnote{Number of IEROs with good photometry in the $i_{775}$, $z_{850}$, $K_s$ and 3.6$\mu$m bands.}) are also EROs and almost two-thirds (163 out of 258) are also classified as DRGs; there are 212 sources that comply with both the ERO and DRG criteria, and 156 ERGs that are simultaneously classified as ERO, IERO, and DRG (the cERGs). Identified as ``pure'' sources are 194 pure EROs (pEROs), 2 pure IEROs (pIEROs), and 48 pure DRGs (pDRGs).

Figure~\ref{c2fig:venncerg} shows the overlap between the different sub-populations. The initial columns of Table~\ref{c2tab:tabctp} summarise the numbers referred above.

\section{Multi-wavelength AGN identification and classification} \label{c2sec:agnid}

One of the major problems for the characterisation of ERGs, or for any distant galaxy population, is to identify the existence of AGN activity. The many techniques that exist target different AGN types and redshift ranges, and no single technique can guarantee a highly discriminatory success rate. X-rays,  optical, MIR or radio, originating from different regions in the vicinity of the AGN, and differently affected by dust obscuration, provide independent ways to reveal such activity. Extreme examples are indeed found, as for instance, the powerful AGN-like radio sources hosted by normal star-forming galaxies (as seen from the optical and IR data) found in the ATLAS survey \citep{Norris07,Mao08}. As a result, many AGN-selection techniques at different wavelengths have been extensively compared in the literature \citep[e.g.,][]{Donley08,Eckart10,Griffith10}. In this work, we use the multi-wavelength data available in this field to carry out a thorough identification and classification of AGN activity in the ERG population.

\subsection{Optical Spectroscopy}

Since ERGs are intrinsically UV/optically faint, spectroscopy will be of limited use to reveal their nature. While a pure passive evolve system will present a spectrum without spectral lines and dominated by a stellar continuum, in the case of obscured SF and AGN activity, dust may hide any sign of line emission. Overall, there are only 8 spectroscopic AGN identifications (Narrow Line AGN or QSO type-2 classifications), galaxies which are also identified as AGN by the criteria described in the following sections.

Spectroscopy allows, nonetheless, for the rejection of galactic stars selected as EROs. In this sample, two were found and discarded from further study. This indicates that contamination of the ERG population by unidentified galactic stars is likely small in this study.

\subsection{X-Rays} \label{c2sec:xr}

X-ray emission is arguably the most effective discriminator of AGN activity in a galaxy, as such high energies are unachieved by normal stellar populations. Due to the sensitivity levels currently reached with the deepest observations \citep[the 2\,Ms CDF fields:][and recently increased to 4\,Ms, \citealt{Xue11}]{Alexander03,Luo08}, the most powerful AGN ($L_{0.5-10\,keV}\,>\,10^{44}$\,erg\,s$^{-1}$) can be detected beyond the highest redshift currently observed, $z>7$. On the other hand, both low luminosity AGN and vigourous star-forming galaxies ($L_{0.5-10\,keV}\,\sim\,10^{41-42}$\,erg\,s$^{-1}$) can only be detected out to $z\sim1-2$. If enough signal is detected, detailed spectral analysis can be used to distinguish between AGN and SF activity as the origin of the X-ray emission.

In this work, the ERG sample was cross-matched with the catalogues from the 2\,Ms $Chandra$ observations \citep{Luo08}. For the region considered here -- GOODSs ISAAC -- the X-ray observations reach aim-point sensitivity limits of $\approx1.9\times 10^{-17}$ and $\approx1.3\times 10^{-16}$\,erg\,cm$^{-2}$\,s$^{-1}$ for the soft (0.5--2.0\,keV) and hard (2--8\,keV) bands, respectively. 

X-ray detections were searched for within 1.5'' of each ERG position. Counterparts were found for 67 of the 553 EROs (12\%), 39 of the 259 IEROs (15\%), and 40 of the 289 DRGs (14\%) (see Table~\ref{c2tab:tabctp}). These detection fractions are consistent with those found by \citet{Alexander02}, for EROs, and \citet{Papovich06}, for DRGs.

\begin{figure}[t]
  \begin{center}
    \includegraphics[width=0.6\columnwidth]{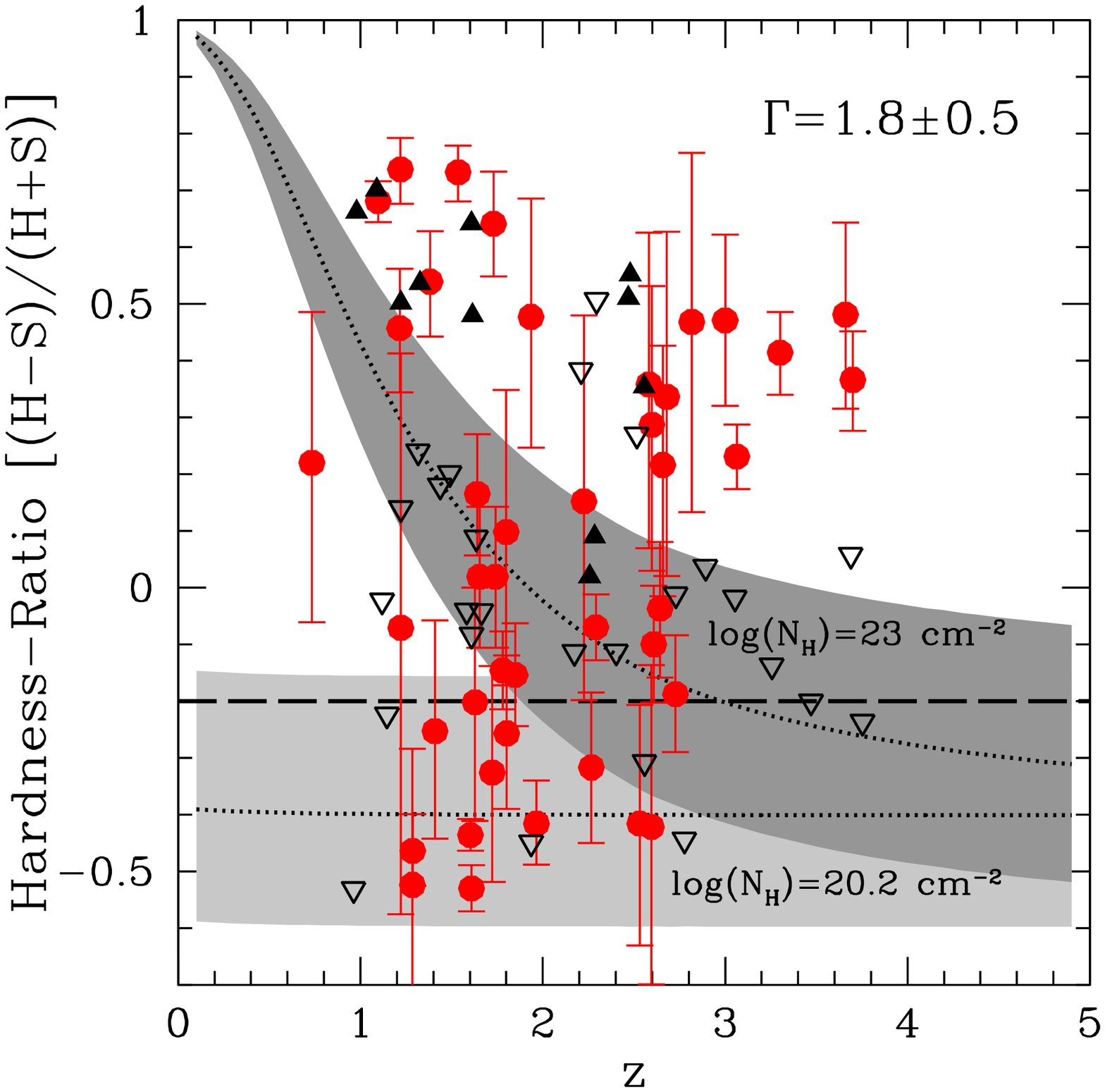}
  \end{center}
  \caption[HR degeneracy at high-$z$]{Figure 3 from \citet{Messias10} showing X-ray HR evolution with redshift for obscured ($\rm{N_H}=10^{23}$cm$^{-2}$, grey shaded region) and unobscured ($\rm{N_H}=10^{20.2}$cm$^{-2}$, light grey shaded region) X-ray power-law emission models ($\Gamma=1.8\pm0.5$), calculated using PIMMS (ver. 3.9k). Filled circles show the distribution of the X-ray detected AGN ERGs with a robust HR estimate. Upper limits (no hard-band detection) appear as empty triangles while filled triangles denote lower limits (no soft-band detection). The dashed horizontal line highlights the HR constraint (HR$=-0.2$) for type discrimination used by \citet{Szokoly04}. It is clear that for high-redshift sources ($z\gtrsim2$) the simple HR criterion becomes degenerate as an obscuration measure.}
  \label{c2fig:hrevol}
\end{figure}

We adopt a similar X-ray classification criteria as \citet{Szokoly04}, who consider both the X-ray Luminosity ($L_X$), estimated from the 0.5--10\,keV flux, and the hardness ratio (HR). The HR is used as an indicator for obscuration and is calculated using the count rates in the hard band (HB, 2--8\,keV) and in the soft band (SB, 0.5--2\,keV): HR = (HB-SB)/(HB+SB). Any source displaying an HR greater than -0.2 (equivalent to column densities of $\log(\rm{N_H[cm^{-2}]})>20$ at $z\sim0$) is considered to be an obscured system. However, the HR is quite degenerate at high redshifts \citep[Figure~\ref{c2fig:hrevol},][but also \citealt{Alexander05} and \citealt{Luo10}]{Eckart06,Messias10}. In this work, a slightly different procedure is used, estimating directly $\rm{N_H}$ from the more robustly determined soft-band (or hard-band) to full-band ratio, as explained below.

In order to estimate $\rm{N_H}$, we have used the Portable, Interactive Multi-Mission Simulator\footnote{http://heasarc.nasa.gov/docs/software/tools/pimms.html} (PIMMS, version 3.9k). The soft-band/full-band (SB/FB) and hard-band/full-band (HB/FB) flux ratios\footnote{As opposed to the commonly used SB/HB flux ratios, the use of ratios based on FB flux allows for an estimate of $\rm{N_H}$ when the source is detected in the FB but no detection is achieved in the SB nor in the HB.} were estimated for a range of column densities ($20<\log(\rm{N_H[cm^{-2}]})<25$, with steps of $\log(\rm{N_H[cm^{-2}]})=0.01$), and redshifts ($0<z<7$, with steps of $z=0.01$), considering a fixed photon index, $\Gamma=1.8$ \citep[$\rm{f(E)\propto{E}^\Gamma}$,][]{Tozzi06}. The comparison with the observed values results in the estimate of $\rm{N_H}$, which can then be used to derive an intrinsic X-ray luminosity referred in the criteria below (for simplicity throughout the text, $\rm{L_{X}}$ refers to an intrinsic luminosity).

The criteria used as equivalent to \citet{Szokoly04} are listed as follows:
\begin{eqnarray}
\begin{array}{ll}
{\rm Galaxy:}~ L_{X}<10^{42}\,{\rm erg\,s}^{-1}~\&~{\rm N_H}\leq10^{22}~\rm{cm}^{-2}\nonumber \\
{\rm AGN-2:}~10^{41} \leq L_{X}<10^{44}\,{\rm erg\,s}^{-1}~\&~{\rm N_H}>10^{22}~\rm{cm}^{-2} \nonumber \\
{\rm AGN-1:}~10^{42} \leq L_{X}<10^{44}\,{\rm erg\,s}^{-1}~\&~{\rm N_H}\leq10^{22}~\rm{cm}^{-2} \nonumber \\
{\rm QSO-2:}~L_{X}\geq10^{44}\,{\rm erg\,s}^{-1}~\&~{\rm N_H}>10^{22}~\rm{cm}^{-2} \nonumber \\
{\rm QSO-1:}~L_{X}\geq10^{44}\,{\rm erg\,s}^{-1}~\&~{\rm N_H}\leq10^{22}~\rm{cm}^{-2} \nonumber 
\end{array}
\end{eqnarray}

The rest-frame X-ray luminosity is calculated as: \[ L_{X} = 4 \pi\,d^{2}_{L}\,f_{X}\,(1+z)^{\Gamma-2}\,\rm{erg}\,\rm{s}^{-1} \] where \emph{f}$_{X}$ is the X-ray flux in the 0.5-10 band and the photon index is the observed $\Gamma$ (when $\log(\rm{N_H[cm^{-2}]})\leq20$) or $\Gamma=1.8$ (when $\log(\rm{N_H[cm^{-2}]})>20$). The luminosity distance, d$_{\rm L}$ is calculated using either the spectroscopic redshift or, if not available, the photometric redshift. The 0.5--8~keV luminosities, derived using \citet{Luo08} catalogued 0.5--8~keV fluxes, were converted to 0.5--10~keV considering the adopted $\Gamma$.

In total, these criteria enable the identification of 72 sources hosting an AGN with 20 X-ray sources powerful enough to be classified as QSOs. The majority of the AGN are classified as type-2 sources: 43 X-ray detections have $\log\rm{(N_H[cm^{-2}])}>22$ (i.e., indicative of large obscuration) while only 17 show lower values (with the remaining 7 having uncertain $\rm{N_H}$ determinations, with no discrimination possible), indicating a possible 2--3:1 obscured to unobscured ratio. However, although in agreement with what is referred in the literature \citep[see the discussion in][]{Donley08}, this value should be taken with care. Although we correct for the redshift effect by considering the $\rm{N_H}$ value instead of HR, the former may still be slightly affected at high redshifts as its calculation relies on flux ratios. Ideally, such high-redshift populations would require observations extending to softer X-ray energies ($<$0.5 keV) below those reliably achieved by $Chandra$.

As a final remark, the reader should note that at $\log(\rm{L_X[erg\,s^{-1}]})>44$ the ratio is even higher, 6:1, close to that found for sub-millimetre galaxies \citep{Alexander05}. This result is relevant for the discussion at the end of Section~\ref{c2sec:morphsec}.

\subsection{Mid-Infrared} \label{c2sec:mir}

Over the last few years with the sensitivity of IRAC and MIPS on board $Spitzer$, several MIR criteria have been developed for the identification of AGN at the centre of galaxies. A power-law MIR spectral energy distribution, for example, is characteristic of AGN emission \citep[e.g.,][]{Donley07}. Somewhat more generic colour-colour diagrams have also been investigated, and AGN loci in such plots defined \citep[e.g.][]{Ivison04,Lacy04,Stern05,Hatziminaoglou05}. This wavelength range is of particular interest for the ERG population, given their red SEDs. Here, we have applied MIR diagnostics to our ERG sample, as described below.

Observational data at X-ray and IR wavelengths provide complementary views of AGN activity. The most obscured AGN may be missed by even the deepest X-ray surveys but can still be identified by their hot-dust emission at IR wavelengths. On the other hand, depending on the amount of dust and its distribution, and on the AGN strength, the MIR emission from X-ray classified AGN may not be dominated by the hot dust in the vicinity of the AGN itself. A detailed comparison of the relative merits of AGN selection by the X-rays and the MIR was performed by \citet{Eckart10}, showing that only a multi-wavelength combination of AGN criteria can help to overcome biases present in single-band selection. However, even the combination of MIR and X-rays will not result in complete AGN samples, as the identification of low power AGN will ultimately depend on the depth of the surveys. By performing this study in GOODSs, with some of the deepest data both at X-ray and MIR wavelengths, we maximise the identification rate of AGN. For a more in depth discussion on this subject, please consult Chapter~\ref{ch:agn}.

IRAC counterparts were found for practically all (98\%) ERG sources. The vast majority (89\%) are detected simultaneously in all IRAC bands: 526 of the 553 EROs, 257 of the 259 IEROs, and 258 of the 289 DRGs. The MIPS 24$\mu$m detection rate is understandably lower 337(55\%), given the lower relative sensitivity: 306/167/171 of the 553/259/289 EROs/IEROs/DRGs are detected (Table~\ref{c2tab:tabctp}).

\subsubsection {Classification: MIR colours} \label{c2sec:kisec}

In recent years, several AGN colour-selection criteria have been developed employing MIR IRAC observations \citep{Ivison04,Lacy04,Stern05,Hatziminaoglou05}. Here we follow the KI criterion proposed in Chapter~\ref{ch:agn} of this thesis. An AGN is considered to present the following colours:
\begin{eqnarray}
\begin{array}{c}
K_s-\left[4.5\right]>0 \nonumber \\
\wedge \nonumber \\
\left[4.5\right]-\left[8.0\right]>0 \nonumber \\
\end{array}
\end{eqnarray}

\begin{figure}
  \begin{center}
    \includegraphics[width=0.9\columnwidth]{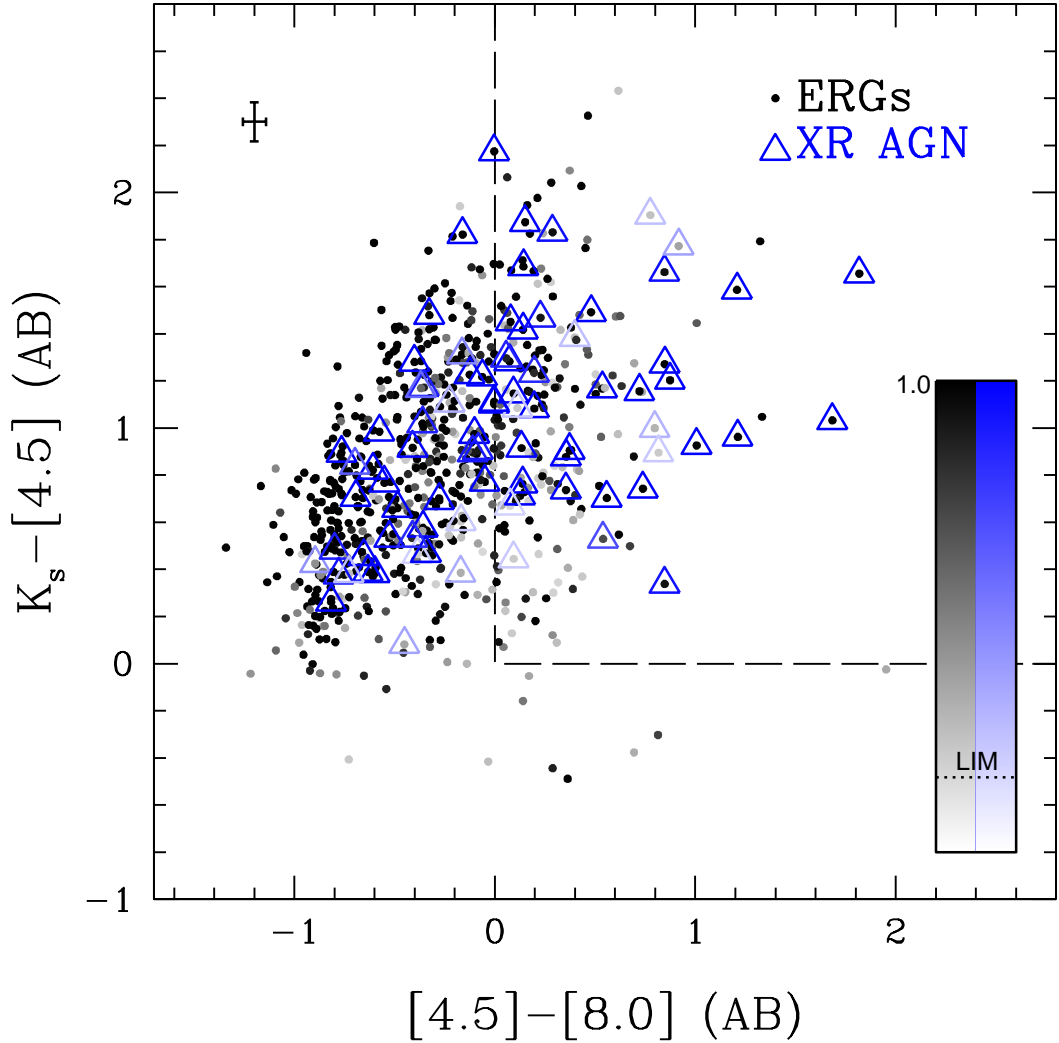}
  \end{center}
  \caption[ERGs on the KI colour-colour space]{The distribution of ERGs in the KI colour-colour diagnostic plot proposed in Chapter~\ref{ch:agn}. The AGN region is delimited by the dashed line. The ERGs (as dots) classified as AGN by the X-rays criterion are highlighted as triangles. The darker the circle the higher is its source probability, while the bluer the triangle, the higher is the probability for an X-ray AGN classification (see the scale on the right-hand side, the dotted line marks the limit probability for a source to be considered in the study, and, hence, in this plot). The error bars on the top left show the average photometric errors.}
  \label{c2fig:ki}
\end{figure}

Figure~\ref{c2fig:ki} shows the distribution of ERGs on the KI colour-colour space. It identifies as AGN 97 (18\%) EROs, 24 of which are also classified as AGN from the X-rays; 78 (30\%) IEROs, 17 of which also have an X-ray AGN classification; and 100 (35\%) DRGs, 19 of which also appear as X-ray AGN. The relatively high number of potential AGN identified, over that revealed by the X-rays, is known and expected \citep[][and references therein]{Donley07,Donley08}. It is worth noting that these results are likely more reliable (i.e., less contaminated by non-AGN sources) than those obtained by traditional MIR colour criteria \citep[e.g.,][]{Lacy04,Stern05}, as shown in Chapter~\ref{ch:agn}.

\vspace{0cm}

\subsubsection{MIR degeneracy at $z>2.5$} \label{c2sec:mir25}

One problem in using MIR photometry to identify AGN (with both power-law and colour-colour criteria) arises at $z \gtrsim 2.5$, as both star-forming galaxies and AGN start to merge into the same MIR colour-colour space (see Chapter~\ref{ch:agn}). The main reason for this is the increasing relative strength of stellar emission in the MIR, as compared to that of an AGN, as redshift increases. At higher redshifts, a prominent 1.6\,$\mu$m stellar bump passes through the IRAC bands, allowing for the detection of a steep spectral index not from AGN emission, but from the stellar emission alone. At z$>$2.5, the KI criterion classifies as AGN 57 EROs, 48 IEROs, and 71 DRGs, with 19, 16, and 21 (respectively) X-ray confirmed at these redshifts.

In the present work, this is not a serious problem, as most of the ERG sample (79\%) lies at $z\leq2.5$ (see Section~\ref{c2sec:reddis}). Nevertheless, it should be noted that at higher redshifts, this could result in a likely overestimate of the presence of AGN. One can attempt to correct for this effect, by using the MIPS 24$\mu$m observations: at $z\sim2.5-5$, the 1.6\,$\mu$m bump will be shifted to the 6--10\,$\mu$m range. Therefore, in the absence of significant AGN emission, one expects a blue [8.0]-[24] colour. In Chapter~\ref{ch:agn} of this thesis, a proper study of this colour is pursued, indicating that sources showing [8.0]-[24]$<$1 at these redshifts are dominated by non-AGN activity.

\begin{figure}
  \begin{center}
    \includegraphics[width=0.85\columnwidth]{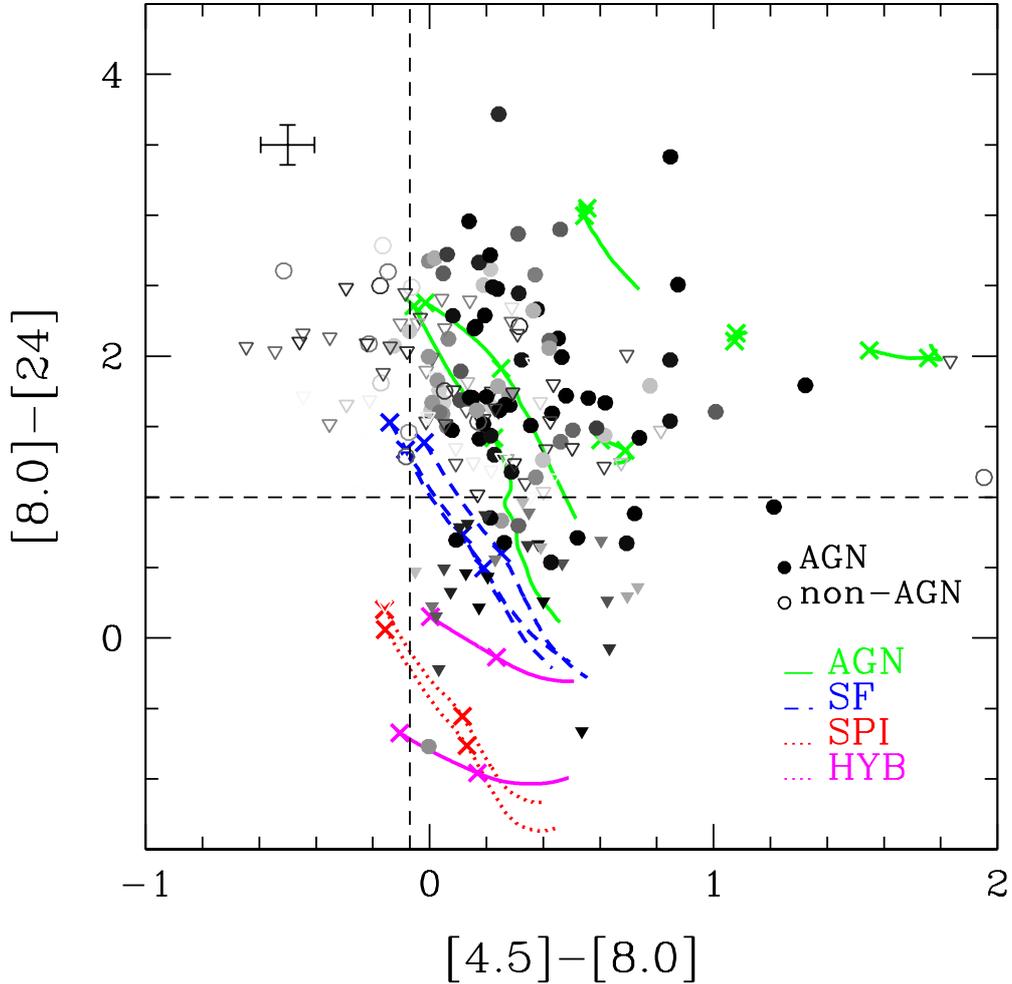}
  \end{center}
  \caption[Correcting KI at $z>2.5$]{Mid-infrared [4.5]-[8.0] vs [8.0]-[24] colour-colour plot for $z>2.5$ ERGs in the current sample. Filled symbols represent the AGN classification from the KI criterion (darker symbols mean higher AGN probability given by KI before correction) and open symbols otherwise (darker symbols mean higher probability not to be KI selected, similar Figure~\ref{c2fig:ki}). Downward pointing triangles indicate $[8.0]-[24]$ upper limits. The error bars on the top left show the average photometric errors. The tracks represent the expected colours of template SEDs where the IR is dominated by star-formation (dotted and dashed tracks, where the latter represent more intense SF activity) or AGN activity (solid tracks), redshifted between $z=2.5$ and $z=4$, with crosses at $z=2.5$ and $z=3$. The templates displayed are (from bottom to top): two Spirals (Sc and Sd, red) and two hybrids (S0+QSO2, magenta), three starbursts (M82, NGC 6240, and Arp220, blue), and six Hybrids and AGN (IRAS 22491-1808, IRAS 20551-4250, QSO-2, \citet{Afonso01} ERO, Mrk 231, and IRAS 19254-7245 South, green).}
  \label{c2fig:z3agn}
\end{figure}

In Figure~\ref{c2fig:z3agn} we present the MIR [4.5]-[8.0] vs [8.0]-[24] colour-colour plot for $z>2.5$ ERGs in the current sample (highlighting those classified as AGN by the KI criterion). The tracks represent the expected colours of template SEDs when redshifted between $z=2.5$ and $z=4$. Templates come from the SWIRE Template Library \citep{Polletta07}, two hybrids\footnote{Hybrids are sources presenting a combination of non-AGN and AGN IR emission. See Chapter~\ref{ch:agn}.} from \citep{Salvato09}, and the extreme ERO of \citet{Afonso01}, which is dominated by an obscured AGN in the MIR. The vertical line indicates the [4.5]-[8.0] colour constraint of the KI criterion, while the horizontal line shows our adopted colour cut separating AGN and star-forming processes at these redshifts. The AGN template that crosses over this [8.0]-[24] threshold at the highest redshifts is IRAS 22491-1808, a possible mixture of AGN and stellar\footnote{By \textit{stellar emission}, we mean in this sentence the emission by pure stellar continuum, dust heated by stellar emission, and polycyclic aromatic hydrocarbons.} MIR emission \citep{Berta05,Polletta07}, where the AGN component is progressively less sampled by the MIR bands as redshift increases. Concerning the current sample, there are 12/10/16 EROs/IEROs/DRGs at $z>2.5$ classified as AGN by the KI criterion which {\it do not require} an AGN SED to explain their MIR emission and, consequently, their KI probability to be AGN is corrected.

We note the presence of two interesting sources in Figure~\ref{c2fig:z3agn}. The one isolated in the upper right, is one of the seven optically unidentified radio sources found in \citet[][their source \#42]{Afonso06}. Inspection of the $K_s$ and 24${\mu}m$ images reveals no signs of blending, strengthening the accuracy of the 24${\mu}m$ flux. This source also has X-ray emission characteristic of a type-2 AGN ($L_{X}=10^{43.3} {\rm erg\,s}^{-1}$ and $\log\rm{(N_H[cm^{-2}])}=23$). The colour-track closest to this source in Figure~\ref{c2fig:z3agn} is that of the highly obscured AGN ERO found by \citet{Afonso01}. The assigned photometric redshift is $z=3.1$ \citep{Luo10}. The spectral index ($\mathcal{S}_\nu=\nu^{-\alpha}$) obtained from 1.4\,GHz and 5\,GHz observations is $\alpha=1.3\pm0.3$ \citep{Kellermann08}, implying an ultra steep spectrum source \citep[e.g.][]{Tielens79,Chambers96}. The high-$z$ obscured AGN scenario postulated in \citet{Afonso06} for this source is thus strengthened.

\begin{figure}
  \begin{center}
    \includegraphics[width=0.5\columnwidth]{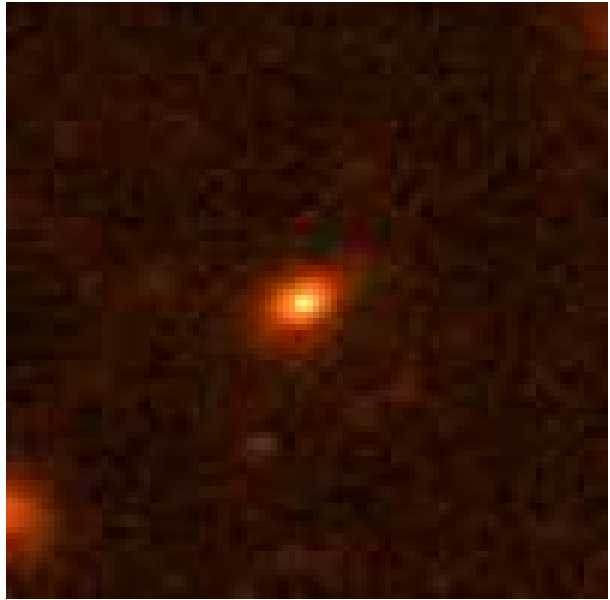}
  \end{center}
  \caption[An evolved disc at $z_{\rm phot}=2.5$]{The GNS-$H160$ image cut-out of the candidate for passively evolved system at high redshift confirms the disc profile expected from the template tracks in Figure~\ref{c2fig:z3agn}.}
  \label{c2fig:discgns}
\end{figure}

The other interesting source is the bluest [8.0]-[24] 24${\mu}m$ detection, with MIR colours characteristic of spiral galaxies or an earlier type with a small AGN contribution. It is also X-ray detected, but has no radio emission. This is a candidate for a high-$z$ evolved system, based on its optical non-detection and extremely blue [8.0]-[24] colours typical of late-type galaxies\footnote{Although these colours may also be seen in low-metalicity sources, the sources in this study are ERGs and, hence, are not low-metalicity systems.}. The redshift assigned to this source, $z_{\rm phot}=2.54$ \citep{Luo10}, is at the highest redshifts in which similar sources have ever been found \citep{Stockton08,vanderWel11}. A NICMOS image taken from the GOODS NICMOS Survey\footnote{http://www.nottingham.ac.uk/astronomy/gns} \citep{Conselice11}, confirms the disc-like nature of this evolved source (Figure~\ref{c2fig:discgns}). A S\'ersic index of $n=1.2$ \citep[][and private communication]{Buitrago08} strengthens the visual disc classification. Its effective radius is equivalent to $r_e=2\,\rm{kpc}$, implying a compact disc. For comparison, considering the scale length of the Milky Way to be $h\sim3\,\rm{kpc}$ \citep{Kent91,Freudenreich98}, its effective radius is $r_e=h\times1.678\sim5\,\rm{kpc}$, when considering the disc profile to be exponential \citep[see, for instance,][]{Graham05}. A few more galaxies fall close to the late-type galaxy colour-colour tracks (Figure~\ref{c2fig:z3agn}), and are also interesting. A discussion on the implications for the existence of passive evolved discs at such high redshifts is presented in Section~\ref{c5sec:pdgs}.

\subsection{Radio}

Radio emission is essentially unaffected by dust obscuration, thus being extremely useful for the estimate of SF activity in ERGs. However, since both star-formation and AGN activity can produce radio emission, it is often difficult or impossible to rely on radio properties alone to reveal the power source in a galaxy. Indications from radio spectral indices are of limited use, as both star-formation and AGN emission usually result from synchrotron radiation with $S_\nu \propto \nu^{-0.8}$, and only some AGN show signs of flat or even inverted radio spectra. Very high resolution \textit{VLBI} radio imaging has also been used with limited success to impose limits on the size of the radio emitting region, identifying star-forming galaxies where the radio emission is resolved, and a possible AGN where not \citep{Muxlow05,Middelberg08,Seymour08}. The only straightforward radio AGN criterion is the radio luminosity itself, as the highest luminosities can only be produced by the most powerful AGN.

\citet{Afonso05} performed a detailed study of the sub-mJy radio population, and found star forming galaxies with radio luminosities up to $L_{\rm 1.4\,GHz} \sim 10^{24.5}\,$W\,Hz$^{-1}$. We thus take this value as the upper limit for SF activity. We note that this value corresponds to a SFR of almost 2000\,M$_{\odot}$\,yr$^{-1}$ \citep[][see Section~\ref{c2sec:sfrfin}]{Bell03}. The existence of galaxies with higher rates of SF activity is unlikely. 

For the current work we have used the 1.4\,GHz Australia Telescope Compact Array observations of this field, which reach a uniform 14--17\,$\mu$Jy rms throughout the GOODSs field \citep[see][for more details]{Afonso06,Norris06}, and the Very Large Array data also in GOODSs \citep{Kellermann08,Miller08}, reaching deep rms levels (typically 8\,$\mu$Jy).

First, the two radio catalogues were matched. Then, with a search radius of 1.5'' and considering the VLA sources coordinates, the radio catalogue was cross-matched with FIREWORKS catalogue, implying 73 (nominal value) sources with a radio counterpart. For ATCA-only radio sources, a larger matching radius of 3'' was considered revealing five (nominal value) more sources with a radio counterpart. Overall, there are 24 (4\%) ERGs detected at radio frequencies: 23 (4\%) EROs, 16 (6\%) IEROs, 14 (5\%) DRGs. Six sources have radio luminosities in excess of $10^{24.5}\,$W\,Hz$^{-1}$. They are also classified as AGN by both the X-ray and KI criteria. On the other hand, nine radio-detected ERGs are not classified as AGN by the X-ray and KI criteria. Only one nominal source with a non-negligible probability to be radio AGN (a 40\% probability to have $\rm{L_{1.4GHz}}>10^{24.5}\,$W\,Hz$^{-1}$) remains unclassified as AGN by the other two AGN criteria.

The sensitivity available even in the current deepest radio surveys limits our radio detected sample to sources with L$_{1.4\rm{GHz}}>10^{23}$\,W\,Hz$^{-1}$, hence, the small detection rate indicates that powerful AGN and the most intense starbursts\footnote{Figure~7 in \citet{Afonso05}, clearly shows that the distribution of star-forming galaxies decreases drastically at L$_{1.4\rm{GHz}}>10^{23}$\,W\,Hz$^{-1}$.} are not common in the ERG population.

\section{Properties of ERGs} \label{c2sec:ergprop}

\subsection{Redshift Distributions} \label{c2sec:reddis}

As noted above, robust spectroscopic redshifts are available for around 22\% of the ERG sample. Photometric redshift estimates are also available from the FIREWORKS and \citet{Luo10} catalogues, covering almost the complete ERG sample. In case only a photometric redshift is available, the redshift probability distribution is taken into account. When separating the sample into redshift bins, only sources with a probability $\mathcal{P}>0.317/2$ (Section~\ref{c2sec:method}) to fall inside a given bin are considered. These sources are weighted by their own probability.

\begin{figure}
  \begin{center}
    \includegraphics[width=0.6\columnwidth]{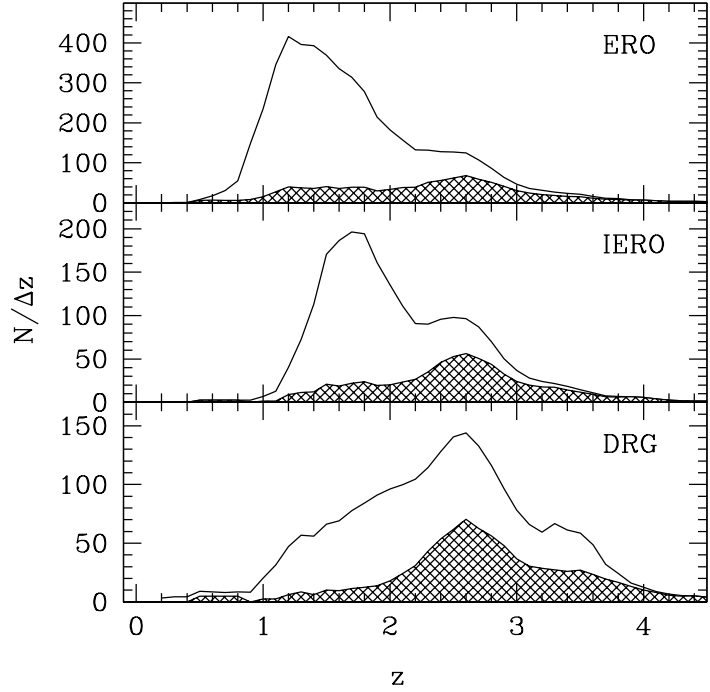}
  \end{center}
  \caption[Redshift distributions of ERGs]{Redshift distributions of different ERG sub-populations: EROs, IEROs, and DRGs. The hatched histograms correspond to ERGs identified as AGN. Note the y-axis units are N/$\Delta{z}$ and different scales are adopted for the individual panels. The distributions were obtained with a moving bin of width $\Delta{z}=0.4$ and adopting steps of $\Delta{z}=0.1$.}
  \label{c2fig:geral}
\end{figure}

The redshift distributions for the ERO, IERO, and DRGs are shown in Figure~\ref{c2fig:geral}. Although the range of redshifts sampled in all ERG classes is similar ($1<z<3$), the average value increases from $z=1.80$ for EROs, to $z=2.11$ for IEROs, and to $z=2.47$ for DRGs populations \citep[the slightly higher values relative to previous works is likely due to the fainter flux cut adopted in this work, e.g.][]{Conselice08,Papovich06}. This is as expected given the source selection, designed to identify objects at such redshifts. 

The AGN in the ERG population follow a similar redshift distribution but the AGN fraction increases rapidly at higher redshifts. This will be addressed in the next section.

\begin{figure}
  \begin{center}
    \includegraphics[width=0.6\columnwidth]{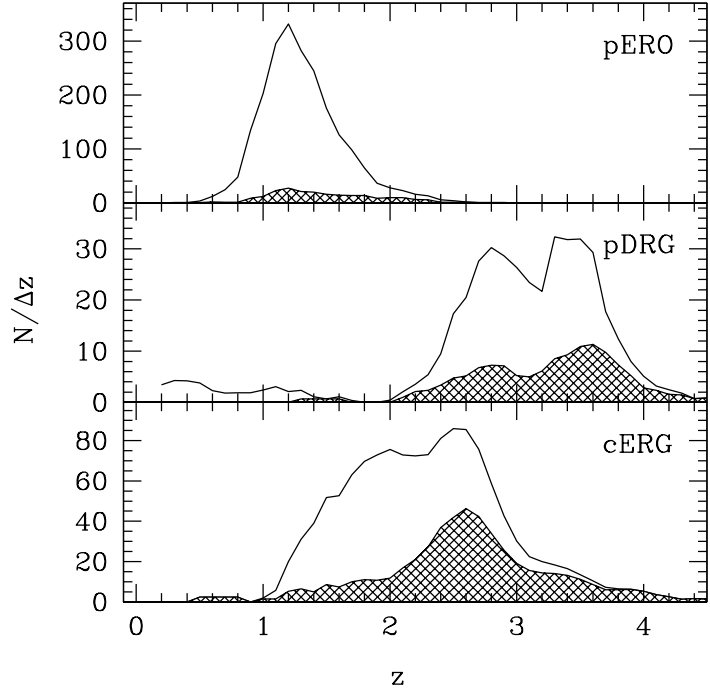}
  \end{center}
  \caption[Redshift distributions of pure and common ERGs]{Redshift distributions for ``pure'' and ``common'' ERG sub-populations: pEROs, pDRGs and cERGs. The hatched histograms correspond to ERGs identified as AGN. Note the y-axis units are N/$\Delta{z}$ and different scales are adopted for the individual panels. The distributions were obtained with a moving bin of width $\Delta{z}=0.4$ and adopting steps of $\Delta{z}=0.1$.}
  \label{c2fig:purecom}
\end{figure}

Figure~\ref{c2fig:purecom} displays the redshift distributions for pEROs, pDRGs, and cERGs. The redshift distribution of pEROs is quite narrow, selecting sources essentially at $z=1-2$ (peaking at $z\sim 1.3$), while the pDRG population is notably less numerous, and at higher redshifts ($z$=2--4). The ``pure'' criteria thus appear to be good and easy techniques to select high-z sources in narrow distinct redshift bins. Sources classified as cERGs, appearing as red in all three ERG selection criteria, cover a broad redshift range, from $z=1$ to $z=4$. There are practically no cERGs at $z<1$ in this particular sample due to the IERO criterion.

\subsection{AGN content of ERGs} \label{c2sec:agncont}

As described in the previous section, several multi-wavelength indicators were used to identify AGN in the ERG population. The indicators have different sensitivities to AGN characteristics, such as distance, dust obscuration, or AGN strength. Their combination will, thus, allow for a more complete census of AGN content in these sources.

We do not find a numerous population of very powerful AGN among the ERGs, as given by the X-rays (only 20 ERGs with $\rm{L_{X}}\geq10^{44}$\,erg s$^{-1}$) and radio (only six ERGs with L$_{\rm 1.4\,GHz}\geq 10^{24.5}\,$W\,Hz$^{-1}$) luminosities. These represent, respectively, only $3.2\pm0.7\,\%$\footnote{This and the following percentage values are approximated to the decimal part and the presented errors are poissonian.} and $1.0\pm0.4\,\%$ of the ERG sample. Such ratio is close to that observed in the complete $K$-selected FIREWORKS sample, where 31 ($0.7\pm0.1\,\%$) QSOs and 9 ($0.2\pm0.1\,\%$) radio-powerful sources are found. However, it is worth noting that a high fraction of these powerful sources are classified as ERG, around 65\% in both samples (20 out of 31 QSOs, and 6 out of 9 radio powerful sources). This apparent contradiction is probably related to the short duty-cycle expected for such kind of sources \citep[e.g.,][]{Hopkins06}, where an AGN will not pass much time as a radio-loud source nor as an X-ray QSO, but will always present an ERG colour before and after the strong on-set of AGN activity.

Overall, we select 154 (25\%) AGN-dominated systems in the ERG sample (23\% for EROs, 33\% for IEROs, and 39\% for DRGs). This fraction increases from low to high redshift, from 10\% at $1\leq{z}<2$ to 45\% at $2\leq{z}\leq3$. Among the X-ray identified AGN, 40\% are also classified as such by the KI criterion. Conversely, 25\% of the KI identified AGN are X-ray detected.

The high AGN fraction and its increase with redshift, might lead one to think that the KI criterion is overestimating the number of AGN at high redshifts, even though a tentative correction was applied (see Section~\ref{c2sec:mir25}). We have investigated the AGN fraction evolution from $1\leq{z}<2$ to $2\leq{z}\leq3$ based, independently, on the X-ray and KI indicator. In both wavebands, the AGN fraction increases significantly from low to high redshifts, rising from 8 to 17\% when the X-ray is considered and from 3 to 37\% when the MIR is considered. Although it seems possible that star-forming galaxies may still be affecting the KI criterion at high redshift (see Section~\ref{c2sec:mir25}), it is shown in Chapter~\ref{ch:agn} of this thesis that KI is still very reliable up to the highest redshifts. This may also partly be an effect of Malmquist bias, with lower luminosity systems, more likely to be dominated by star formation, being progressively lost at higher redshift. In any case, this increase is consistent with the known history of AGN activity in the Universe \citep[e.g.,][]{Osmer04,Richards06}. Section~\ref{c2sec:masssec} also helps understanding this rise in AGN host fraction.

\begin{figure}
  \begin{center}
    \includegraphics[width=0.6\columnwidth]{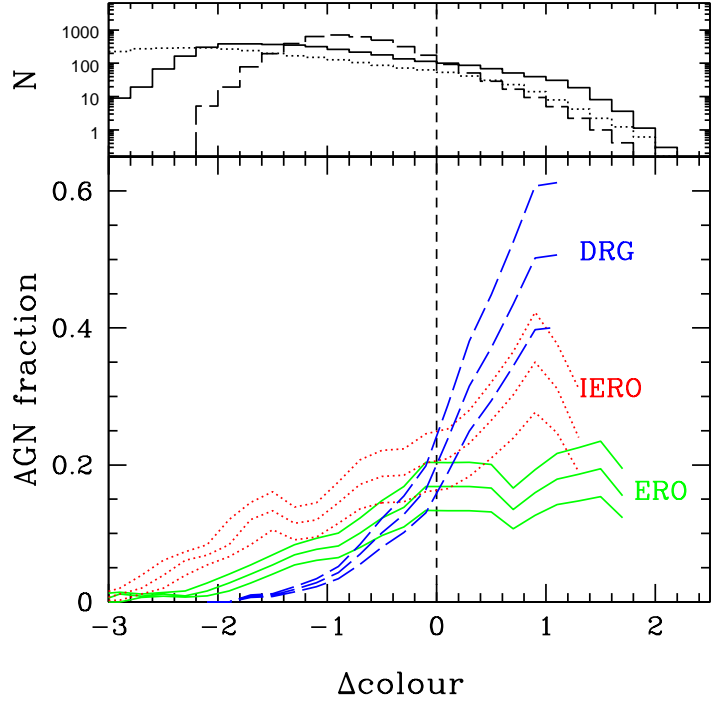}
  \end{center}
  \caption[AGN fraction with colour]{AGN fraction as a function of colour for EROs (green solid line), IEROs (red dotted line), and DRGs (blue dashed line). The trends are computed with a moving bin of 0.4\,mag and steps of 0.2\,mag.  The upper and lower limits consider poissonean errors and the bin width interval. The x-axis represents the difference in magnitude to the colour threshold adequate for each population: $i_{775}-K_s=2.5$ for EROs, $z_{850}-[3.6]=3.25$ for IEROs and $J-K_s=1.35$ for DRGs. The upper panel shows the $K_s$-selected population distribution in the considered colours (the same colours and line patterns are adopted).}
  \label{c2fig:coragn}
\end{figure}

\begin{figure}
  \begin{center}
    \includegraphics[width=0.9\columnwidth]{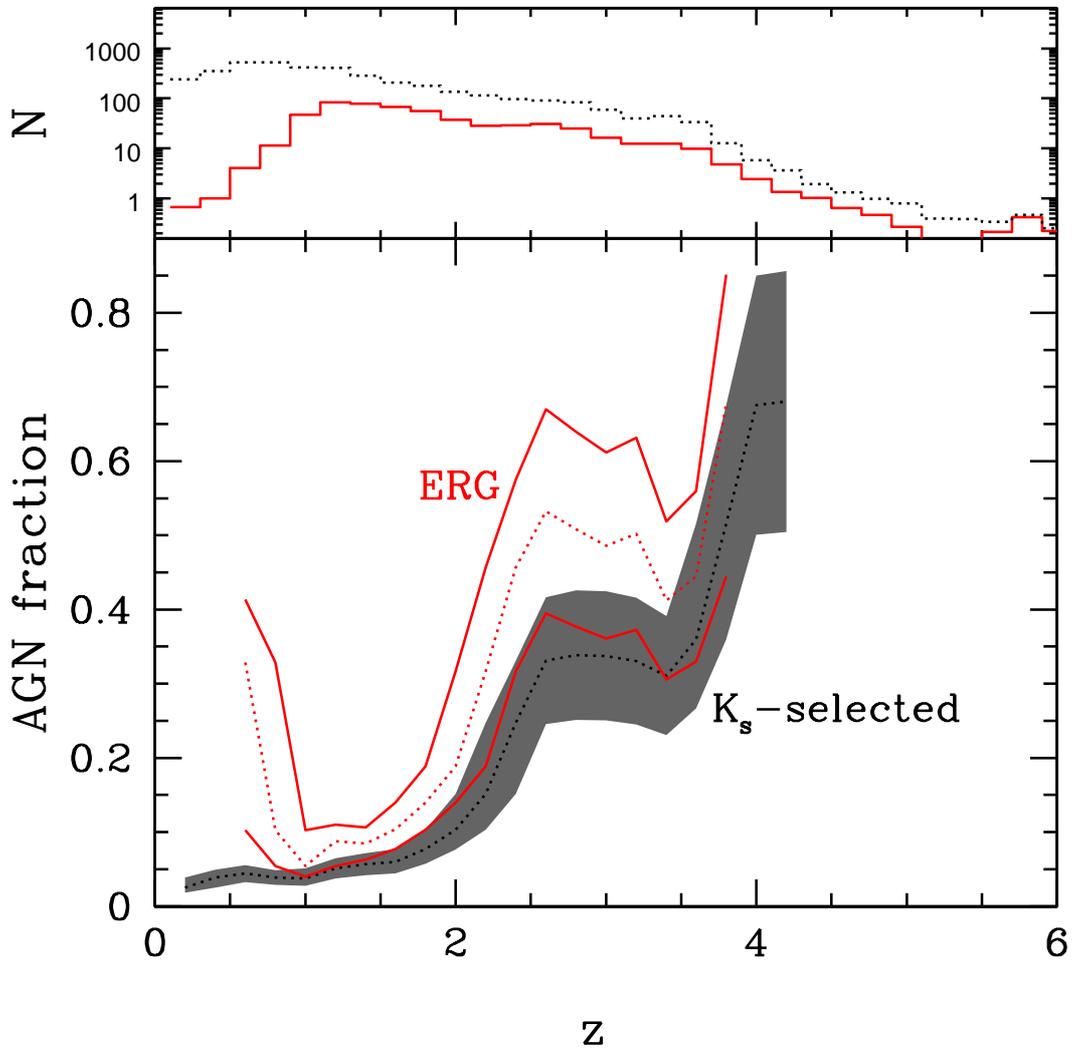}
  \end{center}
  \caption[AGN fraction with redshift]{AGN fraction with redshift. The grey shaded region (dotted line in the upper panel) represents the total $K_s$-selected population, while the open red delimited region (solid line in the upper panel) refers to the overall ERG population. The trends are computed with a moving bin of 0.4\,mag and steps of 0.2\,mag. Dotted lines in lower panel show central trends. Upper and lower limits consider poissonean errors and the bin width interval.}
  \label{c2fig:zagn}
\end{figure}

\begin{figure}
  \begin{center}
    \includegraphics[width=0.9\columnwidth]{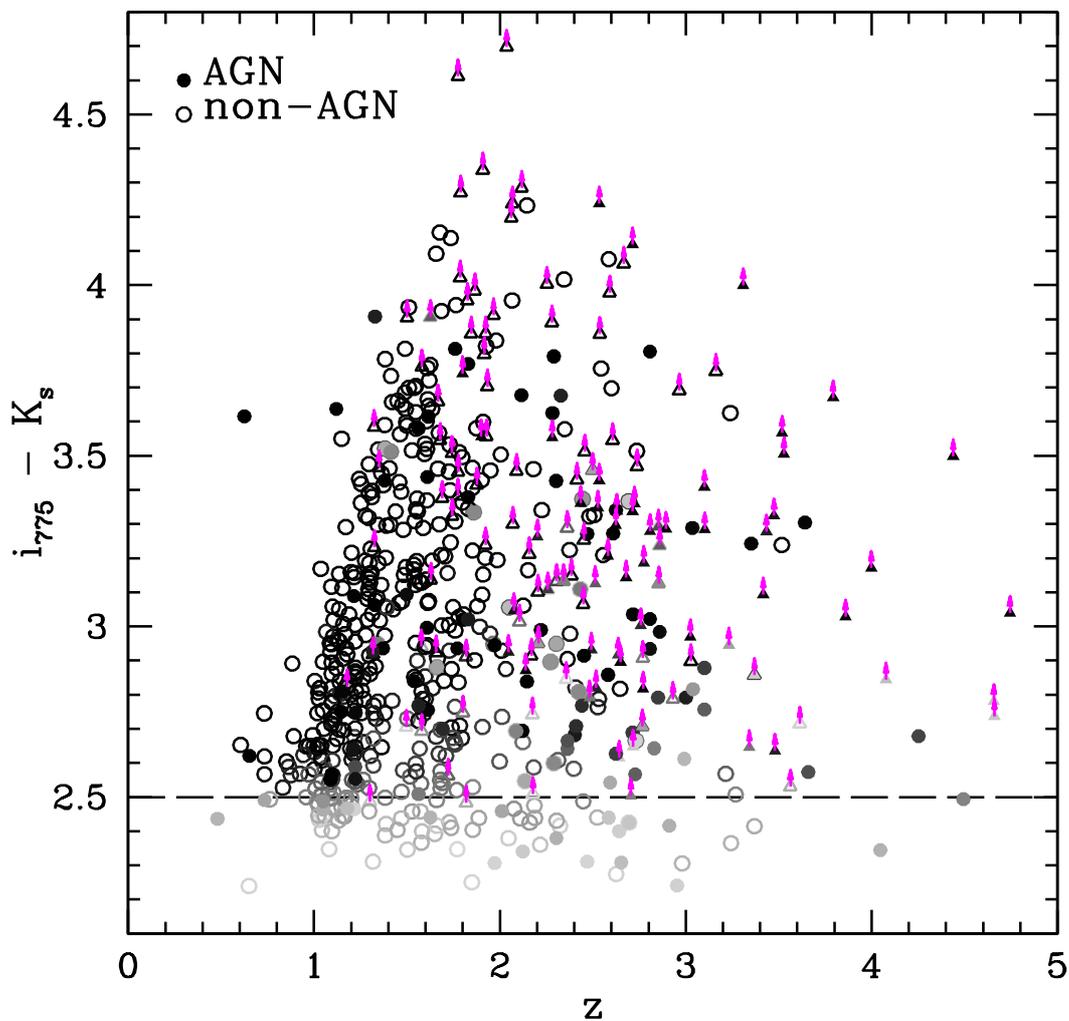}
  \end{center}
  \caption[Variation of $i_{775}-K_s$ colour with redshift]{Variation of $i_{775}-K_s$ colour with redshift. The ERO criterion colour cut is shown as a dashed line. The dot intensity refers to the source probability (see Figure~\ref{c2fig:ki}). AGN appear as filled symbols, while non-AGN as open symbols. Sources undetected in the $i_{775}$-band appear as triangles.}
  \label{c2fig:ikvsz}
\end{figure}

\begin{figure}
  \begin{center}
    \includegraphics[width=0.9\columnwidth]{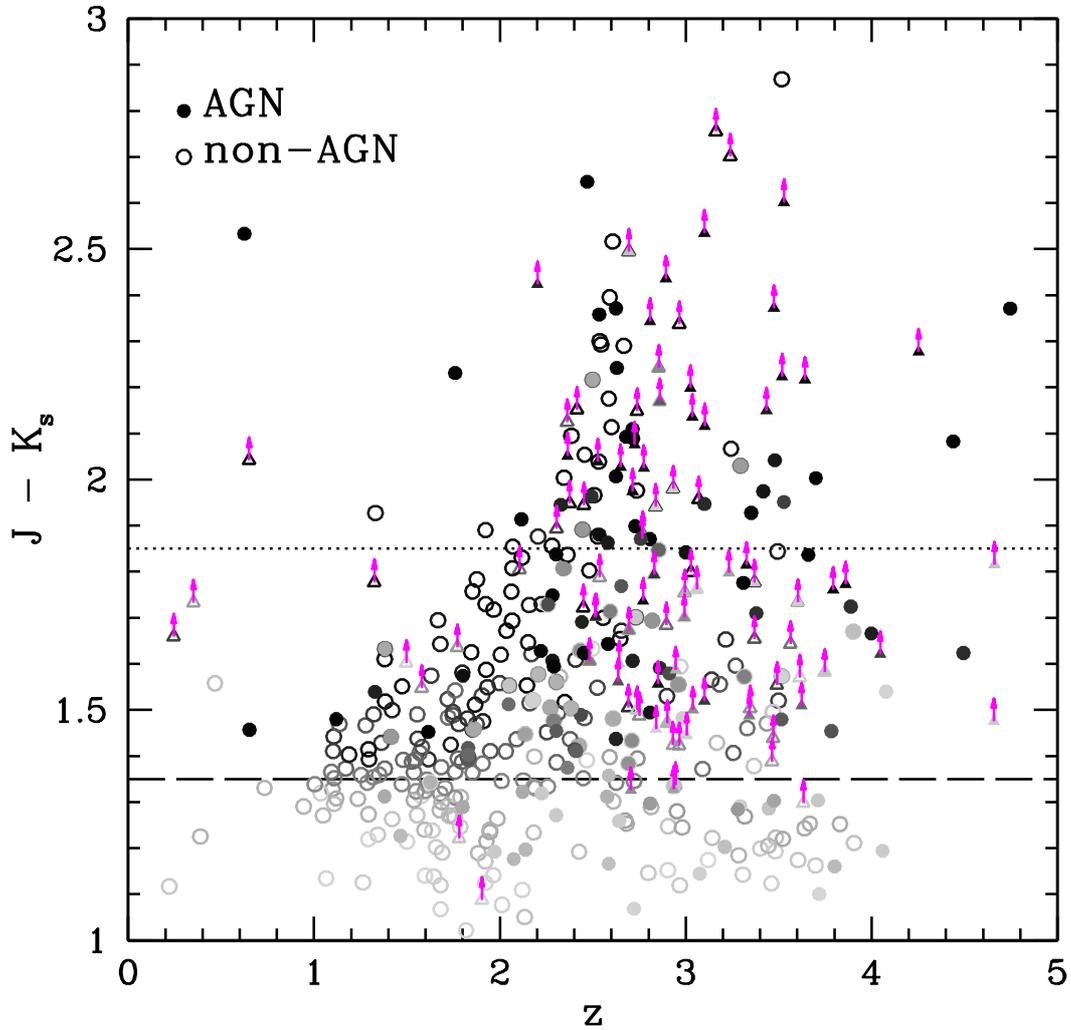}
  \end{center}
  \caption[Variation of $J-K_s$ colour with redshift]{Variation of $J-K_s$ colour with redshift. The DRG criterion colour cut is shown as a dashed line. A 0.5\,mag redder J-K$_s$ cut (dotted line) selects almost no $z<2$ DRGs. Given the data cloud trend, objects at $z<2$ with $J-K_s>1.8$ are believed to have catastrophic $z_{\rm phot}$ estimates. The dot intensity refers to the source probability (see Figure~\ref{c2fig:ki}). AGN appear as filled symbols, while non-AGN as open symbols. $J$-undetected sources appear as triangles.}
  \label{c2fig:jkvsz}
\end{figure}

The ERG populations do tend to include a higher fraction of AGN hosts than the non-ERG population. This is clear from Figure~\ref{c2fig:coragn}, where ERG AGN fractions (found in the positive side of the x-axis) have AGN fractions of 15--50\%. Note the AGN fraction is not a simple function of colour, as shown in Figure~\ref{c2fig:coragn}. All three colours ($i-K_s$, $z_{850}-[3.6]$, and $J-K_s$) imply similar AGN fractions around the colour threshold ($\sim20\%$), but show different behaviour with increasing colours. The more extreme colours among the ERGs do not necessarily correspond to a significantly higher fraction of AGN identifications, and in fact, that appears to be true for EROs, always around or below $\sim20\%$, and maybe IEROs, which the respective trend drops at the most extreme colours (although already being affected by small number statistics). This has implications for some of the works selecting compton-thick AGN at IR wave-bands down to the faintest limits. For instance, \citet{Fiore08} considers a fainter MIPS$_{24\mu\rm{m}}$ flux cut providing that an extreme $R-K>5$ (Vega) colour -- quite similar to $i-K$ -- selects a higher fraction of AGN. However, Figure~\ref{c2fig:coragn} indicates that probably the $J-K_s$ colour will be much more efficient for such task. The difference between the ERO and DRG trends result from each criterion itself and the fact that AGN fraction rises toward higher redshifts (Figure~\ref{c2fig:zagn}). Redder $i_{775}-K_s$ colours will always select a population mostly at low-$z$ ($1\leq{z}<2$, Figure~\ref{c2fig:ikvsz}), where the AGN fraction is shown to be smaller. On the other hand, redder $J-K_s$ constraints imply a higher fraction of high-$z$ ($2\leq{z}\leq3$) sources, where the AGN fraction is higher. For example, essentially no low-$z$ source has $J-K_s\gtrsim1.8$ (Figure~\ref{c2fig:jkvsz}). The AGN fraction versus colour trend of the IEROs (Figure~\ref{c2fig:coragn}) lies between that of the EROs and DRGs, which could be driven by the redshift distribution of the IERO population, which also lies between that of the EROs and DRGs.

As a final remark, two features should be highlighted in Figure~\ref{c2fig:zagn}. Note how at $z<1$ the ERG AGN fraction (solid line) increases dramatically up to the 30\% level. The reader should recall that $i-K_s$, $z_{850}-[3.6]$ or $J-K_s$ colours are sensitive to a prominent 4000\AA\ break in a galaxy SED only beyond $z\sim1$. Any source having ERO-, IERO- or DRG-like colours at $z<1$, has to be quite an obscured source. One of the most impressive examples for such type of object is that found by \citet{Afonso01}: a $z_{spec}=0.67$ obscured dusty starburst dominated in the IR by the AGN in its core. At higher redshifts, at $z\sim$3 both ERG and total trends present a noticeable rise in the AGN fraction. This is expected to be linked to the known peak of AGN activity at these redshifts \citep[see for instance][]{Osmer04,Richards06}.

\subsection{Radio Stacking} \label{c2sec:stack}

Another important aspect necessary to understanding the properties of ERGs is their SFR and the contribution of these populations to the overall $\dot{\rho}_{\ast}$ of the Universe. Dust obscuration is a serious source of uncertainty in estimating SFRs in ERGs from rest-frame ultraviolet luminosities. Radio emission is not affected by dust obscuration and can be used as a SF diagnostic. However, these galaxies are distant enough that even the deepest radio surveys are only sensitive to the brightest star forming systems (detection limits corresponding to several hundred M$_{\odot}$\,yr$^{-1}$ at $z\gtrsim1$). Instead, stacking methods can be used to evaluate the statistical star-forming properties of ERGs. Stacking, as used here, is simply an ``image stacking'' procedure, where image sections (called stamps) centred at each desired source position are combined. The aim is to reach much lower noise levels, possibly providing a statistical detection of samples whose elements are individually undetected in the original image.

For the radio stacking analysis we have used the 1.4\,GHz Australia Telescope Compact Array observations of this field\footnote{The VLA data was not preferred due to its high resolution. Although it may seem an advantage, it is more likely to be affected by bad source registration, producing flux loss in the stacking procedure.}, reaching a uniform 14--17\,$\mu$Jy rms throughout the GOODSs field \citep[see][for more details]{Afonso06,Norris06}. Our adopted stacking methodology can be summarised in the following steps. 

First, using the radio image of the field, stamps of 60 by 60 pixels (equivalent to 120'' by 120'') were considered, allowing for a good sampling of the vicinity of each source, necessary to identify strong neighbouring sources that can bias the stacking.

Every stamp containing a radio source within an 18'' radius from the central (ERG) position, was rejected, as the wings of a neighbouring radio detection can extend to the central part of the stamp. This rejects actual radio counterparts. However, the stacking is only used to estimate the average flux of the unidentified ERGs in the radio image, as the inclusion of radio detections would likely bias the final result. In this context the term ``detection'' does not only apply to the robust detections (roughly at a $>4.5\sigma$ level), but also to ``possible'' detections (all remaining candidate radio sources at a $>3\sigma$ level).

The remaining stamps for each sample of ERGs can then be stacked. Previous work often uses median stacking \citep[e.g.,][]{White07}, in an attempt to be robust to radio detections and high/low pixels. The penalty for this is the loss of sensitivity. Having removed all detections and possible detections from the list of stamps, a weighted average ($weight=rms^{-2}$) stacking procedure is followed. At each pixel position a rejection for outliers is implemented, rejecting high (low) pixels above (below) the $3\sigma$ ($-3\sigma$) value {\it for that pixel position}. The number of rejected pixels in the central region is always zero confirming that previous rejection steps work efficiently. 

The final flux and the noise level are measured in the resulting stacked image. To evaluate the reliability of detections in the stacked images we performed Monte Carlo (MC) simulations. Random positions in the radio image were selected and stacked, following the procedure described above. Each of these positions were required to be farther than 6'' from the known $K_s$ sources, as we are interested in evaluating systematics of the radio image alone. Appropriate numbers of stacked stamps were used, to compare to the actual numbers of the ERG (sub-)samples. The procedure was repeated 10000 times for a given number of stamps\footnote{The number of stamps chosen for each set of 10'000 tries are: 10, 20, 30, 40, 50, 75, 100, 125, 150, 175, 200, 300, 400, 500, and 1000.}. A stacked sample will be considered to have produced a reliable detection only if no MC simulation (among 10000) has resulted in a higher S/N value.

\subsection{Star formation activity in ERGs} \label{c2sec:sfrfin}

Following the procedure outlined above, we have performed a radio stacking analysis for different sub-groups within the ERG population. The radio data was stacked for each of the populations of EROs, IEROs, DRGs, pEROs, pDRGs, and cERGs. Within these samples, stacking of the radio images was also performed separately for the total and non-AGN sub-populations. Since redshift estimates exist for the vast majority of the ERGs, stacking is performed separately for both low and high redshifts ($1\leqslant{z}<2$ and $2\leqslant{z}\leqslant$3, respectively). Besides minimising biases in the stacking signal, due to different populations and different (radio) luminosities being sampled at different redshifts, this also allows us to search for a hint of any evolutionary trend. Given the incompleteness of the sample at the highest redshifts no attempt was made to perform a specific radio stacking analysis for $z>3$ ERGs. Table~\ref{c2tab:tabnum} lists the number of sources considered in each of the sub-populations and those in each of the stacking steps referred in the previous section.

\ctable[
   cap     = Robust Radio stacking of ERG populations,
   caption = Robust Radio stacking of ERG populations,
   label   = c2tab:tabnum
]{crrrrr}{}{ \FL
POP & $\rm{N_{NOM}}$\tmark[a] & $\rm{N_{TOT}}$\tmark[a] & $\rm{N_{<18''}}$\tmark[b] & $\rm{N_{3\sigma}}$\tmark[c] & $\rm{N_{fin}}$\tmark[d] \ML
\bf{$K_s$} \NN 
$z12$ &  1803 & 1429 & 170 & 14 & 1230 \NN
$z12;\rm{nAGN}$ &  1646 & 1307 & 147 & 11 & 1134 \NN
$z23$ &  781 & 512 & 63 & 7 & 435 \NN
$z23;\rm{nAGN}$ &  518 & 318 & 32 & 4 & 276 \NN
\bf{EROs} \NN 
$z12$ &  451 & 357 & 51 & 7 & 294 \NN
$z12;\rm{nAGN}$ &  396 & 316 & 42 & 6 & 264 \NN
$z23$ &  197 & 124 & 20 & 3 & 101 \NN
$z23;\rm{nAGN}$ &  94 & 56 & 6 & 2 & 48 \NN
\bf{IEROs} \NN 
$z12$ &  188 & 133 & 20 & 4 & 106 \NN
$z12;\rm{nAGN}$ &  160 & 114 & 15 & 4 & 92 \NN
$z23$ &  147 & 93 & 16 & 2 & 75 \NN
$z23;\rm{nAGN}$ &  71 & 42 & 5 & 1 & 36 \NN
\bf{DRGs} \NN 
$z12$ &  148 & 73 & 9 & 2 & 61 \NN
$z12;\rm{nAGN}$ &  126 & 61 & 6 & 2 & 52 \NN
$z23$ &  227 & 130 & 18 & 1 & 111 \NN
$z23;\rm{nAGN}$ &  103 & 57 & 5 & 1 & 50 \LL\NN
}

\ctable[
   caption = { },
   continued,
   nosuper,
   mincapwidth = 15cm
]{crrrrr}{
  \tnote[Note.]{---The $z12$ and $z23$ abbreviations stand for $1\leq{z}<2$ and $2\leq{z}\leq3$, respectively.}
  \tnote[$^a$]{$\rm{N_{NOM}}$ and $\rm{N_{TOT}}$ are, respectively, the nominal counts and the effective total sources found in the sample. All the other columns take the source probability into account as does $\rm{N_{TOT}}$.}
  \tnote[$^b$]{Number of stamps with a radio detection within 18'' of the ERG position, consequently rejected from the final stacking.}
  \tnote[$^c$]{Number of stamps with a possible radio detection at the ERG position (signal between 3$\sigma$ and $\sim4.5\sigma$), also removed from the final stacking.}
  \tnote[$^d$]{Final number of stamps included in the stacking.}
}{ \FL
POP & $\rm{N_{NOM}}$\tmark[a] & $\rm{N_{TOT}}$\tmark[a] & $\rm{N_{<18''}}$\tmark[b] & $\rm{N_{3\sigma}}$\tmark[c] & $\rm{N_{fin}}$\tmark[d] \ML
\bf{cERGs} \NN
$z12$ &  97 & 49 & 6 & 2 & 40 \NN
$z12;\rm{nAGN}$ &  81 & 40 & 3 & 2 & 34 \NN
$z23$ &  120 & 77 & 12 & 1 & 64 \NN
$z23;\rm{nAGN}$ &  58 & 35 & 3 & 1 & 31 \NN
\bf{pEROs} \NN 
$z12$ &  287 & 199 & 29 & 2 & 166 \NN
$z12;\rm{nAGN}$ &  258 & 179 & 24 & 1 & 152 \NN
$z23$ &  25 & 8 & 2 & 0 & 6 \NN
$z23;\rm{nAGN}$ &  12 & 4 & 1 & 0 & 3 \NN
\bf{pDRGs} \NN 
$z12$ &  6 & 2 & 0 & 0 & 2 \NN
$z12;\rm{nAGN}$ &  5 & 1 & 0 & 0 & 1 \NN
$z23$ &  64 & 21 & 1 & 0 & 19 \NN
$z23;\rm{nAGN}$ &  23 & 8 & 0 & 0 & 7 \LL\NN
}

While the stacking procedure enables the average flux to be estimated from the radio-undetected sample ($<3\sigma$ signal), the entire population should be considered when measuring the ERG contribution to the global $\dot{\rho}_{\ast}$ of the Universe. The approach adopted here was to consider all radio-undetected ERGs as having a radio flux given by the average signal from the stacking analysis, and all radio-detected ERGs\footnote{For this purpose, radio-detections refer to signals above $3\sigma$ in the radio map; see Section~\ref{c2sec:stack}.} to contribute with their measured flux density. The conversion from radio flux to radio luminosity is performed by using the assumed redshift (spectroscopic or photometric) and a radio spectral index of $\alpha=0.8$ ($S_\nu \propto \nu^{-\alpha}$, characteristic of a synchrotron dominated radio spectrum at 1.4\,GHz): \[ \rm{L_{1.4\,GHz} = 4 \pi\,d^{2}_{L}\,S_{1.4\,GHz}\,10^{-33}(1+z)^{\alpha-1}\,W\,Hz^{-1}} \] where $d_{L}$ is the luminosity distance (cm) and $\rm{S_{1.4GHz}}$ is the 1.4 GHz flux density (mJy). The corresponding SFR is obtained using the calibration from \citet{Bell03}:
\begin{eqnarray}
\rm{SFR~(M_{\odot}~yr^{-1})}=\left\lbrace
\begin{array}{lcc}
5.52\times10^{-22}\rm{L_{1.4\,GHz}} &,&\rm{L>L_c} \nonumber \\
\frac{5.52\times10^{-22}}{0.1+0.9\left(\rm{\frac{L}{L_c}}\right)^{0.3}}\rm{L_{1.4\,GHz}} &,&\rm{L{\leq}L_c} \nonumber
\end{array} \right.
\end{eqnarray}
where $\rm{L_c=6.4\times10^{21}\,W\,Hz^{-1}=10^{21.81}\,W\,Hz^{-1}}$. The contribution to $\dot{\rho}_{\ast}$ was estimated for individual galaxies using the $\rm{1/V_{max}}$ method \citep{Schmidt68}: 

\[\dot{\rho}_{\ast}=\sum{\frac{\rm{SFR^i}}{\rm{V^i_{max}}}}\]

\noindent where $\rm{V_{max}}$ is the volume in which a given source $i$ would be possible to detect:

\[{\rm V_{max}}=\Omega\frac{\rm c}{\rm H_0}\int_{z_1}^{z_2}\frac{\rm d_L^2\prod_i\eta_i}{(1+z)^2\sqrt{\Omega_{\rm M}(1+z)^3+\Omega_\Lambda}}\rm{d}z\]

The solid angle is given by $\Omega$, while \textit{c} stands for the speed of light, and H$_0$ for the Hubble constant. The luminosity distance again appears as $\rm{d^2_L}$, and $\prod_i\eta_i$ is the product of every incompleteness factors affecting the sample (e.g., sources rejected due to bright neighbours affecting their flux estimates). The value of $z_1$ is the lowest redshift probed (set to 1 in the lower redshift bin and 2 in the upper redshift bin). The value of $z_2$ is the minimum between the maximum redshift probed ($z_{\rm max}$, set to 2 in the lower redshift bin and 3 in the upper redshift bin) and the redshift at which a given source would be detected, in the survey selection band, with the minimum source flux observed in the sample in which that same source is considered: $z_2={\rm min}\{z_{\rm max},z(f_{\rm min})\}$. Hence, the final value of ${\rm V_{max}}$ gives the volume in which a given type of galaxy would be detected, this is, $1/{\rm V_{max}}$ is the contribution to the density of sources by a given galaxy in a given redshift bin. It is expected a certain bias if strong clustering is observed between the sources in a sample, when compared to other galaxy samples. Even if $z_{\rm min}$ and $z_{\rm max}$ are set to be far apart, if the sample is physically restricted to a small volume, their fluxes will be comparable, implying a small difference between $z_1$ and $z_2$, hence small volumes (large 1/V values). However, as seen in Figure~\ref{c2fig:geral}, ERGs are well spread over the full $1\leq{z}\leq3$ redshift range, and they spread for almost four and three magnitudes (in observed $K_s$) at $1\leq{z}<2$ and $2\leq{z}\leq3$, respectively.

\begin{figure}
  \begin{center}
    \includegraphics[width=0.5\columnwidth]{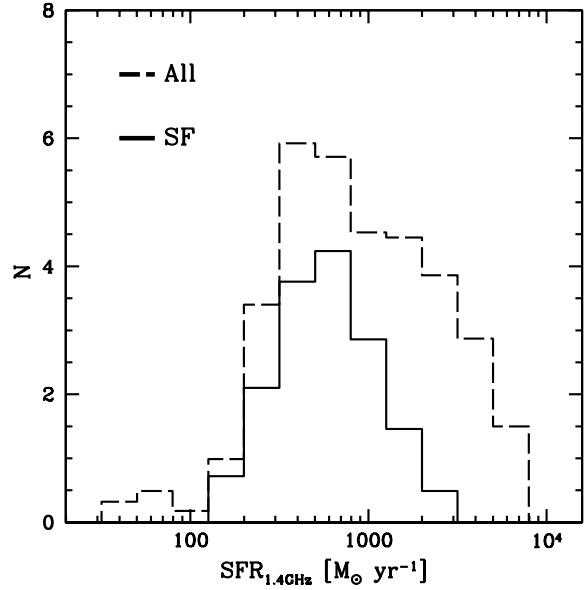}
  \end{center}
  \caption[SFR distribution of the radio detected ERGs]{SFR distribution of the radio detected ERGs (considering any signal in the radio map with $>3\sigma$). Dashed histograms show the overall distribution, while solid histograms refer to the sources considered as star-forming systems.}
  \label{c2fig:sfrdist}
\end{figure}

The ${\rm V_{max}}$ for each galaxy is estimated by using a k-correction derived from the galaxy's own SED (through interpolation of the observed multi-wavelength photometry). Again, radio detected ERGs, contributed with their estimated intrinsic luminosity and SFR, derived with the assigned redshift estimate and its detected flux. In Figure~\ref{c2fig:sfrdist} the SFR distribution of these sources is presented. Those classified as star-forming (16 in total), range from $\sim$100 to $\sim$2000\,M$_{\odot}$ yr$^{-1}$ \citep[in reasonable agreement with][]{Georgakakis06}. On the other hand, the luminosity and SFR estimates of radio-undetected ERGs were based on the resulting stacking signal of the sample and, likewise, the individual ERG redshift value.

The results are given in Table~\ref{c2tab:tabsta}. For each ERG sub-population we list: (1) the ERG sub-population; (2) the total number of sources in the sample; (3) the final number of stamps included in the stacking; (4) the rms of the final stacked image; (5) the measured flux in the central region of the stacked image; (6) the respective S/N; (7) number of Monte-Carlo simulations (out of 10000) that resulted in higher S/N values, a measure of the reliability of the ERG detection (conservatively, whenever $\rm{N_{MC}}>0$ the stacking signal is considered spurious); (8) the average redshift for the sub-population; (9) the average radio luminosity for the radio non-detected sources -- taking into account the stacking signal only -- and, in parenthesis, the median for the entire sub-population (including radio detected sources); (10) the average SFR, for non-AGN samples, corresponding to the radio luminosities in column (9); (11) the resulting radio luminosity density ($\mathcal{L}_{1.4\,{\rm GHz}}$); (12) the corresponding $\dot{\rho}_{\ast}$. For columns (9) to (12), the upper limits, corresponding to non-detections of the stacked signal ($\rm{N_{MC}}>0$), are estimated using the maximum S/N found on the corresponding MC simulations. No stacking is attempted for populations with less than 10 stamps.

\footnotesize
\ctable[
   cap     = Radio properties of ERGs,
   caption = \normalsize{Properties of Extremely Red Galaxy populations from radio stacking analysis},
   label   = c2tab:tabsta,
   sideways,
   nosuper
]{crrrrrrrrrrrr}{}{ \FL
POP & $\rm{N_{TOT}}$ & $\rm{N_{fin}}$\tmark[a] & $rms$ & S$_{\rm 1.4\,GHz}$ & S/N & $\rm{N_{MC}}$\tmark[b] & $\overline{z}$ & $\overline{\log\,(\rm{L_{1.4\, GHz}})}$\tmark[c] & $\overline{\rm SFR}$\tmark[c] & $\log$ ($\mathcal{L}_{1.4\,{\rm GHz}}$)\tmark[c] & $\dot{\rho}_{\ast}$\tmark[d] \NN
 &  &  & [$\mu$Jy] & [$\mu$Jy] &  &  &  & [W$\,$Hz$^{-1}$] & [M$_{\odot}\,$yr$^{-1}$] & [W$\,$Hz$^{-1}\,$Mpc$^{-3}$] & [M$_{\odot}\,$yr$^{-1}\,$Mpc$^{-3}$] \ML
\bf{K$_s$} \NN  
$z12$ &  1429 & 1230 & 0.501 & 1.102 & 2.200 & 0 & 1.39 & 22.0(22.1) & 6(3) & 19.8(20.2) & 3.4e-02(8.0e-02) \NN 
$z12;nAGN$ &  1307 & 1134 & 0.533 & 1.118 & 2.096 & 0 & 1.38 & 22.0(22.0) & 6(4) & 19.8(20.1) & 3.1e-02(6.1e-02) \NN 
$z23$ &  512 & 435 & 0.790 & 2.746 & 3.477 & 0 & 2.46 & 23.0(23.0) & 57(49) & 20.2(20.4) & 8.7e-02(1.3e-01) \NN 
$z23;\rm{nAGN}$ &  318 & 276 & 0.932 & 1.646 & 1.766 & 0 & 2.40 & 22.8(22.8) & 32(27) & 19.8(20.0) & 3.2e-02(5.0e-02) \NN 
\bf{EROs} \NN  
$z12$ &  357 & 294 & 0.984 & 1.327 & 1.349 & 0 & 1.44 & 22.1(23.8) & 8(5) & 19.2(19.9) & 9.5e-03(3.9e-02) \NN 
$z12;\rm{nAGN}$ &  316 & 264 & 1.081 & 0.926 & 0.856 & 0 & 1.44 & 22.0(24.4) & 6(3) & 19.0(19.7) & 5.9e-03(2.6e-02) \NN 
$z23$ &  124 & 101 & 1.705 & 6.610 & 3.877 & 0 & 2.44 & 23.4(23.4) & 136(121) & 20.0(20.2) & 5.0e-02(8.3e-02) \NN 
$z23;\rm{nAGN}$ &  56 & 48 & 2.186 & 5.029 & 2.301 & 1 & 2.36 & $<$23.2(23.2) & $<$82(70) & $<$19.4(19.7) & $<$1.4e-02(2.6e-02) \NN 
\bf{IEROs} \NN  
$z12$ &  133 & 106 & 1.648 & 6.346 & 3.851 & 0 & 1.64 & 23.0(23.0) & 52(42) & 19.6(19.9) & 2.1e-02(4.3e-02) \NN 
$z12;\rm{nAGN}$ &  114 & 92 & 1.716 & 5.193 & 3.026 & 0 & 1.64 & 22.9(23.0) & 42(36) & 19.4(19.8) & 1.5e-02(3.2e-02) \NN 
$z23$ &  93 & 75 & 1.900 & 6.726 & 3.540 & 0 & 2.45 & 23.4(23.4) & 139(124) & 19.9(20.0) & 3.9e-02(6.1e-02) \NN 
$z23;\rm{nAGN}$ &  42 & 36 & 2.503 & 4.116 & 1.644 & 21 & 2.37 & $<$23.2(23.3) & $<$92(76) & $<$19.3(19.5) & $<$1.2e-02(1.8e-02) \NN 
\bf{DRGs} \NN  
$z12$ &  73 & 61 & 1.893 & 7.701 & 4.068 & 0 & 1.57 & 23.0(23.1) & 59(37) & 19.4(19.5) & 1.3e-02(1.9e-02) \NN 
$z12;\rm{nAGN}$ &  61 & 52 & 2.034 & 7.018 & 3.450 & 0 & 1.57 & 23.0(23.0) & 53(32) & 19.2(19.4) & 9.7e-03(1.4e-02) \NN 
$z23$ &  130 & 111 & 1.555 & 5.295 & 3.405 & 0 & 2.50 & 23.31(23.33) & 115(105) & 19.9(20.0) & 4.2e-02(6.0e-02) \NN 
$z23;\rm{nAGN}$ &  57 & 50 & 2.225 & 4.964 & 2.231 & 7 & 2.43 & $<$23.2(23.2) & $<$89(81) & $<$19.4(19.6) & $<$1.4e-02(2.0e-02) \LL\NN
}

\ctable[
   caption = { },
   sideways,
   continued,
   nosuper
]{crrrrrrrrrrrr}{
  \tnote[Note.]{ --- The $z12$ and $z23$ abbreviations stand for $1\leq{z}<2$ and $2\leq{z}\leq3$, respectively. The upper limits for Luminosity and SFR estimates, whenever $\rm{N_{MC}}>0$, are calculated considering the maximum S/N obtained in the respective set of MC simulations.}
  \tnote[$^a$]{Final number of stamps included in the stacking, after the various rejection steps described in Section~\ref{c2sec:stack}.}
  \tnote[$^b$]{Number of MC simulations (out of 10000) that resulted in higher S/N values.}
  \tnote[$^c$]{In parenthesis, the median value also taking into account radio detections ($>3\sigma$) excluded from the stacking procedure (see Section~\ref{c2sec:stack}).}
  \tnote[$^d$]{In parenthesis, the estimated value of $\dot{\rho}_{\ast}$ taking into account radio detections ($>3\sigma$) excluded from the stacking procedure (see Section~\ref{c2sec:stack}).}
}{ \FL
POP & $\rm{N_{TOT}}$ & $\rm{N_{fin}}$\tmark[a] & $rms$ & S$_{\rm 1.4\,GHz}$ & S/N & $\rm{N_{MC}}$\tmark[b] & $\overline{z}$ & $\overline{\log\,(\rm{L_{1.4\, GHz}})}$\tmark[c] & $\overline{\rm SFR}$\tmark[c] & $\log$ ($\mathcal{L}_{1.4\,{\rm GHz}}$)\tmark[c] & $\dot{\rho}_{\ast}$\tmark[d] \NN
 &  &  & [$\mu$Jy] & [$\mu$Jy] &  &  &  & [W$\,$Hz$^{-1}$] & [M$_{\odot}\,$yr$^{-1}$] & [W$\,$Hz$^{-1}\,$Mpc$^{-3}$] & [M$_{\odot}\,$yr$^{-1}\,$Mpc$^{-3}$] \ML 
\bf{cERGs} \NN  
$z12$ &  49 & 40 & 2.420 & 8.203 & 3.390 & 0 & 1.64 & 23.1(23.2) & 69(52) & 19.3(19.5) & 1.0e-02(1.7e-02) \NN 
$z12;\rm{nAGN}$ &  40 & 34 & 2.626 & 7.095 & 2.702 & 0 & 1.66 & 23.0(23.1) & 59(48) & 19.1(19.3) & 7.4e-03(1.2e-02) \NN 
$z23$ &  77 & 64 & 2.073 & 6.685 & 3.225 & 0 & 2.46 & 23.4(23.4) & 140(125) & 19.8(19.9) & 3.2e-02(4.5e-02) \NN 
$z23;\rm{nAGN}$ &  35 & 31 & 2.714 & 4.939 & 1.820 & 22 & 2.39 & $<$23.3(23.3) & $<$104(90) & $<$19.3(19.4) & $<$1.1e-02(1.5e-02) \NN 
\bf{pEROs} \NN  
$z12$ &  199 & 166 & 1.311 & -1.139 & -0.869 & 2532 & 1.32 & $<$22.2(22.3) & $<$10(7) & $<$19.1(19.4) & $<$6.4e-03(1.3e-02) \NN 
$z12;\rm{nAGN}$ &  179 & 152 & 1.403 & -1.366 & -0.974 & 1924 & 1.32 & $<$22.2(22.2) & $<$10(7) & $<$19.1(19.2) & $<$6.2e-03(8.5e-03) \NN 
$z23$ &  8 & 6 & \ldots & \ldots & \ldots & \ldots & \ldots & \ldots & \ldots & \ldots & \ldots \NN 
$z23;\rm{nAGN}$ &  4 & 3 & \ldots & \ldots & \ldots & \ldots & \ldots & \ldots & \ldots & \ldots & \ldots \NN 
\bf{pDRGs} \NN  
$z12$ &  2 & 2 & \ldots & \ldots & \ldots & \ldots & \ldots & \ldots & \ldots & \ldots & \ldots \NN 
$z12;\rm{nAGN}$ &  1 & 1 & \ldots & \ldots & \ldots & \ldots & \ldots & \ldots & \ldots & \ldots & \ldots \NN 
$z23$ &  21 & 19 & 2.957 & 2.231 & 0.754 & 1666 & 2.65 & $<$23.5(23.5) & $<$176(174) & $<$19.3(19.3) & $<$1.1e-02(1.1e-02) \NN 
$z23;\rm{nAGN}$ &  8 & 7 & \ldots & \ldots & \ldots & \ldots & \ldots & \ldots & \ldots & \ldots & \ldots \LL\NN
}

\normalsize

In the table, two rows appear for each sub-population. The reason for this is the controversial inclusion of AGN sources when estimating the SFRs and $\dot{\rho}_{\ast}$. As referred above, AGN are potentially non-star-forming emitters at radio frequencies, thus being a strong source of bias. Yet, studies at the sub-mJy level point to a probable dominance of star-forming systems \citep{Muxlow05,Simpson06,Kellermann08,Smolcic08,Seymour08,Ibar09,Padovani09}. Adding to that, radio-selected AGN tend to appear in a whole different population from that of X-ray and IR-selected AGN (considered in this work), probably meaning a different accretion mode in radio-selected AGN \citep[][and references therein]{Hickox09,Griffith10}, implying that the AGN selection in this work is actually too strict. Also, in this ERG sample, there is not a major presence of strong AGN (although the strongest do tend to show ERG colours) and \citet{Dunne09} believe, based both in radio spectral indexes and comparison with sub-mm-derived SFRs, there may be no significant bias when including AGN sources in the stacking of a sub-mJy radio population. However, in Section~\ref{c2sec:agncont} it is shown that AGN are common in this sample, hence, even if not dominant, there might exist a significant bias when computing the contribution $\dot{\rho}_{\ast}$ of ERGs to the overall star-formation history of the universe. The two extreme scenarios for this are: ($i$) the non-AGN population presents the best estimate possible, or ($ii$) the combined non-AGN/AGN provides an upper limit for the contribution of ERGs, while the non-AGN indicates the lower limit of $\dot{\rho}_{\ast}$. Option ($ii$) also provides an upper limit for the star-formation happening in AGN hosts, which has been proven to occur at significant levels \citep[e.g.,][]{Shi07,Shi09,Silverman09,Xue11,Mullaney11} and even at rates up to thousands of M$_\odot$\,yr$^{-1}$ \citep[e.g.,][]{Dunlop94,Ivison95,Hughes97,Shao10}). It should be stressed, nevertheless, that no sources with $\log(\rm{L_{1.4\,GHz}[W\,Hz^{-1}])>24.5}$ were included in the calculations of the values presented in Table~\ref{c2tab:tabsta}.

The analysis suggests that the bulk of the ERO population have modest SF activity. At $1~{\leq}~z<2$, where most EROs are found, the average SFR is below a few\footnote{The significantly greater value of the population median SFR is a result of the adopted weighted median, applied to both stacked and radio detected samples. The values resulting from the stacking will have much greater relative errors, resulting in significantly smaller weights when compared to the radio detected sources at $>300$\,M$_\odot$\,yr$^{-1}$.} M$_{\odot}$ yr$^{-1}$. Only at $2~{\leq}~z\leq3$, EROs -- many (81\%) being simultaneously classified as DRGs -- reveal intense average SFRs, up to 140 M$_{\odot}$ yr$^{-1}$, entering the Luminous IR Galaxies (LIRG) regime. This suggests that at low-$z$ the passive/evolved systems represent a significant fraction of the ERO population (56\%, see pEROs discussion ahead), as opposed to the high-$z$ regime where the dusty systems dominate. DRGs and IEROs at $1\leq{z}<2$ show substantial SFRs, $\sim50$\,M$_\odot$\,yr$^{-1}$. It should be noted that practically all IEROs and DRGs at these redshifts are also classified as EROs, explaining the similar results for the cERGs. This also supports previous claims of a dusty starburst nature for these sources \citep{Smail02,Papovich06,Wuyts09c}. At $2\leq{z}\leq3$, none of the non-AGN ERG populations is successful in achieving a stacking signal, being indicative of $\lesssim80$\,M$_\odot$\,yr$^{-1}$ SFRs.

The overall SFR for the DRG population is comparable to what is found in the literature \citep{Rubin04,Forster04,Knudsen05,Reddy05}. \citet{Papovich06} studied 153 DRGs selected also in the GOODSs to a limiting magnitude of K$_{s,TOT}<23$. They find an average SFR for the DRG population at $1\lesssim{z}\lesssim3$ of $200-400\,\rm{M_{\odot}\,yr^{-1}}$, which is higher than our result. However, the SFR estimate is based in the $24\,\mu$m flux alone, method which has been shown, using $Spitzer$ \citep{Papovich07} and \textit{Herschel Space Observatory} data \citep{Nordon10,Rodighiero10}, to overestimate the actual SFR values.

The low average SFR for EROs at $1~{\leq}~z<2$ is due to the numerous pEROs (199, 56\% of the $1~{\leq}~z<2$ EROs): the stacking analysis of pEROs found in this redshift range fails to produce any signal. This population likely corresponds to the passively evolving component of EROs. On the other hand, pDRGs at $2~{\leq}~z\leq3$ must also be characterised by relatively low SFRs: although the stacking analysis is unable to give such indication (only limiting the average SFR to $\lesssim170$\,M$_{\odot}$ yr$^{-1}$), pDRGs are the sources responsible for the observed difference of the average SFR of cERGs and that of DRGs in this redshift range. Having this, although the SFR upper limit for pDRGs is rather high, one can adopt $\sim100$\,M$_{\odot}$ yr$^{-1}$ based on the DRG stacking. This is more likely to be close to the real SFR value.

\begin{figure}
  \begin{center}
    \includegraphics[width=0.9\columnwidth]{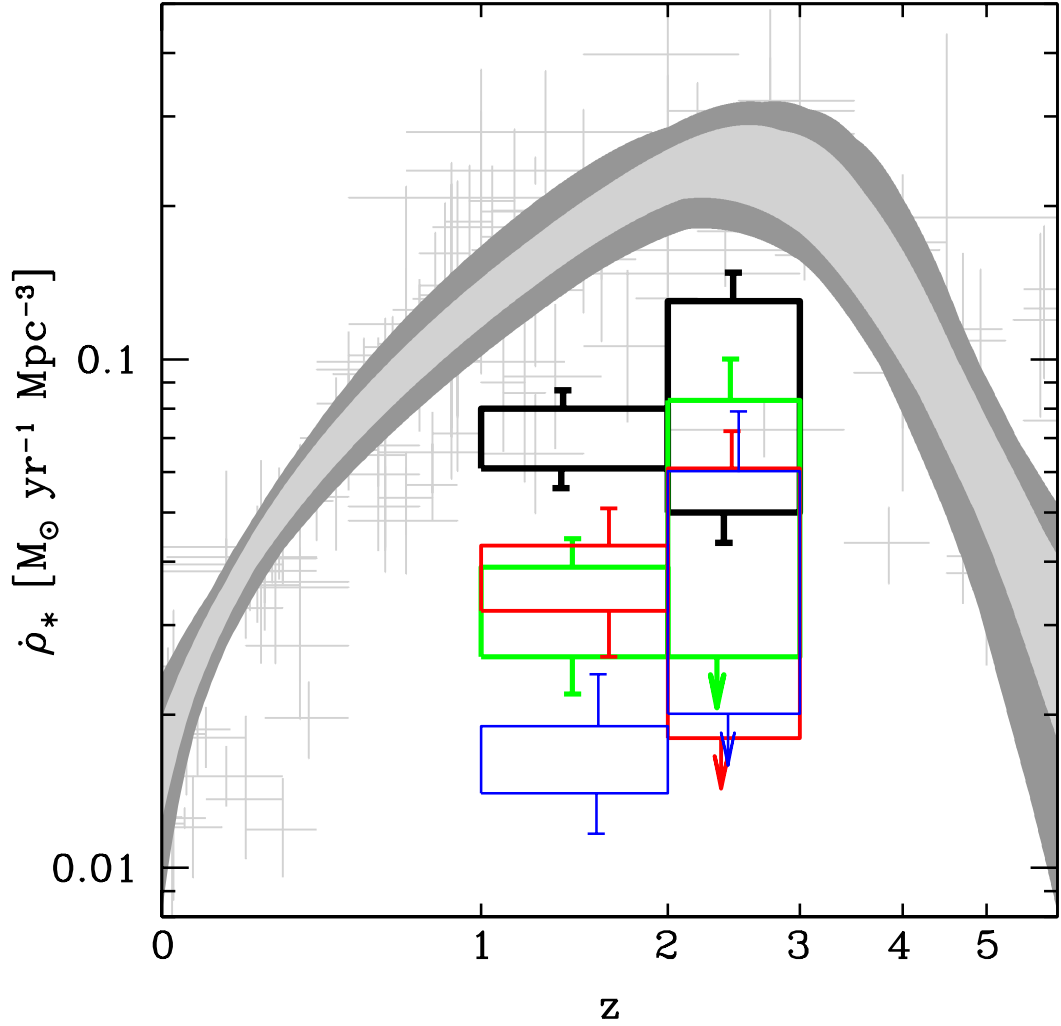}
  \end{center}
  \caption[ERGs SFR densities with redshift]{Contribution of ERG populations to the total $\dot{\rho}_{\ast}$ at $1\leq{z}<2$ and $2\leq{z}\leq3$. $K_s$-selected sources are denoted by black boxes, EROs by green boxes, IEROs by red boxes, and DRGs by blue boxes. The compilation of \citet{HopkinsBeacom06} is displayed for reference (grey crosses and shaded region, correspond to the $\dot{\rho}_{\ast}$ $1\sigma$ and $3\sigma$ confidence regions). Down pointing arrows indicate upper limits.}
  \label{c2fig:sfh}
\end{figure}

The $\dot{\rho}_{\ast}$ behaviour for ERGs roughly follows the general trend for star-forming galaxies, increasing from $1~{\leq}~z<2$ to $2~{\leq}~z<3$ (Figure~\ref{c2fig:sfh}\footnote{The error bars in the figure take into account cosmic variance as calculated in:\\http://casa.colorado.edu/$\sim$trenti/CosmicVariance.html}). Overall, the ERG contribution to the total $\dot{\rho}_{\ast}$ jumps from $\sim10\%$ in the low redshift bin (IEROs contribution), up to $\sim40\%$ at $2~{\leq}~z\leq3$, where EROs are the highest contributors (up to $\dot{\rho}_{\ast}\sim$0.09\,M$_{\odot}$\,yr$^{-1}$\,Mpc$^{-3}$). The range in $\dot{\rho}_{\ast}$ values for the ERO population clearly makes the point on whether one should include the AGN population on not, as the overall $\dot{\rho}_{\ast}$ is $\sim3$ times higher than the upper limit for the non-AGN population. IEROs are the population on which it is impossible to draw any conclusion on evolution, yet they are clearly the biggest contributors at low redshift. DRGs tend to be the ERG population to contribute the least in the full $1\leq{z}\leq3$ range.

\subsection{Dust content} \label{c2sec:dustcont}

Knowing that radio is unaffected by dust obscuration and UV is, both regimes are compared to give an estimate of the amount of dust present in these sources. Adopting the radio SFR estimates as the true values, we estimate how much obscuration is affecting the UV based results. The time gap between the emission at these two spectrum regimes is considered negligible (a few Myr at most). The hot stars strongly emitting in the UV will quickly reach the SNe stage, at which synchrotron emission is produced.

The calibration used to calculate UV SFRs was that given by \citet{Dahlen07} based on the rest-frame 2800\,\AA\ Luminosity. This was obtained through interpolation of the photometry bands available in the FIREWORKS catalogue. The ratio of the observed UV luminosity ($\rm{L_{OBS}}$) and that necessary to justify the radio luminosity ($\rm{L_{INT}}$, intrinsic luminosity) provides the obscuration affecting the UV: 
\[\rm{A_{2800}=-2.5\times\log(L_{OBS}/L_{INT})}\]
Having this, we can now obtain E(B-V) knowing that:
\[\rm{A_{2800}=E(B-V)_{stellar}\times\textit{k}_{2800}}\]
where $\rm{E(B-V)_{stellar}=0.44\times}$E(B-V)$_{\rm gas}$ and $k_{2800}$ is the extinction coefficient at 2800\,\AA. This can be obtained from an equation like that provided in \citet{Calzetti00}:
\[k_{\lambda}=2.659\times(-2.156+\frac{1.509}{\lambda}-\frac{0.198}{\lambda^2}+\frac{0.011}{\lambda^3})+\rm{R_V}\]
with $\lambda=0.28\,{\mu}$m and $\rm{R_V}=4.05$, as the Absolute to Relative Attenuation Ratio. The extinction coefficient is estimated to be $k_{2800}=7.26$.

The results are presented in columns 2--4 of Table~\ref{c2tab:tabdust}. Average values of E(B-V)$\sim$0.5--0.6 are in agreement with the literature \citep[e.g.,][]{Cimatti02a,Bergstrom04,Papovich06,Georgakakis06}, although slightly lower. This owes to the fact that we are using rest-frame UV detected, thus biasing toward less obscured sources. Nonetheless, they already show significant dust content. In this sample \citep[as what happens in][]{Georgakakis06}, the highest level of obscuration is observed for a radio detected source: E(B-V)$\sim1$.

\ctable[
   cap     = The dust content of ERG populations,
   caption = The dust content of ERG populations,
   label   = c2tab:tabdust,
]{crrrr}{}{ \FL
POP & $\overline{\rm{SFR_{UV}}}$\tmark[a] & $\overline{\rm{SFR_{1.4\,GHz}}}$ & $\overline{\rm{A_{2800}}}$\tmark[b] & $\overline{{\rm E}(B-V)}$\tmark[c] \NN
& [$\rm{M_\odot\,yr^{-1}}$] & [$\rm{M_\odot\,yr^{-1}}$] & [AB] & [AB] \ML
\bf{K$_s$} \NN  
$z12$ &  4(2) & 6(3) & 0.9(1.3) & 0.1(0.2) \NN 
$z12;\rm{nAGN}$ &  4(2) & 6(4) & 0.9(1.3) & 0.1(0.2) \NN 
$z23$ &  9(4) & 57(49) & 2.2(2.1) & 0.3(0.3) \NN 
$z23;\rm{nAGN}$ &  10(5) & 32(27) & 1.5(1.7) & 0.2(0.2) \NN 
\bf{EROs} \NN  
$z12$ &  2(1) & 8(5) & 1.7(4.8) & 0.2(0.7) \NN 
$z12;\rm{nAGN}$ &  2(2) & 6(3) & 1.3(5.3) & 0.2(0.7) \NN 
$z23$ &  6(5) & 136(121) & 3.7(3.6) & 0.5(0.5) \NN 
$z23;\rm{nAGN}$ & 5(3) & $<$82(70) & $<$3.2(3.4) & $<$0.4(0.5) \NN 
\bf{IEROs} \NN  
$z12$ &  2(1) & 52(42) & 3.7(4.0) & 0.5(0.6) \NN 
$z12;\rm{nAGN}$ &  2(1) & 42(36) & 3.5(3.7) & 0.5(0.5) \NN 
$z23$ &  5(6) & 139(124) & 3.8(3.6) & 0.5(0.5) \NN 
$z23;\rm{nAGN}$ & 5(4) & $<$92(76) & $<$3.2(3.4) & $<$0.4(0.5) \NN 
\bf{DRGs} \NN  
$z12$ &  2(1) & 59(37) & 3.9(3.9) & 0.5(0.5) \NN 
$z12;\rm{nAGN}$ &  2(1) & 53(32) & 3.8(3.6) & 0.5(0.5) \NN 
$z23$ &  6(5) & 115(105) & 3.5(3.5) & 0.5(0.5) \NN 
$z23;\rm{nAGN}$ &  5(4) & $<$89(81) & $<$3.2(3.4) & $<$0.4(0.5) \LL\NN 
}

\ctable[
   caption = { },
   continued,
   nosuper,
   mincapwidth = 15cm
]{crrrr}{
  \tnote[Note.]{ --- The $z12$ and $z23$ abbreviations stand for $1\leq{z}<2$ and $2\leq{z}\leq3$, respectively.}
  \tnote[$^a$]{Using the conversion from \citet{Dahlen07}.}
  \tnote[$^b$]{Estimated directly from the comparison between UV and radio SFR estimates.}
  \tnote[$^c$]{Using the conversion from \citet{Calzetti00}.}
}{ \FL
POP & $\overline{\rm{SFR_{UV}}}$\tmark[a] & $\overline{\rm{SFR_{1.4\,GHz}}}$ & $\overline{\rm{A_{2800}}}$\tmark[b] & $\overline{{\rm E}(B-V)}$\tmark[c] \NN
& [$\rm{M_\odot\,yr^{-1}}$] & [$\rm{M_\odot\,yr^{-1}}$] & [AB] & [AB] \ML
\bf{cERGs} \NN 
$z12$ &  2(1) & 69(52) & 4.1(4.4) & 0.6(0.6) \NN 
$z12;\rm{nAGN}$ &  2(1) & 59(48) & 3.9(4.2) & 0.5(0.6) \NN 
$z23$ &  5(6) & 140(125) & 3.7(3.6) & 0.5(0.5) \NN 
$z23;\rm{nAGN}$ &  6(4) & $<$104(90) & $<$3.3(3.3) & $<$0.5(0.5) \NN 
\bf{pEROs} \NN 
$z12$ &  2(2) & $<$10(7) & $<$1.9(2.7) & $<$0.3(0.4) \NN 
$z12;\rm{nAGN}$ &  2(2) & $<$10(7) & $<$2.0(1.9) & $<$0.3(0.3) \NN 
$z23$ & \ldots & \ldots & \ldots & \ldots \NN 
$z23;\rm{nAGN}$ & \ldots & \ldots & \ldots & \ldots \NN 
\bf{pDRGs} \NN 
$z12$ & \ldots & \ldots & \ldots & \ldots \NN 
$z12;\rm{nAGN}$ & \ldots & \ldots & \ldots & \ldots \NN 
$z23$ & 6(2) & $<$176(174) & $<$4.0(3.8) & $<$0.5(0.5) \NN 
$z23;\rm{nAGN}$ & \ldots & \ldots & \ldots & \ldots \LL\NN
}

\subsection{Mass Functions}\label{c2sec:masssec}

ERGs are known to be among the most massive objects at high redshifts \citep[$>5\times10^{10}$\,M$_\odot$,][]{Georgakakis06,vanDokkum06}. However, not only mass estimates are very model dependent (models which improve with time), but previous work was based in shallower and/or less numerous samples and/or different redshift ranges \citep{vanDokkum06,Georgakakis06,Marchesini07,Grazian07}. Here, recent estimates for the FIREWORKS sample are considered in order to assess the mass distributions of these ERG populations and their contribution to the total galaxy $\rho_{\rm M}$ at high redshift. The mass estimates are those referred in \citet{Marchesini09}, and follow the prescription described in \citet{Wuyts07}. Briefly, \citep[][BC03]{Bruzual03} models were fitted to the observed optical-to-8\,$\mu$m SED with the HYPER$z$\footnote{http://webast.ast.obs-mip.fr/hyperz/} stellar population fitting code, version 1.1 \citep{Bolzonella00}. Different star formation histories (SFHs) were considered (single stellar population without dust, a constant star formation history with dust, and an exponentially declining SFH with an e-folding time-scale of 300 Myr with dust). $\rm{A_V}$ values ranged from 0 to 4 in step of 0.2\,mag, and the attenuation law of \citet{Calzetti00} is considered. In this work a Salpeter initial mass function\footnote{\citet{Marchesini09} adopt a pseudo-Kroupa IMF by scaling down the stellar masses by a factor of 1.6.} (IMF) is adopted for consistency with the work done in the previous sections. The values of the galaxy stellar mass consider the masses of living stars plus stellar remnants instead of the total mass of stars formed, thus discarding the mass returned to the ISM by evolved stars via stellar winds and supernova explosions. For a detailed study on the systematic uncertainties obtained by adopting different set of parameters and models see \citet{Muzzin09} and \citet{Marchesini09}. For example, using Charlot \& Bruzual (in preparation) models instead of BC03, differing on the treatment of TP-AGB stellar phase, the estimated galaxy masses are 25\% lighter \citep[$\rm{M_\odot[BC03]=0.75{\times}M_\odot[CB09]}$,][]{Muzzin09}. Mass estimates were considered only when obtained with reliable photometry (a pixel weight of w$>0.3$ from UV to 8\,$\mu$m) and if the source has less than 50\% probability to be associated with an unobscured AGN, as higher probabilities may imply a high contribution from non-stellar emission to the galaxy SED, thus resulting in misleading mass estimates.

\begin{figure}
  \begin{center}
    \includegraphics[width=0.9\columnwidth]{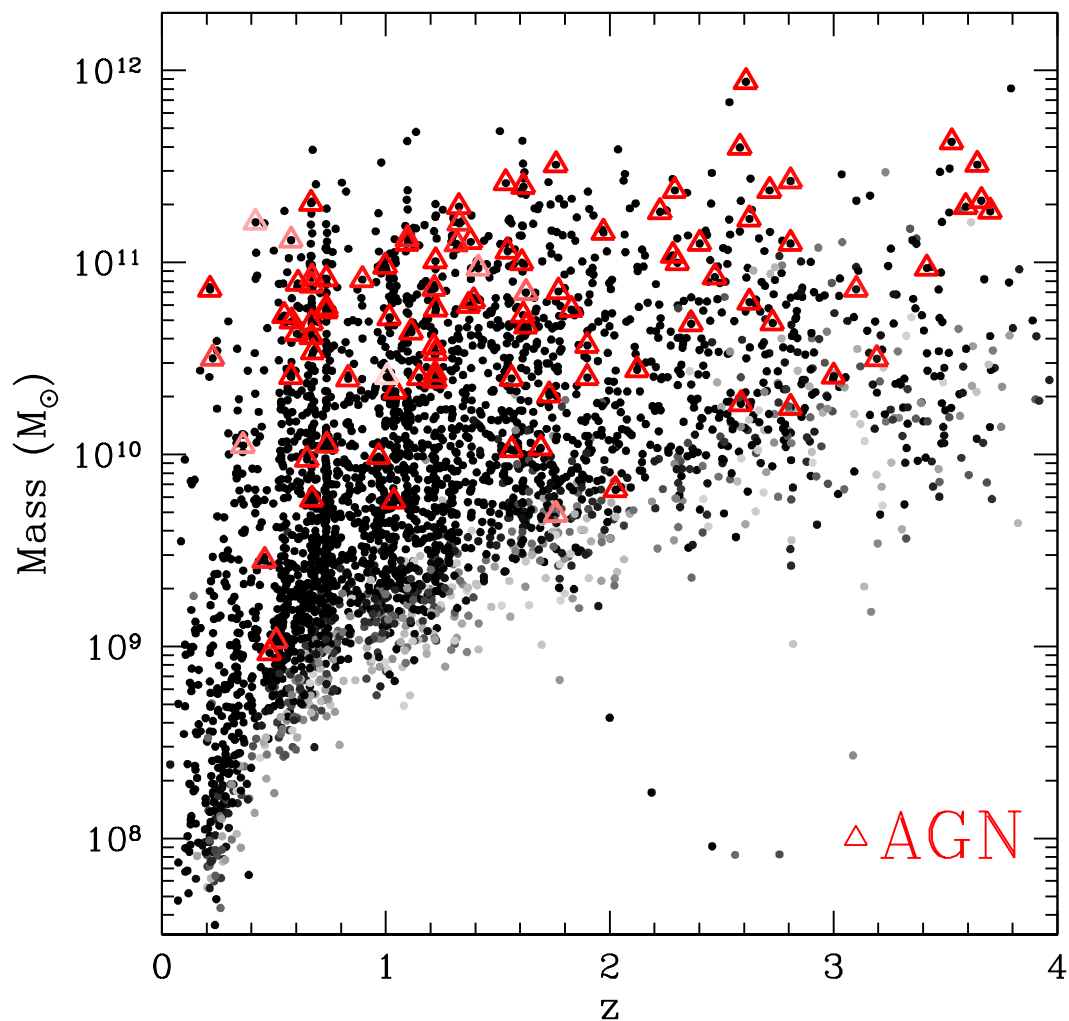}
  \end{center}
  \caption[$K_s$-sample $z$-mass distribution]{Distribution of sources in the $z$-mass space. X-ray identified AGN (as red triangles) are overlaid for reference, showing that they are mostly hosted by $2\times10^{10}$\,M$_\odot$ galaxies. The darker the points, the higher is the probability to have $K_s<23.8$, while the redder the triangle, the higher is the probability to be a source with an X-ray AGN with $K_s<23.8$ (similar Figure~\ref{c2fig:ki}).}
  \label{c2fig:zagnvsmass}
\end{figure}

Again, the ERGs are separated into sub-populations and redshift intervals, and AGN and non-AGN populations. The AGN/non-AGN separation is important. Although \citet{Marchesini09} stress that AGN IR emission does not significantly alter the mass estimates, their conclusion is based on a comparison with re-computed mass estimates without considering the 5.8 and 8.0\,$\mu$m IRAC channels. However, in an error-weighted SED fitting procedure, these channels will unavoidably count less due to their tendentiously higher photometric errors. Also, the higher number of optical filters, and their tendentiously smaller photometric errors, imply that the nIR and IR filters will tend to be less considered when compared to optical ones \citep[see][for a tentative correction]{Rodighiero10}. In Chapter~\ref{ch:lfs}, we show as well that, depending on the source redshift, $H$ to 4.5\,$\mu$m bands may also be affected by AGN emission. Although it may not produce a scatter in the stellar mass estimate, a dangerous upward scaling bias may happen. Finally, it is known that the fraction of AGN increases both with redshift and stellar mass \citep{Papovich06,Kriek07,Daddi07}. This is seen in Figures~\ref{c2fig:zagn} and \ref{c2fig:zagnvsmass}, where in the latter X-ray identified AGN tend to be hosted by $\gtrsim10^{10}$\,M$_\odot$ massive galaxies. As one considers higher redshifts, the sample is restricted to higher mass galaxies, hence producing an apparent rise in the AGN fraction of the sample (again supporting the high AGN hosts fractions of 25--40\% found for ERG populations).

\begin{figure}
  \begin{center}
    \includegraphics[width=0.9\columnwidth]{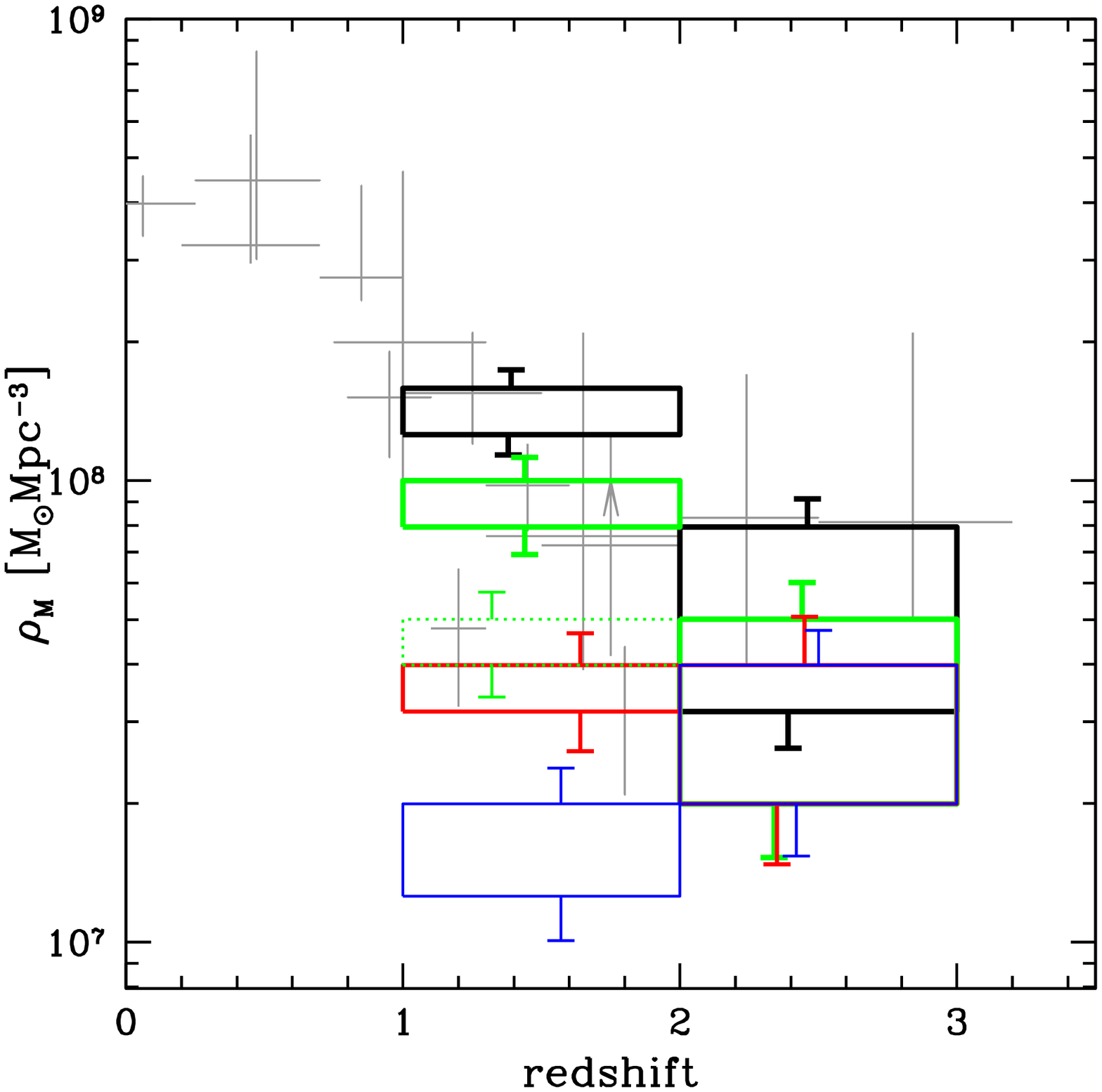}
  \end{center}
  \caption[ERGs mass densities with redshift]{Contribution of ERG populations to the total $\dot{\rho}_{\rm M}$ at $1\leq{z}<2$ and $2\leq{z}\leq3$. EROs are denoted by green boxes, IEROs by cyan boxes, and DRGs by blue boxes. Also, pEROs are denoted by the dotted green box at $1\leq~z<2$. The compilation from the literature \citep{Cole00,Fontana03,Fontana04,Glazebrook04} is displayed for reference (grey crosses).}
  \label{c2fig:massdens}
\end{figure}

Mass densities are obtained considering the 1/Vmax method as previously described. The results are presented in Table~\ref{c2tab:masstab}, and compared to the overall tendency observed in the universe in Figure~\ref{c2fig:massdens}. The $\rho_{\rm M}$ for the total $K_s$-selected sample are also estimated in the considered redshift intervals and are in agreement with those presented by \citet{Marchesini09}. ERGs may constitute up to 60--70\% of the total mass of the $1<z<3$ universe, although they represent only 25\% of the $1<z<3$ $K_s$ selected sample. The average and median mass estimates are roughly equal among all three ERG populations in the full $1\leq{z}\leq3$ range. At $1\leq{z}<2$, one can consider the ERO population to be complementary composed mainly by IEROs and pEROs. Figure~\ref{c2fig:massdens} shows that both populations comprise comparable $\rho_{\rm M}$ values, representing together $\sim60\%$ of the Universe mass at $1\leq{z}<2$. These mass densities estimates are in agreement with what is seen for $1<z<2$ EROs \citep{Georgakakis06} and $2<z<3$ DRGs \citep{vanDokkum06, Rudnick06,Grazian06}.

\ctable[
   cap     = Mass and sSFRs of the ERGs,
   caption = Mass and Specific SFRs of the Extremely Red Galaxy,
   label   = c2tab:masstab,
]{crrr}{}{ \FL
POP & $\log(\overline{\rm M})$\tmark[a] & $\log(\rho_{\rm M})$\tmark[b] & $\log(\overline{\rm sSFR})$\tmark[c] \NN
 & [M$_{\odot}$] & [M$_{\odot}$\,Mpc$^{-3}$] & [yr$^{-1}$] \ML
\bf{K$_s$} \NN
$z12$ &  10.1(10.6) & 8.1(8.2) & -9.3(-9.3) \NN
$z12;\rm{nAGN}$ &  10.1(10.6) & 8.1(8.1) & -9.3(-9.3) \NN
$z23$ &  10.4(10.4) & 7.8(7.9) & -8.6(-8.5) \NN
$z23;\rm{nAGN}$ &  10.3(10.4) & 7.5(7.5) & -8.8(-8.7) \NN
\bf{EROs} \NN
$z12$ &  10.7(11.1) & 7.9(8.0) & -9.9(-8.9) \NN
$z12;\rm{nAGN}$ &  10.7(11.1) & 7.9(7.9) & -10.0(-8.4) \NN
$z23$ &  11.0(11.1) & 7.6(7.7) & -8.8(-8.9) \NN
$z23;\rm{nAGN}$ &  11.0(11.1) & 7.3(7.3) & $<$-9.1(-9.0) \NN
\bf{IEROs} \NN
$z12$ &  10.8(11.3) & 7.6(7.6) & -9.1(-9.0) \NN
$z12;\rm{nAGN}$ &  10.8(11.5) & 7.5(7.5) & -9.2(-9.0) \NN
$z23$ &  11.0(11.4) & 7.6(7.6) & -8.9(-8.9) \NN
$z23;\rm{nAGN}$ &  11.1(11.1) & 7.3(7.3) & $<$-9.1(-9.1) \NN
\bf{DRGs} \NN
$z12$ &  10.7(11.1) & 7.2(7.3) & -9.0(-9.1) \NN
$z12;\rm{nAGN}$ &  10.7(11.1) & 7.1(7.1) & -9.0(-9.1) \NN
$z23$ &  10.9(11.0) & 7.6(7.6) & -8.9(-8.9) \NN
$z23;\rm{nAGN}$ &  11.0(11.1) & 7.3(7.3) & $<$-9.0(-9.0) \LL\NN
}

\ctable[
   caption = { },
   continued,
   nosuper,
   mincapwidth = 16cm
]{crrr}{
  \tnote[Note.]{ --- The $z12$ and $z23$ abbreviations stand for $1\leq{z}<2$ and $2\leq{z}\leq3$, respectively. The upper limits for Luminosity and SFR estimates, whenever $\rm{N_{MC}}>0$ (Table~\ref{c2tab:tabsta}), are calculated considering the maximum S/N obtained in the respective set of MC simulations.}
  \tnote[$^a$]{The number in parenthesis indicates the median value. Errors at the $1\sigma$ level reach 0.2--0.4.}
  \tnote[$^b$]{The number in parenthesis indicate the estimates when accounting for the radio detected sources. Errors are at the 0.1--0.2 level and account for cosmic variation).}
  \tnote[$^c$]{In parenthesis, the median value also taking into account radio detections ($>3\sigma$) excluded from the stacking procedure (see Section~\ref{c2sec:stack}).}
}{ \FL
POP & $\log(\overline{\rm M})$\tmark[a] & $\log(\rho_{\rm M})$\tmark[b] & $\log(\overline{\rm sSFR})$\tmark[c] \NN
 & [M$_{\odot}$] & [M$_{\odot}$\,Mpc$^{-3}$] & [yr$^{-1}$] \ML
\bf{cERGs} \NN
$z12$ &  10.9(11.2) & 7.1(7.2) & -9.0(-9.0) \NN
$z12;\rm{nAGN}$ &  10.8(11.1) & 7.0(7.1) & -9.0(-9.1) \NN
$z23$ &  11.1(11.1) & 7.5(7.5) & -8.9(-9.0) \NN
$z23;\rm{nAGN}$ &  11.1(11.2) & 7.2(7.2) & $<$-9.1(-9.1) \NN
\bf{pEROs} \NN
$z12$ &  10.7(11.1) & 7.7(7.7) & $<$-9.8(-9.6) \NN
$z12;\rm{nAGN}$ &  10.7(11.1) & 7.6(7.6) & $<$-9.75(-9.8) \NN
$z23$ &  10.7(11.0) & 6.2(6.2) & \ldots \NN
$z23;\rm{nAGN}$ &  10.6(10.7) & 5.8(5.9) & \ldots \NN
\bf{pDRGs} \NN
$z12$ &  9.7(9.6) & 5.1(5.1) & \ldots \NN
$z12;\rm{nAGN}$ &  9.5(9.6) & 4.5(4.5) & \ldots \NN
$z23$ &  10.6(10.8) & 6.5(6.5) & $<$-8.4(-8.4) \NN
$z23;\rm{nAGN}$ &  10.5(10.7) & 6.0(6.0) & \ldots \LL\NN
}

\begin{figure}
  \begin{center}
    \includegraphics[width=0.9\columnwidth]{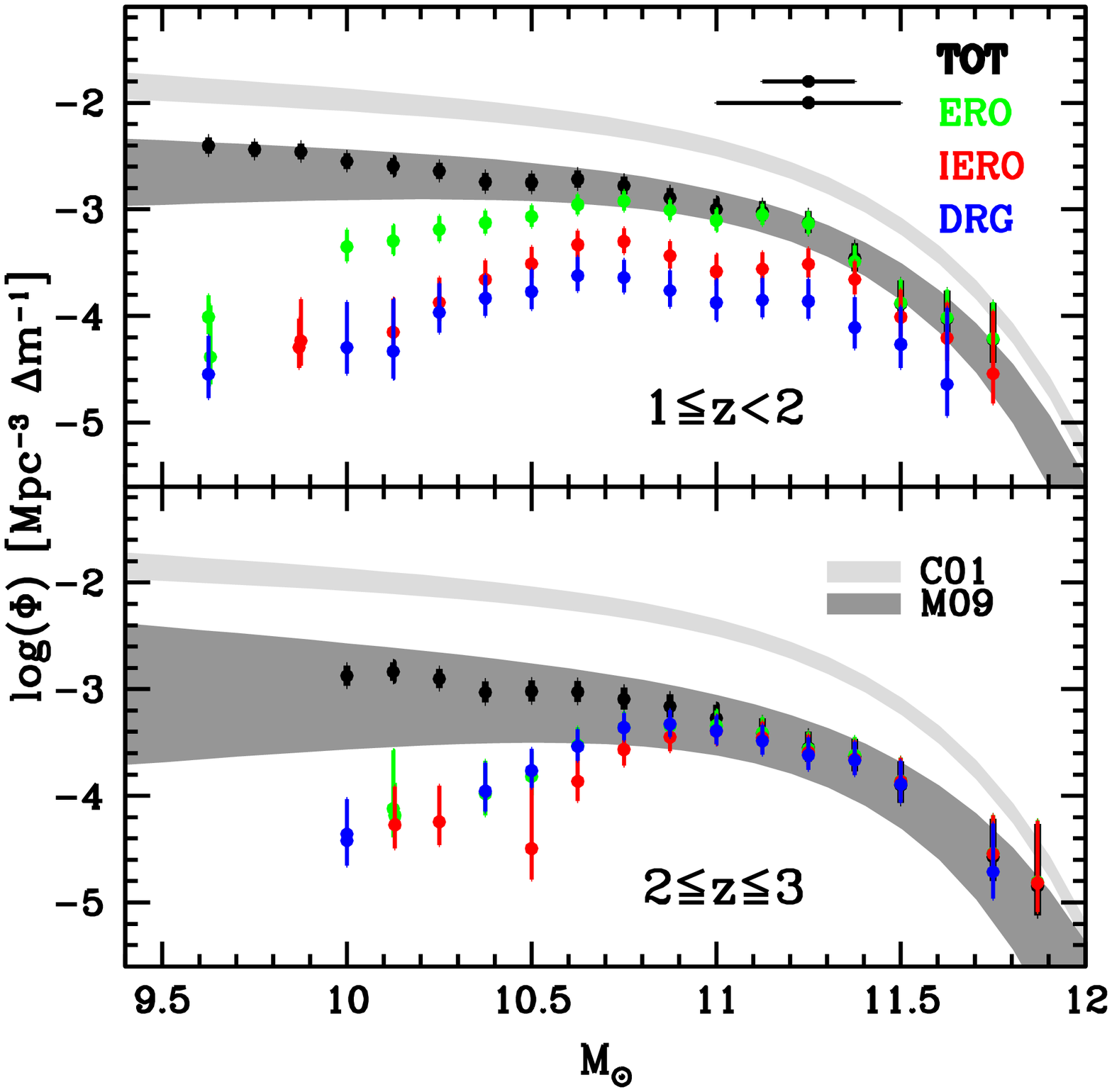}
  \end{center}
  \caption[ERGs mass functions with redshift]{Mass functions for the $K_s$ population (in black), EROs (green), IEROs (red), and DRGs (blue), at $1\leq~z~<2$ (upper panel) and $2\leq~z~\leq3$ (lower panel). A moving bin of width 0.25\,$\log$(M$_\odot$) was used with steps of 0.125\,$\log$(M$_\odot$). Whenever the number of sources in each bin was small, a large bin was used (0.5\,$\log$(M$_\odot$)). The bin widths are show at the top. The shaded regions are mass functions derived from \citet[][at $z\sim0.1$]{Cole01} and \citet[][at $1.3<z<2$ and $2<z<3$]{Marchesini09}, respectively, light-grey and dark-grey regions.}
  \label{c2fig:massf}
\end{figure}

Figure~\ref{c2fig:massf} shows the mass functions (MFs) of the overall $K_s$-selected galaxy population and for the different ERG populations. On the overall $K_s$-selected MF, one can distinguish a dip at $\log(\rm{M/M_\odot})\sim10.4$ referred in the literature at $z<1$ \citep[e.g.,][]{Drory09,Pozzetti10}. This confirms that the feature is present at even higher redshifts. Also, there seems to exist another dip at even higher masses ($\log(\rm{M/M_\odot})\sim11$), and it seems to be stronger for IEROs and DRGs, although this result is still preliminary. Comparing to what we see at lower redshifts \citep[e.g.,][and Chapter~\ref{ch:lfs}]{Drory09,Pozzetti10}, where both early and late type galaxies contribute at a comparable level to the high mass end, a possible interpretation may be the on-going mass build up of late type galaxies (comprising dusty starbursts), while a significant amount of early type galaxies has already reached an advanced stage of mass build up (producing the second hump).

\begin{figure}
  \begin{center}
    \includegraphics[width=0.6\columnwidth]{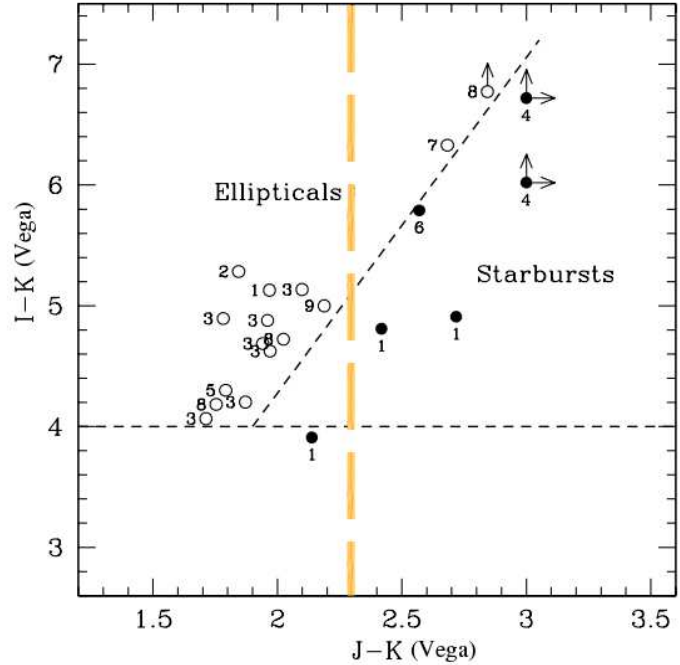}
  \end{center}
  \caption[$I-K$ versus $J-K$: disentangling old and passive populations]{The original Figure~4 from \citet{PozzettiMannucci00} with a vertical dashed line over-plotted representing the adopted DRG cut in this thesis. Note the colour axis are in Vega system. The region referred in the text is the triangle to the left of the long dashed line.}
  \label{c2fig:pm00}
\end{figure}

Note that at the highest mass bins (M$>10^{11}$\,M$_\odot$), the contribution of ERGs to the overall mass densities reaches 100\%. There are evidences for an evolutionary trend. Note that, while at high-$z$, all three populations equally dominate the high mass bins (due to the overlap between them), at low-$z$, only the EROs maintain their strong contribution to the total mass function of $K_s$ selected sources. However, all three present comparable stellar masses (Table~\ref{c2tab:masstab}). The reader should recall that low-$z$ DRGs and low-$z$ IEROs are also mostly classified as EROs. What is likely to be happening is that part of the star-forming population seen at high redshifts and selected by all three criteria, have extinguished their fuel at low-$z$ and are gradually missed by the IERO and DRG criteria\footnote{The DRG criterion is even more affected because, at $z>2$, it relies on the 4000\,\AA\ break being redshifted into the spectral range between $J$ and K$_s$ bands, which, of course, does not happen at $z<2$.}, turning into passive evolved systems (becoming pEROs). The remainder still have some obscured star-formation happening, producing the characteristic red colours in all three ERG criteria, enabling the selection as IEROs and/or DRGs. This effect was first explored at $z<2$ by \citet{PozzettiMannucci00} using an $i-K$ versus $J-K$ colour-colour space to separate early-type galaxies from dusty starbursts. However, they found necessary to have a diagonal cut extending the selection of dusty starbursts to bluer $J-K$ colours (Figure~\ref{c2fig:pm00}), which \citet{Georgakakis06} confirmed \citep[and also][through a similar $R-K$ versus $J-K$ diagram, but see \citealt{Pierini04}]{Mannucci02,Cimatti03}. Hence, some star-forming systems are expected to be included in the pERO population, yet, from the radio estimates, they are not dominant in the pERO population. The next section will reveal more details on this subject.



\subsection{Morphology} \label{c2sec:morphsec}

Galaxy morphology can be an efficient way to break the photometric degeneracy between the passively evolved and dusty starburst populations. This field of research has improved enormously, specially with the observations provided by the 20-year old \textit{HST}. Its high resolution and sensitivity enabled the scientific community to improve our knowledge of galaxy evolution through means of morphology studies up to high redshifts. Still, it is a field where a lot is still left to be discovered. In a time where the number of galaxies per survey reaches millions, the community turned its efforts to develop morphological (non-)parametric criteria, avoiding the time-consuming and subjective visual inspection. A set of morphological criteria, frequently used by the astronomy community, rely on the Concentration, Asymmetry, and Smoothness \citep[CAS,][]{Conselice03}, Gini and M$_{20}$ \citep{Lotz06} coefficients \citep[see][for a comparison between these parameters]{Lotz04,Conselice08}.

Many are the combinations between the five coefficients which are believed to separate different types of galaxy systems, from early-type galaxies to highly disturbed systems. However, for this high-$z$ ERG sample, the simple criterion applied to $z\sim1.5$ and $z\sim4$ galaxy samples by \citet{Lotz06} will be the one adopted: M$_{20}>-1.1$ for merger candidates, and M$_{20}<-1.8$ and G$>5.7$ for bulge dominated galaxies. The second-order moment values of the 20\% brightest galaxy pixels, M$_{20}$, owes its name to the way it is computed: it is the product between the flux and the squared distance (to galaxy centre) of the 20\% brightest pixels in a galaxy light profile \citep[normalised by the second-order moment for the entire, 100\%, galaxy pixels,][]{Lotz04}. Hence, M$_{20}$ traces any off-centre bright distributions in a galaxy profile (either bright star-formation clumps, bars, spiral arms, or star clusters). In the high redshift universe, high M$_{20}$ values are thus expected to be related to star-formation clumps, features taken as a hint of merger activity. The Gini index, G, has its origins in demographic studies to provide the degree of wealth distribution within a population. Smaller Gini values indicate a more uneven distribution of pixels \citep{Lotz04}.

The morphology code used in this work was that developed by \citet{Lotz04,Lotz06} (the reader is referred to these works to fully understand the concepts adopted ahead). The images considered for the study are those from the latest Great Observatories Origin Deep Survey South (GOODSs) ACS-\textit{HST} release (v1.9). The total drizzled image was divided into overlapping cells to avoid the loss of galaxies at the boundaries. The value adopted for those galaxies with more than one measurement were the estimates with the best signal-to-noise ratio (S/N). SExtractor was used to provide the segmentation files and the input catalogues to the code. Only sources with a signal to noise (S/N) detection of $S/N>2.5$ and effective radius ($\rm{R_{eff}}$) of $\rm{R_{eff}}>2\times$FWHM (Full Width at Half Maximum) are considered for the morphology study, as done by \citet{Lotz04,Lotz06}.

\begin{figure}
  \begin{center}
    \includegraphics[angle=-90,width=0.9\columnwidth]{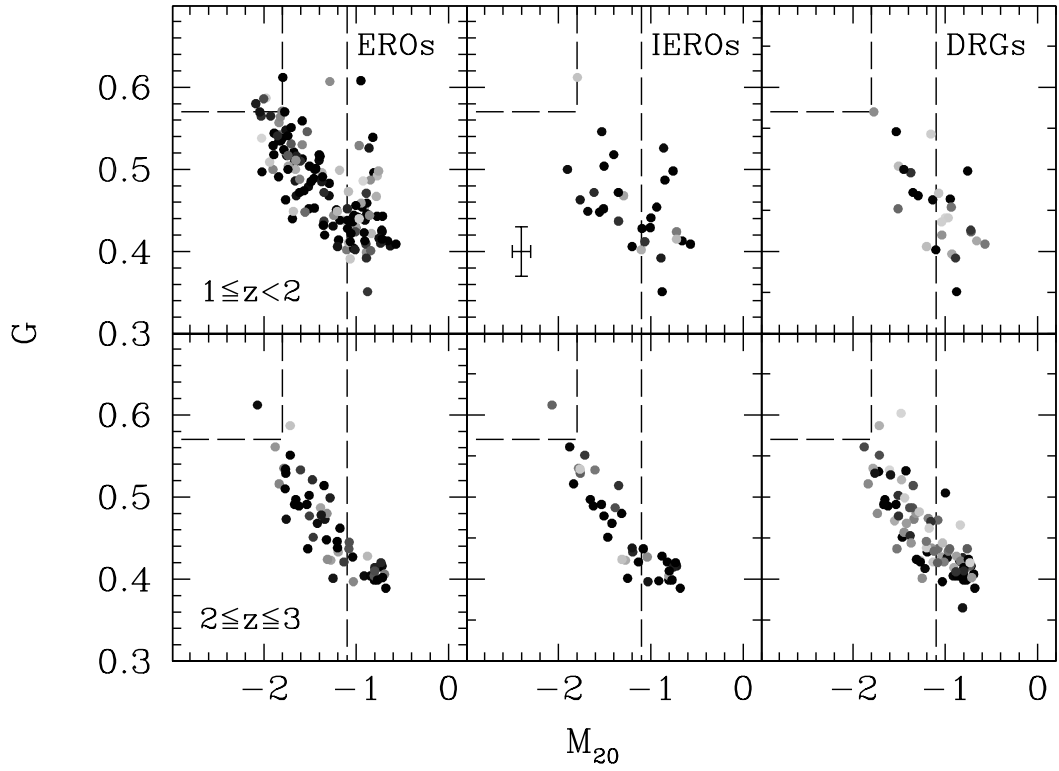}
  \end{center}
  \caption[ERGs on the Gini-M$_{20}$ space]{Distribution of sources in the Gini-M$_{20}$ space. The panels refer to different redshift ranges (upper panels for $1\leq~z<2$ and lower panels for $2\leq~z~\leq3$), and different populations (EROs on the left-hand side, IEROs in the middle, and DRGs on the right-hand side). The error-bars in the top middle panel show typical errors for a source with S/N=2.5. The darker the point, higher is the source probability (see Figure~\ref{c2fig:ki}).}
  \label{c2fig:gm20geral}
\end{figure}

\begin{figure}
  \begin{center}
    \includegraphics[angle=-90,width=0.9\columnwidth]{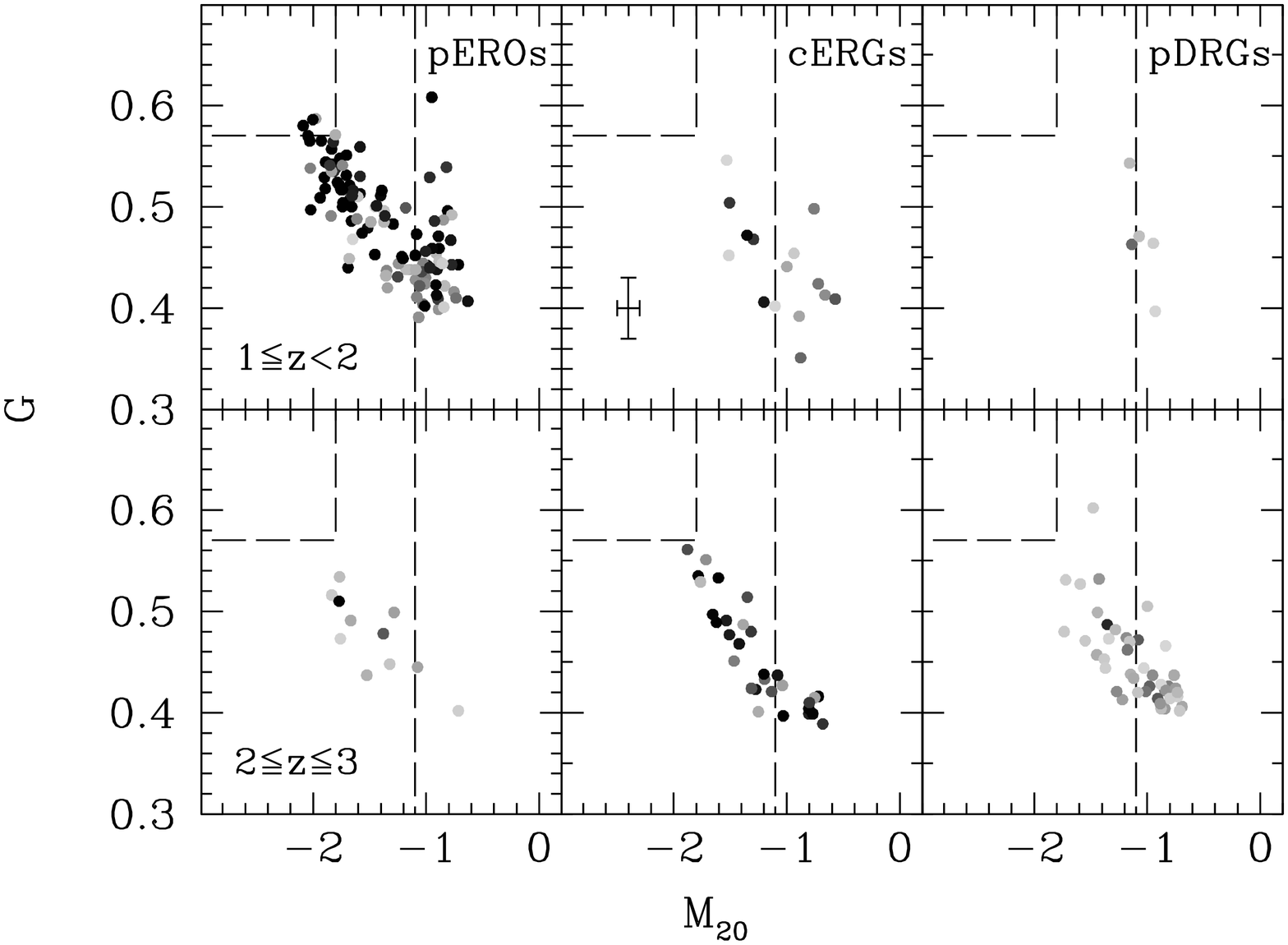}
  \end{center}
  \caption[Pure and common ERGs on the Gini-M$_{20}$ space]{Distribution of sources in the Gini-M$_{20}$ space. The panels refer to different redshift ranges (upper panels for $1\leq~z<2$ and lower panels for $2\leq~z~\leq3$), and different populations (pEROs on the left-hand side, cERGs in the middle, and pDRGs on the right-hand side). The error-bars in the top middle panel show typical errors for a source with S/N=2.5. The darker the point, higher is the source probability (see Figure~\ref{c2fig:ki}).}
  \label{c2fig:gm20purecom}
\end{figure}

\begin{figure}
  \begin{center}
    \includegraphics[width=0.8\columnwidth]{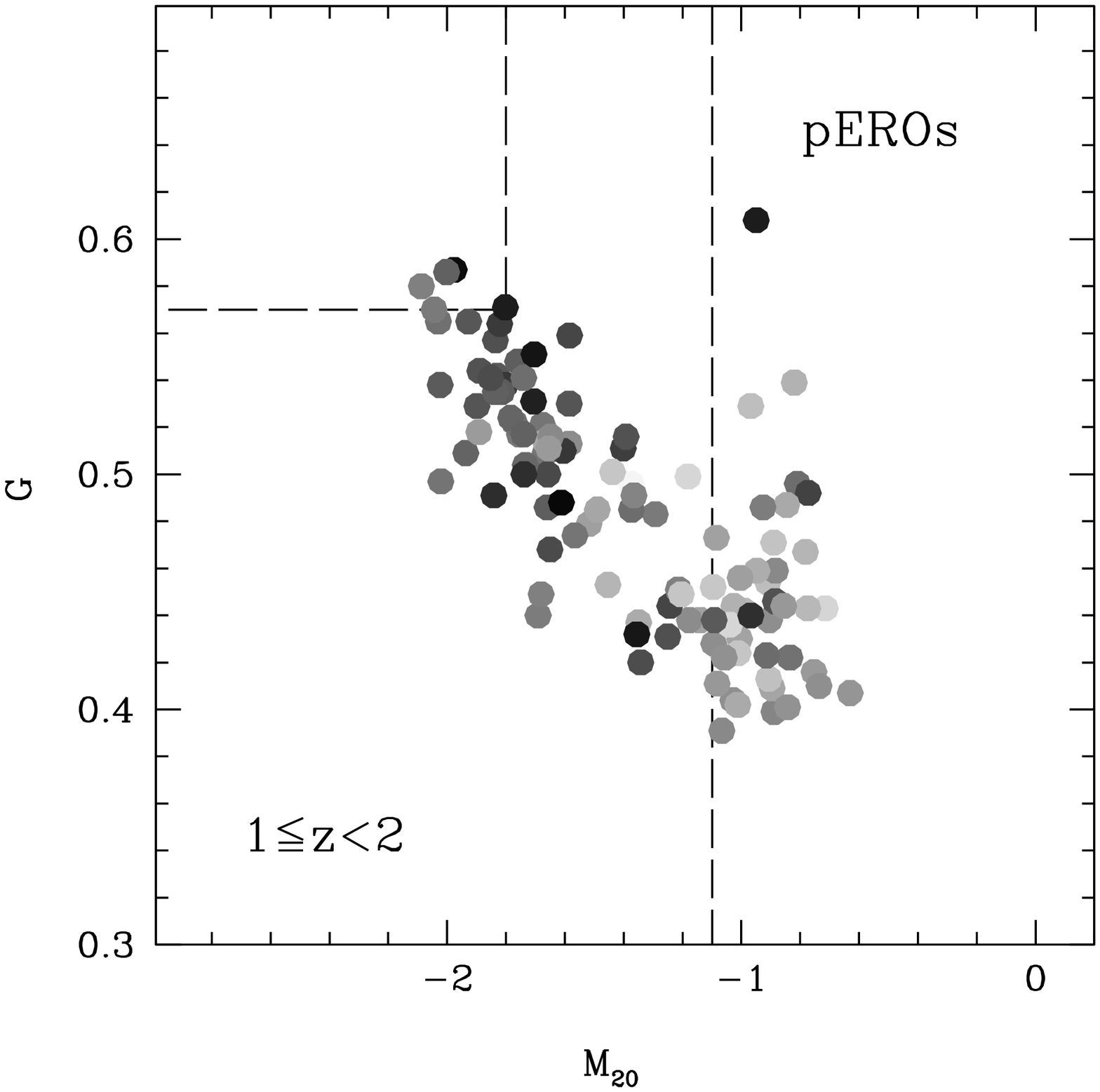}
  \end{center}
  \caption[pEROs on the Gini-M$_{20}$ space: dependency on $J-K$]{Pure-EROs in the Gini-M$_{20}$ space. Point intensity indicates how red in $J-K$ a given source is, where lighter dots mean $J-K$ colours close to verify the DRG criterion (similar to Figure~\ref{c2fig:ki}). It is visible the gradual change toward less red colours from the lower right to the upper left.}
  \label{c2fig:gm20pero}
\end{figure}

In order for a fair comparison between the lower and upper redshift intervals, two bands are considered to constraint the same observed rest-frame wavelength: the $V_{606}$ band for the $1\leq{z}<2$ redshift bin and $z_{850}$ band for the $2\leq{z}\leq3$ redshift bin\footnote{This redshift--band combination is based on the rest-frame $\sim$2800\,\AA\ wavelength being redshifted into these specific bands at these redshift intervals.}. Figures~\ref{c2fig:gm20geral} and \ref{c2fig:gm20purecom} show the Gini-M$_{20}$ space and the difference between low and high redshift sources in each of the ERG populations. One of the main features to point out is the ERO distribution, clearly showing a wide range of values in both redshift ranges. Note that, at low redshift, there is a significant part of the ERO population close to or in the upper left region \citep[reserved for bulge dominated sources,][]{Lotz06}, whereas IEROs and DRGs fall at higher M$_{20}$ with fractions of almost 50\% inside the region where merger candidates are expected \citep[M$_{20}$>-1.1,][]{Lotz06}. Again, the strong similarity between all samples is observed at the highest redshifts. This figure strongly supports the scenario proposed in the previous section, where part of the high redshift population (seen here with smaller M$_{20}$ and higher Gini values) becomes less active, thus becoming pEROs at low redshifts. The fact that the Gini-M$_{20}$ values of pEROs are comparable to those for the high-$z$ ERG population also agrees with recent studies defending the presence of already settled early-type galaxies at $z\sim2$ or higher \citep[e.g.,][]{Pozzetti03,Papovich06,vanDokkum06,Wuyts09b,Marchesini09}. Note, however, from Figure~\ref{c2fig:gm20purecom} that not all the pERO population is strongly bulge dominated, as expected from the discussion at the end of the previous section. Yet the presence of pEROs is stronger close to the upper left region (and the radio data reveals a SFR upper limit of the order of unity). Furthermore, Figure~\ref{c2fig:gm20pero} shows there is a gradual increase in $J-K$ colours for sources with increasing M$_{20}$ value, probably meaning higher obscuration and star-formation. Again, this points to an evolutionary scenario: extremely red systems at high redshifts present significant SFRs and already spheroidal type morphologies, but fuel is exhausted as they evolve to $z\sim1$ and they become more passive, thus being gradually missed by the IERO and DRG criteria.

In an attempt to link the results inferred from radio SFRs, mass estimates, and morphology in this section, we propose that practically all ERGs comprise the same population, but seen in different evolution stages. This has been proposed before. In the review by \citet{McCarthy04}, for instance, evidences are presented for a link between high redshift DRGs to low redshift EROs, and refer sub-millimetre galaxies (SMGs) as the probable extreme star forming ancestors of evolved ERGs at $z\sim1$ \citep[see also][]{McCarthy04b}. ERGs in general are known to be found in dense environments \citep[e.g.,][]{Georgakakis05,Kim11}, with a higher degree of clustering for the passive-evolved component \citep[e.g.,][]{Daddi02,Roche02,Foucaud07,Kong09}. The reader should recall that, the more massive the host dark matter halo is, the sooner baryonic mass is expected to collapse \citep{Baugh99,Tanaka05,deLucia06,Neistein06}. Hence, as one probes lower density fields more likely it is to find younger and bursty galaxies, as the baryonic mass assembly started later than in galaxies found in denser regions. For instance, \citet{Roche02} believe that both passive-evolved and dusty starburst ERG components end as old ellipticals in the local Universe, and \citet{Fontanot10} find no passively evolved versus dusty starburst population bimodality in EROs. Also, \citet{Bussmann09} and \citet{Narayanan10} propose that dusty obscured galaxies (DOGs), in case a merger scenario is considered, are candidates for galaxies in an evolution phase between SMGs and quiescent DRGs \citep[but see][for an alternative Lyman break galaxy origins scenario]{Shapley05,Stark09}. Adding to that, an X-ray analysis reveals comparable obscured AGN fractions for SMGs and DRGs when considering the most X-ray luminous sources (Section~\ref{c2sec:xr}). Knowing that the QSO duty-cycle is expected to be short \citep[e.g.,][and Section~\ref{c2sec:agncont} in this work]{Hopkins06}, in order for such property to hold, the transition between the SMG and DRG phases may actually be quite fast. There is evidence that support this scenario, where SMGs are believe to be rapid, highly dissipative, gas-rich major mergers \citep{Narayanan09} with short-lived ($\sim1$\,Gyr) starbursts \citep{Tacconi06,Tacconi08}.


\subsubsection{The case of pDRGs} \label{c2sec:pdrgs}

Most pure-DRGs (85\%) are found in a very interesting epoch of the universe, when both star-formation and AGN activity peak \citep{Osmer04,HopkinsBeacom06,Richards06,Hopkins07}, $2<z<4$. In the literature, one can find scarce indirect references to this population \citep[][who also study those DRGs with bluer rest-frame $U-V$ colours]{Forster04,Wuyts07}, sometimes even regarded as a result of photometric errors \citep{Papovich06}. \citet{Wuyts07} refer these galaxies as the least massive among the DRGs.

The colours and redshift distribution of pDRGs imply an evolved stellar population (bands $J$ and $K$ straddle the 4000\AA\ break) and an excess of flux at rest-frame ultra-violet (based on the less extreme red $i-K_s$ or $z-[3.6]$ colours). Such colours can be produced either by exponentially decaying or constant star-formation histories. While the former applies to passively evolving galaxies, the latter scenario is considered for merging systems \citep{Forster04,Wuyts09a,Wuyts09b}. This calls for a morphological inspection in order to break such degeneracy.

\begin{figure}[t]
  \begin{center}
    \includegraphics[width=0.7\columnwidth]{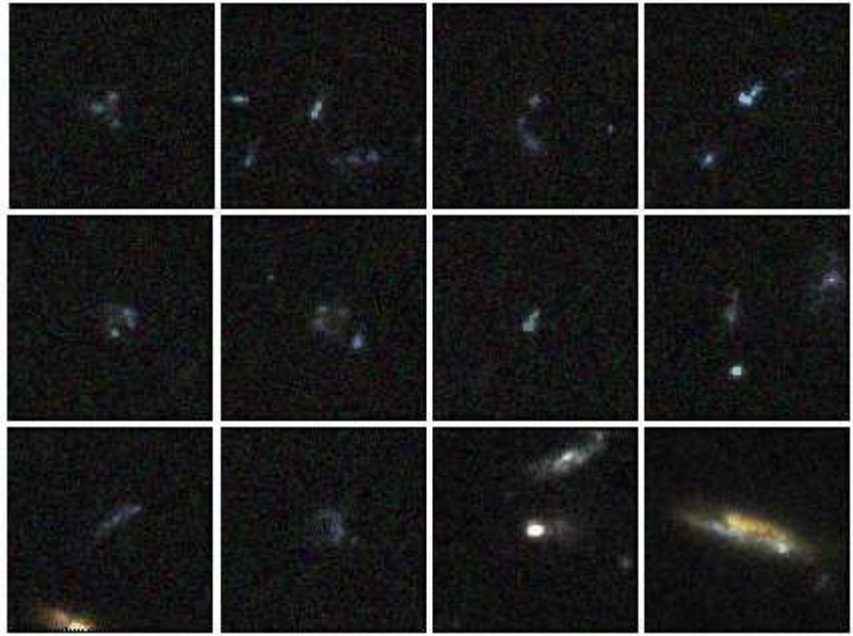}
  \end{center}
  \caption[ACS imaging of pDRGs]{Twelve examples of pDRGs found in GOODSs. The cut-outs are taken from the MAST cut-out service and are 5'' wide ACS-$Viz$ band combinations. The galaxy seen in the bottom right corner is at $z_{spec}=0.5$, while all the remainder are at $2<z<4$.}
  \label{c2fig:pdrgacs}
\end{figure}

To maximise the statistics in this section, the requirement for a pDRG to be non-IERO is discarded\footnote{The reader should recall that most IEROs are also EROs, so this is a plausible assumption.}, the magnitude cut is extended to fainter fluxes, down to the catalogue limiting magnitude of $K_s=24.3$, and the $2<z<4$ redshift range is considered. There are 237 DRGs found in this way, 88 of which are pDRGs. The total $K_s$ population amounts to 894 sources under these constraints. All 88 pDRGs were visually inspected. Figure~\ref{c2fig:pdrgacs} shows a few examples of the selected pDRGs with disturbed light profiles. The example on the lower right corner is a $z_{\rm spec}=0.5$ galaxy that clearly shows why the pDRG selection can identify galaxies at such low redshifts. Due to its disturbed morphology, some star-forming regions of the galaxy are not obscured by dust (producing the UV excess), while the rest of the galaxy is strongly obscured by a dust lane originating red $J-K$ colours. The source to its left is at $z_{\rm spec}=2.2$, showing an extended low surface brightness feature to the right (West). This is the source seen in Figure~\ref{c2fig:gm20purecom} with the highest Gini coefficient (and M$_{20}<-1.1$) in the upper redshift bin of pDRGs. Interestingly, a proper flux contrast scale reveals two nuclei separated by 0.1''. This reveals that automated algorithms will not always select merger candidates, and visual inspection should be pursued whenever possible.

However, are pDRGs more disturbed than the remainder galaxy population at $2<z<4$? To allow for a fair comparison, three complementary samples are considered: pDRGs, DRGs not pDRGs, and $K_s$-selected non-DRG sources. Considering the above mentioned M$_{20}>-1.1$ cut to select merging system candidates (estimated in the $i_{775}$ band when at $2<z<3$ and in the $z_{850}$ band when at $3<z<4$), one obtains fractions of 31\%, 32\%, and 33\% for non-DRG, DRGs not pDRGs, and pDRG samples, respectively. Although the conclusion is that pDRGs are as disturbed as the remainder galaxy population, the M$_{20}$ parameter shows a consistent value in agreement in recent estimates using other morphology criteria. In Figure 10 by \citet{ConseliceArnold09}, the general trend for the expected evolution of galaxy merger fraction with redshift (considering estimates based on both visual and non-parametric classifications) shows a peak at $z\sim3$ and at the 30\% level.

\begin{figure}
  \begin{center}
    \includegraphics[width=0.75\columnwidth]{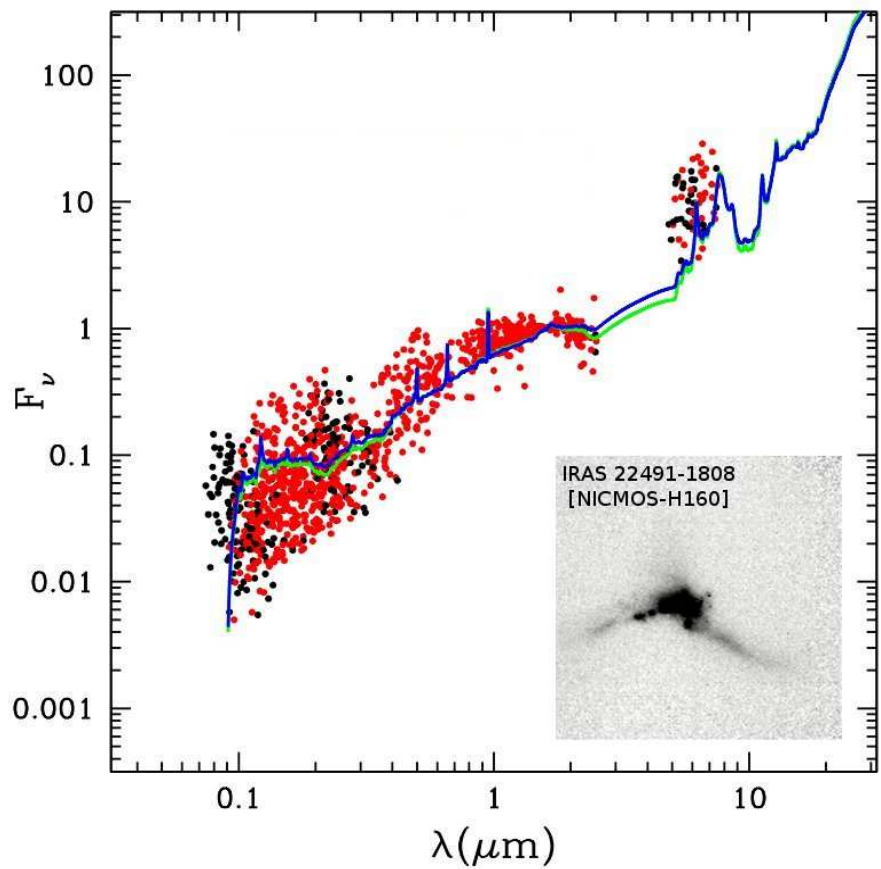}
  \end{center}
  \caption[SED fitting to pDRG photometry]{The data points are rest-frame photometry normalised at 1.6$\mu$m (black dots indicate upper-limits). The two resulting best $\chi^2$ fits are based in a template of IRAS 22491-1808 with a contribution of 20\% (green) and 30\% (blue) from AGN emission. A NICMOS-$H160$ image of IRAS 22491-1808 \citep{Scoville00} is displayed in the lower right corner.}
  \label{c2fig:rffit}
\end{figure}

\begin{figure}
  \begin{center}
    \includegraphics[width=0.75\columnwidth]{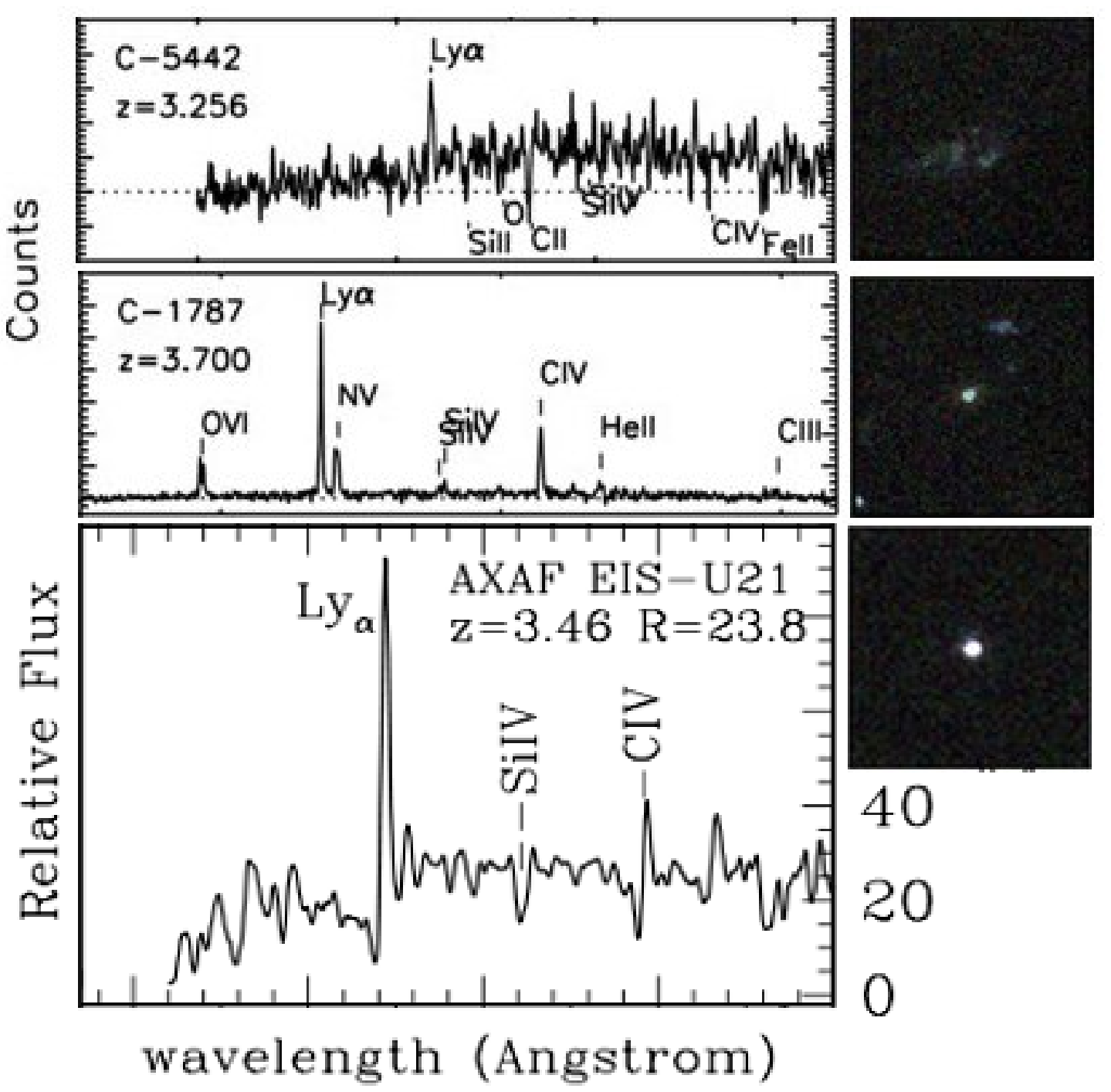}
  \end{center}
  \caption[Spectroscopy of AGN pDRGs]{Three examples of candidate AGN spectra found in the pDRG sample. The top one is the FeLoBAL candidate, the middle panel shows the type-2 QSO from \citet{Norman02}, while the bottom panel shows the evident P-cygni profiled emission lines of AXAF EIS-U21 \citep{Cristiani00}. To the right of each spectrum, are the respective ACS-$Viz$ stacked imaging (5'' wide) from MAST cut-out service.}
  \label{c2fig:specpdrg}
\end{figure}

One last procedure is used to confirm the active nature of the pDRG population. In Figure~\ref{c2fig:rffit}, the best $\chi^2$ fit to the photometric data of pDRGs wavelengths is presented. Two type of fits are shown: Fit-1 considers the optical-nIR-IR data and Fit-2 the nIR-IR-MIPS$_{24\mu\rm{m}}$ data. Among an extremely rich template library (the same used in Chapter~\ref{ch:agn}), the two best fits are hybrids based in the SED of IRAS 22491-1808. The overlaid image stamp \citep{Scoville00} shows the morphology of this local ULIRG to be characteristic of a merger system. It should be pointed out that the AGN contribution in each template is of the order of 20\% (Fit-1) to 30\% (Fit-2). This is in agreement with the literature, which hints to a strong co-existence of star-forming and AGN activity at such high redshifts \citep[e.g.,][]{Hopkins07,Lotz08}. In fact, five X-ray detections are observed, with an average X-ray luminosity of $\log(\rm{L_X\,[erg\,s^{-1}]})\sim43.8$ and column density $\log(\rm{N_H\,[cm^{-2}]})\sim23$\footnote{However, the two sources with the highest source $\mathcal{P}$ are distinct. One \citep[that referred by][]{Norman02}, with 100\% $\mathcal{P}$, has estimated intrinsic $\log(\rm{L_X\,[erg\,s^{-1}]})\sim44.8$ and $\log(\rm{N_H\,[cm^{-2}]})\sim24$ \citep[in agreement with][]{Norman02}. The other, with 96\% $\mathcal{P}$, has $\log(\rm{L_X\,[erg\,s^{-1}]})\sim43.4$ and $\log(\rm{N_H\,[cm^{-2}]})<20$.}. Also, from the available spectroscopy (seven sources), two narrow line AGN are found. Nominally, one is the type-2 QSO announced by \citet{Norman02} as the farthest object of such type found at the time, another shows P-cygni profile emission lines (characteristic of expanding shells of material). A third object (classified as AGN in the X-rays) is a candidate for a Fe Low-ionization broad absorption line system \citep[FeLoBAL,][]{Gregg02,Hall02,Farrah07,Farrah10}. This population of galaxies, is maybe transiting between AGN and star-burst dominated phases. The three spectra are displayed in Figure~\ref{c2fig:specpdrg}, together with the respective optical image cut-out, showing that compact systems do appear in the pDRG population. The obscured nature of these AGN hosts reinforces the idea that the UV flux comes mostly from star-formation processes.

These results thus support that pDRGs are an appropriate population for the study of the co-existence of star-forming regions with AGN activity in the epoch of greatest activity in the universe. We aim to assess questions such as ``Which came first? The starburst or the AGN phase?'', ``What is the mechanism behind such transition?'', and ``What is the time range for the transition from a starburst to AGN dominated phase or vice-versa?''. In order to do so, the spectral coverage of the pDRGs must be improved (currently $\sim$10\% of the population is found to have a measured spectrum). For this purpose, an observational proposal has been recently submitted to FORS2 at the VLT-UT1 telescope.

\section{Conclusions} \label{c2sec:conc}

We have presented a multi-wavelength analysis of the properties of the ERG population in the GOODSs field. EROs, IEROs, DRGs -- and various combinations between these groups -- are considered, their AGN content identified and their contribution to the global $\dot{\rho}_{\ast}$ and $\rho_{\rm M}$ estimated. A new approach is adopted where each source contribution is weighted upon the uncertainties of the estimated parameters (e.g., photometric redshifts, fluxes). All this, together with the estimated masses and rest-frame UV morphologies, leads to the following conclusions: 

\begin{itemize}

\item the different criteria for the selection of red galaxies select, as previously known, sources at different redshift ranges: while the bulk of EROs and IEROs can be found at $1<z<2$, DRGs are mostly found at $2<z<3$. Different combinations of the three criteria result in samples with distinct redshift properties: while cERGs are observed in a wide redshift range, $1<z<3$, and have no low-$z$ ($z<1$) interlopers, pEROs and pDRGs appear in distinct redshift intervals, at $1<z<2$ and $2<z<4$, respectively. The ``pure'' criteria appear, thus, to result suitable and simple techniques to select high-$z$ sources in well constrained redshift intervals. See Section~\ref{c2sec:reddis}.

\item the ERG population does not include a large number of {\emph {powerful}} AGN, as indicated by the X-rays and radio observations. One fourth of the ERG sample hosts potential AGN activity, with the fraction of AGN increasing from EROs to IEROs to DRGs (resp., 23\%, 33\%, and 39\%). Among ERGs, and according to the X-ray properties, Type-2 sources dominate (a 2--3:1 ratio, up to 6:1 for $\log(\rm{L_X[erg\,s^{-1}]})>44$ sources). An X-ray estimate of the Type-2 to Type-1 AGN ratio among the ERG population is, however, indeterminate, requiring observations extending to lower X-ray energies (higher wavelengths). See Section~\ref{c2sec:xr} and Section~\ref{c2sec:agncont}.

\item The multi-wavelength AGN identification confirms that AGN tend to be found in more massive galaxies, and the AGN fraction increases with redshift, presenting a peak at $z\sim3$, in agreement with the literature. We also note the rise of the AGN fraction at $z<1$, supporting the findings that sources presenting ERG colours at low-$z$, tend to be dusty systems hosting an AGN. Also, the AGN fraction evolves differently with colour, showing that the $J-K$ colour is more efficient to select a higher fraction of AGN with the advantage that the observed population will mostly be at z>2. This is important for the selection of faint IR-excess sources as AGN candidates \citep[as in][]{Fiore08}. See Section~\ref{c2sec:agncont}.

\item EROs at $z<2$ are often pEROs ($\sim$60\%), which are mostly passively evolved systems without strong SFR activity, on average below $\sim10$\,M$_{\odot}$\,yr$^{-1}$. On the other hand, essentially all EROs at $2<z<3$ are classified as DRGs and may show up to $\sim140$\,M$_{\odot}$\,yr$^{-1}$. See Section~\ref{c2sec:sfrfin}.

\item The overlapping population, the cERGs, displays an intense average SFR at $1\leq{z}<2$ (up to $\rm{69\pm15\,M_{\odot}\,yr^{-1}}$), supporting previous claims of a dusty starburst nature for these sources \citep{Smail02,Papovich06}. See Section~\ref{c2sec:sfrfin}.

\item The contribution of ERGs to the $\dot{\rho}_{\ast}$ increases with redshift: from up to $\sim25\%$ at $1<z<2$ to up to $\sim40\%$ at $2<z<3$. IEROs show the highest contribution to the global star formation history among the three ERG population at low-$z$. See Section~\ref{c2sec:sfrfin}.

\item SFR densities from ERG populations were estimated for SF-dominated and total populations separately after a thorough AGN multi-wavelength identification. Although the inclusion of AGN ERGs in the stacking would only slightly increase the average radio luminosities shown by the non-AGN samples, that inclusion for the $\dot{\rho}_{\ast}$ estimate may lead to significant (and, at this point, undetermined) biases. See Section~\ref{c2sec:sfrfin}.

\item The use of a [8.0]-[24] colour diagnostic allows for a tentative separation between AGN and star-forming galaxies at $z>2.5$, where mid-IR (2 to 8\,$\mu$m) diagnostics become degenerate. In particular, the use of this diagnostic enables the identification of a $z\sim2.5$ massive ($5\times10^{11}\,$M$_\odot$) evolved system with MIR colours and morphology typical of a disc galaxy. See Section~\ref{c2sec:mir25}, Chapters~\ref{ch:agn} and \ref{ch:fut}.

\item A direct comparison between rest-frame UV light and radio emission from ERGs, points to a higher dust obscuration in the common population (cERG), up to E$(B-V)\sim0.6$. The lowest obscuration level is found for pEROs, which are believed to be mostly passively evolved systems. See Section~\ref{c2sec:dustcont}.

\item ERGs comprise high fraction of the Universe stellar mass at $1<z<3$, $\sim$60\%, although they represent only 25\% (at $1\leq{z}<2$) and 30\% (at $2\leq{z}\leq3$) of the total galaxy population. Mass functions show that at the highest masses, ERGs may comprise practically 100\% of the Universe stellar mass. The use of a moving bin allows the tentative discovery of a dip in the mass function at $\sim10^{11}$\,M$_\odot$ (probably the result of different contributions of early, producing the higher-mass hump, and late type galaxies, producing the lower-mass hump), and confirms the existence at $z>1$ of the lower-mass (at $\sim10^{10.4}$\,M$_\odot$) dip referred in the literature at $z<1$. See Section~\ref{c2sec:masssec} and Chapter~\ref{ch:lfs}.

\item The morphology analysis reveals bulge dominated galaxies at $2\leq{z}\leq3$ and shows an heterogeneous ERG population. The separation into pure and common populations does not point to any bimodality. This evidence, together with the remainder results and work in the literature, supports the scenario that EROs, IEROs, and DRGs are a sequence of galaxy evolution phases (showing a significant overlap). Also, one of the possibilities for the high-$z$ progenitor Population of ERGs may be indeed sub-millimetre galaxies after a fast transition into a DRG phase (based in literature evidences and X-ray information).

\item The peculiar population of pDRGs at $2<z<4$ is also studied, showing that they are indeed a mix between old and young stellar populations. Peculiar cases are found, including a the type-2 QSO found by \citep{Norman02} and a galaxy revealing P-cygni-shaped emission-lines. Although the spectral coverage is small and X-ray detections are not numerous, there are tentative evidences for a transition scenario between AGN and star-forming phases for pDRGs, as similarly defended for FeLoBALs.

\end{itemize} 
\chapter{Selecting $0<z<7$ AGN}
\label{ch:agn}
\thispagestyle{empty}

\section{Introduction}

Following the steps of its space-based predecessors (Infra-red Astronomical Satellite and Infra-red Space Observatory), the successful mission of the \emph{Spitzer Space Telescope} (\textit{Spitzer}) has opened a new window to the scientific community, by unveiling a deeper infra-red (IR) universe. Examples include mass estimates of high-$z$ galaxies \citep{Wuyts07,Ilbert10}, star formation history of galaxies \citep{leFloch05,Perezgonz05} and black hole growth and demographics throughout the age of the universe \citep{Lacy04,Stern05,Donley07,Fiore08,Fiore09}. A major accomplishment has been the development of purely photometrical techniques, in the 3--8~$\mu$m range, for the efficient selection of sources with enhanced IR emission redward of the 1.6\,${\mu}$m stellar peak, characteristic of an active galactic nucleus \citep[AGN; e.g.,][hereafter L07 and S05, respectively]{Lacy04,Lacy07,Stern05}. Long known since the 70's \citep[with ground-based telescopes,][and references therein]{KleinmannLow70,Rieke78} and 80's \citep[with the start of IR space-based observations,][]{deGrijp85,Miley85,Neugebauer86,Sanders89}, active galaxies are prone to show intense emission at IR wavelengths. This is a powerful tool as it allows the selection of AGN sources not revealed at other wavelengths. This is mostly due to dust obscuration hiding AGN signatures at optical and even X-ray wavelengths. The absorbed energy is subsequently reprocessed by the enshrouding dust and emitted at IR wavelengths, producing an IR emission excess beyond 1.6\,${\mu}$m\footnote{Blueward of this wavelength, the contribution of AGN emission through this reprocessed light mechanism diminishes significantly due to dust sublimation.}.

These MIR criteria have been repeatedly compared with those in the X-rays, arguably more reliable despite missing a high fraction of the obscured AGN population \citep{Barmby06,Donley08,Eckart10}. But \emph{reliability} and \emph{completeness} are highly dependent on the characteristics of the sample, and often difficult, if not impossible, to quantify. The combined effect of the survey depth, the wavelength coverage and, as a result, sensitivity to different physical processes as a function of redshift affect the AGN selection process. For example, the MIR \emph{wedge} type criteria (S05, L07) become increasingly affected by stellar dominated systems beyond $z\sim2.5$. If, however, one applies these criteria to shallow MIR samples, where high-$z$ star-forming (SF) galaxies are unlikely to be detected, then one finds these criteria reasonably reliable.

But if one's purpose is to obtain a truly complete and reliable AGN sample, then relying on MIR criteria alone is of course inappropriate. As put by \citet{Barmby06}, \emph{``no proposed MIR colour AGN selection will identify them all''}. The same can obviously be said about the other wavelength regimes: \emph{no individual AGN criterion -- in any spectral regime! -- will identify all AGN}. Furthermore, no single waveband criteria will be 100\% reliable. For example, high-mass X-ray binaries, if abundant in a galaxy, may mimic obscured AGN properties due to their hard X-ray spectra and high X-ray luminosities \citep[$\Gamma$=0.5--1, $\rm{L_X}=10^{42-43}$\,erg\,s$^{-1}$;][]{Colbert04,Alexander05}; Wolf-Rayet galaxies having compact optical profiles, extremely blue colours \citep{Kewley01}, high ionization emission lines (NV, SiIV, and CIV stellar wind features) and broad emission features \citep[$\sim2000$\,km\,s$^{-1}$;][]{Beals29,Schulteladbeck95,Herald00,Crowther07} may be misclassified as having an AGN dominated optical spectral energy distribution (SED); and, finally, extremely obscured starbursts at high-$z$ can mimic AGN characteristic IR red colours \citep{Donley08,Narayanan10}.

A high completeness (the fraction of the true AGN host population selected by a given criterion) and reliability (fraction of correct AGN classifications within the selected sample) can only be attained by combining different wavelength criteria, thus sampling different physical conditions and processes indicative of the existence of an AGN. Following this reasoning, \citet{Richards09} investigated a 6 to 8 dimensional criterion based in the optical and MIR regimes to present a sample of $>5000$ AGN candidates using wide, deep fields. While this method (gradually improved as one adds X-rays, radio or even morphological information) and that of SED fitting \citep{Walcher11} will likely provide the best results, the intrinsic degree of complexity and the difficulty to apply in anything but the most intensively observed fields on the sky make the simpler IR colour-colour criteria stand out. Considering the high redshift Universe, for example, where sources will be difficult to detect at most wavelengths, one would aim to develop the most reliable and complete criterion \emph{possible} that solely requires the use of a single observational facility and the minimum number of observations.

With the approaching launch of the \emph{James Webb Space Telescope} (\textit{JWST}), optimised to near and MIR wavelengths ($1-25\,\mu$m) and with a particular emphasis on the high-redshift Universe, it is fundamental to investigate AGN selection criteria that can be directly applied to the resulting deep surveys. In this work we will present several near-to-mid-infrared \textit{JWST}-suited colour criteria aiming to select a variety of AGN populations. Resulting from the use of a large set of observed and theoretical SEDs, these colour criteria are defined and tested against several control samples (selected from X-rays to radio frequencies) existing in deep galaxy surveys covered by $Spitzer$. Reliability and completeness are estimated for the proposed criteria and compared to those of existing MIR AGN diagnostics.

In Section~\ref{c3sec:samp} the different possibilities for the mechanisms behind the IR emission are discussed and the new criteria are presented. A test bench will be explored in Section~\ref{c3sec:ctrlsp} using the above-referenced broad set of control samples. We discuss the sensitivity of IR colour-colour criteria toward specific types of AGN and the conceptual improvements of the new proposed IR AGN diagnostics in Section~\ref{c3sec:disc}. The implications to $JWST$ surveys will be highlighted in Section~\ref{c3sec:impjwst}, followed by the final conclusions of this work in Section~\ref{c3sec:concagn}.

\section{Distinguishing AGN from Stellar/SF IR contributions} \label{c3sec:samp}

\begin{figure}
  \begin{center}
    \includegraphics[width=1\columnwidth]{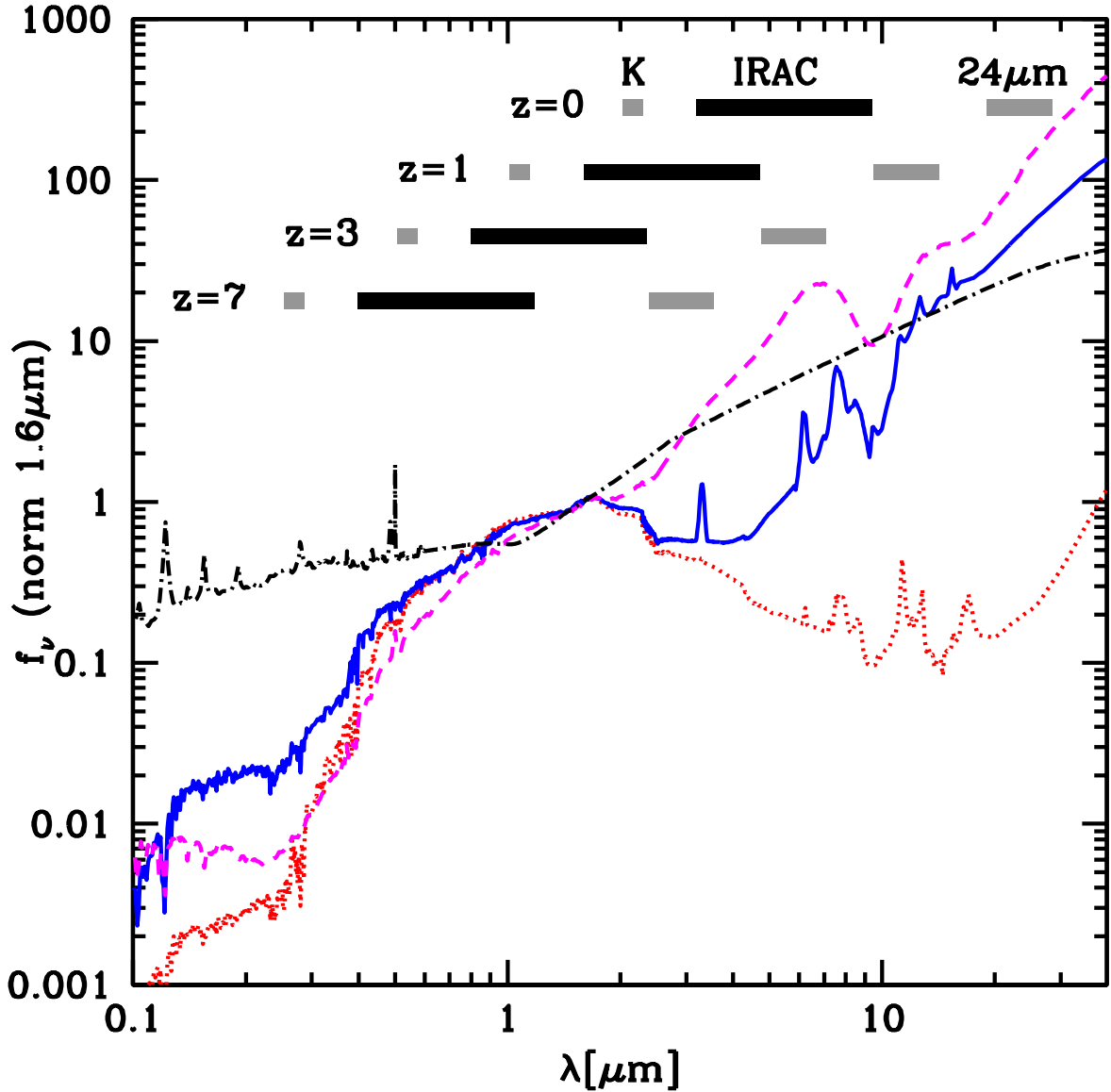}
  \end{center}
  \caption[Infrared spectral energy distributions]{Four examples of galaxy templates taken from the SWIRE Template Library \citep{Polletta07} and flux normalised to 1.6$\mu$m: S0 (early type galaxy, red dotted line), M82 (starburst galaxy, blue line), IRAS 19254-7245 (a hybrid source, magenta dashed line), and type-1 QSO (AGN, black dot-dashed line). The shaded regions show what rest-frame wavelength the K, IRAC, and MIPS 24${\mu}m$ filters will be observing depending on the redshift.}
  \label{c3fig:sed}
\end{figure}

The SEDs of stellar/SF dominated systems have some distinctive characteristics, allowing the separation of this population from AGN host galaxies through IR colours alone. Figure~\ref{c3fig:sed} illustrates a few examples using galaxy templates taken from the SWIRE Template Library \citep{Polletta07}. In stellar and/or SF dominated SEDs, henceforth referred to as normal galaxy SEDs, the overall blackbody emission from the stellar population, caused by the minimum in the opacity of the H$^{-}$ ion, produces an emission peak at 1.6\,${\mu}$m, which clearly stands out, as does the CO absorption at 2.35--2.5\,${\mu}$m from red supergiants. Furthermore, the strength of the Polycyclic Aromatic Hydrocarbons (PAH) features, seen mostly beyond 6\,$\mu$m, increases with star formation activity. It is in this spectral region (1--6\,${\mu}$m) that the difference between normal galaxies and AGN dominated SEDs is the greatest. The existence of an AGN is frequently accompanied by a rising power-law continuum ($f_{\nu}~\propto~{\nu}^{\alpha}$) redward of $\sim$1$\,\mu$m, as a result of reprocessed X-ray, UV, and optical light emitted in the MIR by the hot dust surrounding the central region of an active galaxy \citep{Sanders89,Sanders99,PierKrolik92}.

This feature is unique for AGN hosts and is revealed in IRAC colour-colour spaces \citep{Lacy04,Stern05,Hatziminaoglou05}, by power-law techniques \citep{AlonsoHerrero06,Polletta06,Donley07} or by IR emission excess diagnostics \citep{Daddi07,Dey08,Fiore08,Polletta08}. The latter are particularly sensitive to the reddest, most obscured types. In these cases, the AGN MIR emission is even more obvious when compared to a severely obscured UV-Optical emission. Each of these criteria has its own advantages and problems. While colour-colour \emph{wedges} tend to select more complete AGN samples, the power-law and IR-excess (IRxs) techniques have a higher reliability in the selection of specific AGN types \citep{Donley08}. However, as one probes more distant galaxy samples, the identification becomes more complicated \citep{Barmby06,Donley08}.

In the following sections, having the wide near-to-mid IR range of \textit{JWST} in mind, $K$-band-to-IRAC (KI) and $K$-band-to-IRAC/MIPS (KIM) colour-colour spaces will be explored as diagnostics for AGN identification at low and high redshifts. These are further tested against other AGN diagnostics, making use of a wide set of galaxy model SEDs, and a broad variety of control samples.

\subsection{The template set} \label{c3sec:templates}

The templates used throughout this paper come from published work as follows: 10 templates covering early to late galaxy types, three starbursts, six hybrids\footnote{By \textit{hybrids} we refer to SEDs simultaneously showing stellar/SF and AGN emission.}, and seven AGN, all from \citet{Polletta07}; nine starburst ULIRGs from \citet{Rieke09}; one blue starburst and 18 hybrid SEDs from \citet{Salvato09}; and one extremely obscured hybrid from \citet{Afonso01}. Except for early type and blue starburst model templates, all SEDs are derived from mixed model and observational information. The latter either comes from broad band photometry, SDSS optical spectra, Infrared Space Observatory (ISO) 5--12\,${\mu}$m or \textit{Spitzer}-IRS 5--36\,$\mu$m spectra. The hybrid SEDs from \citet{Salvato09} were obtained by the combination of stellar/SF dominated SEDs with AGN dominated SEDs: IRAS 22491-1808 SED with that characteristic of a QSO type-1 object, and an S0 template with one characteristic of a QSO type-2 object \citep[all four SEDs from][]{Polletta07}.

With such a varied template library, the galaxy colour-$z$ space is expected to be adequately sampled. High redshift extreme examples are considered \citep[such as the Torus template used to fit the heavily obscured type-2 QSO at $z=2.54$, SWIRE\_J104409.95+585224.8,][]{Polletta06} and hybrid templates, shown to be efficient up to high redshifts ($z<5.5$) and down to faint fluxes \citep[$i<24.5$,][]{Salvato09,Salvato11}, are also taken into account. It is worth noting nevertheless that even local templates are successful in fitting some of the most extreme high redshift sources (for instance, the case of Arp220 as a local analogue of HR10, an extremely red galaxy at $z=1.44$, \citealt{HuRidgway94} and \citealt{Elbaz02}, or M82 as an analogue for star-formation dominated sub-millimetre galaxies, \citealt{Pope08}).

The SED templates are organised in four groups: (a) Early to Late-type galaxies, (b) Starbursts, (c) Hybrids, and (d) AGN. The following investigation will focus on how these groups populate near-to-mid IR colour-colour diagnostic plots, aiming to separate the AGN/Hybrid population, (c) and (d) above, from that for normal galaxies, i.e., (a) and (b).

\section{The new approach}

\subsection{An enhanced wedge diagram: the KI criterion}\label{c3sec:ki}

\begin{figure}
  \begin{center}
    \includegraphics[width=1\columnwidth]{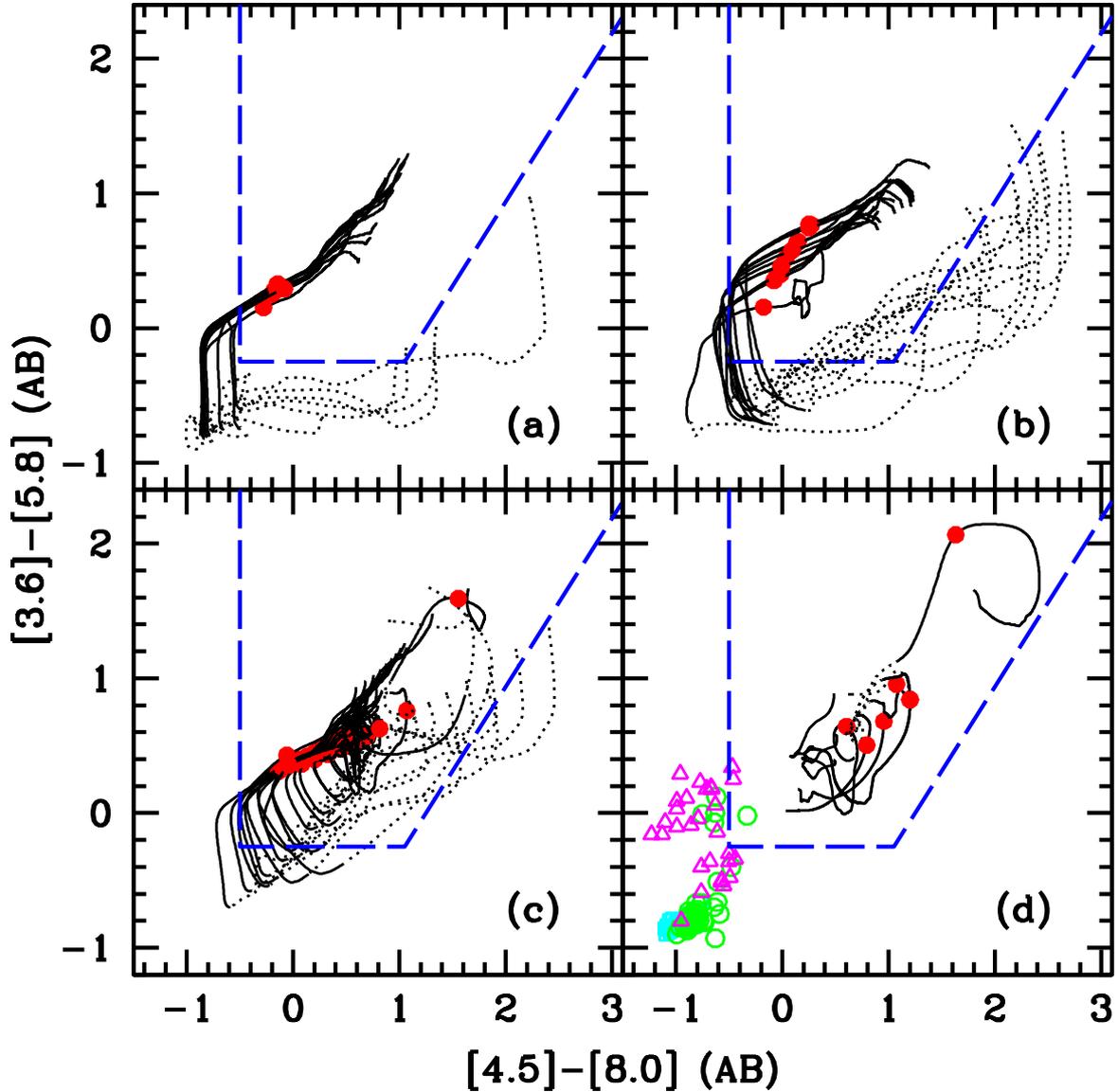}
  \end{center}
  \caption[The \citet{Lacy04,Lacy07} AGN criterion]{Model colour tracks displayed in the L07 criterion colour-colour space. Dashed blue line refers to the boundaries proposed in that work for the selection of AGN. Each panel presents a specific group: (a) Early/Late, (b) starburst, (c) Hybrid and (d) AGN. The dotted lines refer to the $0<z<1$ redshift range, and solid line to $1\leq{z}\leq7$. Red circles along the lines mark $z=2.5$. Dwarf stars \citep{Patten06} are shown for reference. M-dwarfs appear as open cyan squares, L-dwarfs as open green circles, and T-dwarfs as open magenta triangles to show where these red point-like cool stars appear.}
  \label{c3fig:c1324}
\end{figure}

\begin{figure}
  \begin{center}
    \includegraphics[width=1\columnwidth]{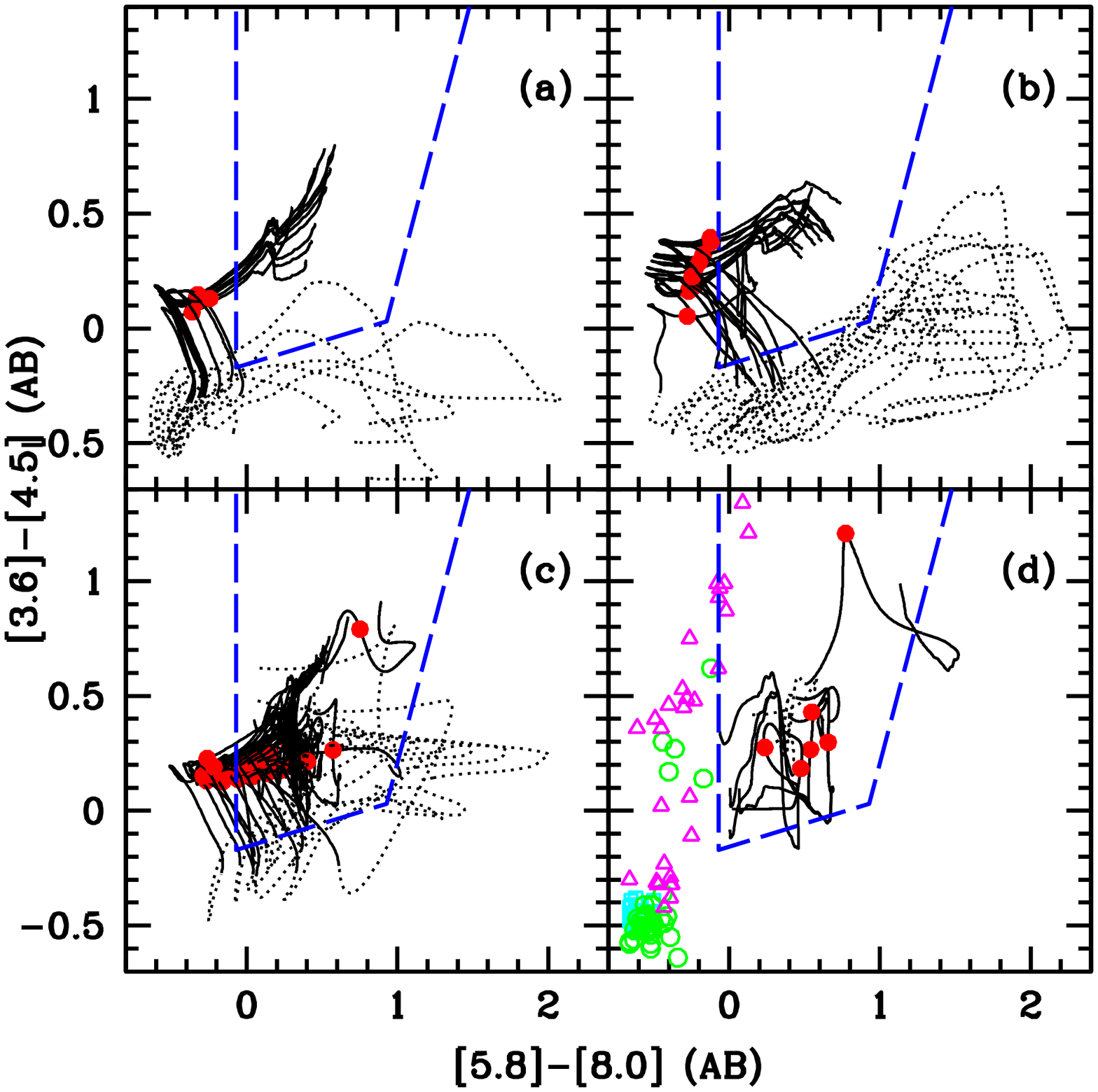}
  \end{center}
  \caption[The \citet{Stern05} AGN criterion]{Model colour tracks displayed in the S05 criterion colour-colour space. Symbols and panel definition as in Figure~\ref{c3fig:c1324}.}
  \label{c3fig:c3412}
\end{figure}

In Figures~\ref{c3fig:c1324} and \ref{c3fig:c3412} the colour tracks (spanning the range $0<z<7$) for the template SEDs considered are presented on the L07 and S05 criteria colour-colour spaces, respectively. In both, the nominal AGN regions encompass most of the AGN and hybrid tracks for a large range of redshifts, as they were built to do. We note, however, that the use of a very diverse SED template set already shows some shortcomings of these diagnostic plots. In both Figures, the two upper panels (early/late and starburst galaxies) show a significant contamination of the nominal AGN region by normal galaxies (i.e., non-AGN) not only at high redshifts ($z\gtrsim2-3$) but also much closer ($z\lesssim1$), as already noted by previous studies \citep{Barmby06,Donley08}. The fact that some hybrid templates fall, at some point, out of the selection regions is expected as the SF or AGN emission contribute differently to the observed bands at different redshifts. Again we point out that colour-colour criteria will only successfully identify AGN whose emission dominates in at least some of the observed bands, which won't be the case for many AGN \citep{Rigopoulou99,Maiolino03,Treister06}. Cool dwarf stars may fall close to the boundaries or inside the selecting regions, thus being also potential (point-like) contaminants.

In order to enhance these wedge diagrams, one can extend the wavelength coverage to shorter wavebands, out of the IRAC range. This is obviously outside the IRAC framework behind the original definition of such wedge diagrams, but suits the larger \textit{JWST} wavelength coverage. By considering shorter wavelengths, one is of course probing a spectral region mostly dominated by stellar emission (see Figure~\ref{c3fig:sed}). Such a scenario is an advantage as we now compare a stellar dominated wave-band with one that has contribution either from stellar or AGN light. Such a comparison will yield a useful colour dispersion ideal for the separation of the two types of system.

\begin{figure}
  \begin{center}
    \includegraphics[width=1\columnwidth]{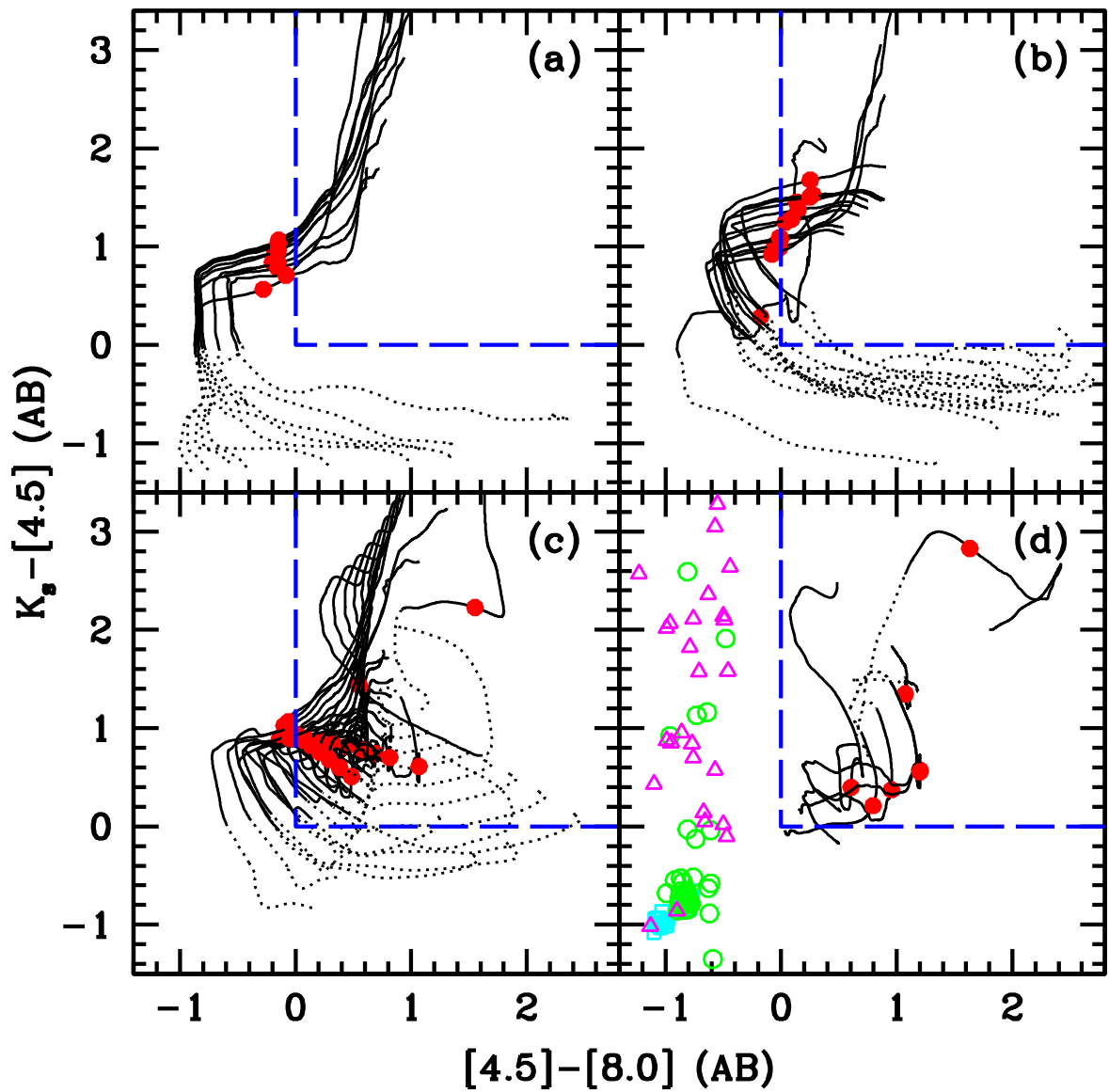}
  \end{center}
  \caption[The KI AGN criterion]{The proposed KI criterion. Symbols and panel definition as in Figure~\ref{c3fig:c1324}.}
  \label{c3fig:kc24}
\end{figure}

A particularly relevant combination of colours is $K-[4.5]$ versus [4.5]-[8.0] (Figure~\ref{c3fig:kc24}). This, henceforth called the KI (K+IRAC) criterion, is defined by the following simple conditions ($\wedge$ denotes the ``AND'' condition): 
\begin{eqnarray}
\begin{array}{c}
K_s-\left[4.5\right]>0 \nonumber \\
\wedge \nonumber \\
\left[4.5\right]-\left[8.0\right]>0 \nonumber \\
\end{array}
\end{eqnarray}

Comparing with the L07 and S05 criteria, the contamination by normal galaxies at $z\gtrsim2.5$ seems similar. However, the contamination by normal galaxies at $z\lesssim2.5$ appears significantly reduced (higher reliability), with no effect on the ability to select AGN (completeness). Another conceptual improvement of KI, adding to the simplicity of its definition, is the unbounded upper right AGN region. This avoids the loss of heavily obscured AGN (with extremely red colours). This is in opposition to what is seen in S05, for example, where the Torus template moves out from the selecting region at the highest redshifts ($z\gtrsim4$).

One can also note the usefulness of the simple $K-[4.5]$ colour in excluding low redshift normal galaxies: the condition $K-[4.5]>0$ is able to reject a large fraction of the $z<1$ non-AGN galaxies. Such a property also makes this simple colour-cut of great use to the study of AGN and star-formation co-evolution in the last half of the history of the universe.

\begin{figure}[t]
  \begin{center}$
  \begin{array}{cc}
    \includegraphics[width=0.5\columnwidth]{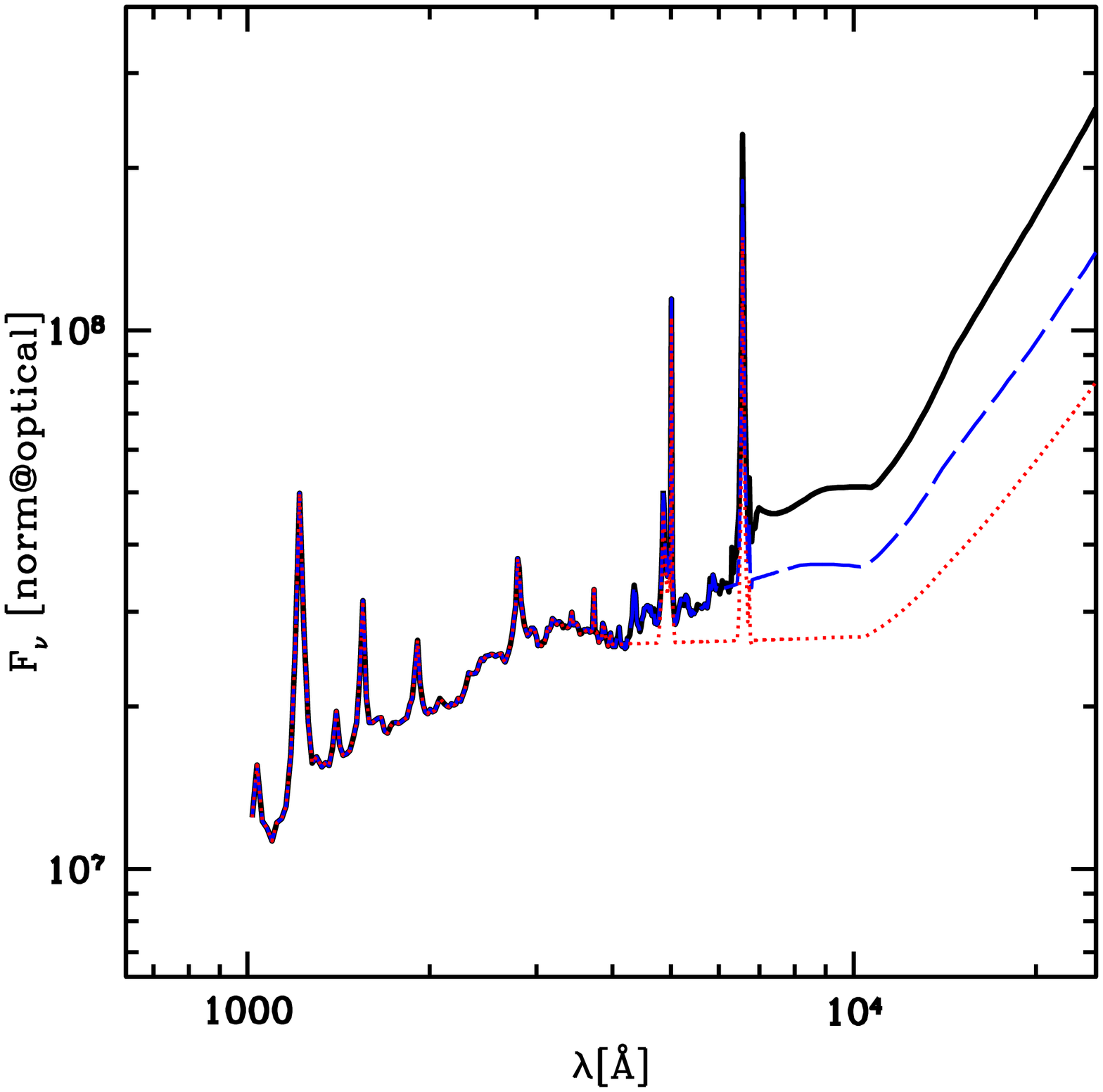} &
    \includegraphics[width=0.5\columnwidth]{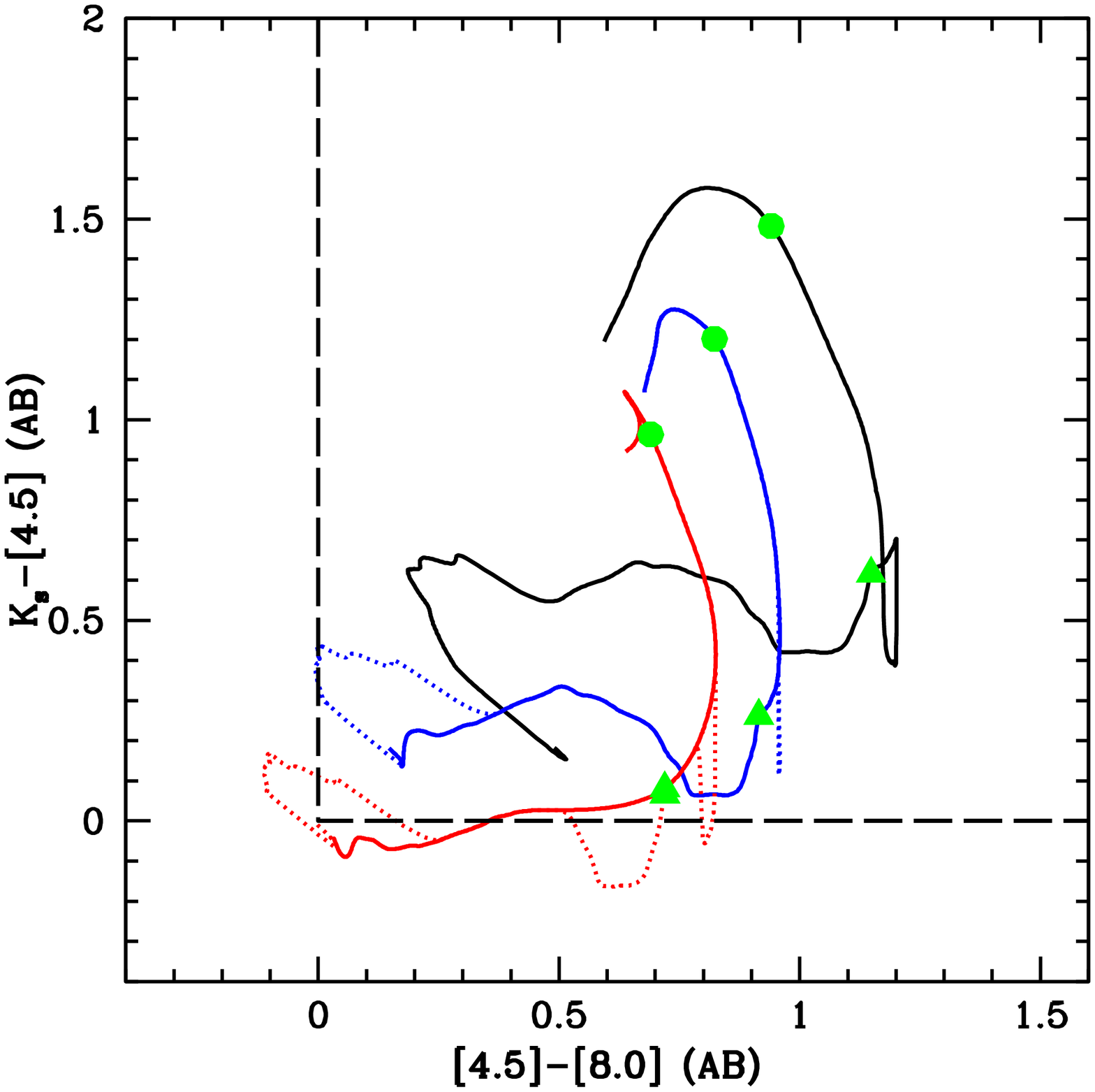}
  \end{array}$
  \end{center}
  \caption[Line emission effect on Photometry]{The effects of considering prominent lines in type-1 QSO SED models. The left panel shows three such SEDs originally from \citet{Polletta07}: two updated versions of QSO1 (dashed blue line, now with an H$\alpha$ line) and BQSO1 (red dotted, now with both H$\alpha$ and OIII lines), and the unchanged TQSO1 (solid black). On the right panel the original SED model tracks are shown as solid lines (from top to bottom: TQSO1, QSO1, and BQSO1), whereas the inclusion of strong emission lines produces the deviations given by the dotted segments. Circles and triangles show $z=1$ and $z=3$, respectively. The tracks extend from $z=0$ to $z=7$.}
  \label{c3fig:lines}
\end{figure}

It should be noted that not all the templates considered take into account prominent emission lines. These may affect the photometry and produce some degree of scatter in the colour-colour tracks. This is visible in Figure~\ref{c3fig:lines} where updated versions of the QSO1 and BQSO1 templates \citep{Polletta07} were considered, now with both H$\alpha$ and OIII lines included (visible in the unchanged TQSO1 template). Although in specific redshift intervals (when a certain emission line is redshifted into a given filter), AGN sources with smaller AGN contribution in the IR (like BQSO1) fall out of the AGN region of the KI criterion, the bulk of the AGN population is expected to remain inside the KI boundaries nonetheless. Also, the presence of emission lines in starburst SEDs (e.g., H$\alpha$, Pa$\alpha$) will improve the results by producing colours that will place a given starburst further away from the AGN region. This reasoning equally applies to S05 and L07, although the effect on the latter is expected to be smaller due to its larger selection region.

\subsection{Extending to high redshifts: the KIM criterion} \label{c3sec:imhighz}

One serious problem is the contamination by normal galaxies at high redshifts ($z\gtrsim2.5$). All three criteria (L07, S05, and KI) fail to disentangle AGN dominated systems from normal galaxies at those redshifts. To avoid this problem, the longer MIR wavelength range to be available in the \textit{JWST} ($>20\,{\mu}$m) will be considered. For this purpose, we extend the criterion to the MIPS-$24\,{\mu}m$ band.

The use of this waveband for AGN selection has always been peculiar. Given the degeneracy between AGN and non-AGN when exploring IRAC-MIPS colours \citep{Lacy04,Hatziminaoglou05,Cardamone08}, other authors have tended to use the MIPS-$24\,{\mu}m$ band for unique, extreme objects (like the IRxs techniques) or in single, unconventional situations. For instance, while \citet{Garn10} use [8.0]-[24] against [5.8]-[8.0] for a $z\sim0.8$ sample to identify those sources showing AGN activity, \citet{Treister06} and \citet{Messias10} use a single [8.0]-[24] colour cut at, respectively, $z\sim2$ and $z>2.5$ for the same purpose\footnote{\citet{Ivison04} and \citet{Pope08} also address [8.0]-[24] against [4.5]-[8.0] to distinguish AGN from normal galaxies, but, in those works, only the [4.5]-[8.0] colour is effectively used for that purpose.}. Colours involving the 24\,$\mu$m band are usually avoided due to the large wavelength gap between this band and other commonly available MIR bands (usually the \textit{Spitzer}-IRAC bands). At high-$z$ (e.g., $z\sim3$), however, the sampled rest-frame wavebands (2 and 6\,$\mu$m corresponding to observed 8 and 24\,$\mu$m, respectively) are not much more separated than the 3.6 and 8.0${\mu}m$ IRAC bands for nearby galaxies. A different issue is the lower sensitivity and larger point spread function of the MIPS 24\,$\mu$m images, compared with those for the IRAC channels, which affects the accuracy of colour measurements using this longer wavelength band.

\begin{figure}
  \begin{center}
    \includegraphics[width=1\columnwidth]{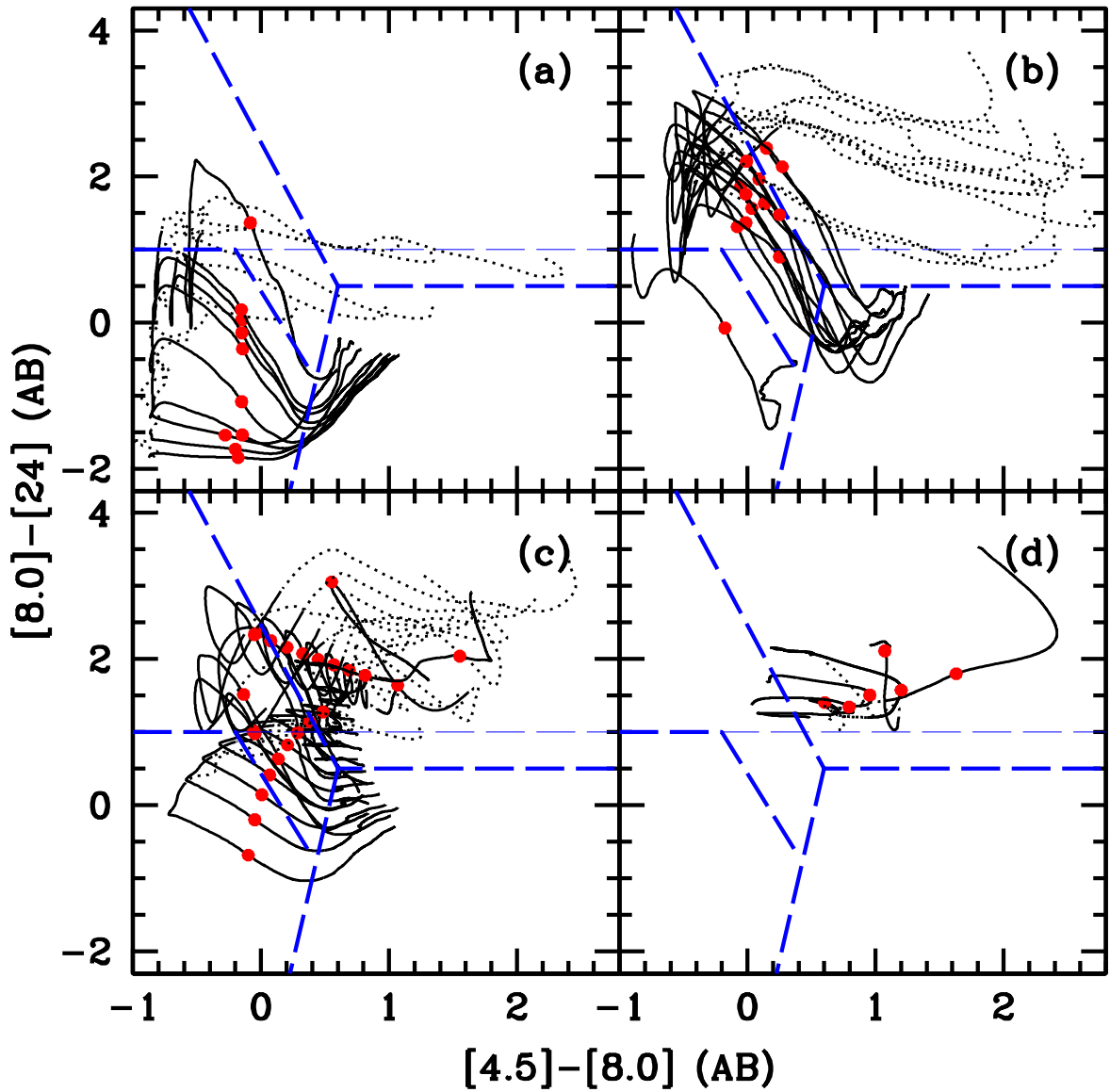}
  \end{center}
  \caption[The IM colour-colour space]{The IRAC-MIPS colour-colour space and the proposed criterion. Symbols and panel definition as in Figure~\ref{c3fig:c1324}. The thin dashed blue line refers to a simpler criterion valid at $z\gtrsim3$, as detailed in the text.}
  \label{c3fig:c24m24}
\end{figure}

Furthermore, normal galaxies show a wide [8.0]-[24] colour range (as a result of different PAH and dust emission between galaxies), which is further increased by redshift (in a normal galaxy, while at low-$z$ the 8.0\,$\mu$m filter is dominated by PAH emission, at higher redshifts only dust and stellar continua emission contribute to this band). This results in a considerable colour overlap with AGNs, limiting this single colour usefulness to separate both populations \citep{Lacy04,Cardamone08}. Nonetheless, adding a shorter wavelength MIR colour helps to break this degeneracy. Figure~\ref{c3fig:c24m24} illustrates a proposed colour-colour separation diagnostic efficient at high redshifts. One can see that beyond $z\sim1$, AGN (lower panels) and normal galaxies (upper panels) occupy essentially different regions in the [8.0]-[24] versus [4.5]-[8.0] space. This is of great interest for the characterisation of high redshift galaxy populations, such as Lyman Break Galaxies (LBGs) and equivalents at $z\gtrsim2$ \citep{Steidel03,Steidel04,Adelberger04}.

Furthermore, normal galaxies show a wide [8.0]-[24] colour range, which is further increased by redshift. This results in a considerable colour overlap with AGNs, limiting this single colour usefulness to separate both populations \citep{Lacy04,Cardamone08}. Nonetheless, adding a shorter wavelength MIR colour helps to break this degeneracy. Figure~\ref{c3fig:c24m24} illustrates a proposed colour-colour separation diagnostic efficient at high redshifts. One can see that beyond $z\sim1$, AGN (lower panels) and normal galaxies (upper panels) occupy essentially different regions in the [8.0]-[24] versus [4.5]-[8.0] space. This is of great interest for the characterisation of high redshift galaxy populations, such as Lyman Break Galaxies (LBGs) and equivalents at $z\gtrsim2$ \citep{Steidel03,Steidel04,Adelberger04}.

However, at low redshifts a high degree of degeneracy exists, with AGN and normal galaxies occupying the same colour-colour region. A rejection of low-redshift ($z<1$) normal/star forming galaxies would, however, remove this overlap, allowing for a powerful AGN-selection criteria to be built. This can be achieved, as noted in the previous section, by using the $K-[4.5]>0$ colour cut, which will allow for the rejection of a large fraction of the $z<1$ non-AGN galaxies, with AGN and hybrid galaxies in this redshift range remaining mostly unaffected.

\begin{figure}
  \begin{center}
    \includegraphics[width=0.5\columnwidth]{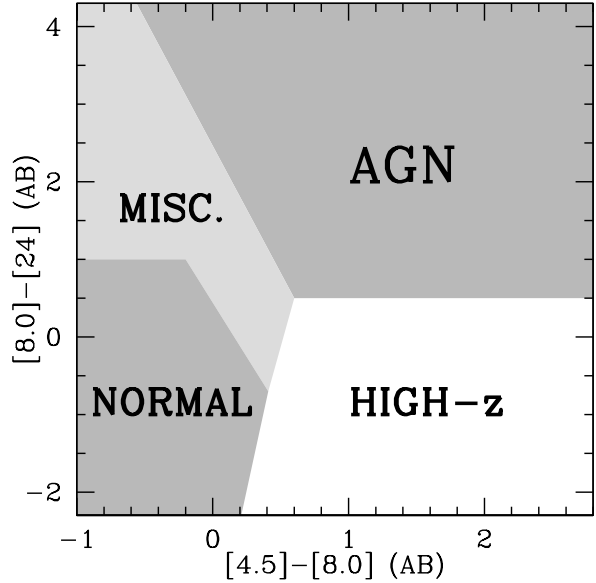}
  \end{center}
  \caption[The IM regions of interest]{The IRAC-MIPS colour-colour space and the proposed IM criterion regions.}
  \label{c3fig:kimreg}
\end{figure}

Under these conditions (either at $z>1$ or having excluded $z<1$ normal galaxies by applying a criterion such as $K-[4.5]>0$), one is then able to define four regions in the IRAC-MIPS (IM) colour-colour space as shown in Figures~\ref{c3fig:c24m24} and \ref{c3fig:kimreg}: AGN dominated, miscellaneous (where both pure starburst and hybrid systems with a reasonable AGN contribution appear), normal galaxies and, finally, a region occupied by non-AGN sources at higher redshifts ($z\gtrsim3-4$). The boundaries of each of these regions are set by the following conditions ($\vee$ and $\wedge$ denote the ``OR'' and ``AND'' conditions, respectively):
\small
\begin{eqnarray}
\begin{array}{l}
{\rm (i)~AGN:} \nonumber \\
\left[8.0\right]-\left[24\right] > -3.3\times(\left[4.5\right]-\left[8.0\right])+2.5 ~\wedge \nonumber \\
\left[8.0\right]-\left[24\right] \geq 0.5 \nonumber \\
{\rm (ii)~Miscellaneous:} \nonumber \\
\left(\right.\left[8.0\right]-\left[24\right] \geq 1 ~\vee \nonumber \\
\;\;\,\left[8.0\right]-\left[24\right] > -2.8\times(\left[4.5\right]-\left[8.0\right])+0.4\left.\right) ~\wedge \nonumber \\
\left[8.0\right]-\left[24\right] \leq -3.3\times(\left[4.5\right]-\left[8.0\right])+2.5 ~\wedge \nonumber \\
\left[8.0\right]-\left[24\right] > 7.5\times(\left[4.5\right]-\left[8.0\right])-4 \nonumber \\
{\rm (iii)~Normal:} \nonumber \\
\left[8.0\right]-\left[24\right] < 1 ~\wedge \nonumber \\
\left[8.0\right]-\left[24\right] < -2.8\times(\left[4.5\right]-\left[8.0\right])+0.4 ~\wedge \nonumber \\
\left[8.0\right]-\left[24\right] > 7.5\times(\left[4.5\right]-\left[8.0\right])-4 \nonumber \\
{\rm (iv)~High\!-z:} \nonumber \\
\left[8.0\right]-\left[24\right] < 0.5 ~\wedge \nonumber \\
\left[8.0\right]-\left[24\right] \leq 7.5\times(\left[4.5\right]-\left[8.0\right])-4 \nonumber \\
\end{array}
\end{eqnarray}
\normalsize

These IM conditions, when considered together with the $K-[4.5]>0$ cut, which implements the rejection of $z<1$ normal galaxies, define what we will henceforth call the KIM (K+IRAC+MIPS) criterion.

\begin{figure}
  \begin{center}
    \includegraphics[width=1\columnwidth]{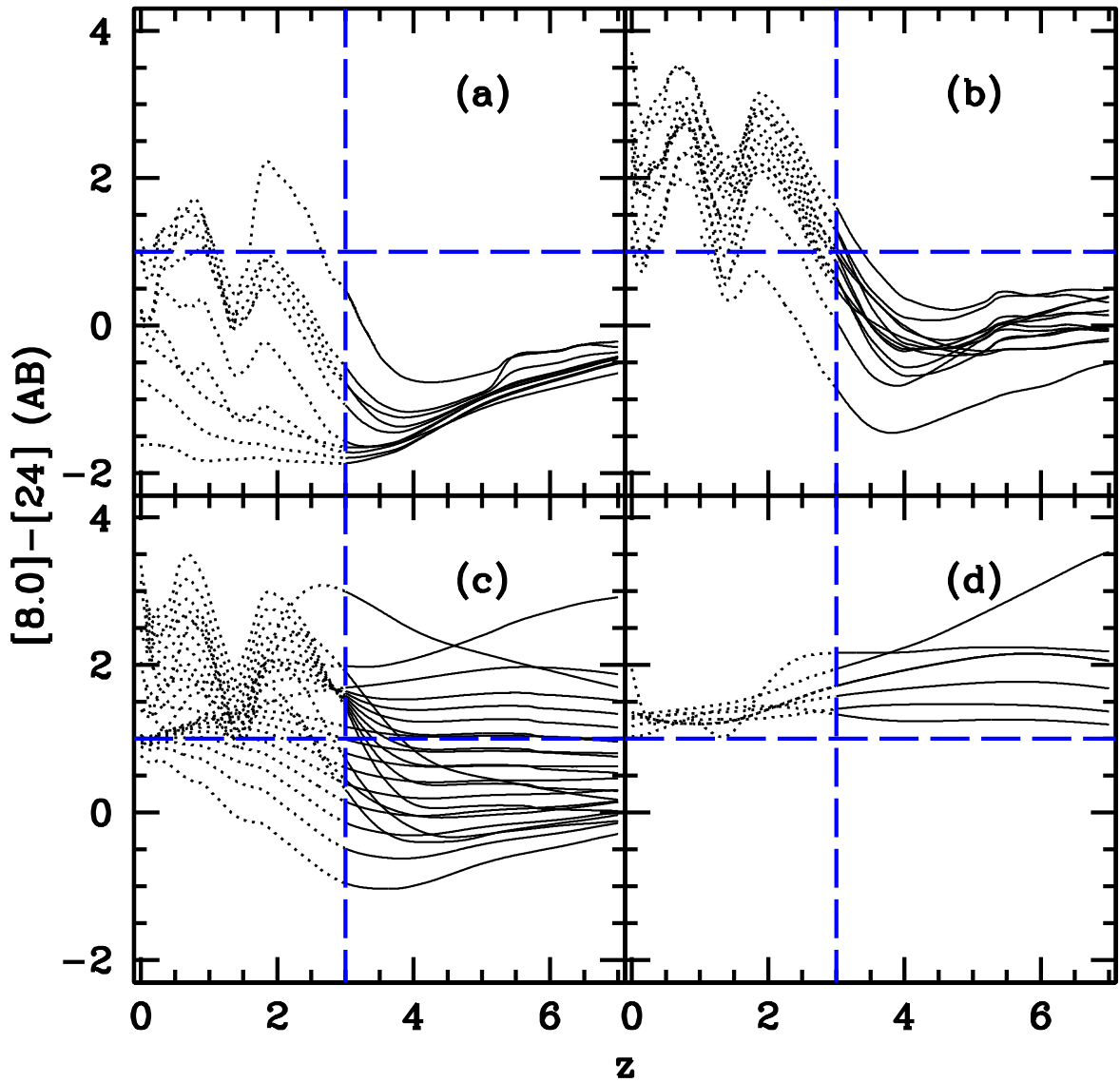}
  \end{center}
  \caption[The $8.0-24$ colour evolution with redshift]{The $[8.0]-[24]$ colour evolution with redshift. Panel definition as in Figure~\ref{c3fig:c1324}. The horizontal line shows $[8.0]-[24]=1$, while the vertical one indicates $z=3$. At $z>3$, only AGN dominated galaxies show $[8.0]-[24]>1$ colours.}
  \label{c3fig:z824}
\end{figure}

We further note from Figure~\ref{c3fig:c24m24} that for $z\gtrsim3$, essentially all SEDs with $[8.0]-[24]>1$ are dominated by AGN emission. Figure~\ref{c3fig:z824} details this behaviour, clearly showing that stellar dominated galaxies at $z\gtrsim3$ show $[8.0]-[24]<1$ colours, as described by \citet{Messias10}.

\section{Test bench} \label{c3sec:ctrlsp}

In the previous section we proposed: $K-[4.5]$ as an useful colour for the efficient segregation of the galaxy population into AGN-dominated and normal SEDs at $z<1$; the KI criterion as an alternative to L07 and S05; and KIM (a 4 band, 3 colour criterion), as a diagnostic which, according to the colour tracks of the templates used, enables the selection of AGN sources at $0<z<7$ with little contamination by normal galaxies. This is of great interest as it enables the tracking of AGN activity since the epoch of reionization to the current time. The usefulness of these criteria can only be evaluated, however, by pursuing a test with well characterised control samples. By using different galaxy samples, and considering other available AGN criteria, based on distinct spectral regimes, we can obtain some estimate of the reliability and completeness of the new proposed diagnostics in comparison with commonly used ones. Again, one must keep in mind that any AGN criteria will be complete and reliable only at some level, so caution must be exercised when comparing the results.

We will perform these tests with five control samples. Firstly, we use a sample of galaxies from the Great Observatories Origin Deep Survey South \citep[GOODSs,][]{Giavalisco04} and another from the Cosmic Evolution Survey \citep[COSMOS,][]{Scoville07}, both with available AGN/non-AGN classification from X-rays and/or optical spectroscopy. Secondly, we assemble samples of IRxs sources found in GOODSs and COSMOS fields. The QSO sample from the Sloan Digital Sky Survey \citep[SDSS,][]{Schneider10}, reaching $z\sim6$, is also considered for the testing, as well as the High-\textit{z} Radio Galaxy (H\textit{z}RG) sample from \citet{Seymour07}. The first two samples allow for an indication of the completeness and reliability of the IR AGN selection criteria, while the AGN samples (IRxs sources, SDSS QSOs and H\textit{z}RGs) will allow for independent measures of their completeness up to the highest redshifts, with the caveat that the AGN samples are, themselves, incomplete.

In the following subsections, Completeness ($\mathcal{C}$) is defined as the fraction of the AGN population that a given IR criterion is able to select (AGN$_{SEL}$/AGN$_{TOT}$), while Reliability ($\mathcal{R}$) refers to the fraction of the IR sources selected by a given criterion which are part of the ``true'' AGN population ($\rm{AGN_{SEL}/N_{SEL}}$, where $\rm{N_{SEL}=AGN_{SEL}+non-AGN_{SEL}}$). We again stress that the true AGN population is unknown, and we are always limited to a fraction of it as unveiled by other selection methods, which can themselves be more or less biased.

\subsection{The GOODSs and COSMOS samples} \label{c3sec:goodssample}

The ideal sample to test the MIR-AGN selection criteria would be a sample of galaxies with complete AGN/non-AGN characterisation for all of its members. Such a thorough characterisation is at this stage impossible, this being precisely one of the reasons for the development of MIR AGN-selection criteria. As such, one can only aim to assemble a sample of galaxies where both AGN and non-AGN populations are represented, and keep in mind that the comparison between the MIR criteria being tested will only be indicative of relative performance.
 
For this first test sample, we have selected 2288 galaxies from MUSIC/GOODSs catalogue \citep{Grazian06,Santini09} and 7180 from COSMOS \citep{Ilbert09} with an X-ray classification and/or a good quality\footnote{Spectra flagged as 0 (very good) or 1 (good) in the MUSIC catalogue, and with 90\% probability in the COSMOS catalogue.} optical spectroscopic classification. Whenever a spectroscopic redshift was not available, the photometric estimates by \citet[][in GOODSs]{Luo10} and \citet[][in COSMOS]{Salvato09} were adopted.

The IR data used for the MUSIC catalogue comes from \citet{Vandame02} and Dickinson et al. (in prep.), and that for the COSMOS comes from \citet{Sanders07}, \citet{Lefloch09}, and \citet{McCracken10}. Regarding the X-rays, the 2\,Ms $Chandra$ Deep Field South \citep[CDFs,][]{Luo08} data was used, as well as the XMM data in COSMOS \citep{Cappelluti09,Brusa10}. The X-ray AGN classification is similar to that of \citet{Szokoly04}. There, the X-ray luminosity and hardness-ratio (HR) are used to identify the AGN population. The HR is a measure of the source obscuration and is defined as HR$\equiv$(H-S)/(H+S) with H and S being, respectively, the net counts in the hard, 2--8\,keV, and soft, 0.5--2\,keV, X-ray bands. However, this ratio becomes degenerated with redshift (\citealt{Eckart06} and \citealt{Messias10}, but also \citealt{Alexander05} and \citealt{Luo10}). Hence we compute for each source the respective column densities ($\rm{N_H}$) using the Portable, Interactive Multi-Mission Simulator\footnote{http://heasarc.nasa.gov/docs/software/tools/pimms.html} (PIMMS, version 3.9k). The soft-band/full-band (SB/FB) and hard-band/full-band (HB/FB) flux ratios\footnote{The use of ratios based on FB flux instead of the commonly used SB/HB flux ratios, allows for an estimate of $\rm{N_H}$ when the source is detected in the FB but no detection is achieved in either the SB or HB.} were estimated for a range of column densities ($20<\log(\rm{N_H[{\rm cm}^{-2}]})<25$, with steps of $\log(\rm{N_H[{\rm cm}^{-2}]})=0.01$), and redshifts ($0<z<7$, with steps of $z=0.01$), considering a fixed photon index, $\Gamma=1.8$ \citep{Tozzi06}. The comparison with the observed values results in the estimate of $\rm{N_H}$, which can then be used to derive an intrinsic X-ray luminosity. The HR constraint used by \citet{Szokoly04} ($\rm{HR}=-0.2$) is equivalent to $\log(\rm{N_H[{\rm cm}^{-2}]})=22$ at $z\sim0$, and this is the value considered throughout the whole redshift range. Hence, an X-ray AGN is considered to have ($\vee$ and $\wedge$ denote the ``OR'' and ``AND'' conditions, respectively):
\[\rm{L_X^{int}}>10^{41}\,\rm{erg\,s^{-1}}~\wedge~\rm{N_H}>10^{22}\,{\rm cm}^{-2}\]\[\vee\]\[\rm{L_X^{int}}>10^{42}\,\rm{erg\,s^{-1}}\]

The remaining X-ray detections are hence regarded as non-AGN sources. The intrinsic X-ray (0.5--10\,keV) luminosities are estimated as: \[ \rm{L_X^{int}} = 4 \pi\,d^{2}_{L}\,f_{X}^{int}\,(1+z)^{\Gamma-2}\,\rm{erg}\,\rm{s}^{-1} \] where $\rm{f_{X}^{int}}$ is the obscuration-corrected X-ray flux in the 0.5--10 keV band and $\Gamma$ is the observed photon index (when $\log(\rm{N_H[{\rm cm}^{-2}]})\leq20$) or $\Gamma=1.8$ (when $\log(\rm{N_H[{\rm cm}^{-2}]})>20$). The luminosity distance, d$_{\rm L}$ is calculated using either the spectroscopic redshift or, if not available, the photometric redshift. The 0.5--8~keV luminosities, derived using \citet{Luo08} catalogued 0.5--8~keV fluxes, were converted to 0.5--10~keV considering the adopted $\Gamma$. For simplicity, the luminosity `int' label is dropped from now on, as we will always be referring to intrinsic luminosities, unless stated.

Regarding the spectroscopic sample, the AGN sources are those which display broad line features or narrow emission lines characteristic of AGN (BLAGN or NLAGN). The remaining sources with a spectroscopic classification are regarded as part of the non-AGN population (e.g., SF galaxies, stars). The NLAGN classification comes from MUSIC catalogue in GOODSs, and from \citet{Bongiorno10} in COSMOS.

In both the GOODS and COSMOS final AGN samples, most sources have an X-ray AGN classification (82\% and 80\%, respectively), and a significant fraction also has a spectroscopic AGN classification (21\% in GOODS and 55\% in COSMOS).

\subsubsection{KI/KIM efficiency in GOODSs}

\begin{figure}
  \begin{center}
    \includegraphics[width=0.9\columnwidth]{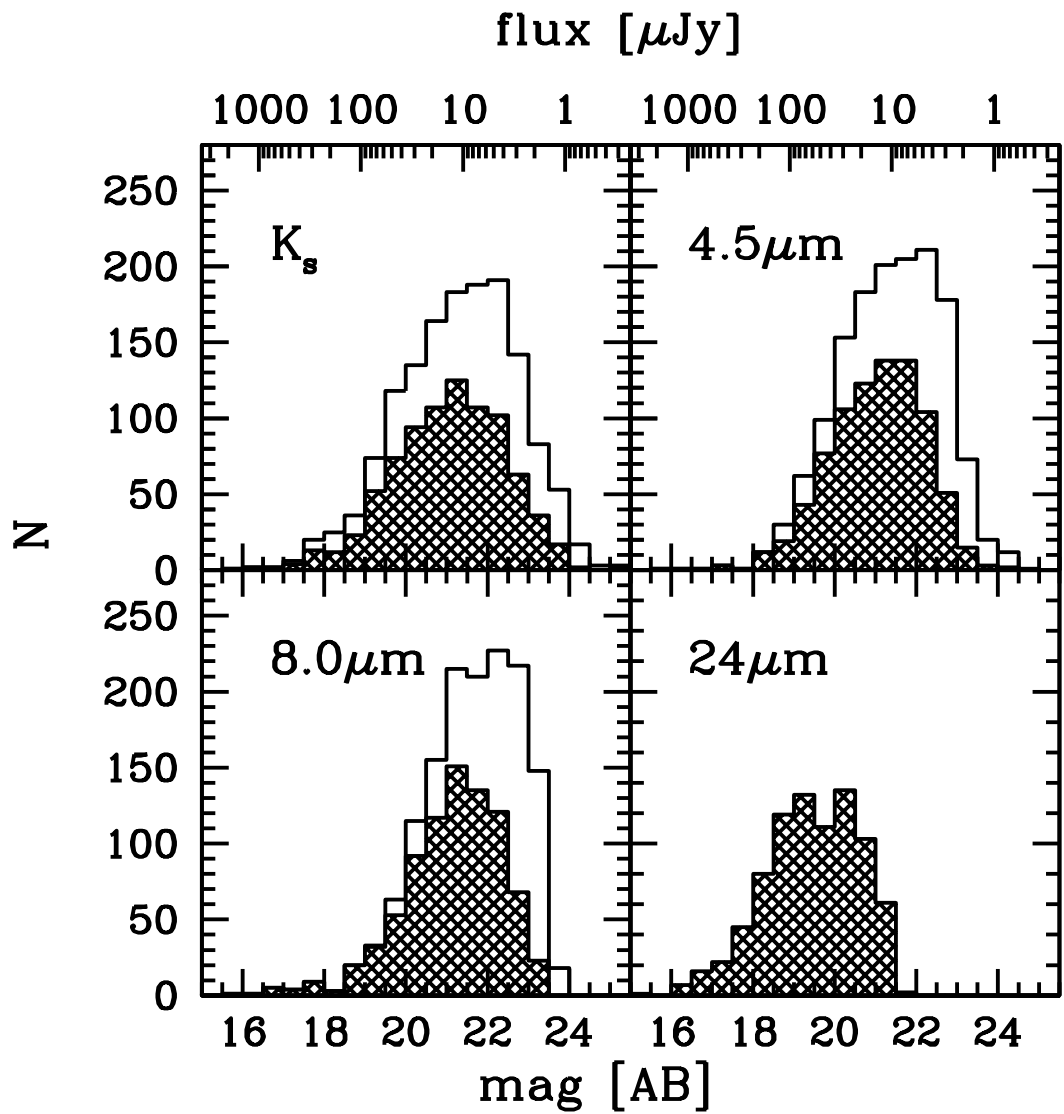}
  \end{center}
  \caption[Magnitude distribution of GOODSs sample]{The distribution in magnitude of the final GOODSs sample with reliable photometry in $K$-IRAC bands (open histogram) and $K$-IRAC-MIPS$_{24\mu{\rm m}}$ (hatched histogram). Each panel refers to the magnitude distribution in he following bands: $K_s$ (upper left), 4.5\,$\mu$m (upper right), 8.0\,$\mu$m (lower left), and 24\,$\mu$m (lower right). Note that in the latter both histograms coincide.}
  \label{c3fig:magdist}
\end{figure}

For consistency, we only consider sources with photometry estimates with a flux error smaller than a third of the flux value (equivalent to an error in magnitude smaller than 0.36) in all $K_s$-IRAC bands when testing L07, S05, and KI. This requirement will remove many of the fainter objects, but the final sample is still among the deepest ever used to test these IR criteria. The magnitude distribution of the sources considered is shown in Figure~\ref{c3fig:magdist}. Among the 1441 sources composing the final sample, 171 (12\%) are classified as AGN hosts (141 in X-rays and 38 through spectroscopy). The sample is further separated into redshift ranges ($0\leq{z}<1$, $1\leq{z}<2.5$, $2.5\leq{z}<4$). The adopted threshold of $z=2.5$, is the redshift beyond which L07, S05, and KI are believed to be strongly contaminated by SF systems as shown in section~\ref{c3sec:ki}. This results in 801, 536, and 94 sources with $K_s$-IRAC photometry at $0\leq{z}<1$, $1\leq{z}<2.5$, and $2.5\leq{z}<4$, respectively. When testing KIM we also require reliable 24\,$\mu$m photometry (see Figure~\ref{c3fig:magdist}). However, this requirement restricts the sample to the brightest sources, unavoidably increasing the probability of finding AGN dominated sources \citep{Brand06,Treister06,Donley08} and resulting in an unfair comparison with the remainder criteria (L07, S05, and KI). Hence, when comparing KIM to L07, S05, and KI, we consider the sample of 835 sources (460/325/47 in the respective redshift bins) with reliable $K_s$-IRAC-MIPS$_{24\,\mu{\rm m}}$ photometry, of which 139 (17\%) are classified as AGN hosts.

Tables~\ref{c3tab:xrspcs}, \ref{c3tab:xrspcs125}, and \ref{c3tab:xrspcs254} summarise the final statistics for the application of each of the IR criteria to the GOODSs control sample at different redshift ranges. L07 reaches the highest levels of completeness ($\mathcal{C}$) in the full redshift range covered, yet at the expense of its reliability ($\mathcal{R}$), this is, it selects too many sources as AGN (higher $\mathcal{C}$), L07 boundaries include consequently higher fractions of both AGN and non-AGN (lower $\mathcal{R}$). However, at $1\leq{z}<2.5$, L07 presents an $\mathcal{R}$ value comparable to that of S05. At $z<2.5$, both KI and KIM present the best $\mathcal{R}$ levels. While at $z<1$ KI and KIM present similar $\mathcal{R}$ to S05, at $1\leq{z}<2.5$, KI and KIM reach an impressive level of improvement over L07 and S05.

\ctable[
   cap     = GOODSs $0\leq{z}<1$ control sample,
   caption = GOODSs X-ray and Spectroscopic $0\leq{z}<1$ control sample test.,
   label   = c3tab:xrspcs,
   nosuper,
   mincapwidth = 15cm
]{cccccc}{
  \tnote[Note.]{ --- This table is restricted to the $0\leq{z}<1$ GOODSs sample. While in the upper set of rows it is required reliable photometry --- a magnitude error below 0.36 --- in $K$+IRAC bands, in the lower set of rows we also require reliable 24\,$\mu$m photometry.}
  \tnote[$^a$]{Number of sources selected by a given criterion with a AGN/non-AGN classification from X-rays and/or spectroscopy.}
  \tnote[$^b$]{Number of selected sources with an AGN classification, from either the X-rays or optical spectroscopy.}
  \tnote[$^c$]{Completeness calculated as AGN$_{\rm{SEL}}$/AGN$_{\rm{TOT}}$.}
  \tnote[$^d$]{Reliability calculated as AGN$_{\rm{SEL}}$/N$_{\rm{SEL}}$.}
  \tnote[$^e$]{The first row in each group refers to the total number of sources with reliable $K$+IRAC (upper group) and $K$+IRAC+24\,$\mu$m (bottom group) photometry. For reference, the value in parenthesis in $\cal{R}$ column gives the overall fraction of identified AGN hosts, equivalent to the $\cal{R}$ of a criterion selecting all sources with reliable photometry in the considered bands.}
}{ \FL
Sample & Criterion & N$_{\rm SEL}$\tmark[a] & AGN\tmark[b] & {$\cal C$}\tmark[c] & {$\cal{R}$}\tmark[d] \ML
& [none]\tmark[e] & 801 & 47 & \ldots & (6) \NN
K+IRAC & L07 & 105 & 22 & 47 & 21 \NN
& S05 &  26 & 12 & 26 & 46 \NN
& KI  &  24 & 12 & 26 & 50 \NN
\hline
& [none]\tmark[e] & 460 & 42 & \ldots & (9) \NN
K+IRAC+ & L07 & 76 & 19 & 45 & 25 \NN
MIPS$_{24\mu\rm{m}}$ & S05 & 20 & 11 & 26 & 55 \NN
& KI  & 17 & 10 & 24 & 59 \NN
& KIM & 15 &  8 & 19 & 53 \LL\NN
}

\ctable[
   cap     = GOODSs $1\leq{z}<2.5$ control sample,
   caption = GOODSs X-ray and Spectroscopic $1\leq{z}<2.5$ control sample test.,
   label   = c3tab:xrspcs125,
   nosuper
]{cccccc}{
  \tnote[Note.]{This table is restricted to the $1\leq{z}<2.5$ GOODSs sample. Table structure and columns definitions as in Table~\ref{c3tab:xrspcs}.}
}{ \FL
Sample & Criterion & N$_{\rm SEL}$ & AGN & {$\cal C$} & {$\cal{R}$} \ML
& [none] & 536 & 80 & \ldots & (15) \NN
$K$+IRAC & L07 & 171 & 50 & 63 & 29 \NN
& S05 & 104 & 28 & 35 & 27 \NN
& KI  &  60 & 32 & 40 & 53 \NN
\hline
& [none] & 325 & 61 & \ldots & (19) \NN
$K$+IRAC+ & L07 & 111 & 39 & 64 & 35 \NN
MIPS$_{24\mu\rm{m}}$ & S05 &  70 & 25 & 41 & 36 \NN
& KI  &  40 & 26 & 43 & 65 \NN
& KIM &  37 & 23 & 38 & 62 \LL\NN
}

\ctable[
   cap     = GOODSs $2.5\leq{z}<4$ control sample,
   caption = GOODSs X-ray and Spectroscopic $2.5\leq{z}<4$ control sample test.,
   label   = c3tab:xrspcs254,
   nosuper
]{cccccc}{
  \tnote[Note.]{This table is restricted to the $2.5\leq{z}<4$ GOODSs sample. Table structure and columns definitions as in Table~\ref{c3tab:xrspcs}.}
}{ \FL
Sample & Criterion & N$_{\rm SEL}$ & AGN & {$\cal C$} & {$\cal{R}$} \ML
& [none] & 94 & 40 & \ldots & (43) \NN
$K$+IRAC & L07 & 93 & 40 & 100 & 43 \NN
& S05 & 54 & 29 & 73 & 54 \NN
& KI  & 73 & 36 & 90 & 49 \NN
\hline
& [none] & 47 & 33 & \ldots & (70) \NN
$K$+IRAC+ & L07 & 47 & 33 & 100 & 70 \NN
MIPS$_{24\mu\rm{m}}$ & S05 & 32 & 24 & 73 & 75 \NN
& KI  & 41 & 30 & 91 & 73 \NN
& KIM & 33 & 24 & 73 & 73 \LL\NN
}

At high-$z$ ($2.5\leq{z}<4$), the fraction of identified AGN hosts is already high (40\%, increasing to 70\% when restricting to the MIPS$_{24\,\mu{\rm m}}$ detected sample). All but one object in the sample fall inside the L07 region, while S05 and KI show yet again higher reliability. Note, however, that KI is significantly more complete than S05. This incompleteness was shown for high-$z$ QSOs by \citet{Richards09}, who consequently extended S05 boundaries to bluer [5.8]-[8.0] colours. The result is the same when restricting to the MIPS$_{24\,\mu{\rm m}}$ detected sources, where KIM presents equal efficiencies as S05. This is easily explained with the necessary constraints applied to the sample. By requiring reliable detections in the full $K_s$-IRAC(-MIPS$_{24\,\mu{\rm m}}$) range, the sample is consequently restricted to the most luminous objects, which at the highest redshifts tend to be AGN hosts. Hence, with the current sample, no conclusion can be drawn on the efficiency of these IR criteria at such high redshifts.

\begin{figure}
  \begin{center}
    \includegraphics[width=1\columnwidth]{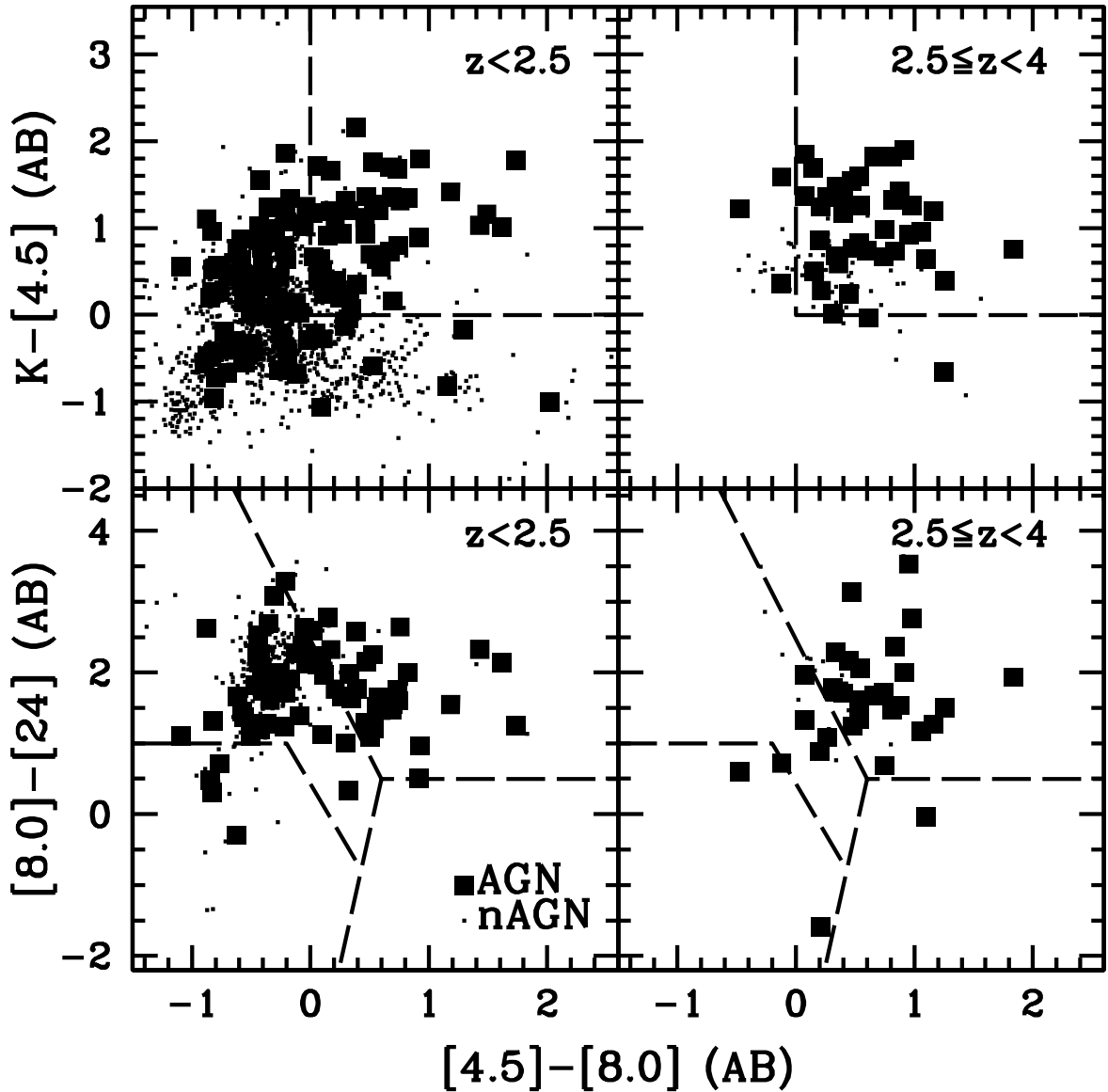}
  \end{center}
  \caption[MUSIC sources on KI and KIM colour-colour spaces]{The MUSIC sources on KI (upper panels) and KIM (lower panels) colour-colour spaces, divided into low-$z$ ($z<2.5$, left panels) and high-$z$ ($2.5\leq{z}<4$, right panels) groups. Squares represent AGN hosts, while dots highlight non-AGN sources. The dashed lines in the upper panels refer to the KI criterion, while the dashed lines in the lower panels refer to the adopted region boundaries from Figure~\ref{c3fig:c24m24}. All sources displayed in the lower panels have $K-[4.5]>0$ as required by the KIM criterion.}
  \label{c3fig:goodsamp}
\end{figure}

Figure~\ref{c3fig:goodsamp} details the application to GOODSs data of KI (upper panels) and KIM (lower panels). For this exercise, we have required reliable photometry in the bands needed for KI (upper panels) or KIM (lower panels). In KIM panels, the boundaries for each of the regions defined in Section~\ref{c3sec:imhighz} are shown. One of the main results from the KIM panels is the extremely low number of sources in the high-\textit{z} region (lower right-hand side of the diagram). This can be seen as a result of the limiting flux at both X-rays and spectroscopic observations, as a detection is required to have enough S/N for a proper classification with either indicator. Under such requirements, high redshift sources are, with the current existing data, likely AGN hosts, thus falling in the AGN region\footnote{The only source found in the high-\textit{z} region is a type-2 QSO ($\log(\rm{L_X[erg\,s^{-1}]})>44$) and indeed shows a redshift estimate of $z_{\rm phot}=3.1.$}. Also, the KIM-normal region is worthy of note. The galaxies that appear here are expected to be, as seen in Figure~\ref{c3fig:c24m24}, either early-to-late type, blue dust-free starbursts or hybrid sources at high redshift (due to the $K-[4.5]>0$ cut), with the IR colours becoming redder with AGN strength. The bluest [8.0]-[24] AGN source at high-$z$ in this region, with an X-ray AGN classification and a faint optical SED ($BViz>26$--$27$), has $z_{\rm phot}=2.54$. Already noted by \citet{Messias10}, it seems to be a very interesting source as its IR colours are compatible with a spiral Sa--Sc galaxy or, if an AGN is contributing to the IR, a galaxy of an earlier type (see Figure~\ref{c3fig:c24m24}). In either case, its optical flux and blue [8-0]-[24] colour hint to one of the most distant known objects of such evolved nature \citep[e.g.,][]{Stockton08,vanderWel11}. A proper discussion on this source and a whole sample of similar objects is differed to a future work (Messias et al., in preparation, but see Sections~\ref{c2sec:mir25} and \ref{c5sec:pdgs}), where the disc-like nature is confirmed. The numbers of GOODSs sources falling in each region of the KIM criterion (Section~\ref{c3sec:imhighz}) are summarised in Table~\ref{c3tab:kimmus}.

\ctable[
   cap     = KIM classification of GOODSs sample.,
   caption = KIM classification of GOODSs sample.,
   label   = c3tab:kimmus,
   mincapwidth = 11cm
]{crr}{
  \tnote{Number of sources with good photometry in all relevant bands ($K_s$, 4.5\,$\mu$m, 8.0\,$\mu$m, and 24\,$\mu$m), pre-selected with $K_s-[4.5]>0$, and with a AGN/non-AGN X-ray or spectroscopic classification.}
}{ \FL
Region & N\tmark[a] & AGN \ML
Total & 387 & 109 \NN
KIM-AGN & 90 & 58 \NN
KIM-Misc & 270 & 43 \NN
KIM-Normal & 26 & 7 \NN
KIM-High-$z$ & 1 & 1 \LL\NN
}

\subsubsection{KI/KIM efficiency in COSMOS}

The same comparison is now performed for COSMOS. No redshift segregation is applied as there is no classified SF system at $z\gtrsim1.6$ in this COSMOS sample. Among the 7180 sources with either a spectral or X-ray classification and adequate $K$-IRAC photometry, 1404 are flagged as AGN hosts. There are 2643 sources with MIPS$_{24\mu{\rm m}}$ detection (844 AGN hosts). Table~\ref{c3tab:Cxrspcs} reports the final statistics on the application of the various considered diagnostics. Having 84\% of this COSMOS sample at $z<1$, the statistics of this sample will be dominated by those of the $z<1$ population. Hence, it is fair to compare with the GOODSs sample at $0\leq{z}<1$ (Tables~\ref{c3tab:Cxrspcs} and \ref{c3tab:xrspcs}). Both imply the same conclusions, where the relative performances between each of the criteria agree between the two samples. L07 is the most complete, yet the least reliable. S05 and KI provide comparable $\mathcal{C}$ and $\mathcal{R}$. KIM is slightly less complete, presenting, however, comparable $\mathcal{R}$ levels to S05 and KI. Together with the results from Table~\ref{c3tab:xrspcs}, this likely means that many AGN dominating the SED at $<8\,\mu$m do not significantly dominate the IR regime at 12--24$\mu$m at least up to $z\sim1$. Table~\ref{c3tab:kimcsm} summarises the results from the application of each of the KIM criteria (Section~\ref{c3sec:imhighz}) to COSMOS sample.

\ctable[
   cap     = COSMOS control sample,
   caption = COSMOS X-ray and Spectroscopic control sample test.,
   label   = c3tab:Cxrspcs,
   nosuper
]{cccccc}{
  \tnote[Note.]{Table structure and columns definitions as in Table~\ref{c3tab:xrspcs}. No redshift range is adopted as there is no classified SF systems at $z\gtrsim1.6$ in the COSMOS sample.}
}{ \FL
Sample & Criterion & N$_{\rm SEL}$ & AGN & {$\cal C$} & {$\cal{R}$} \ML
& [none] & 7180 & 1404 & \ldots & (20) \NN
$K$+IRAC & L07 & 2032 & 1108 & 79 & 55 \NN
& S05 & 1101 &  919 & 65 & 83 \NN
& KI  &  965 &  879 & 63 & 91 \NN
\hline
& [none] & 2643 & 844 & \ldots & (32) \NN
$K$+IRAC & L07 & 1089 & 730 & 86 & 67 \NN
MIPS$_{24\mu\rm{m}}$ & S05 &  700 & 630 & 75 & 90 \NN
& KI  &  644 & 590 & 70 & 92 \NN
& KIM &  529 & 485 & 57 & 92 \LL\NN
}

\ctable[
   cap     = KIM classification of COSMOS sample.,
   caption = KIM classification of COSMOS sample.,
   label   = c3tab:kimcsm,
   mincapwidth = 10cm
]{crr}{
  \tnote{Table structure and columns definitions as in Table~\ref{c3tab:kimmus}.}
}{ \FL
Region & N\tmark & AGN \ML
Total & 838 & 648 \NN
KIM-AGN & 529 & 485 \NN
KIM-Misc & 304 & 158 \NN
KIM-Normal & 1 & 1 \NN
KIM-High-$z$ & 4 & 4 \LL\NN
}

It is difficult to directly compare in absolute value the results achieved with the GOODSs and COSMOS samples, since many survey characteristics differ between the two. As an example, by applying the COSMOS (IR and X-rays) flux limits to the GOODSs sample, the $\mathcal{C}$ and $\mathcal{R}$ values are closer to those of COSMOS. Other issues may contribute to this, such as (a) different photometry extraction methods (ConvPhot in GOODSs and aperture photometry in COSMOS), (b) different spectral coverage depth and procedures for spectral classification, (c) the different photon indices used for the GOODSs and COSMOS samples to convert from count rates to X-ray fluxes, (d) difference in relative sensitivity between soft and hard bands of \textit{Chandra Space Telescope} (\textit{Chandra}, in GOODSs) and XMM-\textit{Newton} (in COSMOS), and (e) cosmic variance. We again stress, however, that relative efficiency between the criteria is the same in the two samples.

\subsection{IR-excess sources} \label{c3sec:irxs}

Also known as IR bright galaxies (IRBGs), IRxs sources are believed to be part of an extreme IR population, the compton-thick (type-2) AGN, frequently missed by optical/X-ray surveys. The selection criteria vary in the literature, but it is accepted that all IRxs diagnostics are quite reliable in selecting this type of source \citep[$>80\%$;][]{Donley08,Treister09b,Donley10}. The diagnostics considered below rely on optical-to-IR colour cuts, more specifically, $R-K$ and $R-[24]$. However, $R$-band photometry is not available in the MUSIC catalogue. We thus convert those colours to equivalent ones using $i$-band ($i-K$ and $i-[24]$) considering a power-law spectrum ($f_{\nu}\propto\nu^{\alpha}$). We highlight three criteria. \citet[][D08]{Dey08} select sources with $S_{24}/S_R>1000$ and $S_{24}>300\,\mu{\rm Jy}$ (equivalent to $i-[24]>7$ and $[24]<17.5$), \citet[][F08]{Fiore08} with $S_{24}/S_R>1000$ and $(R-K)_{\rm VEGA}>4.5$ ($i-[24]>7$ and $i-K>2.5$\footnote{It should be noted, however, that a $J-K$ colour is likely more efficient than $i-K$ in selecting AGN sources \citep[][and Section~\ref{c2sec:agncont}]{Messias10}.}), allowing a fainter flux cut at $S_{24}>40\,\mu{\rm Jy}$ ($[24]<20$). Finally, we also consider the brightest $S_{24}/S_R>1000$ sources by adopting the flux cut of \citet[][P08]{Polletta08}, $S_{24}>1\,{\rm mJy}$ (corresponding to $[24]<16.5$).

These criteria were applied to the MUSIC and COSMOS catalogues and Table~\ref{c3tab:irxs} details the numbers of the selected sources by each of the IR colour criteria. Similar results are achieved in both GOODSs and COSMOS fields: S05 is the criterion selecting fewer IRxs sources, and KIM is always more complete than both KI and S05. KIM is even more complete than L07 when selecting the brightest IRxs sources (P08), emphasising its great potential.

\ctable[
   cap     = Selection of IRxs sources.,
   caption = Selection of IRxs sources.,
   label   = c3tab:irxs,
   nosuper
]{cccc}{
  \tnote[Note.]{ --- The numbers in parenthesis give the equivalent fractions.}
}{ \FL
Region & F08 & D08 & P08 \ML
\multicolumn{4}{c}{GOODSs} \NN
\ldots & 77 & 10 & 1 \NN
L07 & 72 (94\%) & 9 (90\%) & 1 (100\%) \NN
S05 & 29 (38\%) & 5 (50\%) & 1 (100\%) \NN
KI  & 40 (52\%) & 7 (70\%) & 1 (100\%) \NN
KIM & 41 (53\%) & 8 (80\%) & 1 (100\%) \NN
\hline \NN
\multicolumn{4}{c}{{\rm COSMOS}} \NN
\ldots & 991 & 256 & 51 \NN
L07 & 909 (92\%) & 244 (95\%) & 47 (92\%) \NN
S05 & 381 (38\%) & 137 (54\%) & 39 (76\%) \NN
KI  & 493 (50\%) & 179 (70\%) & 46 (90\%) \NN
KIM & 618 (62\%) & 212 (83\%) & 50 (98\%) \LL\NN
}

\subsection{SDSS QSOs} \label{c3sec:sdssqso}

QSOs present in the Sloan Digital Sky Survey Quasar Catalogue Data Release 7 \citep[SDSS-DR7,][]{Schneider10} were cross-matched (2'' radius) with the $Spitzer$ IR catalogues from the COSMOS (S-COSMOS), Lockman Hole, ELAIS-N1, and ELAIS-N2 \citep[SWIRE,][]{Lonsdale03} fields using GATOR\footnote{http://irsa.ipac.caltech.edu/applications/Gator/} at IRSA-NASA/IPAC. The final number of sources is 293. $K$-band photometry comes from 2MASS \citep[for 21\% of the sample,][]{Skrutskie06}, UKIDSS-DXS DR8\footnote{UKIDSS uses the UKIRT Wide Field Camera \citep[WFCAM;][]{Casali07} and a photometric system described in \citet{Hewett06}. The pipeline processing and science archive are described in Irwin et al. (in preparation) and \citet{Hambly08}. We have used data from the 8th data release.} \citep[][29\%]{Lawrence07}, and COSMOS \citep[][23\%]{Ilbert09}. Overall, there are 186 QSOs with reliable photometry in all IRAC channels. Of which, 140 have also MIPS$_{24\mu\rm{m}}$ photometry, and 142 have $K$-band photometry. We find 107 with full $K$-IRAC-MIPS$_{24\mu\rm{m}}$ coverage. To enhance the high-$z$ regime sampling, we further include 13 SDSS-DR3 QSOs at $z\sim6$ \citep{Jiang06}. Of these, 12 are detected in all IRAC and MIPS$_{24\mu\rm{m}}$ channels, while only five have 2MASS $K$-band data.

\begin{figure}
  \begin{center}
    \includegraphics[width=1\columnwidth]{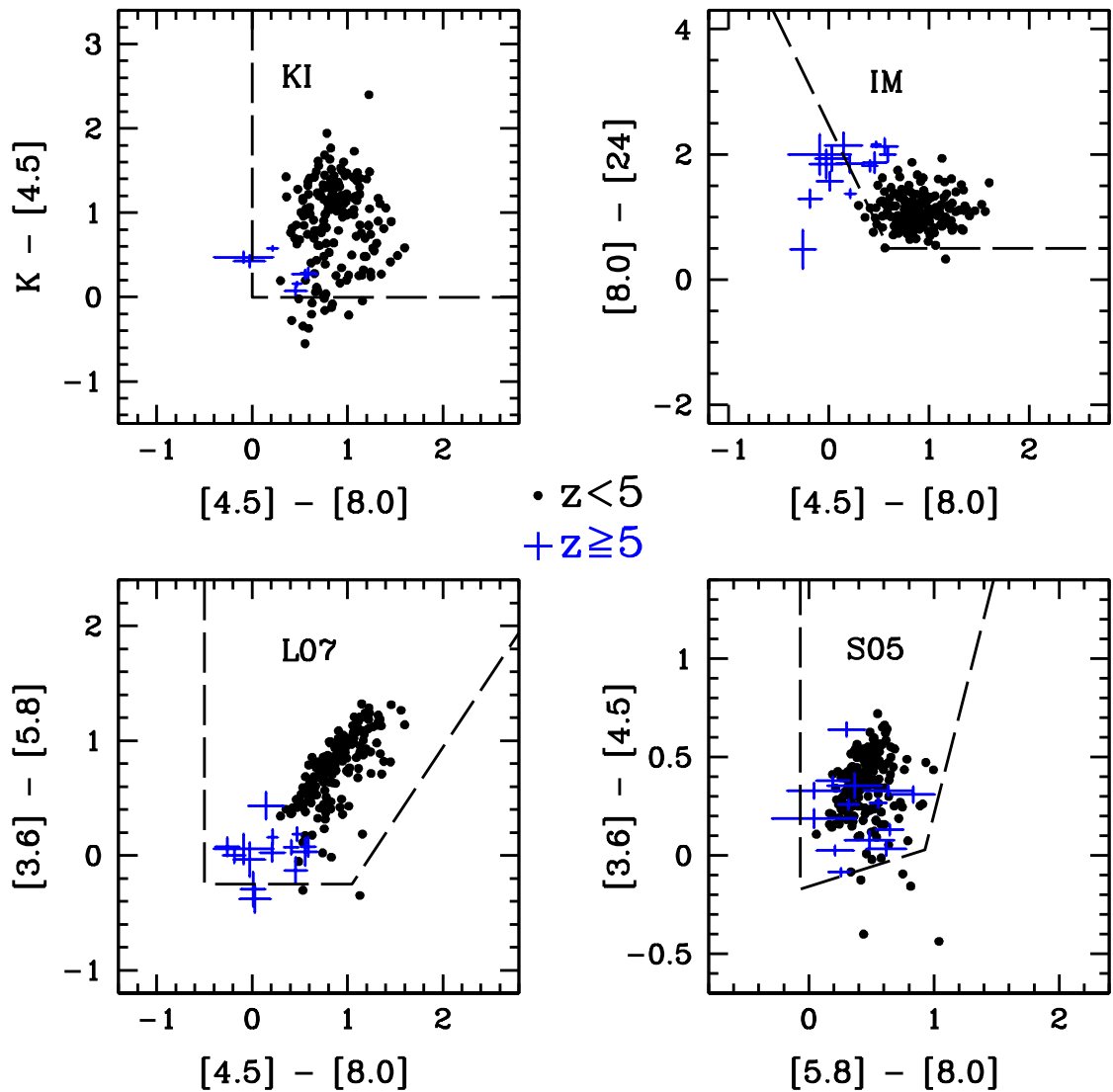}
  \end{center}
  \caption[IR colours for SDSS-DR7 QSOs]{The SDSS-DR7 QSOs found in SWIRE and COSMOS fields together with \citet{Jiang06} sample displayed in KI (top left), IM (top right), L07 (bottom left), and S05 (bottom right) colour-colour spaces. Black dots represent $z<5$ sources (with small photometric errors on average), and blue error bars otherwise.}
  \label{c3fig:sdssPlot}
\end{figure}

Figure~\ref{c3fig:sdssPlot} shows the location of the QSO sample in the KI, IM (Section~\ref{c3sec:imhighz}), L07, and S05 colour-colour spaces. Only sources with reliable photometry in the displayed bands are shown. All four criteria select most of the displayed sample ($>90\%$). For $z>5$ QSOs, the IM completeness drops to 50\%, in agreement with Figure~\ref{c3fig:c24m24} where some QSO templates start to move out of the KIM-AGN region at $z\sim6$. We note, however, that if there is a prior indication for such high redshifts ($z>3$), then the [8.0]-[24] colour can be used by itself and much more efficiently for the identification of AGN (cf. Figure~\ref{c3fig:z824}). For $z>5$ QSOs, all but one show $[8.0]-[24]>1$. The small number of QSOs with blue $K-[4.5]$ colours is explained in light of the discussion in Section~\ref{c3sec:ki}. These are potentially less IR dominant AGN and/or sources possessing strong line emission.

The high completeness levels achieved with this optical selected sample show the eclectic selection of IR criteria. However, optically selected AGN are not the main targets of IR AGN diagnostics, as, by definition, optical surveys \textit{do} detect them. The most interesting use of these criteria is to recover sources undetected at X-ray and optical wavelengths. Sections~\ref{c3sec:irxs} and \ref{c3sec:hzrg} are, in this respect, much more representative of the intended use of IR AGN diagnostics.

\subsection{High redshift Radio Galaxies} \label{c3sec:hzrg}

To test yet another AGN population, we now consider High-\textit{z} Radio Galaxies (H\textit{z}RGs). These are among the most luminous sources in the Universe and are believed to host powerful AGN. We use the sample of 71 H\textit{z}RGs from \citet{Seymour07}. These are all at $z>1$, a redshift range where no normal galaxy is believed to contaminate the AGN IM region proposed in Section~\ref{c3sec:imhighz}. This is a classic example --- such as that of LBGs --- for the direct application of the IM boundaries. Having this, the $K-[4.5]>0$ colour cut is not required to disentangle AGN/non-AGN dominated sources at $z<1$, meaning that one may consider [4.5]-[8.0] and [8.0]-[24] colours alone to determine whether AGN or stellar emission dominates the IR spectral regime.

\begin{figure}
  \begin{center}
    \includegraphics[width=1\columnwidth]{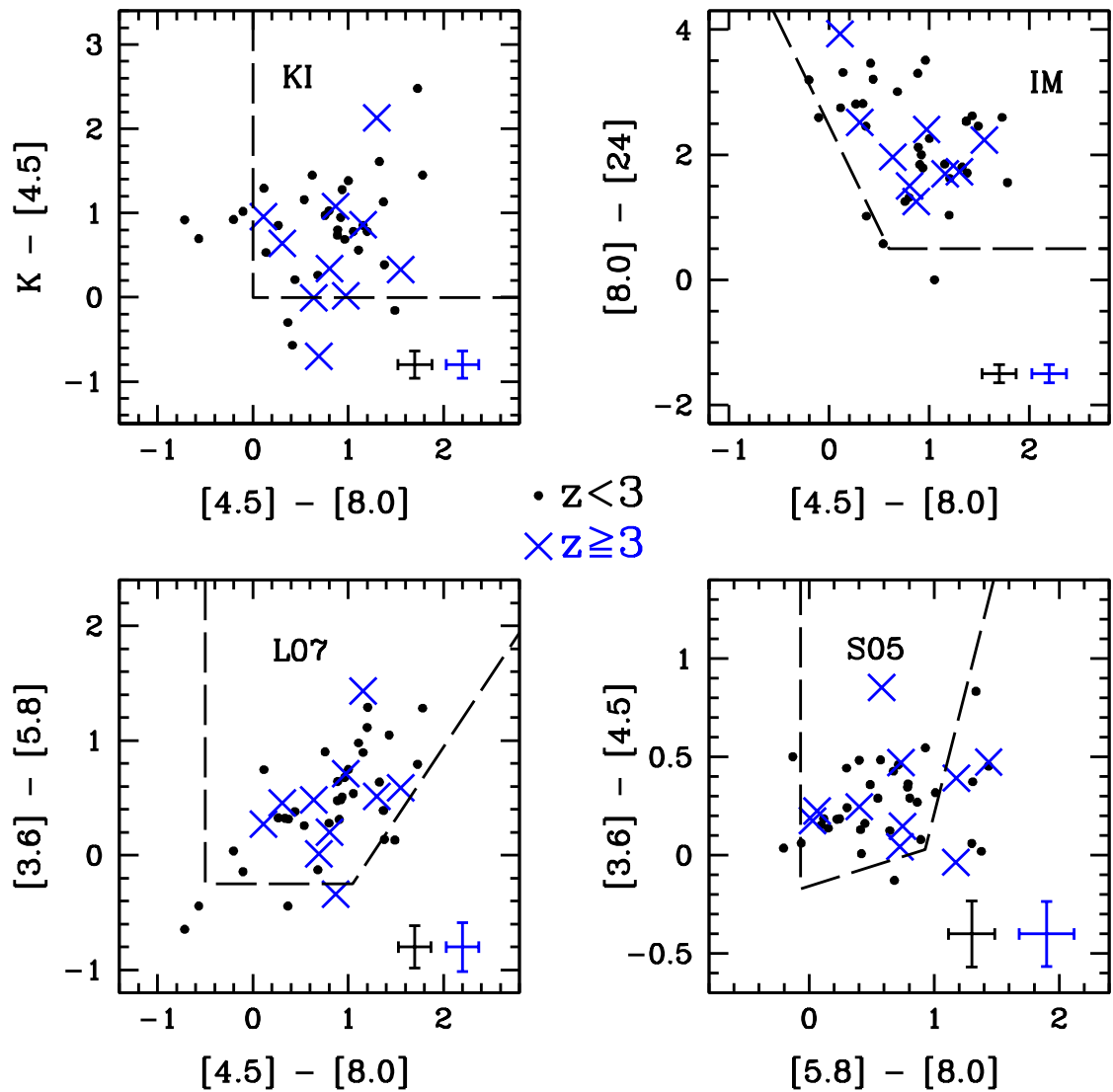}
  \end{center}
  \caption[IR colours of high-$z$ radio sources]{The HzRG ($z>1$) sample from \citet{Seymour07} displayed in the same colour-colour spaces as in Figure~\ref{c3fig:sdssPlot}. Note the objects at $2<[8.0]-[24]<4$ which are even redder than QSOs (Figure~\ref{c3fig:sdssPlot}). Dots show the $z<3$ population, while crosses that at $z\geq3$. Average photometric errors are shown on the lower right for the $z<3$ (left error bar) and $z\geq3$ (right error bar) samples.}
  \label{c3fig:hzrgIM}
\end{figure}

Figure~\ref{c3fig:hzrgIM} shows the location of 62 H\textit{z}RGs in the IM colour-colour diagram. Note the difference to SDSS QSOs (Figure~\ref{c3fig:sdssPlot}), where H\textit{z}RGs show predominantly redder colours. The AGN region correctly selects as AGN 85\% (40 sources) of the sample with adequate photometry (47 sources detected at 4.5, 8.0, and 24$\mu$m). In case no redshift estimate was available, however, one would need the $K-[4.5]>0$ colour cut to apply the IM AGN criterion, i.e., the KIM criterion. The application of KIM would result in a 76\% completeness level. L07 selects 85\% (41 out of 48 sources), and S05 selects 69\% (33 out of 48 sources).

Again we recall that much of the improvement of KI/KIM over the commonly used L07 and S05 will be in terms of reliability, not evaluated with this sample nor those referred in Sections~\ref{c3sec:irxs} and \ref{c3sec:sdssqso}.

\section{Discussion} \label{c3sec:disc}

\subsection{Selection of type-1/2 and low-/high-luminosity sources} \label{c3sec:typesel}

\begin{figure}[t]
  \begin{center}
    \includegraphics[width=0.6\columnwidth]{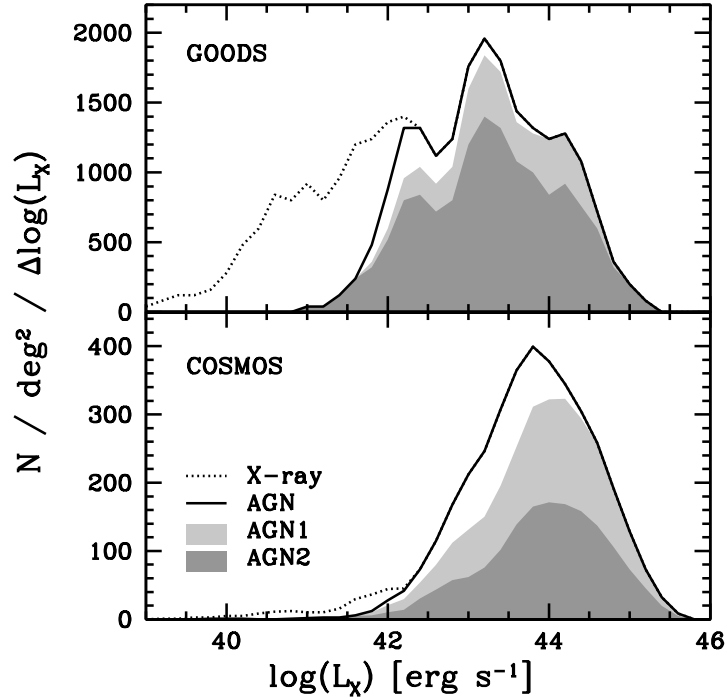}
  \end{center}
  \caption[X-ray luminosity distributions]{The source density distribution with intrinsic X-ray luminosity distribution for GOODSs (upper panel, $\sim140\,\rm{arcmin}^2$) and COSMOS (lower panel, $1.8\,\rm{deg}^2$) samples (note the y-axis are different). The trends were obtained with a moving bin of width $\Delta\log({\rm L}_X)=0.6$, with measurements taken each $\Delta\log({\rm L}_X)=0.2$. The overall X-ray population is represented by the dotted line, the AGN by the continuous line. The AGN population is further separated into the type-1 (light shaded region, $\log(\rm{N_H[\,cm^{-2}]})\leq22$) and type-2 (dark shaded region, $\log(\rm{N_H[\,cm^{-2}]})>22$) sub-populations.}
  \label{c3fig:lxdist}
\end{figure}

Previous studies have claimed that IR colour-colour criteria are biased toward unobscured systems \citep[BLAGN or type-1 AGN;][]{Stern05,Donley07,Cardamone08,Eckart10}, and tend to select the most luminous objects, missing many low-luminosity ones \citep{Treister06,Cardamone08,Donley08,Eckart10}. These tendencies are also assessed in this work. The considered AGN samples are those of GOODSs and COSMOS detailed in Section~\ref{c3sec:goodssample}. X-ray and spectroscopy data are considered in order to separate the samples into type-1 (unobscured) and type-2 (obscured) AGN. The way both regimes were considered and the relevant assumptions for this classification are discussed with more detail in Appendix~\ref{c3app:type12}. The intrinsic X-ray luminosity distribution is shown for GOODSs and COSMOS in Figure~\ref{c3fig:lxdist} for the overall X-ray AGN sample, highlighting the type-1 and type-2 AGN components.

\begin{figure}
  \begin{center}
    \includegraphics[angle=-90,width=0.75\columnwidth]{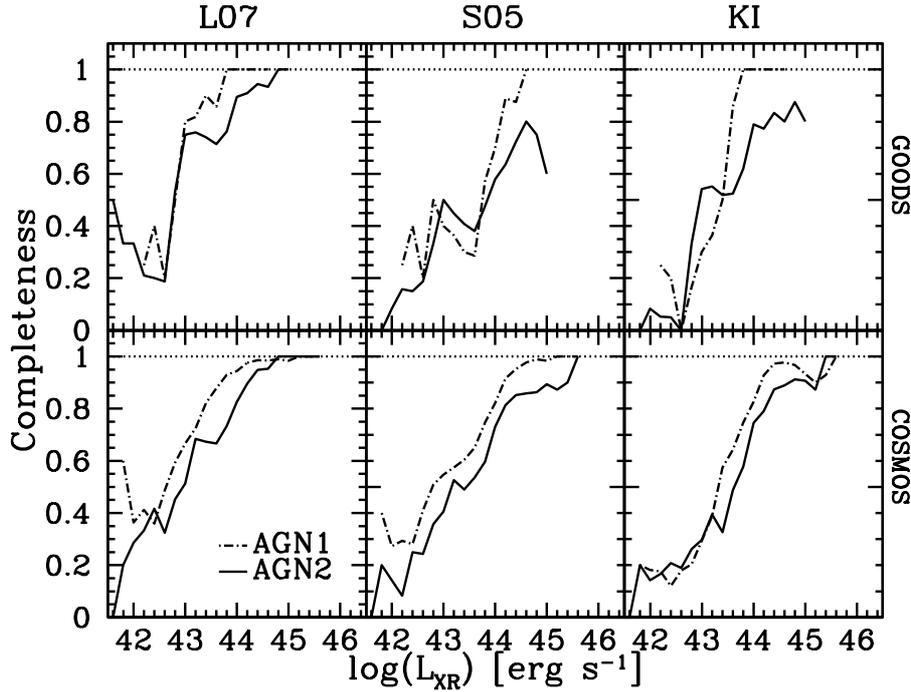}
  \end{center}
  \caption[X-ray AGN completeness of IR AGN criteria]{The AGN completeness for L07, S05, and KI criteria depending on source X-ray luminosity and type-1 (dotted-dashed lines) or type-2 (solid lines) nature.}
  \label{c3fig:s21c}
\end{figure}

We again emphasise that the aim of IR colour-colour criteria is the selection of galaxies with an IR SED dominated by AGN light. However, low-luminosity AGN will often not dominate the IR emission of a galaxy, making their IR selection unlikely. This is clearly seen in Figure~\ref{c3fig:s21c}, where the completeness of the L07, S05, and KI AGN selection criteria increases significantly with source luminosity, in agreement with previous work \citep{Treister06,Cardamone08,Donley08,Eckart10}. One can also note that type-1 and type-2 AGN-selection completeness follow each other quite closely, pointing to a much stronger dependency on source luminosity than on type-1/type-2 nature\footnote{The behaviour would be similar if considering the uncorrected or observed luminosity.}. If, as indicated by previous work, type-1 AGN tend to be more luminous than type-2 AGN (see the discussions in see the discussions in \citealt{Treister09b} and \citealt{Bongiorno10} and references therein), this will then lead to a higher fraction of type-1 objects among IR-selected AGN samples. However, this does not mean the IR criteria are more sensitive to type-1 AGN, as the main dependency is on luminosity (Figure~\ref{c3fig:s21c} points to a very slight preference for the selection of type-1 over type-2, for the same luminosity), but mostly a natural result of luminosity dependence. It is nevertheless worthwhile to note that the classification of AGN as type-1 or type-2 is not always straightforward nor unequivocal (see Appendix~\ref{c3app:type12}). A more robust study of the dependency of the IR AGN-selection criteria on AGN type requires a different approach, which we now present.

Let $\mathcal{S}$ be introduced as the relative sensitivity of a given selection criterion to a certain AGN type over another. Take unobscured (type-1) and obscured (type-2) AGN populations as an example. These sub-populations exist in the overall AGN population at a given proportion. If such a proportion is maintained after applying a given selection criterion (either colour or luminosity based), it means the criterion is equally sensitive to either population, if not, there is a bias. Hence, $\mathcal{S}$ is calculated as the ratio between the proportion estimated using a given criterion and the proportion estimated for the total AGN population. That is, the relative sensitivity regarding type-1 and type-2 AGN is defined as $\mathcal{S}_{\rm 12}=(\rm{A_1/A_2})_{\rm SEL}/(\rm{A_1/A_2})_{\rm TOT}$, where $\rm{A_1}$ and $\rm{A_2}$ are the numbers of type-1 and type-2 objects, respectively. Likewise, the relative sensitivity concerning low ($\log(\rm{L_X[erg\,s^{-1}]})<43.5$) and high X-ray luminosity ($\log(\rm{L_X[erg\,s^{-1}]})\geq43.5$) is defined as $\mathcal{S}_{\rm HL}=(\rm{A_H/A_L})_{\rm SEL}/(\rm{A_H/A_L})_{\rm TOT}$, where $\rm{A_H}$ and $\rm{A_L}$ are the numbers of high and low X-ray luminosity objects, respectively. Values of 1 mean no bias, while, for example, higher values of $\mathcal{S}_{\rm 12}$ or $\mathcal{S}_{\rm HL}$ mean biases favouring the selection of type-1 or high-luminosity AGN, respectively. As an example, in Figure~\ref{c3fig:s21c} the IR criteria clearly show a bias toward the selection of high luminosity sources. This implies by definition $\mathcal{S}_{\rm HL}>1$ for the IR AGN diagnostics. Care should be taken when comparing $\mathcal{S}$ values. For instance, if a given criterion has a lower $\mathcal{S}_{\rm 12}$ value than another criterion, that does not necessarily mean a comparatively higher completeness of type-2 sources, nor lower completeness of type-1 sources. The completeness ought to be estimated separately.

\begin{figure}
  \begin{center}
    \includegraphics[width=0.75\columnwidth]{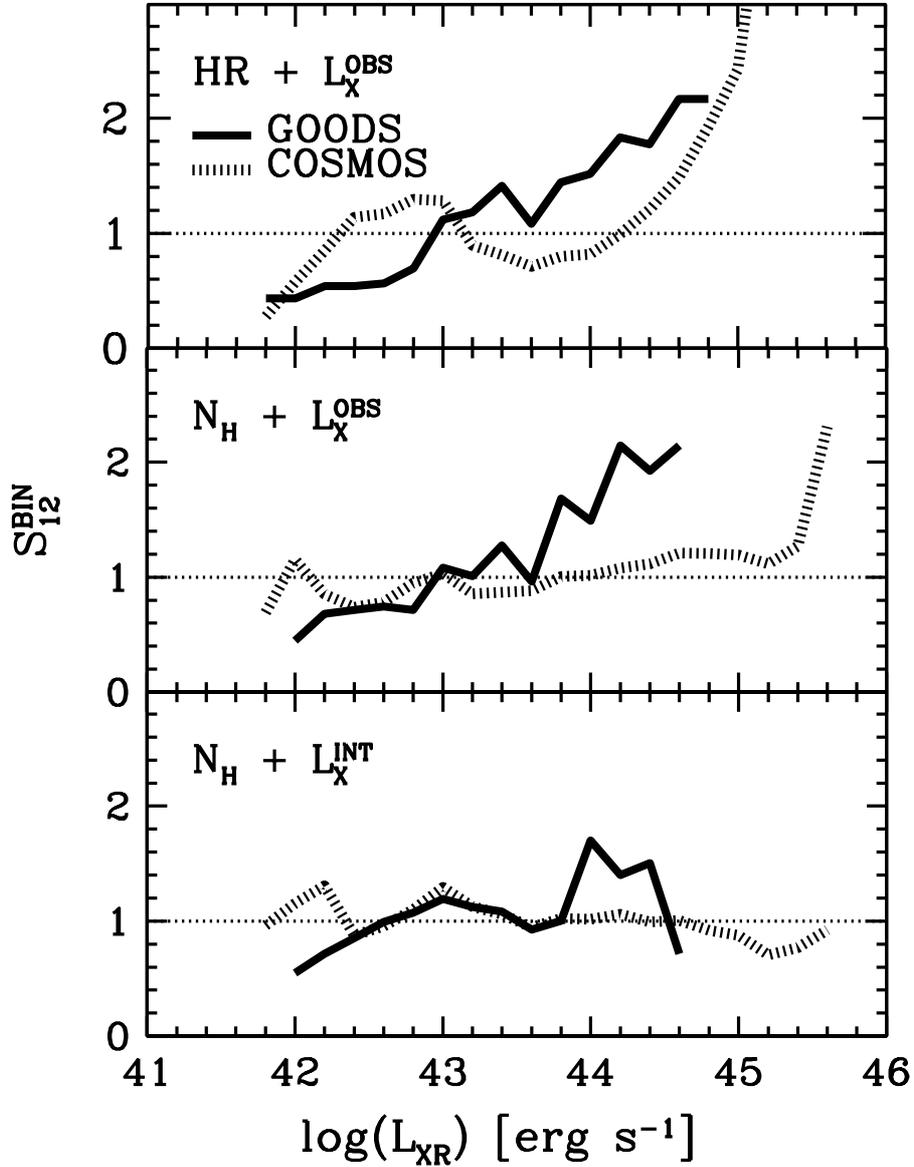}
  \end{center}
  \caption[$\mathcal{S}_{\rm 12}$ dependency on X-ray luminosity]{The variation of $\mathcal{S}_{\rm 12}$ with source X-ray luminosity. The different panels show the effect of different assumptions in assessing the luminosity classes, by considering either the intrinsic or observed luminosities (L$_X^{INT}$ or L$_X^{OBS}$, respectively), and type-1 or type-2 populations, by considering either the HR or $\rm{N_H}$. A moving bin is used as described in Figure~\ref{c3fig:lxdist}. In each bin, $\mathcal{S}_{\rm 12}^{\rm BIN}=(\rm{A_1/A_2})_{\rm BIN}/(\rm{A_1/A_2})_{\rm TOT}$.}
  \label{c3fig:s21lumx}
\end{figure}

Figure~\ref{c3fig:s21lumx} shows the variation of $\mathcal{S}_{\rm 12}$ with luminosity, meaning that in each bin $\mathcal{S}_{\rm 12}=(\rm{A_1/A_2})_{\rm BIN}/(\rm{A_1/A_2})_{\rm TOT}$. The trend is estimated with a moving bin with width $\Delta\log({\rm L_X})=0.6$, with measurements taken every $\Delta\log({\rm L_X})=0.2$ step (similar to the moving average method). The three panels show the difference when considering intrinsic or observed luminosities ($\rm{L_X^{INT}}$ or $\rm{L_X^{OBS}}$, respectively), and N$_{\rm H}$ or HR (two different alternatives for the AGN type-1 and type-2 classifications). In the upper panel, the use of HR and $\rm{L_X^{OBS}}$ imply a bias favouring the selection of type-1 AGN at the highest luminosities, in agreement with, e.g., \citet{Hasinger08} and \citet{Bongiorno10}. Yet, if one considers N$_{\rm H}$ instead (middle panel) this bias appears to decrease. The trend disappears over the full luminosity range if both N$_{\rm H}$ and $\rm{L_X^{INT}}$ are considered instead (lower panel).



\begin{figure}[t]
  \begin{center}
    \includegraphics[width=0.5\columnwidth]{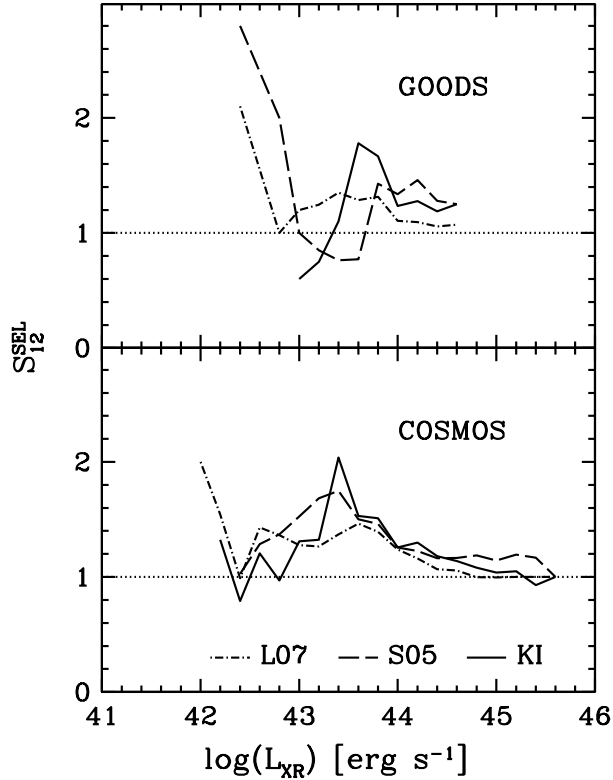}
  \end{center}
  \caption[$\mathcal{S}_{\rm 12}$ dependency on intrinsic X-ray luminosity for IR criteria]{The variation of $\mathcal{S}_{\rm 12}$ with source intrinsic X-ray luminosity for L07, S05, and KI. $\rm{N_H}$ is considered to identify type-1 and type-2 AGNs. A moving bin is used as described in Figure~\ref{c3fig:lxdist}. In each bin $\mathcal{S}_{\rm 12}^{\rm SEL}=(\rm{A_1/A_2})_{\rm SEL}/(\rm{A_1/A_2})_{\rm BIN}$.}
  \label{c3fig:s21f}
\end{figure}

The dependency of the type-1 to type-2 ratio on luminosity (or redshift) affects the evaluation of the type-1/type-2 bias of the IR criteria. So, assuming that IR criteria are clearly dependent on source luminosity (presenting high $\mathcal{S}_{\rm HL}$, Figure~\ref{c3fig:s21c}) and the type-1/type-2 AGN ratio is equal throughout the full range of intrinsic luminosities (Figure~\ref{c3fig:s21lumx}), does our sample imply nevertheless a bias toward type-1 sources, as referred to in the literature? Figure~\ref{c3fig:s21f} helps to clarify this point. Restricting the estimate of $\mathcal{S}_{\rm 12}$ to each luminosity bin ($\mathcal{S}_{\rm 12}=(\rm{A_1/A_2})_{\rm SEL}/(\rm{A_1/A_2})_{\rm BIN}$ for each IR criterion) any possible luminosity dependency seen in Figure~\ref{c3fig:s21lumx} is avoided. Although the GOODSs sample (upper panel) does not allow one to draw any conclusion due to the high scatter, in COSMOS (lower panel) it is clear the IR criteria are biased toward type-1 AGN at intermediate luminosities ($43<\log(\rm{L_X[erg\,s^{-1}]})<44$), while both types seem equally selected at the highest luminosities. This is in agreement with the findings of \citet{Treister09b}, who noticed a lack of IR excess emission in intermediate luminosity obscured AGN, even though their analysis is mainly spectroscopy based.

In \citet{Treister09b}, effects of self-absorption in a thick torus are evoked as the mechanism behind the lack of IR AGN emission. However, can dust-free X-ray obscuration also account for such behaviour? As discussed in Appendix~\ref{c3app:type12}, the existence of dust-free clouds between the nuclear source and the dust torus is responsible for the bulk of the X-ray obscuration, but it will not emit at IR wavelengths. This results in a weaker radiation field at any given radius when compared to a gas-obscuration-free scenario. The inner radius of the dust torus \citep[set by the sublimation radius, e.g.,][]{Nenkova08,HonigKishimoto10} will be smaller and the dust will still be heated up to the highest temperatures, emitting at short IR wavelengths. However, the existence of a weaker radiation field results in a less intense dust emission, when comparing gas obscured and unobscured AGN with equal intrinsic X-ray luminosities. Hence, dust-free obscuration can indeed be another reason for the observed lack of IR AGN emission of intermediate luminosity AGN.

Also, the scenario where the obscuration material of the X-ray nuclear source is not the circumnuclear torus, but instead material present in the disc of the host galaxy itself may happen. AGN are frequently found in disc-like sources either at low redshifts \citep[e.g.,][]{Griffith10,Cisternas11} or at earlier times \citep[e.g.,][]{Schawinski11,Kocevski11}. Extreme examples are also found in the literature. For example, \citet{Polletta06} find five X-ray compton-thick candidates (sources having $\log(\rm{N_H[cm^{-2}]})>24$). Yet, three of them are not selected as such at IR wavelengths, showing instead normal spiral-type SEDs. Available optical imaging of these sources is however inconclusive regarding the morphology and orientation of these systems. This host galaxy disc obscuration effect is not expected to be a significant contributor to the X-ray compton-thick population, as it has been found that the rotation axis of the central black-hole (a) is randomly align to the galaxy disc in Seyfert galaxies \citep{Clarke98,Nagar99,Kinney00} and (b) it seems to avoid the dust torus plane in radio galaxies \citep[][and references therein]{Schmitt02}. Nevertheless, deeper optical and (near-)IR imaging should be pursued as a fundamental tool to confirm such scenario in these three specific X-ray compton-thick sources.

Table~\ref{c3tab:typecompG} shows the results for GOODSs sample, while Table~\ref{c3tab:typecompC} those for COSMOS, both considering the full redshift range. Both show a clear bias towards more X-ray luminous sources (as implied by Figure~\ref{c3fig:s21c}) and, at a lower level, towards type-1 objects.

\ctable[
   cap     = AGN-type selection in GOODSs.,
   caption = AGN-type selection comparison in GOODSs.,
   label   = c3tab:typecompG,
   nosuper,
   mincapwidth = 15cm
]{ccccc|ccc}{
  \tnote[Note.]{ --- While in the upper set of rows it is required reliable photometry ($\delta{\rm mag}<0.36$) in $K$+IRAC bands, in the lower set of rows we also require reliable 24${\mu}m$ photometry. A$_1$ stands for AGN type-1, whereas A$_2$ for type-2 (X-ray or spectroscopic classifications). $\rm{A_L}$ refers to the sources having $\log({\rm L}_{\rm XR}[erg\,s^{-1}])<43.5$, while $\rm{A_H}$ refers to those having $\log({\rm L_X[erg\,s^{-1}])}\geq43.5$. No redshift cut applied to this sample.}
  \tnote[$^a$]{Number of AGN sources selected by the applied MIR criterion.}
  \tnote[$^b$]{Relative sensibility: $\mathcal{S}_{\rm 12}=(\rm{A_1/A_2})_{\rm SEL}/(\rm{A_1/A_2})_{\rm TOT}$ and $\mathcal{S}_{\rm HL}=(\rm{A_H/A_L})_{\rm SEL}/(\rm{A_H/A_L})_{\rm TOT}$. $\mathcal{S}_{\rm 12}$ or $\mathcal{S}_{\rm HL}$ values higher than one mean greater relative sensitivity toward A1 or $\rm{A_H}$ AGN, respectively.}
  \tnote[$^c$]{The first row in each group refers to the total number of sources of a given type with reliable $K$+IRAC (upper group) and $K$+IRAC+24$\mu$m (bottom group) photometry.}
}{ \FL
Sample & Criterion & A$_1$\tmark[a] & A$_2$\tmark[a] & $\mathcal{S}_{\rm 12}$\tmark[b] & $\rm{A_L}$\tmark[a] & $\rm{A_H}$\tmark[a] & $\mathcal{S}_{\rm HL}$\tmark[b] \ML
& [none]\tmark[c] & 33 & 119 & \ldots & 171 & 65 & \ldots \NN
$K$+IRAC & L07 & 26 & 75 & 1.26 & 69 & 58 & 2.25 \NN
& S05 & 17 & 45 & 1.37 & 36 & 38 & 2.78 \NN
& KI  & 20 & 55 & 1.32 & 33 & 51 & 4.07 \NN
\hline
& [none]\tmark[c] & 26 & 101 & \ldots & 137 & 56 & \ldots \NN
$K$+IRAC & L07 & 23 & 61 & 1.47 & 56 & 49 & 2.15 \NN
MIPS$_{24\mu\rm{m}}$ & S05 & 15 & 39 & 1.50 & 30 & 34 & 2.78 \NN
& KI  & 18 & 44 & 1.59 & 27 & 43 & 3.90 \NN
& KIM & 13 & 37 & 1.37 & 25 & 35 & 3.43 \LL\NN
}

\ctable[
   cap     = AGN-type selection in COSMOS.,
   caption = AGN-type selection comparison in COSMOS.,
   label   = c3tab:typecompC,
   nosuper
]{ccccc|ccc}{
  \tnote[Note.]{ --- Table structure and columns definitions as in Table~\ref{c3tab:typecompG}.}
}{ \FL
Sample & Criterion & A$_1$ & A$_2$ & $\mathcal{S}_{\rm 12}$ & $\rm{A_L}$ & $\rm{A_H}$ & $\mathcal{S}_{\rm HL}$ \ML
& [none] & 519 & 629 & \ldots & 468 & 916 & \ldots \NN
$K$+IRAC & L07 & 455 & 445 & 1.24 & 262 & 820 & 1.60 \NN
& S05 & 404 & 354 & 1.39 & 196 & 709 & 1.85 \NN
& KI  & 378 & 339 & 1.36 & 149 & 721 & 2.48 \NN
\hline
& [none] & 371 & 370 & \ldots & 252 & 584 & \ldots \NN
$K$+IRAC & L07 & 346 & 293 & 1.18 & 164 & 549 & 1.45 \NN
MIPS$_{24\mu\rm{m}}$ & S05 & 313 & 246 & 1.27 & 117 & 497 & 1.83 \NN
& KI  & 288 & 229 & 1.26 &  87 & 492 & 2.45 \NN
& KIM & 233 & 190 & 1.23 &  69 & 403 & 2.52 \LL\NN
}

\subsection{Photometric errors}

In the discussion so far, some conceptual advantages of KI and KIM have been presented, such as the open upper right AGN selection region allowing the selection of extremely obscured sources. Also, the use of filters probing widely separated wavelength ranges, such as $K$ and 4.5$\mu$m as opposed to 3.6 and 4.5$\mu$m, for instance. This results in a wider colour-range domain, diminishing the sensitivity to photometric errors, particularly relevant close to the colour-colour space boundaries. This is verified by assessing the errors associated with the numbers in Tables~\ref{c3tab:xrspcs} and \ref{c3tab:Cxrspcs} by varying the data points within the respective photometric errors ($\delta{\rm mag}<0.36$). For instance, in GOODSs, the overall $\mathcal{C}$ of KI can vary between $\sim$46\% and $\sim$59\%. This is a range of $\sim$2\%, which is comparable to that of L07 (4\%), yet significantly smaller than that of S05 (9\%). Restricting to MIPS$_{24\mu\rm{m}}$ detected sources, the ranges are 2, 8, 1, and 7\% for L07, S05, KI, and KIM criteria, respectively. The range for the $\mathcal{R}$ variation is 9\% for KI, again comparable to that of L07, 11\%, and much better than that of S05, 23\%. Again, the MIPS$_{24\mu\rm{m}}$ detected sample holds similar results, with $\mathcal{R}$ variations of 9, 15, 5, and 20\% for L07, S05, KI, and KIM criteria. The same test in COSMOS implies the same conclusion: the boundaries of criteria with filters probing widely separated wavelength ranges are less affected by photometric errors\footnote{Note that KIM may present higher errors only due to the influence of the flux uncertainties at 24$\,\mu$m, while L07 may do as well knowing that the boundaries are set in colour regions already too much populated by non-AGN sources.}.


\subsection{$K-[4.5]$ at $z<1$} \label{c3sec:kc2}

\begin{figure}
  \begin{center}
    \includegraphics[width=0.5\columnwidth]{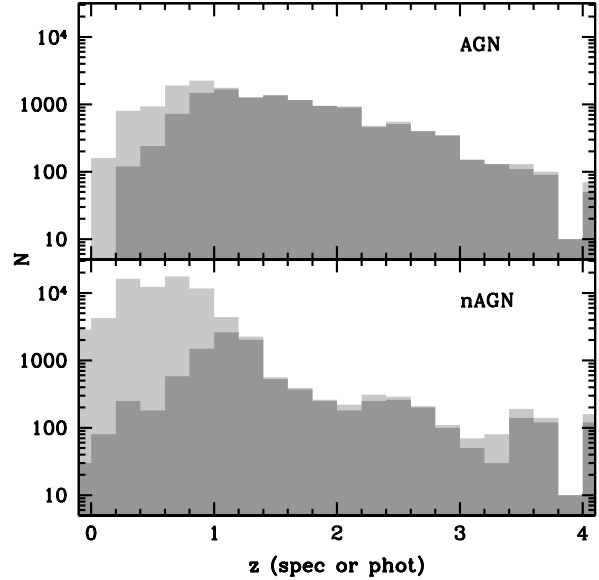}
  \end{center}
  \caption[Application of the $K-4.5>0$ cut]{Applying a $K-[4.5]>0$ cut to the GOODSs and COSMOS samples in order to discard low-z non-AGN systems. Note the logarithmic scale on the ordinate axes.}
  \label{c3fig:zkc20}
\end{figure}

We finally highlight the importance of the $K-[4.5]>0$ colour cut as part of the KI and KIM-AGN criteria. In Figure~\ref{c3fig:zkc20} the redshift distributions for both the AGN and non-AGN populations found in GOODSs and COSMOS are presented. One can see the effect of the $K-[4.5]>0$ cut: at $z<1$ there is a significant rejection of non-AGN galaxies (97\%), while $\sim40\%$ of the AGN population is kept. At $z\geq1$ this colour cut has practically no effect in either galaxy population, selecting 97\% of the AGN population in both fields, and 80\% (54\%) of the non-AGN in GOODSs (COSMOS), not biasing the selection in the IM colour-colour space (Section~\ref{c3sec:imhighz}). As expected, the selection of AGN sources improves with source X-ray luminosity (Figure~\ref{c3fig:s21c}). At low-$z$ the $\mathcal{C}$ of $\log\rm{(L_X[erg s^{-1}])}>43.5$ sources is 60\% (67\%) in GOODSs (COSMOS), and 95\% (97\%) at high-$z$.

\section{Implications for JWST surveys} \label{c3sec:impjwst}

The start of scientific observations of \emph{JWST}, the successor of $Spitzer$ at MIR wavelengths, is expected for 2018. It will be a 6.5m space telescope with the ability to probe the Universe from 1 to 25\,${\mu}$m. As highlighted in this work, this spectral regime has great potential for separating AGN from normal (non-AGN) galaxies.

The sensitivity will of course be better than ever before, and the high-\textit{z} universe will be probed with unprecedented detail. Many galaxies will be studied with MIR spectroscopy, and signs for AGN activity will be naturally found that way \citep[see, for example,][and references therein]{Laurent00}. When dealing with large surveys, however, with thousands of sources and many close to the detection limit, AGN selection will have to rely on photometric diagnostics such as the KI/KIM criteria presented here. By selecting AGN candidates over a broad range of redshifts, $0<z<7$, the KIM criterion will enable the study of AGN phenomena to the earliest epochs. 

\begin{figure}
  \begin{center}
    \includegraphics[width=0.825\columnwidth]{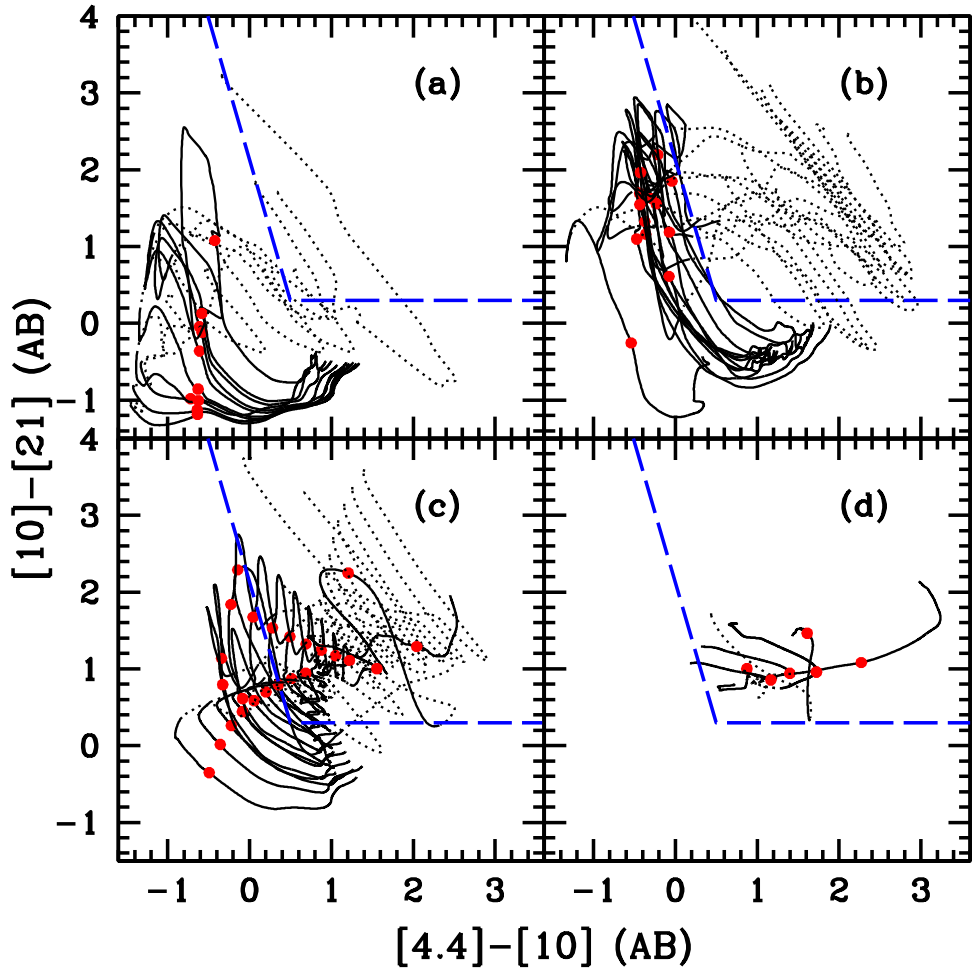}
  \end{center}
  \caption[KIM for $JWST$]{An alternative colour-colour space with \textit{JWST} bands which might improve the AGN selection at $0<z<7$. Symbols and panels definition as in Figure~\ref{c3fig:c1324}.}
  \label{c3fig:jwst}
\end{figure}

\begin{figure}
  \begin{center}
    \includegraphics[width=1\columnwidth]{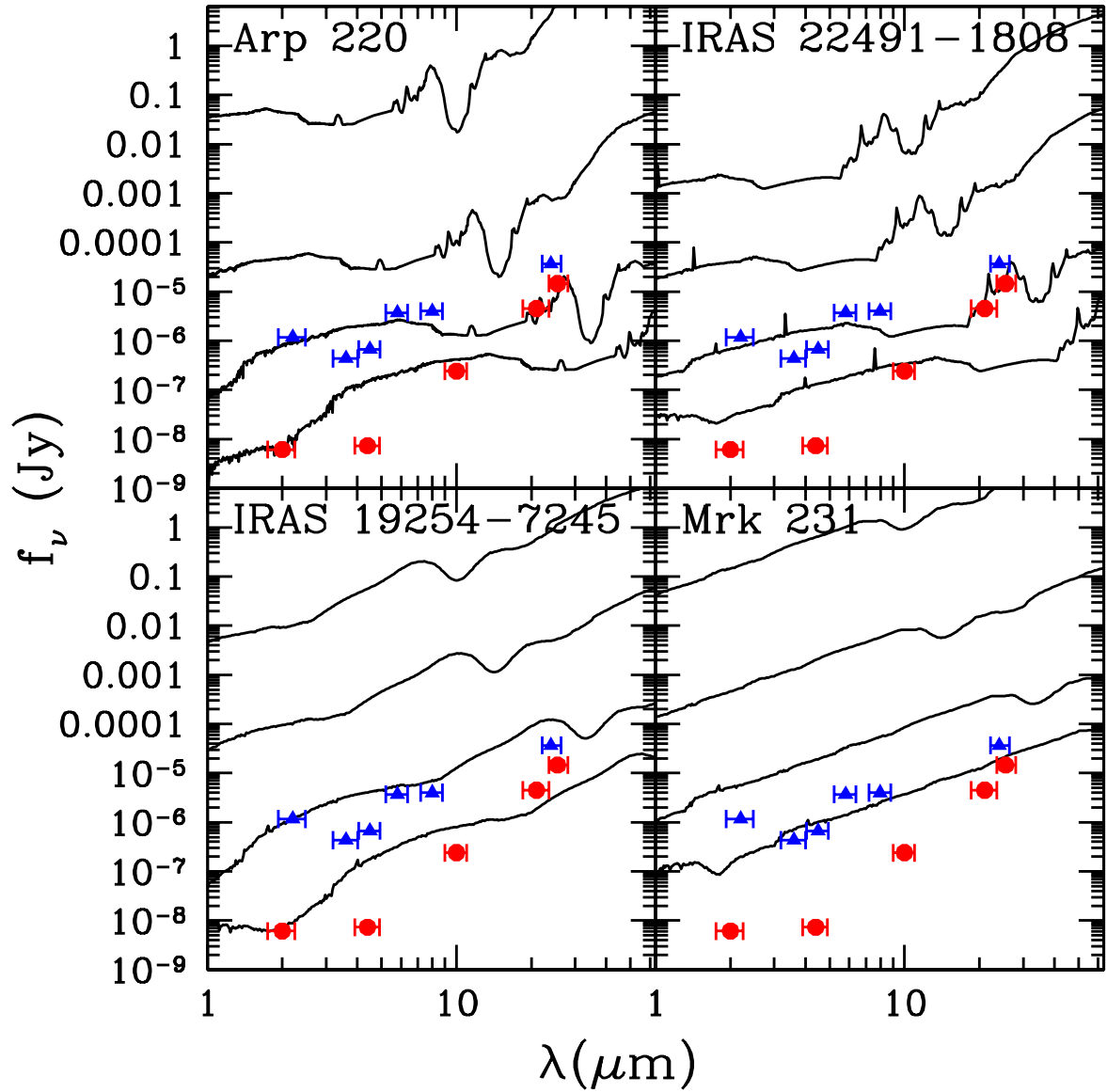}
  \end{center}
  \caption[SED flux evolution with redshift I]{SED flux evolution with redshift for two star-formation dominated systems (Arp220 and IRAS 22491-1908, upper panels), and two AGN dominated systems (IRAS 19254-7245 and Mrk231, lower panels). The redshift steps are $z=z_0,\,0.5,\,2.5,\,7$. The blue dots indicate the $K$-IRAC-MIPS$_{24\mu{\rm m}}$ GOODSs $10\sigma$ total flux level \citep[based in Table 1 of][]{Wuyts08}, red dots give the $10\sigma$ level (at equivalent GOODSs integration times) of the \textit{JWST} filters: 2.0$\mu$m, 4.4$\mu$m, 10$\mu$m, and 21$\mu$m. At longer wavelengths, the gap between $Spitzer$ and \textit{JWST}'s sensitivities is smaller due to the warmer telescope thermal background of \textit{JWST}.}
  \label{c3fig:jwstTp}
\end{figure}

\begin{figure}
  \begin{center}
    \includegraphics[width=1\columnwidth]{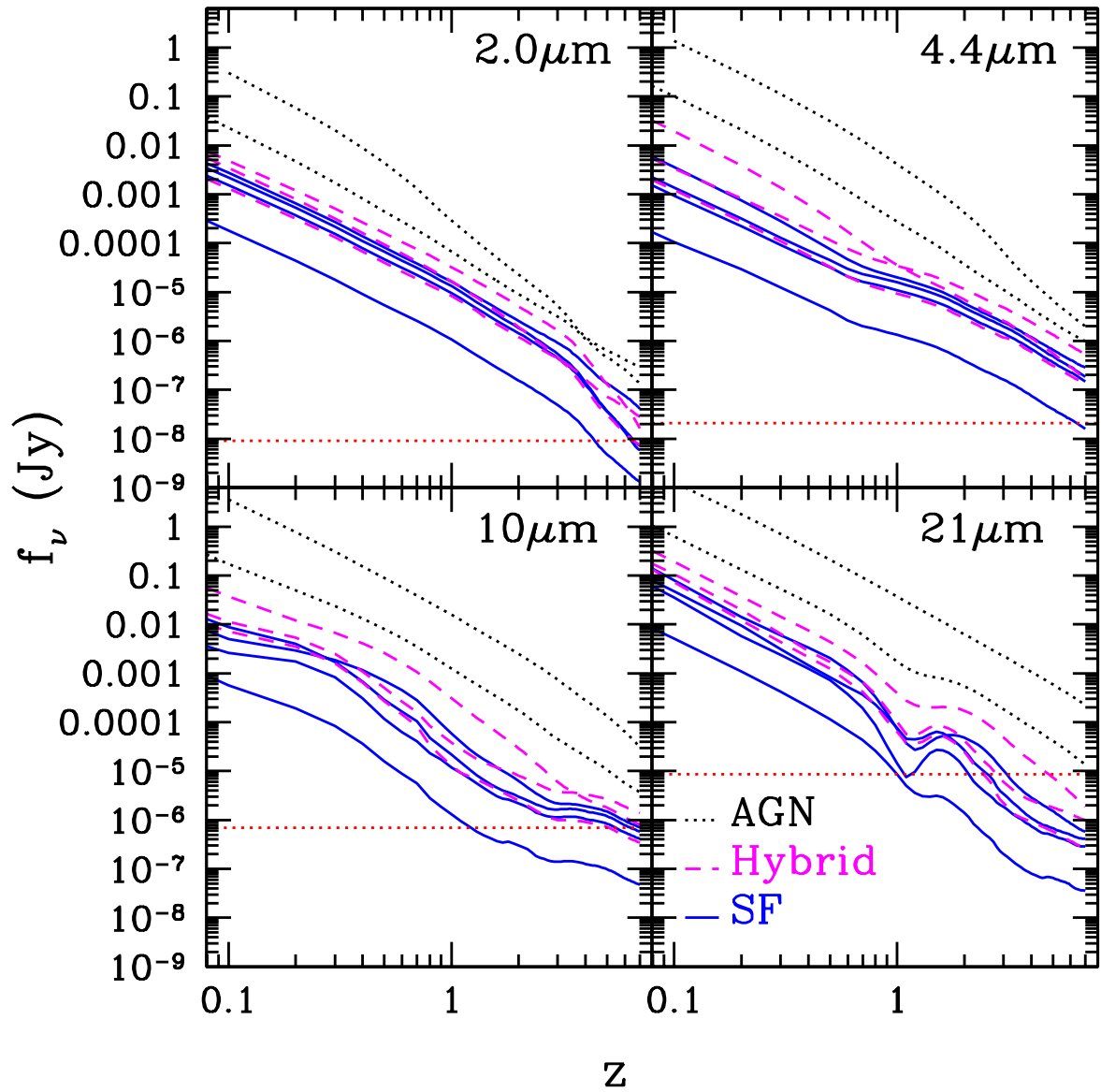}
  \end{center}
  \caption[SED flux evolution with redshift II]{Flux evolution with redshift for starbursts (blue continuous line), hybrids (magenta dashed line), and AGN (black dotted line) in four \textit{JWST} filters. Red dotted horizontal lines mark the $10\sigma$ level (10,000$\:$s integration) of each filter.}
  \label{c3fig:jwstFt}
\end{figure}

While the KI/KIM criteria can already be applied to current data from $Spitzer$, potentially more efficient MIR criteria will be possible with the large wavelength coverage of the \emph{JWST}. Using planned \emph{JWST} filter response curves\footnote{Provided online at:\\ http://www.stsci.edu/jwst/instruments/nircam/instrumentdesign/filters/index\_html \\http://www.stsci.edu/jwst/instruments/miri/instrumentdesign/miri\_glance.html .}, we suggest a possible and promising colour-colour space alternative to that proposed in Section~\ref{c3sec:imhighz}, using the MIRI 10$\mu$m and 21${\mu}$m filters instead of the IRAC 8.0$\mu$m and MIPS 24$\mu$m bands, and the NirCAM 4.4$\mu$m instead of IRAC 4.5$\mu$m \citep[note that these are bands close to those used in Wide-field IR Survey Explorer, WISE; see also][]{Assef10}. In Figure~\ref{c3fig:jwst}, the four panels show that the [4.4]-[10] versus [10]-[21] colour-colour space seems to present a better selection of the AGN/Hybrid model tracks. The AGN model tracks are better delineated by the selection boundaries and, as a bonus, the 21$\mu$m filter is significantly more sensitive than the planed MIRI 25$\mu$m filter (equivalent to the MIPS 24$\mu$m filter), increasing the probability of a detection needed for an AGN classification. This is shown in Figures~\ref{c3fig:jwstTp} and \ref{c3fig:jwstFt}, where AGN dominated sources are detected up to the highest redshift considered in this work ($z\sim7$).

\newpage

\section{Conclusions} \label{c3sec:concagn}

Based on semi-empirical galaxy SED templates, we have developed IR colour based criteria for the selection of a wide variety of AGN in a large redshift range ($0<z<7$). As well as the application to existing data e.g., the recently available WISE data\footnote{http://wise2.ipac.caltech.edu/docs/release/prelim/},][]{Wright10}, these criteria are particularly relevant for the \emph{JWST}, given the wide MIR spectral range considered. We thus propose new AGN IR diagnostics, which select AGN populations at better reliability levels than commonly used IR criteria (e.g., L07 and S05). The $K-[4.5]$ colour is ideal for the $z<1$ universe (Sections~\ref{c3sec:ki} and \ref{c3sec:kc2}); KI is a reliable alternative to the IRAC-based diagnostics (Section~\ref{c3sec:ki}); and KIM (Section~\ref{c3sec:imhighz}), a four band (K, 4.5, 8.0, and 24${\mu}$m), three colour (K-[4.5], [4.5]-[8.0], and [8.0]-[24]) criterion is shown to be more reliable than the L07 and S05 `wedge' criteria on selecting AGN hosts over the full $0<z<7$ range, when tested against some of the deepest IR data available today (Section~\ref{c3sec:ctrlsp}) and based in a template library sampling a broad range of galaxy types. In comparison to S05, these criteria are also shown to be significantly more robust against photometric errors and more complete in the selection of IRxs sources.

We confirm that the completeness of the IR criteria is dependent on both AGN intrinsic luminosity and nature (obscured and unobscured), but strongly on the former. The bias favouring the selection of unobscured sources is seen at intermediate X-ray luminosities ($43<\log(\rm{L_X [erg\,s^{-1}]})<44$).

Nonetheless, the IR selection of AGN will definitely improve with \textit{JWST}. The high resolution provided by \textit{JWST}, will push the detection of low-luminosity AGN in the cores of bright galaxies to higher redshifts. The proposed KI and KIM criteria should be improved as a result of the rich variety of filters to be incorporated in the instruments on board \textit{JWST}. Ultimately, the ability to track AGN activity since the end of the reionization epoch will hold great advantages for the study of galaxy evolution in the future.

\begin{appendices}

\chapter{Obscured/unobscured AGN} \label{c3app:type12}

\section{X-ray versus optical diagnostics}

In this study, both X-rays and optical/nIR spectroscopy are used to identify the unobscured (type-1) and obscured (type-2) AGN populations. However, there are known cases where the two spectral regimes do not agree: either narrow lines are found in the optical spectra of X-ray unobscured AGN hosts, or broad lines are found in the optical spectra of galaxies hosting X-ray obscured AGN.

An optical obscured classification of X-ray unobscured AGN is now believed to be the result of selection effects \citep{Moran02,Severgnini03,Silverman05}, where the light from the host galaxy outshines that of the AGN continuum and broad lines, the latter appearing with apparent smaller equivalent widths (hence classified as narrow line systems) due to the relatively high stellar continuum. This has been further supported by subsequent work \citep{Page06,Cardamone07,Garcet07}. This effect is thought to increase with redshift, as the same slit width will gradually include more light from the host galaxy at higher redshifts.

However, selection effects can not explain an unobscured optical classification together with obscured X-ray emission. Such combination is real and has frequently been found \citep[e.g.,][]{Perola04,Eckart06,Garcet07}. Short time variability on flux and absorption column density \citep{Elvis04,Risaliti07} imply obscuration material at smaller radii than the dusty torus inner radius, which is set by the sublimation radius \citep[R$_{sub}$,][]{Suganuma06,Nenkova08}. These dust-free gas clouds will only absorb X-ray emission, not affecting the optical/nIR broad-line emission. Hence, knowing that the dusty material obscuring optical light also blocks X-ray emission, the X-ray column densities are always larger than those producing the optical obscuration, up to extreme ratios of two orders of magnitude \citep{Maccacaro82,Gaskell07}. This clearly implies that dust-free clouds at $<$R$_{sub}$ frequently represent the bulk of the X-ray obscuration. Being composed by gas, and not dust, these innermost clouds will not re-emit at IR wavelengths. But note that these sources tend to present high luminosities \citep[e.g.,][]{Perola04,Eckart06,Garcet07,Treister09a}. Such property is determinant for an IR AGN classification, as significant initial X-ray flux is needed to still heat the dust torus at the necessary level to overcome the host galaxy light and produce an IR SED dominated by AGN emission. The same should apply to the broad-line emission. If not luminous enough, the host galaxy stellar light continuum will again produce either apparent narrow lines or a spectrum with no AGN emission lines. Note that the former will produce an underestimate of the number of such type of sources, unless adequate spectral observations are done (slit size and orientation set to probe solely the nuclear region). In this scenario, the apparent narrow-line spectral classification agrees with the X-ray obscured classification, inducing the astronomer to account such source as a normal obscured AGN source.

Finally, both optical-colour based AGN criteria and optical spectroscopy classification are known to miss many of the faintest obscured AGN \citep[e.g.,][and references therein]{Treister04}.

\section{$\rm{N_H}$ versus hardness-ratio}

The above discussion leads us to adopt the X-ray spectral regime as the main tool to assess the type-1 and type-2 AGN populations. Such procedure relies on the X-ray spectrum hardness, i.e., how much flux is observed in the hard-band (2--10\,keV) compared to that observed in the soft-band (0.5--2\,keV). The hardness-ratio (HR, Section~\ref{c3sec:goodssample}) is often adopted for such task. However, it is degenerate at high redshifts \citep[][but see also \citealt{Alexander05} and \citealt{Luo10}]{Eckart06,Messias10}, as more energetic rest-frame X-ray wave-bands (less affected by dust obscuration) are observed by both $Chandra$ and XMM-\textit{Newton} telescopes. Also, HR relies on photon counts, which is highly dependent on telescope band efficiency. Flux ratios (used to estimate $\rm{N_H}$) are a more uniform measurement from telescope to telescope. Hence, we adopt the source column density $\rm{N_H}$ (computed as described in Section~\ref{c3sec:goodssample}) as the X-ray diagnostic for obscured/unobscured nature. Although the high-$z$ degeneracy is avoided, the estimate of $\rm{N_H}$ still relies on band-ratios. At high redshift this implies that even small photometric errors correspond to larger uncertainties in the $\rm{N_H}$ value. This is expected to produce a scatter effect instead of a systematic one.

In this work, only when the type-1/type-2 classification is undetermined in the X-rays, is the optical/nIR spectral classification adopted.

\section{Band ratios versus spectral fit} \label{c3sec:brvsspf}

\citet{Akylas06} use an X-ray spectral fit procedure to estimate the column density. They find a tentative trend (as referred by the authors) hinting for a systematic and increasing overestimate of $\rm{N_H}$ with redshift, reaching a 50\% level at $z\sim2.5$, corresponding to a $\Delta\log(\rm{N_H})=0.17$ increase. Also, the quality requirements for a spectral fit induce a bias toward the identification of harder X-ray SEDs, as enough photon counts are needed throughout the full 0.5--8\,keV band to provide a good spectral fit. Both XMM-\textit{Newton} and $Chandra$, however, have soft-bands $\sim$6--7 times more sensitive than the hard-bands. This means that many observed-frame soft sources detected in the soft-band cannot be classified because the hard-band upper limit is too high. The opposite does not hold, as whatever is detected in the hard-band and not in the soft-band already implies (with $Chandra$ and XMM-\textit{Newton} observations) an obscured nature.

The use of band ratios allows reaching to fainter sources, as the flux is summed over two wide bands, instead of the narrow spectral channels. More importantly, the consideration of the full-band flux instead of the hard-band flux (see Section~\ref{c3sec:goodssample}) allows the flux limit to be pushed even deeper and to classify part of the undetermined population missed by spectral fitting or HR diagnostic. Briefly, let two sources have the same full-band flux, which happens to be close to the sensitivity limit. Let one source have a hard SED, while the other has an observed-frame soft SED. The former will be detected in both hard and soft bands allowing for a classification, while the latter is only detected in the soft band, resulting in an undetermined classification. However, if one considers both full and soft bands (where the soft source is detected), a classification is now possible. The full-band ratio method is only possible to apply to the data in GOODSs, as no full-band flux is provided in the XMM-COSMOS catalogue. Note that the $Chandra$ full-band is only 3--4 times less sensitive than the soft band and almost two times more sensitive than the hard-band. This allows us to classify 41 sources more than the HR method, which already classifies 192 (including sources with useful upper-limits). Of this extra sample, 17 (41\%) are classified as unobscured sources ($\log(\rm{N_H\,[cm^{-2}]})<22$), while the remainder are high redshift ($z>$1.5--2) obscured sources.

\section{The adopted classification}

In this work, X-ray column densities are computed to all possible sources (as described in Section~\ref{c3sec:goodssample}). An obscured AGN is considered to have $\log(\rm{N_H\,[cm^{-2}]})>22$, while unobscured AGN have $\log(\rm{N_H\,[cm^{-2}]})\leq22$. If such estimate is indeterminate, the optical/nIR classification is adopted, where broad line features imply unobscured nuclear activity, and high-ionization narrow lines indicate obscured nuclear activity.

Again we stress that no criterion is perfect. The choice of the criteria finally adopted is thought to be, after a careful line of reasoning, the least affected by all the bias inherent to this and similar studies.

\section{Comparison with \citet{Treister09a}}

\citet{Treister09a} present a detailed procedure to estimate the dependency of the fraction of obscured sources ($f_{obs}$) on both X-ray luminosity and redshift. The $f_{obs}$ is found to decrease with increasing luminosity and, after accounting for incompleteness effects \citep[as described in][]{TreisterUrry06}, to increase with increasing redshift. The sample in that work is selected in the extended CDFs (ECDFs, a $Chandra$ survey with an average 230\,ks depth in a region three times larger than CDFs and six times smaller than the COSMOS field).

\begin{figure}
  \begin{center}
    \includegraphics[width=0.75\columnwidth]{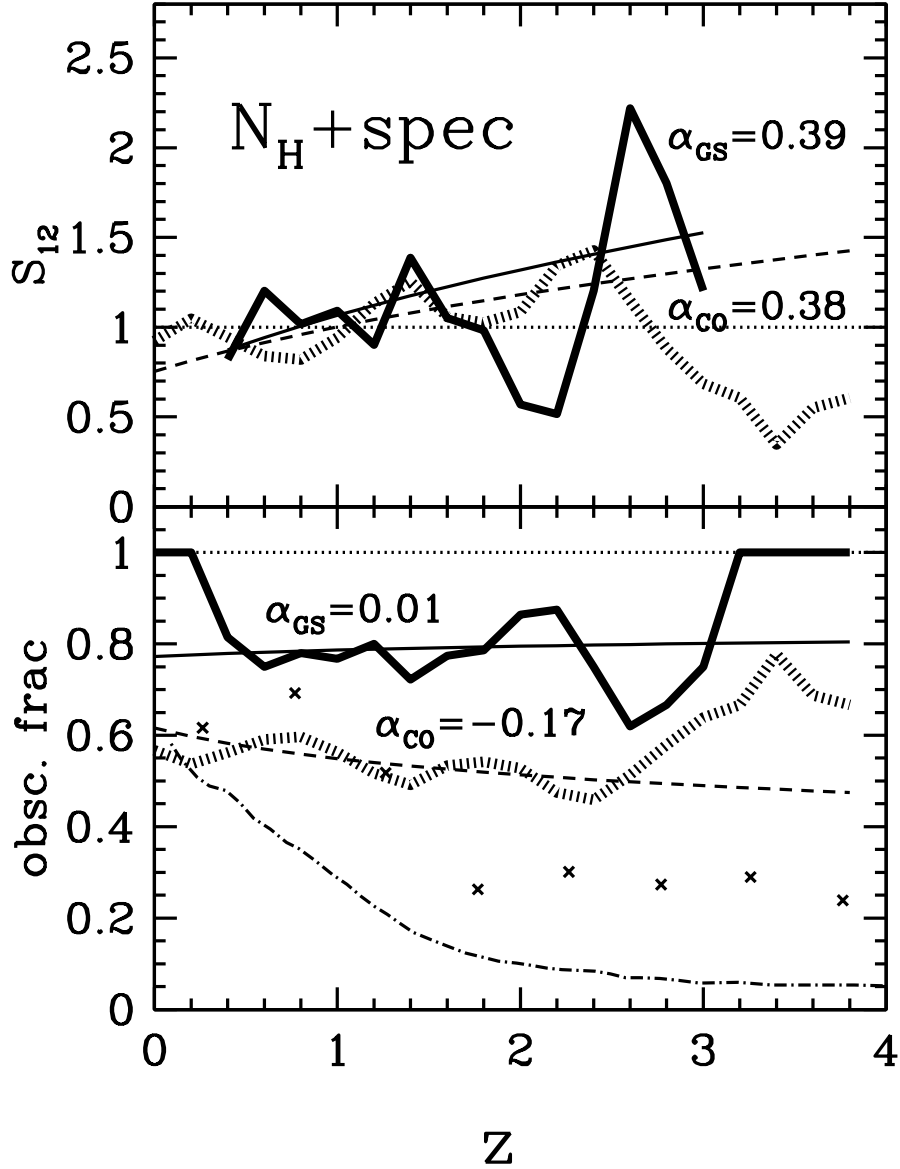}
  \end{center}
  \caption[$\mathcal{S}_{\rm 12}$ and $f_{obs}$ dependency on redshift]{The variation of $\mathcal{S}_{12}$ (upper panel) and the obscured fraction (lower panel) with redshift. The trends for both GOODSs (thick solid lines) and COSMOS (thick dotted lines), are displayed. The power-law ($\propto(1+z)^\alpha$) index $\alpha$ is given for GOODSs ($\alpha_{GS}$) and COSMOS ($\alpha_{CO}$). As a reference, the data points (crosses) and the expected evolution of the obscured fraction induced by sample characteristics in ECDFs (dotted-dashed line) from \citet{Treister09a} are displayed.}
  \label{c3fig:s12z}
\end{figure}

\begin{figure}
  \begin{center}
    \includegraphics[width=0.75\columnwidth]{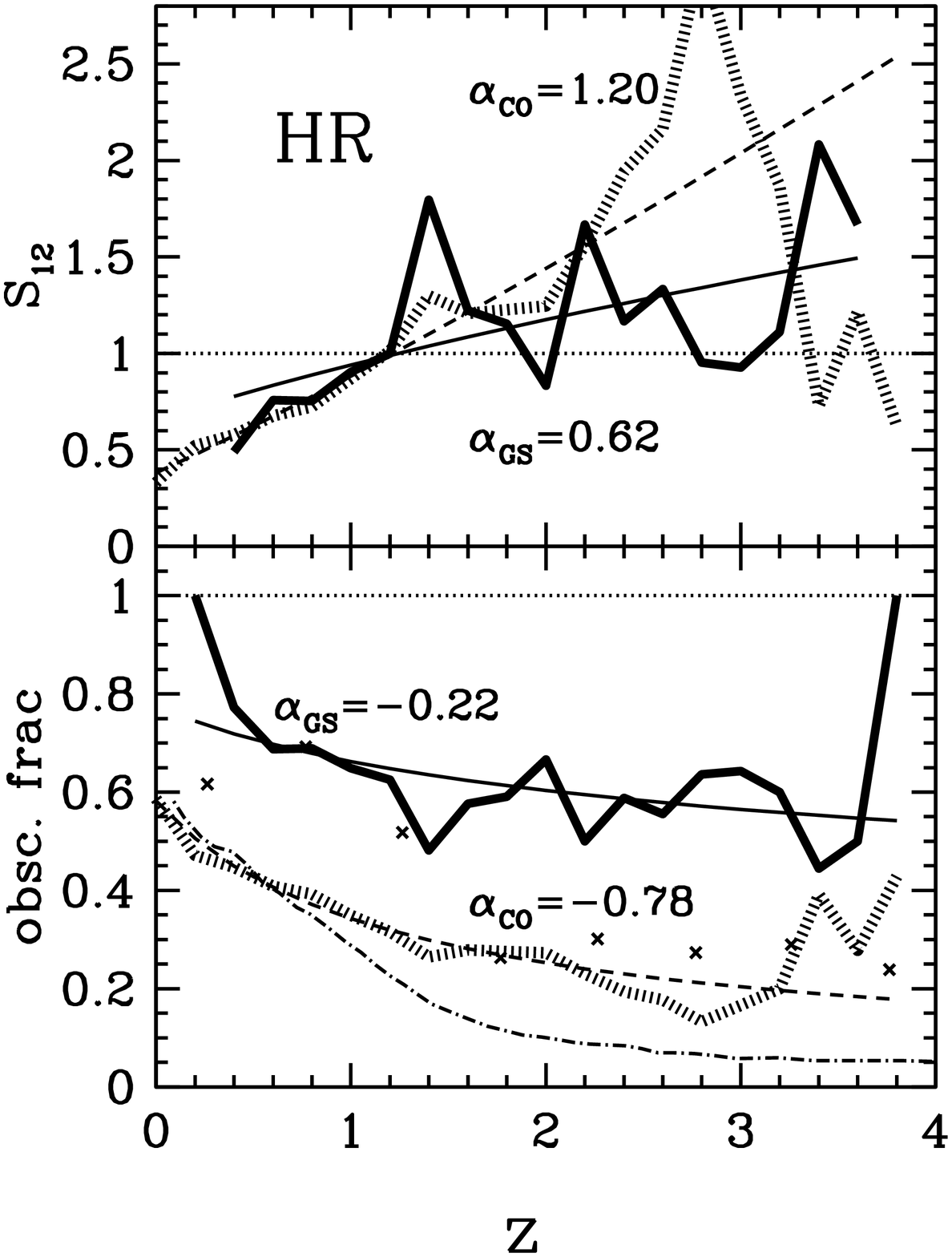}
  \end{center}
  \caption[The high-redshift degeneracy of HR]{The same as in Figure~\ref{c3fig:s12z}, but considering the HR to identify obscured and unobscured sources instead of $\rm{N_H}$ and optical/nIR spectroscopy. Symbols and labelling as in Figure~\ref{c3fig:s12z}.}
  \label{c3fig:s12hr}
\end{figure}

A spectroscopic analysis (including data from 8-m class telescopes) is the main diagnostic between obscured and unobscured AGN (at $z<0.5$ the HR is considered instead). In the lower panel of Figure~\ref{c3fig:s12z} we compare our estimates (using both X-ray and spectroscopic criteria) with the $f_{obs}$ data points (crosses) presented by \citet[][which also consider those from \citealt{TreisterUrry06}]{Treister09a}. The expected $f_{obs}$ trend resulting from incompleteness effects and survey characteristics for the specific ECDFs sample and type-1/type-2 classification was computed (for a constant type-1/type-2 ratio) as described in \citet{TreisterUrry06} and is shown as a dotted-dashed line. As the ECDFs data points are always above the theoretical line, \citet{Treister09a} propose that the $f_{obs}$ is actually increasing. At $z<1.5$ the ECDFs data points fall between the GOODSs and COSMOS trends as expected from an intermediate depth survey, recovering more obscured sources than COSMOS, yet still missing those detected in the 2\,Ms $Chandra$ observations \citep{Luo08}. At $z>1.5$, however, our data points show a higher $f_{obs}$, even when considering the shallower XMM-$Newton$ data in COSMOS. This is due to the different classification methods. Figure~\ref{c3fig:s12hr} clearly proves the statement. If the HR is used instead of $\rm{N_H}$, the lower redshift ($z<1.5$) data points from \citet{Treister09a} still agree with our trends derived from the COSMOS and the deeper GOODSs data, while at higher redshifts the agreement with COSMOS is clear. The reader should recall once again that the HR is increasingly degenerate with increasing redshift \citep[at $z\gtrsim2$, a reasonably obscured source, $\log(\rm{N_H[cm^{-2}]})\sim23$, may be classified as unobscured, Figure~3 in][]{Messias10}. Hence, we deduce that the \citet{Treister09a} analysis (before correction) suffers of equal bias at $z>1.5$ as it is based in spectroscopic classification alone.

Note that the deeper GOODSs/CDFs data imply a non-dependency with redshift (Figure~\ref{c3fig:s12z}), with the caveat that this work may still be missing a significant population of obscured AGN. The recently released 4\,Ms depth $Chandra$ observations \citep{Xue11}\footnote{Data set available at: http://cxc.harvard.edu/cda/Contrib/CDFS.html} will definitely improve this estimate. Hence, while the underlying obscured population remains undetected by current surveys, a procedure like that of \citet{TreisterUrry06} should be pursued for an improved estimate of the dependency of $f_{obs}$ with redshift.

The upper panels in Figures~\ref{c3fig:s12z} and \ref{c3fig:s12hr} show that $\mathcal{S}_{12}$ increases with redshift with a power-law ($\mathcal{S}_{12}\propto(1+z)^\alpha$) index $\alpha\sim0.4$. This means that compared with the overall type-1 to type-2 ratio in the X-ray sample, that ratio increases with redshift, meaning a relatively higher number of type-1 sources.

\begin{figure}
  \begin{center}
    \includegraphics[width=0.5\columnwidth]{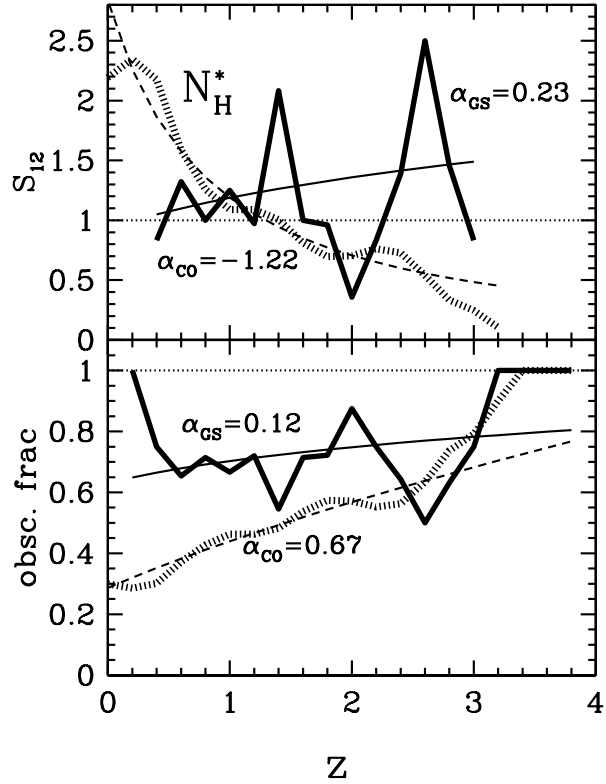}
  \end{center}
  \caption[Induced bias due to hard and soft band relative sensitivities.]{The same as in Figure~\ref{c3fig:s12z}, but considering $\rm{N_H}$ to identify obscured and unobscured sources in the restricted sample composed by sources detected in both soft and hard X-ray bands. Symbols and labelling as in Figure~\ref{c3fig:s12z}.}
  \label{c3fig:s12z2}
\end{figure}

As a final remark to prove the bias induced by the discrepant relative sensitivity between the soft and hard bands (as described in Section~\ref{c3sec:brvsspf}), the same trends are shown in Figure~\ref{c3fig:s12z2} this time considering only the sources detected in both soft and hard bands (as required by spectral fitting analysis). The clear increase of $f_{obs}$ with redshift reflects the referred selection effect.

\end{appendices}
\chapter[Hot dust and its role in shaping the IR LFs]{Hot dust and its role in shaping the infra-red luminosity functions}
\label{ch:lfs}
\thispagestyle{empty}

\section{Introduction}

While FIR/mm studies focus on the cold (T$\lesssim$100\,K) dust re-emission dominating at those wavelengths, in this study, the hot extremes of dust re-emission \citep[T$\sim$500\,--1500\,K, e.g.,][]{Nenkova08} are explored (at $\sim$2--8\,$\mu$m) using observations on the Cosmic Evolution Survey \citep[COSMOS,][]{Scoville07}. The study is mostly based on data from the IR array camera \citep[IRAC,][]{Fazio04} on board the \textit{Spitzer Space Telescope} \citep[\textit{Spitzer},][]{Werner04}, facility which, in less than a decade, has contributed so much to the field of galaxy evolution \citep[for a review, see][]{Soifer08}. The goal is to estimate the hot dust contribution to the SED of the galaxy population at short IR wavelengths, and how it depends on both redshift and galaxy nature. This is intrinsically related to the shape of the overall IR luminosity function and how it evolves with cosmic time.

In Section~\ref{c4sec:samp}, the sample used in this study is described, detailing how each redshift regime was assessed and how each galaxy population --- early and late-types, starbursts, and AGN hosts --- was assembled. Section~\ref{c4sec:estimdust} describes the method used to extract the dust emission in each galaxy SED. In Section~\ref{c4sec:irlf}, we present rest-frame IR LFs and then, in Section~\ref{c4sec:dlfs}, we present the dust luminosity density functions. In this same section, the dust luminosity density is also shown to evolve with redshift and differently between each galaxy population. Finally, Section~\ref{c4sec:conclusions} summarises the conclusions drawn from this study.

\section{The sample} \label{c4sec:samp}

We use observations from the COSMOS field, covering an area of 1.8 deg$^2$, with available multiple-waveband data. The reference catalogue used is the one described in \citet{Ilbert09}. From it, consistent samples of galaxies are extracted depending on the target rest-frame wavelength (3.3 or 6.2\,$\mu$m) and redshift, allowing for an evolution study with cosmic time. Each sample is further separated into: early-type, late-type, starburst, and AGN host populations. While the first three are assembled based on a spectral SED fitting algorithm, the AGN population is estimated by applying a new IR colour-colour diagnostic enabling the selection of AGN host galaxies (Chapter~\ref{ch:agn}). The following sections detail each of these steps.

\subsection{The target rest-frames} \label{c4sec:rfwl}


This study focus on the extreme hot regimes of dust reemission. The target rest-frame wavelengths are at 3.3 m and 6.2 m. These are wavelengths where specific polycyclic aromatic hydrocarbons (PAHs) features are expected to be observed by Spitzer -IRAC filters, and to which hot dust is also known to contribute significantly. Although PAHs are not actual dust particles, they are large molecules (composed by $\sim$50 Carbon atoms) of Carbon rings and Hydrogen, which act as light-blocking small dust grains for UV radiation, producing broad emission features in a galaxy IR SED \citep[for a review, see][]{Tielens11}. These were referred as unidentified IR bands until the 80's, when the PhD work of Kris Sellgren provided a key understanding of such emission features \citep{Sellgren83a,Sellgren83b}. PAHs comprise a significant fraction of the Carbon existing in the universe, they are believed to be closely related to star-formation activity, and to reprocess a substantial fraction of UV-light into the IR wave-bands, hence being a major source of obscuration, which we aim to track in this study. For simplicity throughout this chapter, when referring ``dust'', the PAHs contribution is also included in that class. The IR continuum comes from dust heated by energetic radiation fields. Vigourous obscured star-formation can account for such emission as well as AGN activity \citep[][and references therein]{daCunha08,Nenkova08,HonigKishimoto10,Popescu11}.

\subsection{Redshifts and galaxy populations}

The photometric redshifts assigned to each source found in the sample were estimated with the \textit{Le Phare} code\footnote{http://www.cfht.hawaii.edu/$\sim$arnouts/LEPHARE/cfht\_lephare/lephare.html} (S. Arnouts \& O. Ilbert) \citep{Ilbert09}. The ultraviolet-to-IR coverage (0.15--8\,$\mu$m) is unique and there are, in total, 30 broad, intermediate, and narrow-band filters to be considered \citep[Table 1 in][]{Ilbert09}. The new implemented procedure, which accounts for the contributions from emission lines ([O\,II], H$\beta$, H$\alpha$, and Ly$\alpha$) to the SEDs, and results are described and thoroughly tested in \citet{Ilbert09}. The template library is heterogeneous enough to cover a large range of the colour-$z$ space. Briefly, the considered templates \citep[Figure 1 in][]{Ilbert09} consist of three early and six late-type SEDs from \citet[][and linear interpolations between some of these SEDs for fitting improvement]{Polletta07}, and 12 SEDs generated with \citet[][BC03 henceforth]{Bruzual03} models in order to better match the colours observed in some of the bluest sources found in the field. The BC03 models cover a wide age range (0.03 to 3\,Gyr\footnote{Different metalicities are used depending on the template age:  Z=0.004 for $\leq1.1$\,Gyr templates, Z=0.008 for $1.43<\rm{age\,[Gyr]}<2.1$, and Z=0.02 (solar) for $\geq2.5$\,Gyr (where Z is the mass fraction of all elements heavier than Helium).}). The choice of the extinction curve was SED template-dependent and according to a grid of colour-excess values\footnote{$\rm{E(B-V)}=0,0.05,0.1,0.15,0.2,0.25,0.3,0.4,0.5$}. The spectral types adopted throughout this chapter (early, late, and starbursts) are a result from the fitting procedure of the SED templates to the observed galaxy SEDs \citep{Ilbert09}.

The testing done in \citet{Ilbert09} indicates a fitting quality dependency on both source flux and redshift. The redshift intervals considered in this study are set by the target rest-frame wavelengths --- 3.3 and 6.2\,$\mu$m --- and the spectral responses of the IRAC filters. Table~\ref{c4tab:zrange} shows the redshift ranges considered throughout this study, resulting from the specific redshifts where 3.3 or 6.2\,$\mu$m wavelengths enter or leave the 50\% throughput limits of an IRAC filter. For the first three redshift bins, Figure 9 of \citet{Ilbert09} shows a constant $z_{\rm phot}$ quality with distance, with $\sigma_{\Delta{z}}/(1+z_{\rm spec})\lesssim0.04$ at an $i$-band\footnote{Subaru Telescope: http://www.naoj.org/} magnitude of $i^+<25$ (where $\Delta{z}=z_{\rm spec}-z_{\rm phot}$). For the farthest redshift bin considered in this study, larger errors are expected, with $\sigma_{\Delta{z}}/(1+z_{\rm spec})\lesssim0.05$ for $i^+<24$ and $\sigma_{\Delta{z}}/(1+z_{\rm spec})\lesssim0.1$ for $i^+<25$. When computing the errors of the dust LF estimates, we consider in quadrature the poissonian and the $z_{\rm phot}$-induced errors ($\sigma_{\rm poi}$ and $\sigma_{z\rm{p}}$, respectively): $\sigma_{\rm tot}=\sqrt{\sigma_{\rm poi}^2+\sigma_{z\rm{p}}^2}$.

\ctable[
   cap     = Redshift ranges,
   caption = The adopted redshift ranges and equivalent observing bands for rest-frame 3.3 and 6.2$\mu$m,
   label   = c4tab:zrange,
   pos     = ht
]{ccrr}{}{ \FL
Rest-Frame & IRAC$_{\lambda}$ & $z_{\rm LOW}$ & $z_{\rm HIGH}$ \NN
$[\mu\rm{m}]$ & $[\mu\rm{m}]$ &  & \ML
3.3 & 3.6 & 0.05 & 0.19 \NN
 & 4.5 & 0.21 & 0.52 \NN
 & 5.8 & 0.52 & 0.94 \NN
 & 8.0 & 0.97 & 1.86 \ML
6.2 & 8.0 & 0.05 & 0.52 \LL
}

\begin{figure}
  \begin{center}
    \includegraphics[width=0.5\columnwidth]{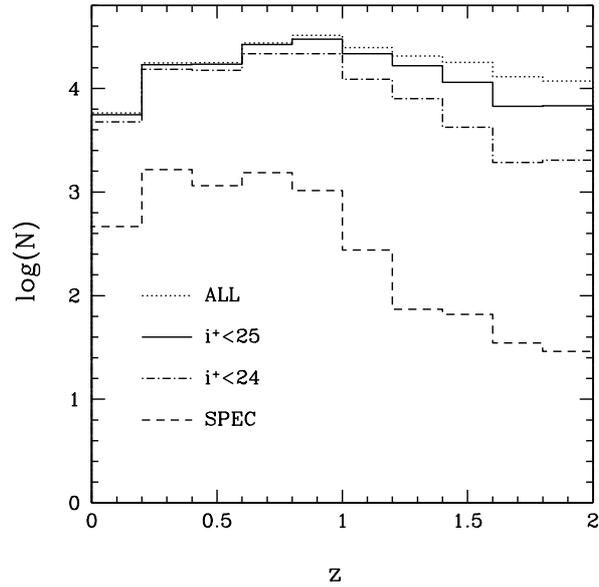}
  \end{center}
  \caption[Sample redshift distribution]{The redshift distribution of the COSMOS sample. Highlighted are the distributions of sources with available good quality spectroscopy (dashed histogram), with $i^+<24$ (dotted-dashed histogram) and $i^+<25$ (solid histogram), while the overall population is denoted by the dotted histogram. Note the logarithmic scale on the y-axis.}
  \label{c4fig:zdist}
\end{figure}

\begin{figure}
  \begin{center}
    \includegraphics[width=0.5\columnwidth]{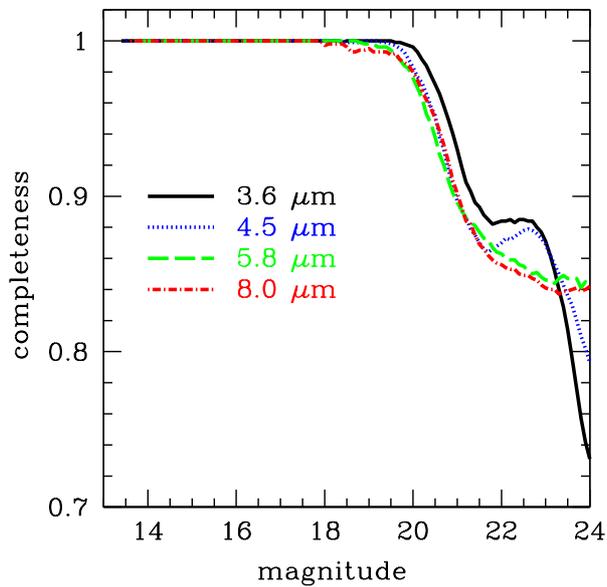}
  \end{center}
  \caption[Redshift completeness with magnitude]{Completeness of reliable redshift estimates depending on source magnitude in each of the IRAC channels: 3.6\,$\mu$m (solid black line), 4.5\,$\mu$m (dotted blue line), 5.8\,$\mu$m (dashed green line), and 8.0\,$\mu$m (dot-dashed red line).}
  \label{c4fig:zincomp}
\end{figure}

When available, the spectroscopic redshift estimate from $z$COSMOS \citep{Lilly09} is considered only if with high probability ($>90\%$ confidence, 8562 sources were found with such constraint). Also, in \citet{Salvato09} the COSMOS team re-computed photometric redshifts for XMM-detected sources. Variability effects in X-ray AGN hosts and AGN emission contribution is properly accounted for the computation of $z_{\rm phot}$. If available, this improved $z_{\rm phot}$ estimate from \citet{Salvato09} is considered instead of that from \citet{Ilbert09}. For the remainder of the sample, only sources with good quality photometric redshifts ($(z_{68\%}^{\rm up}-z_{68\%}^{\rm low})/(1+z_{\rm phot})<0.4$) of sources with $i^+<25$ are considered for the study. The redshift distribution is shown in Figure~\ref{c4fig:zdist}, highlighting the fraction of the sources with available $z_{\rm spec}$, with $i^+<24$, $i^+<25$, and the total population. The incompleteness caused by the quality constraints is accounted for while computing the source densities (Sections~\ref{c4sec:irlf} and \ref{c4sec:dlfs}). Figure~\ref{c4fig:zincomp} shows the variation of the fraction of sources having a reliable redshift estimate depending on observed magnitude in each of the IRAC-channels.

\begin{figure}
  \begin{center}
    \includegraphics[width=1\columnwidth]{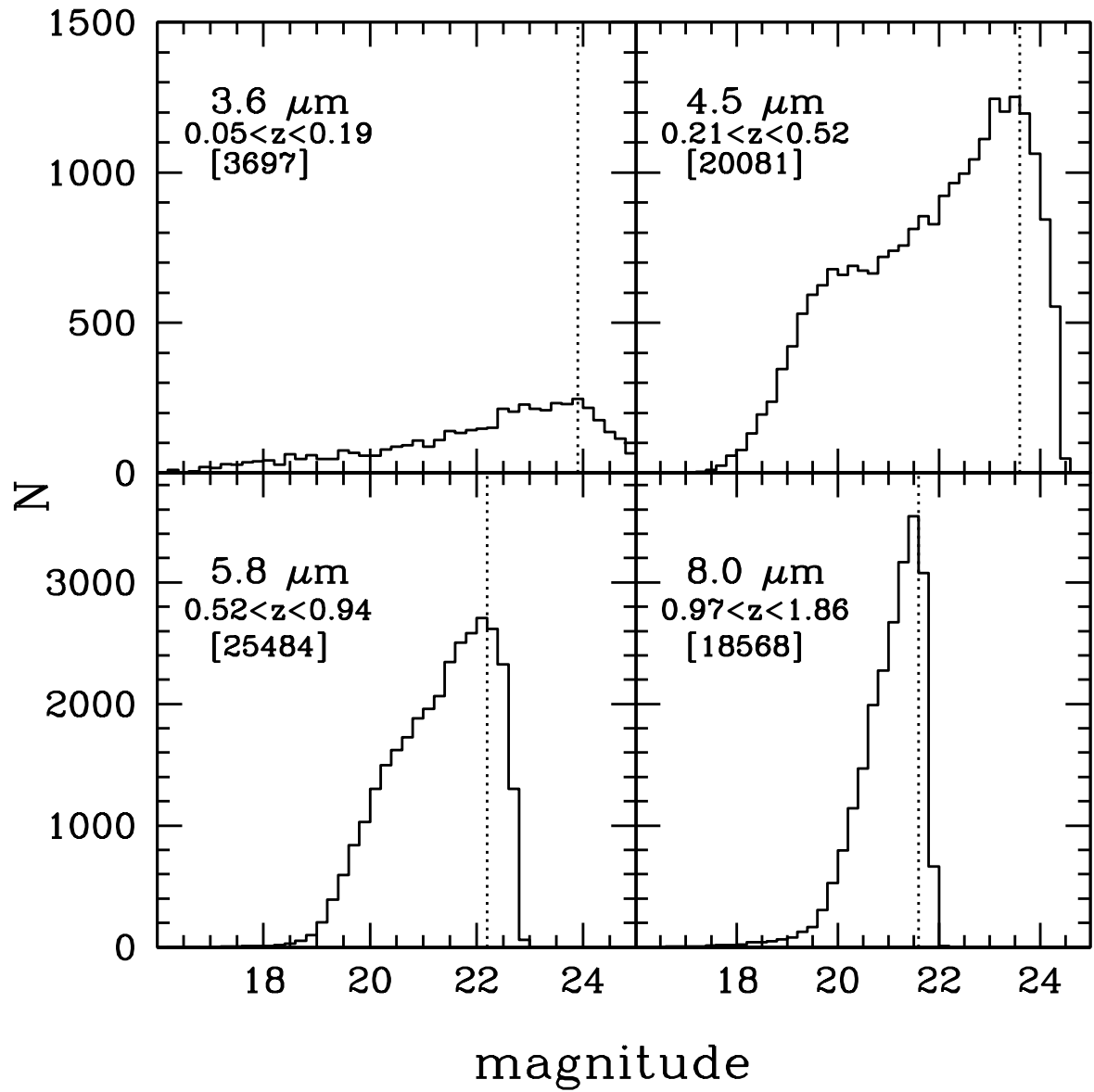}
  \end{center}
  \caption[Sample completeness]{Magnitude cuts to constrain the completeness depending on redshift interval and observed band (see Table~\ref{c4tab:zrange}): 3.6\,$\mu$m (upper left), 4.5\,$\mu$m (upper right), 5.8\,$\mu$m (lower left), and 8.0\,$\mu$m (lower right). The numbers inside squared brackets give the number of selected galaxies, i.e., those found to the left of the vertical line (the magnitude cut). Note that the y-axis scale in the upper panels is different from that in the lower panels.}
  \label{c4fig:magcut}
\end{figure}

The final samples are selected in the four IRAC channels (when following the rest-frame 3.3\,$\mu$m wavelength with redshift) and at 8\,$\mu$m, when studying the rest-frame 6.2\,$\mu$m wavelength in the nearby universe. Hence, for the estimate of dust LFs at rest-frame 3.3\,$\mu$m we consider all $0.05<z<0.19$ sources with $[3.6]<23.9$ \citep[for consistency with][]{Ilbert09}, all $0.21<z<0.52$ sources with $[4.5]<23.6$, all $0.52<z<0.94$ sources with $[5.8]<22.2$, and all $0.97<z<1.86$ sources with $[8.0]<21.6$. For the estimate of dust LFs at rest-frame 6.2\,$\mu$m, we consider all $0.05<z<0.52$ sources with $[8.0]<21.6$. These completeness magnitude cuts are set as the magnitude value at which the magnitude distribution starts to drop (Figure~\ref{c4fig:magcut}). The use of apparent magnitude as a completeness cut instead of absolute magnitude, is not a problem as redshift constraints are also applied. Within a narrow redshift, the apparent magnitude can be used as a proxy of the absolute magnitude. In our case, even in the highest redshift bin, which is not narrow in any way, this proxy is valid, as we are already limited to the brightest objects. The final numbers are detailed in Table~\ref{c4tab:numb}.

\ctable[
   cap     = Sample statistics,
   caption = The numbers of each population with redshift,
   label   = c4tab:numb,
   pos     = ht,
   nosuper
]{ccrrrrr}{
 \tnote[Note.]{ --- Numbers in parenthesis give the fraction (in \%) of the total population each population represents at each redshift interval}
}{ \FL
Rest-Frame & $z_{\rm BIN}$ & TOTAL & EARLY & LATE & STARB. & AGN \NN
$[\mu\rm{m}]$ &  &  &  &  &  & \ML
3.3 & $0.05<z<0.19$ &  3697 &  704\,(19) &  437\,(12) &  2479\,(67) &  77\,(2) \NN
& $0.21<z<0.52$ & 20081 & 3027\,(15) & 3645\,(18) & 12780\,(64) &  629\,(3) \NN
& $0.52<z<0.94$ & 25484 & 3548\,(14) & 4825\,(19) & 13672\,(54) & 3439\,(13) \NN
& $0.97<z<1.86$ & 18568 & 1619\,(9) & 4188\,(23) &  7796\,(42) & 4965\,(27) \ML
6.2 & $0.05<z<0.52$ & 11759 & 1948\,(17) & 3637\,(31) &  5953\,(51) &  221\,(2) \LL\NN
}

\subsection{IR selection of AGN} \label{c4sec:agnsamp}

\begin{figure}
  \begin{center}
    \includegraphics[width=0.5\columnwidth]{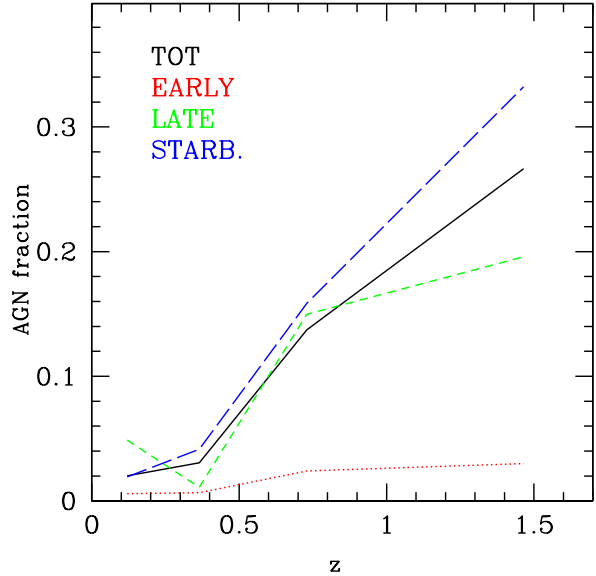}
  \end{center}
  \caption[AGN fraction evolution with redshift]{The evolution with redshift of the AGN fraction in the total IR (black solid line), early-type (red dotted line), late-type (green short dashed line), and starbursts (blue long dashed line) populations.}
  \label{c4fig:agnfrac}
\end{figure}

Both X-ray and spectroscopic observations can be used for the identification of AGN hosts. However, we are just interested in identifying those galaxies whose nuclear emission dominates the IR regime, which frequently is not the case for either optical or X-ray identified AGN \citep[e.g.,][and Chapter~\ref{ch:agn} of this thesis]{Treister06,Donley08,Eckart10}. We adopt the AGN diagnostics described in Chapter~\ref{ch:agn} of this thesis involving $K-[4.5]$ and $[4.5]-[8.0]$ colours. AGN hosts are considered to have $K-[4.5]>0$ at $z<1$, and $K-[4.5]>0$ and $[4.5]-[8.0]>0$ at $z\geq1$. Table~\ref{c4tab:numb} details the number of sources selected as AGN in each redshift interval. Figure~\ref{c4fig:agnfrac} shows the evolution of the AGN fraction depending on both redshift and galaxy type. At high redshift, most AGN are found in starburst and late spectral-type galaxies.

\section{Estimating rest-frame IR luminosities and dust contribution} \label{c4sec:estimdust}

This study focus on two specific rest-frame spectral regimes: 3.3 and 6.2\,$\mu$m (Section~\ref{c4sec:rfwl}). However, at these wavelengths, both stellar and hot dust continuum emission contribute to the galaxy SED. In order to disentangle these, we first estimate the stellar emission at 3.3 and 6.2\,$\mu$m for each galaxy. This is done in two steps. First we consider a reference wavelength which we expect to be solely due to stellar emission. With that, we then use a pure stellar emission model to estimate the stellar emission at 3.3 and 6.2\,$\mu$m. The remaining flux is thus assumed to be due to dust emission alone.

\begin{figure}
  \begin{center}
    \includegraphics[width=0.8\columnwidth]{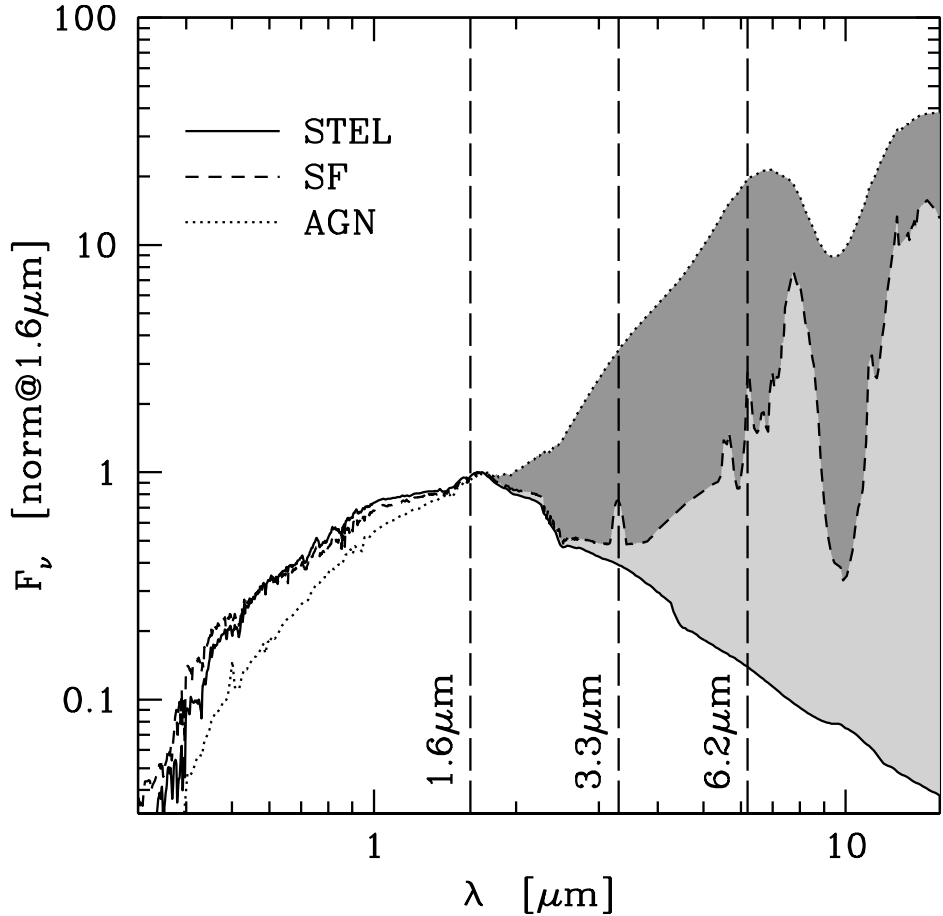}
  \end{center}
  \caption[Stellar and dust IR emission]{Separating the IR emission into stellar and dust contributions. The solid line shows an elliptical template, dominated by stellar emission alone, used for the conversions from rest-frame 1.6\,$\mu$m luminosities to 3.3 and 6.2\,$\mu$m stellar luminosities. Together with the solid line, the dashed and dotted lines delimit, respectively, the dust contribution to the IR SED of Arp220 (a dusty starburst) and IRAS 19254-7245$_{\rm south}$ (an AGN host) galaxies \citep[templates from][]{Polletta07}.}
  \label{c4fig:ellmod}
\end{figure}

The reference wavelength to estimate the stellar emission from is that of the peak of the stellar bump at 1.6\,$\mu$m ($H$-band). This emission bump is frequently observed in galaxy SEDs (Figure~\ref{c4fig:ellmod}), and, at $\lesssim2\,\mu$m, no significant dust emission is expected. Either because SF UV/optical (stellar or AGN) emission is not enough to heat dust for it to re-emit and dominate at such low wavelengths, or because, and most importantly, any dust particle in the radiation field is dissociated. Hence, at wavelengths below $\sim2\,\mu$m, only emission from the Wien tail by the hottest dust grains (around thermally pulsating asymptotic giant branch, TP-AGB, stars or in a dust torus around an AGN) and from scattered AGN light is expected, which is expectable not to be substantial (however, see discussion in Section~\ref{c4sec:dlfs}).

\begin{figure}
  \begin{center}
    \includegraphics[width=0.45\columnwidth]{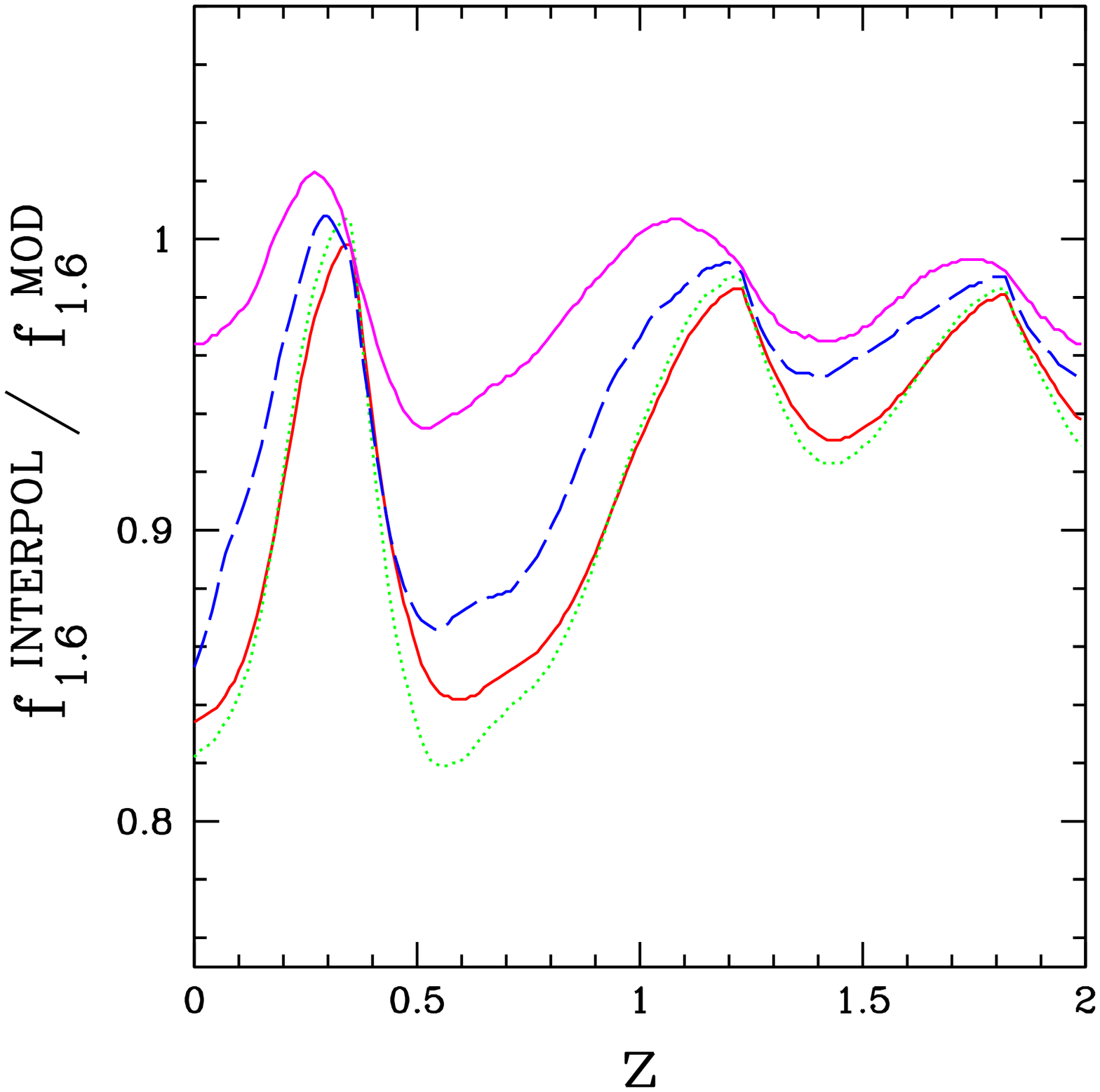}
  \end{center}
  \caption[Interpolating the 1.6\,$\mu$m]{The redshift induced effect of interpolating the galaxy SED in order to estimate the flux at rest-frame 1.6\,$\mu$m. The y-axis shows the ratio between the interpolated flux ($\rm{f^{INTERPOL}_{1.6}}$) and the actual model value ($\rm{f^{MOD}_{1.6}}$) at $1.6\,\mu$m. Different types are shown: early (solid red line) late (dotted green line), blue starburst (dashed blue line), and AGN (solid magenta line).}
  \label{c4fig:interpol}
\end{figure}

The source flux at rest-frame 1.6\,$\mu$m is obtained through interpolation between the two wavebands which straddle this rest-frame wavelength at the source's redshift. However, although necessary, interpolation will generally underestimate the true rest-frame 1.6\,$\mu$m flux value depending on the source redshift and SED shape. This is evident from Figure~\ref{c4fig:interpol} where discrepancies between estimated and true value (always below the 20\% level) are shown for typical early (red) and late (green) galaxies, blue starbursts (blue), and AGN hosts (magenta). These trends were used to correct the interpolated 1.6\,$\mu$m flux for each galaxy at its redshift.

With the estimated stellar flux at 1.6\,$\mu$m, the corresponding stellar contribution at 3.3 and 6.2$\mu$m is obtained. This is done with a pure stellar model \citep[solid line in Figure~\ref{c4fig:ellmod}, 13\,Gyr elliptical from][]{Polletta07}. The conversion from 1.6\,$\mu$m stellar flux to that at 3.3 and 6.2\,$\mu$m is slightly redshift dependent, because the
considered filters will probe slightly different rest-frame
wavelength ranges. Hence, using the pure stellar model (Figure~\ref{c4fig:ellmod}), a conversion table was produced by convolving that stellar model with the NIR filters (from $J$-band to 8\,$\mu$m) at each redshift step of $\Delta{z}=0.01$.

\begin{figure}
  \begin{center}
    \includegraphics[width=1\columnwidth]{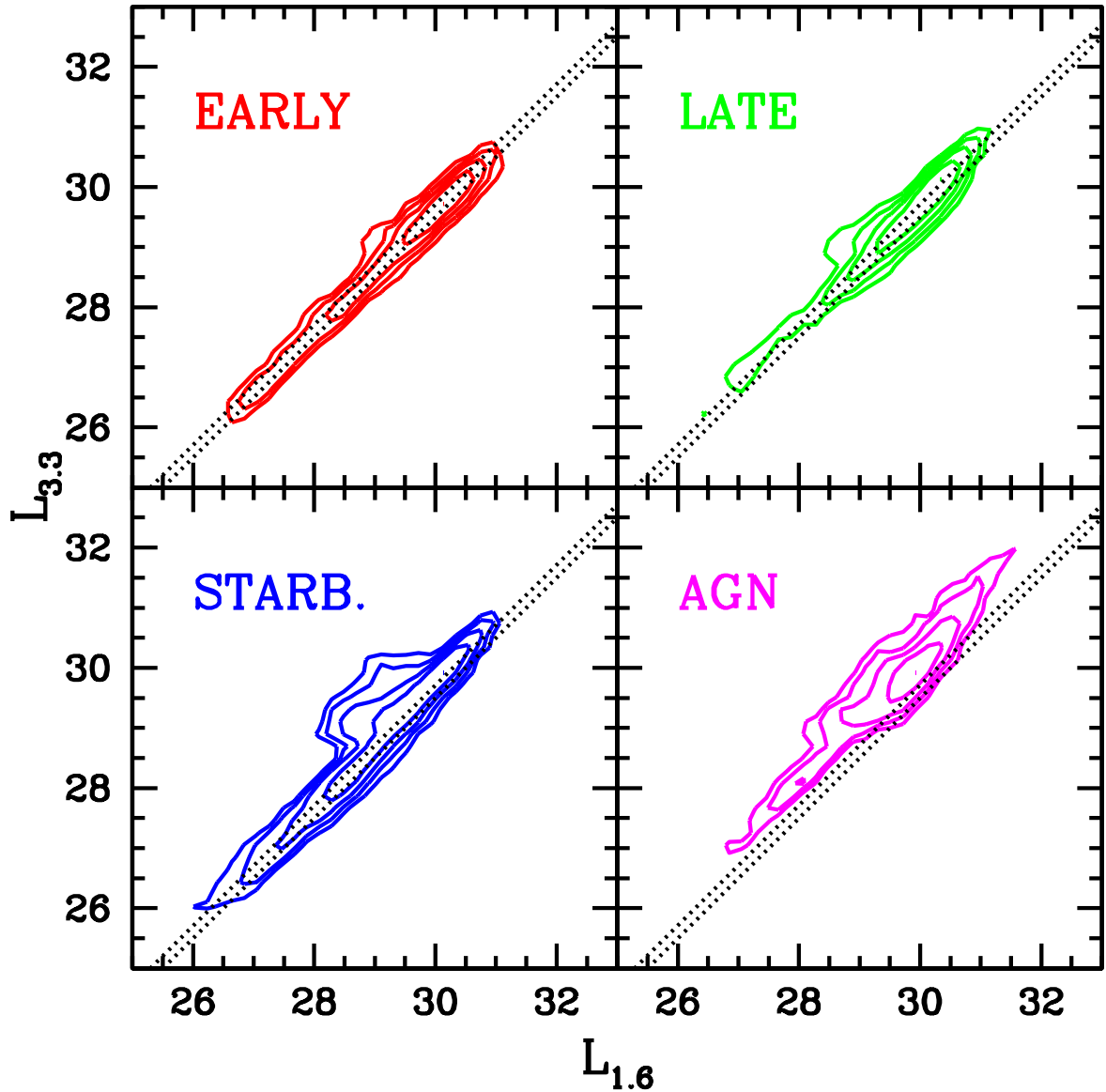}
  \end{center}
  \caption[Rest-frame luminosities: 1.6 versus 3.3\,$\mu$m]{Luminosities at rest-frames 1.6\,$\mu$m (L$_{1.6}$) and 3.3\,$\mu$m (L$_{3.3}$). The dotted lines delimit the region where pure stellar emission should fall. Each panel is reserved to a different population: early-type (upper left), late-type (upper right), starburst (lower left), and AGN host (lower right). The contours are simply demonstrative of the sample distribution and are defined based on the maximum source density in each plot, hence the isocontour levels differ between panels.}
  \label{c4fig:cont33}
\end{figure}

\begin{figure}
  \begin{center}
    \includegraphics[width=1\columnwidth]{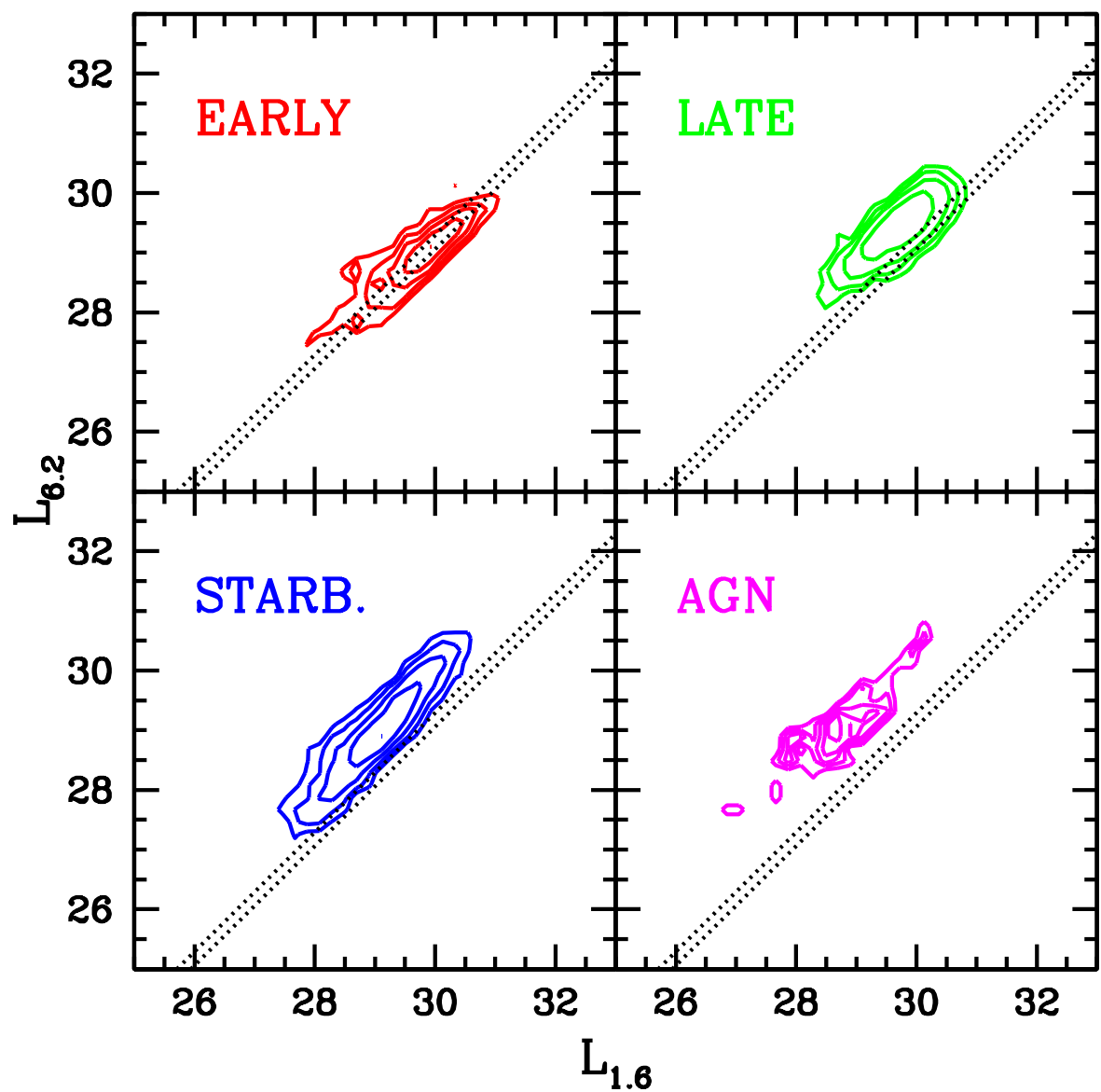}
  \end{center}
  \caption[Rest-frame luminosities: 1.6 versus 6.2\,$\mu$m]{Luminosities at rest-frames 1.6\,$\mu$m (L$_{1.6}$) and 6.2\,$\mu$m (L$_{3.3}$). The dotted lines delimit the region where pure stellar emission should fall. Panel and contour definition as in Figure~\ref{c4fig:cont33}.}
  \label{c4fig:cont62}
\end{figure}

Figures~\ref{c4fig:cont33} and \ref{c4fig:cont62} show the estimated 1.6\,$\mu$m luminosities versus the observed --- hence including stellar plus dust emission --- 3.3 and 6.2\,$\mu$m luminosities for each galaxy population considered in this study. The regions between the dotted lines represent the locus where SEDs dominated by stellar emission alone are expected to fall. These two figures already show that the dust contribution at 6.2$\mu$m is more significant than at 3.3$\mu$m in AGN hosts, starbursts, and some late-type galaxies. As expected, the elliptical data ``cloud'' tends to fall in the stellar region. Figure~\ref{c4fig:cont33} also shows the distinct feature in the luminosity distribution of the starburst population (and slightly in that of the late-type population). This behaviour is assigned to strong hot-dust emission producing a migration out of the pure stellar region. This is the emission excess we aim to extract when subtracting the stellar emission.

We note that the underlying shape of the galaxy IR SED, due to stellar emission alone, is considered to be common to all galaxy populations referred in this study. This is a fair assumption --- for an universal initial mass function --- knowing that stellar emission in this spectral regime originates in cold stars, which live longer, hence producing a constant SED shape over a wide range of ages. Such assumption may be affected by strong obscuration factors between rest-frame 1.6$\mu$m and 3.3 or 6.2$\mu$m. However, this will only occur in rare extremely obscured systems \citep[da Cunha, private communication, and][]{daCunha08}.

\section{The effect of dust and AGN in IR Luminosity Functions} \label{c4sec:irlf}

\begin{figure}
  \begin{center}
    \includegraphics[width=1\columnwidth]{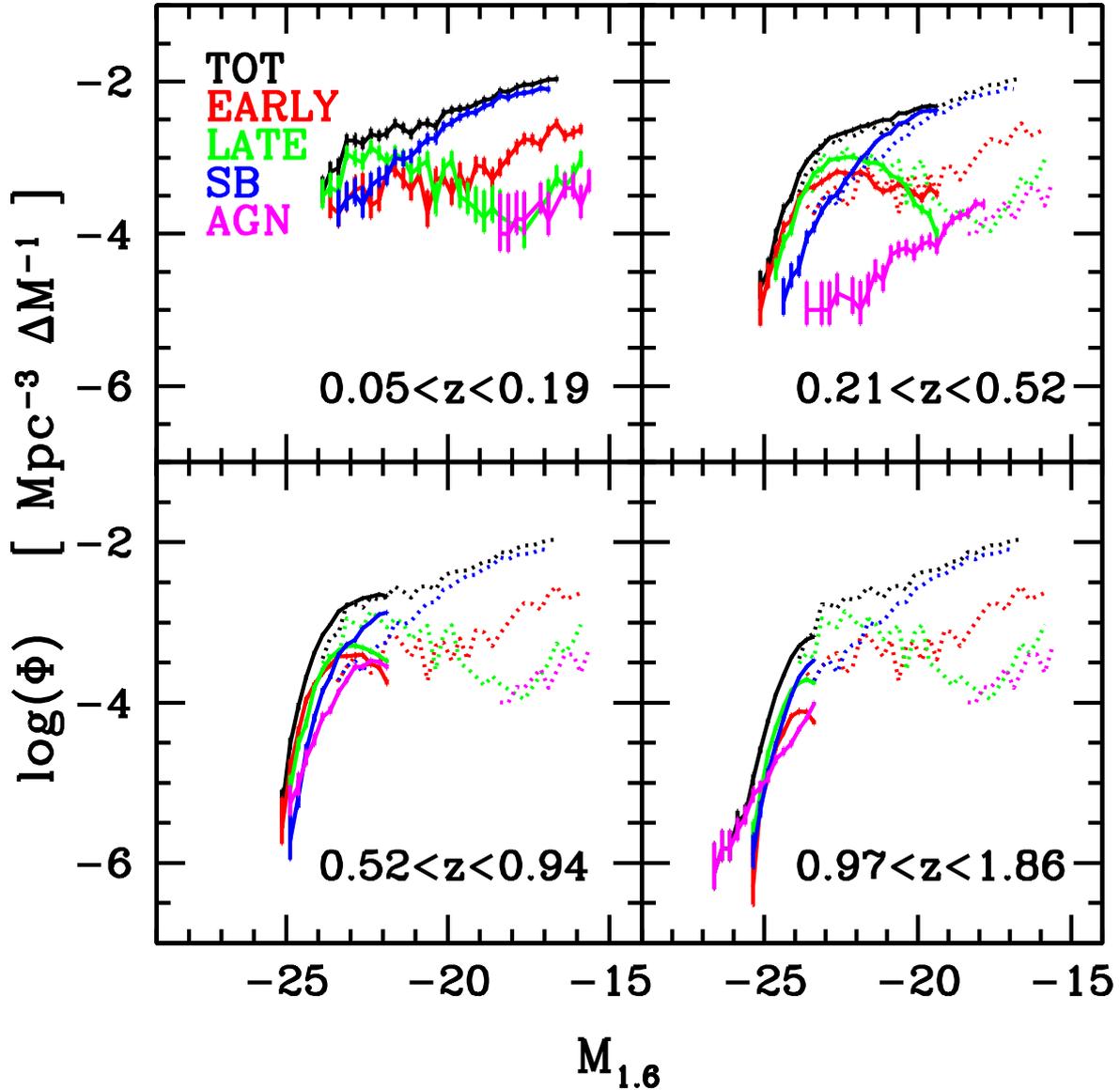}
  \end{center}
  \caption[Rest-frame 1.6\,$\mu$m LFs]{Rest-frame 1.6\,$\mu$m LFs depending on redshift (each panel is reserved to a different redshift range) and galaxy type: Total population (black), Early (red), Late (green), Starburst (blue), and AGN (magenta). The trend shown by each LF in the lowest redshift panel is displayed (dotted LFs) in the subsequent panels for comparison. The LFs of each population were trimmed according to the luminosity below which a significant drop in the source densities (due to incompleteness) is observed in the total population LF or in the sub-population LF. The bin size is of 0.25\,mag.}
  \label{c4fig:lfh}
\end{figure}

\begin{figure}
  \begin{center}
    \includegraphics[width=1\columnwidth]{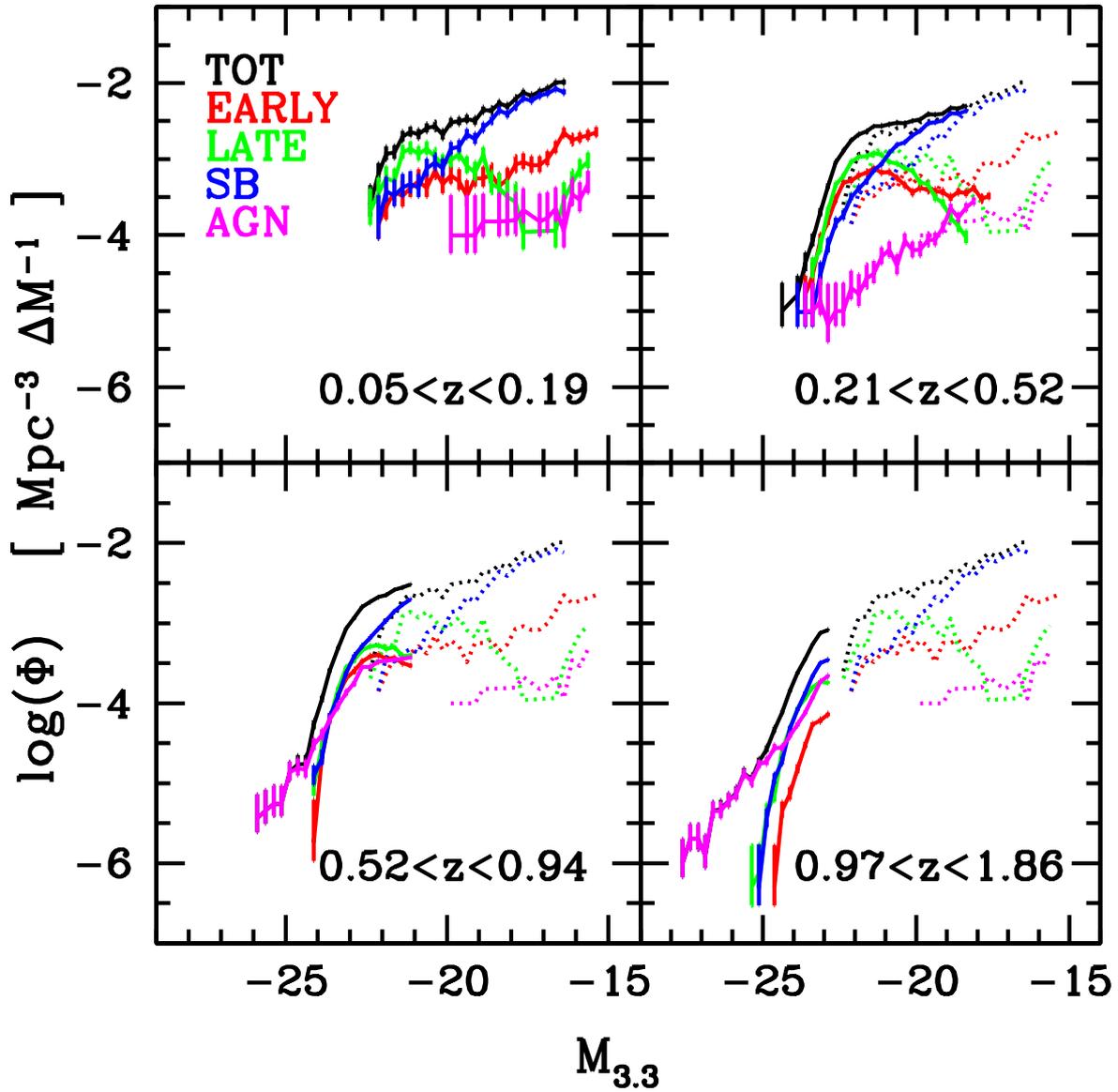}
  \end{center}
  \caption[Rest-frame 3.3\,$\mu$m LFs]{LFs at rest-frame 3.3$\mu$m depending on redshift and galaxy type: Total population (black), Early (red), Late (green), Starburst (blue), and AGN (magenta). The trend shown by each LF in the lowest redshift panel is displayed in the subsequent panels for comparison. The LFs of each population were trimmed as described in Figure~\ref{c4fig:lfh}.}
  \label{c4fig:lf33}
\end{figure}

\begin{figure}
  \begin{center}
    \includegraphics[width=0.5\columnwidth]{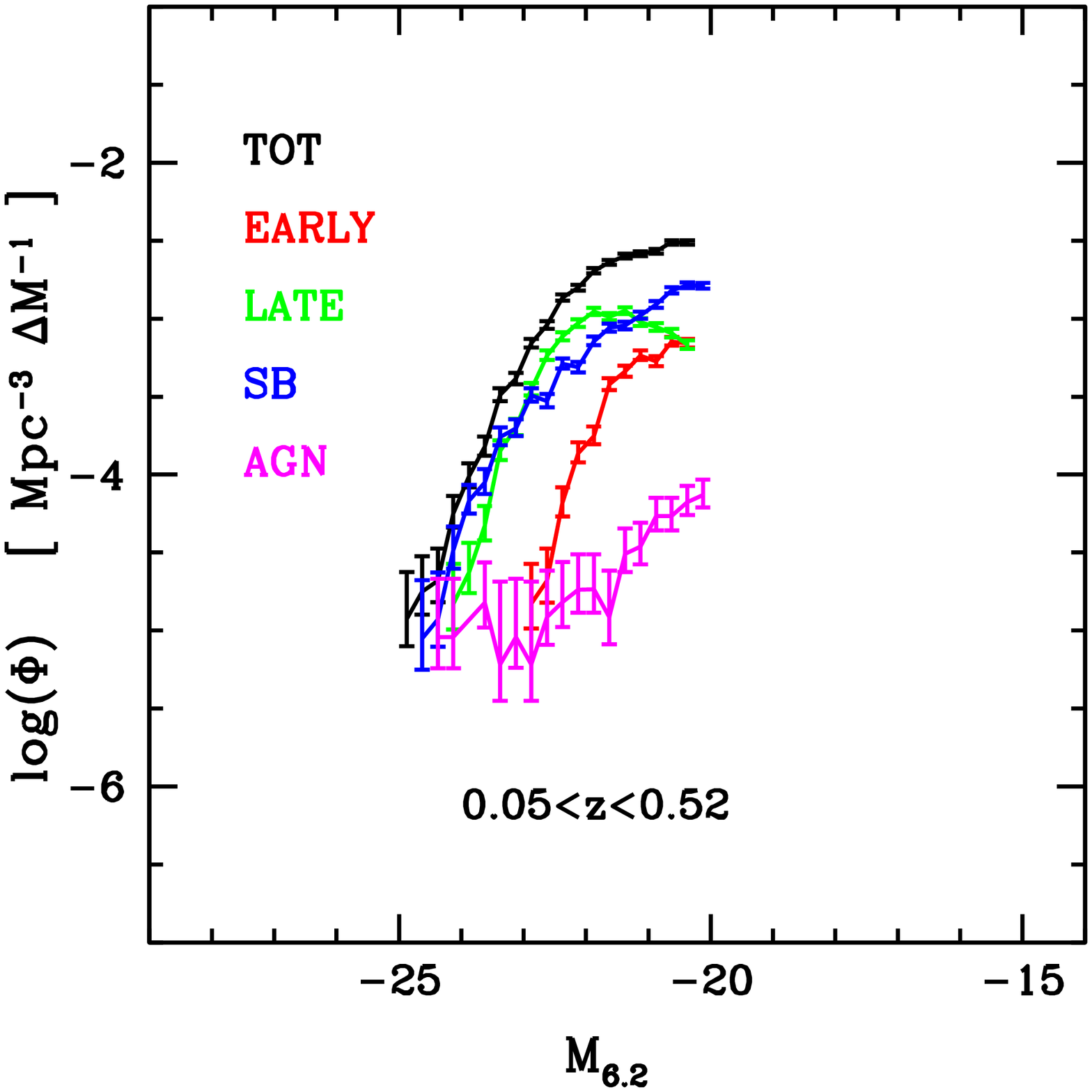}
  \end{center}
  \caption[Rest-frame 6.2\,$\mu$m LFs]{Local 6.2\,$\mu$m LFs depending on galaxy type: Total population (black), Early (red), Late (green), Starburst (blue), and AGN (magenta). The LFs of each population were trimmed as described in Figure~\ref{c4fig:lfh}.}
  \label{c4fig:lf62}
\end{figure}

The total IR LFs are now presented at rest-frames 1.6\,$\mu$m (Figure~\ref{c4fig:lfh}), 3.3\,$\mu$m (Figure~\ref{c4fig:lf33}), and 6.2\,$\mu$m (Figure~\ref{c4fig:lf62}). In each figure, different populations are considered: total (black), early-type (red), late-type (green), starburst (blue), and AGN hosts (magenta). Each panel refers to different redshift intervals (except for 6.2$\mu$m in Figure~\ref{c4fig:lf62} where only one redshift interval is accessible with IRAC bands, Table~\ref{c4tab:zrange}). LFs were obtained through the $1/V_{max}$ method described in Chapter~\ref{ch:erg}. The volume associated with each galaxy is based on the flux limit of the sample and the $k$-correction, derived from the galaxy's own SED (as given by the observed multi-wavelength photometry).

Although here we present rest-frame 1.6\,$\mu$m LFs, this wavelength (at which the stellar continuum peaks) is correlated with the galaxy mass. Therefore, one can attempt to compare the mass functions in the literature \citep[e.g.,][]{Ilbert09,Drory09,Pozzetti10} with our 1.6\,$\mu$m LFs (and in some situations 3.3\,$\mu$m as well). In this study we are unable to follow the features referred in the literature up to the highest redshifts, as in \citet{Drory09}. This is due to the completeness constraints we have applied through magnitude cuts in the IRAC bands. Note that the IRAC bands used for sample selection at the highest redshift bins are also the shallowest.

Nonetheless, many features and trends can be highlighted, some of them new as a result from our approach. Here, we also observe the early and late spectral-type galaxies dominating the bright end, while starbursts dominate the faint-end. At the highest redshifts, however, all three populations seem to contribute at comparable levels to the bright end. Also observed at the lowest redshifts, is the upturn at the faintest magnitudes in the LFs of the early and late populations. These features have been extensively studied and interpreted in previous work \citep[e.g.,][]{Ilbert09,Drory09,Pozzetti10}. Here, we study and highlight the role of the AGN population in shaping the total IR LF.

At the lowest redshift intervals, the 1.6 and 3.3\,$\mu$m power-law shaped LFs of the AGN population seem to follow the upturn found for the LFs of the early and late spectral-type populations at the faintest magnitudes. \citet{Drory09} noted that the blue (starburst) population also presents a comparable steepness at fainter magnitudes, and propose the scenario where they may actually be related. This is supported by \citet{Kormendy85} who proposed a connection between dwarf spheroidal and dwarf irregular galaxies. Adding to that, the fact that dwarf galaxies tend to cluster around massive galaxies \citep{Zehavi05,Haines06,Haines07,Carlberg09}, implies that tidal interactions or ram pressure stripping may be behind the quenching necessary to turn dwarf irregular galaxies into dwarf spheroidal galaxies \citep[see also, e.g., ][]{Boselli08,Henriques08}. The onset of AGN activity seen here may then be understood, as these perturbations and torques on a dwarf galaxy may drive material to the nuclear engine \citep[as it happens in merger events, ][]{diMatteo05,Springel05a,Springel05b} making the AGN ``visible''. In its turn, the nuclear activity may then also act as a quenching mechanism, producing an even faster switch from a star-forming dwarf to a passive dwarf galaxy. Even though most galaxy evolution models do not predict AGN feedback to be a significant (or even existent) quench mechanism in low-mass systems \citep[e.g.,][]{Croton06a,Somerville08}, massive black holes have been found and are expect to exist in dwarf galaxies \citep{Barth04,Reines11,Bellovary11}. Being dwarf systems, they will have a smaller gravitational potential, meaning that a less intense feedback wind may still be efficient as a quench mechanism.

At high-$z$ ($z>0.52$), the power-law shaped AGN LFs produce on the total LF bright-end a deviation from a classical Schechter function \citep[seen better in Figure~\ref{c4fig:lf33}; see also some evidences for this in][although not assigned to AGN emission in that work]{Cirasuolo10}. This behaviour has also been observed in the work based on \textit{Spitzer}-IRS data by \citet{Fu10}. The reader should note that the AGN population is known to present such power-law LFs in the X-rays \citep[e.g.,][]{Aird10}, optical \citep[e.g.,][]{Croom04,Richards05}, and near-IR \citep{Assef11}. These findings thus support the reliability of the AGN selection considered in this work.

The evidence for an AGN flux boost (as a result from hot dust emission) even at rest-frame 1.6\,$\mu$m has important implications for the estimate of stellar mass based on $K$-band luminosity (as does the inclusion of the TP-AGB phase in the stellar models used for such science). Although \citet{Marchesini09} show that AGN emission induces uncertainties in the stellar mass estimate smaller than the photometric mass estimate uncertainty itself, their conclusion is based on a comparison with re-computed stellar masses without considering the 5.8 and 8.0\,$\mu$m IRAC channels, while we show that the flux boost can happen down to $H$-band (probed by the 4.5\,$\mu$m band up to $z\sim2$). Also, the problem is not the scatter which AGN emission may induce, but an upward systematic boost instead, just like the effect produced by emission from TP-AGB stars. It is known that the fraction of AGN is higher both at higher redshifts and high stellar mass \citep{Papovich06,Kriek07,Daddi07}, regimes at which, coincidentally, emission from TP-AGB stars is mostly evoked. \citet{Eminian08} and \citet{Muzzin09} show, however, that the TP-AGB phase does not provide a complete understanding of nIR colours. A thorough study on the systematic boost by hot dust from AGN activity should thus be pursued in every study on stellar mass build up (specially at high redshift where host and nuclear dust emission are frequently blended) with the use of hybrid galaxy models like those of \citet{Salvato09,Salvato11}, or considering the procedure adopted by \citet{Hainline10}.

\begin{figure}
  \begin{center}
    \includegraphics[width=1\columnwidth]{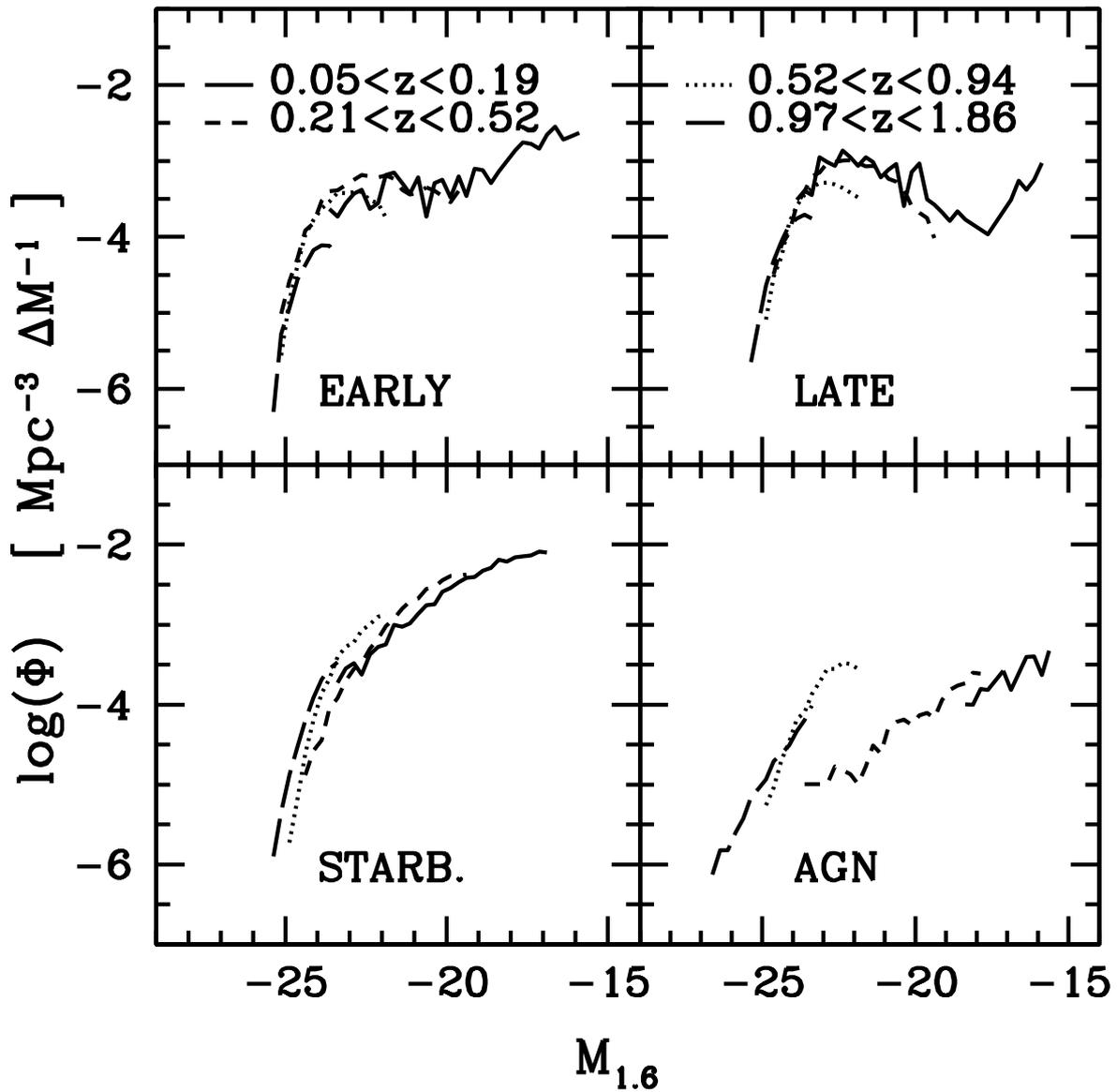}
  \end{center}
  \caption[Evolution of 1.6\,$\mu$m LFs with redshift]{Comparing the rest-frame 1.6\,$\mu$m LFs for each galaxy population between redshift bins: $0.05<z<0.19$ as solid line, $0.21<z<0.52$ as short dashed line, $0.52<z<0.94$ as dotted line, and $0.97<z<1.86$ as long dashed line.}
  \label{c4fig:lfhz}
\end{figure}

\begin{figure}
  \begin{center}
    \includegraphics[width=1\columnwidth]{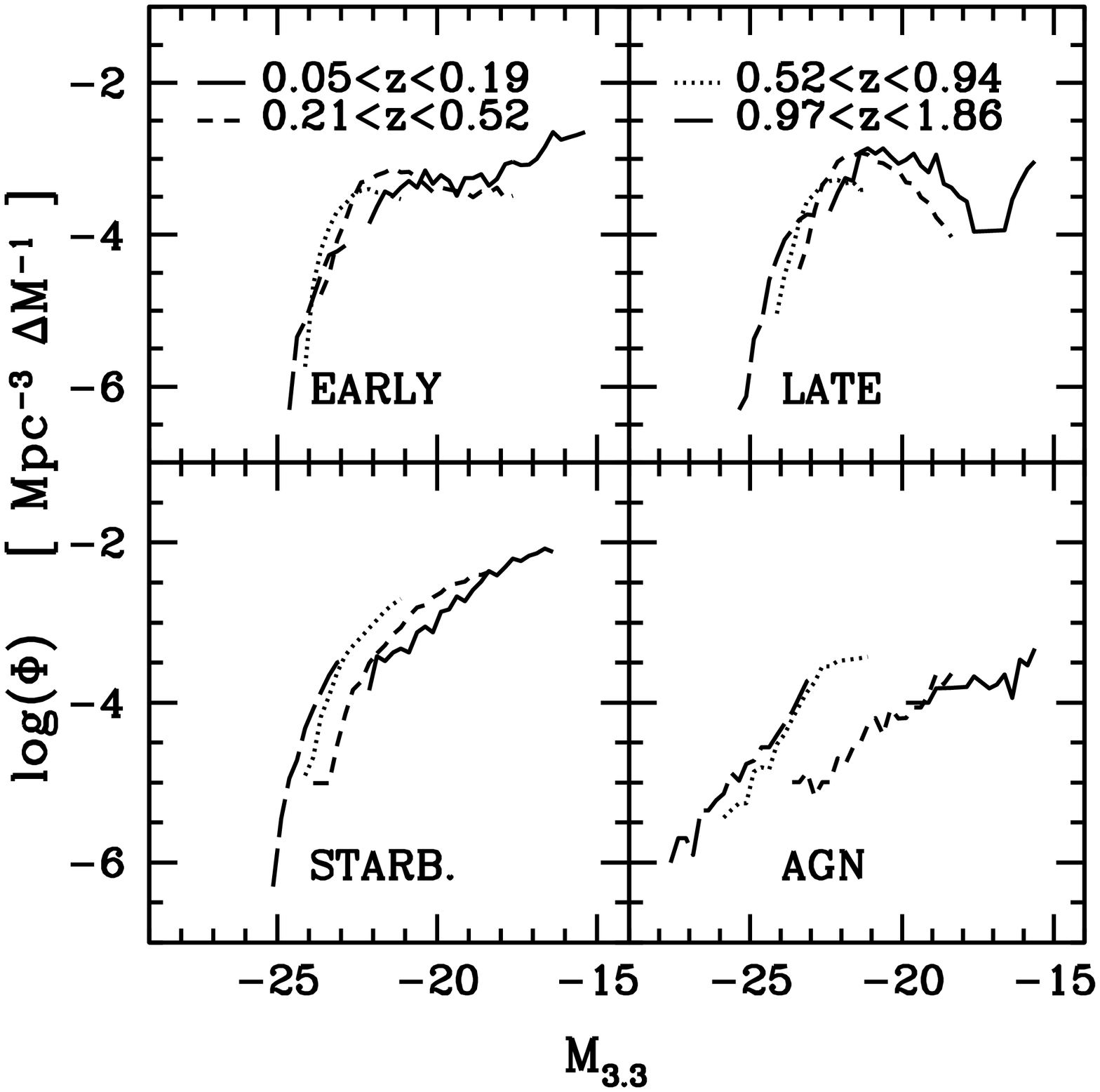}
  \end{center}
  \caption[]{Comparing the rest-frame 3.3\,$\mu$m LFs for each galaxy population between redshift bins. Line coding as in Figure~\ref{c4fig:lfhz}.}
  \label{c4fig:lf3z}
\end{figure}

Both Figures~\ref{c4fig:lfhz} and \ref{c4fig:lf3z} show that the AGN population is the one to show the greatest difference between the low ($z<0.5$) and high ($z>0.5$) redshift intervals. While at high-$z$, one can see the AGN population as the largest contributor to the bright-end of the total population LF (Figures~\ref{c4fig:lfh} and ~\ref{c4fig:lf33}), its activity is completely altered as one moves to low-$z$. At low-$z$, the AGN activity is restricted to the faintest objects. This can thus be understood as some kind of AGN downsizing, where at higher redshifts, AGN activity is seen more in brighter galaxies as opposed to the low redshift regime. However, the data used here is unable to confirm whether the fainter AGN sample seen at low-$z$ is present at high-$z$ or not. In fact, \citet{Cardamone10} find a bimodal AGN host population at $z\sim1$ after correcting for obscuration, with AGN activity found in equal numbers of passive evolving and dust-reddened young galaxies \citep[but see][]{Xue10}. This would imply that the downsizing effect is a selection result and not real, yet it is clear that the AGN activity shuts down first in more luminous galaxies between $0.52<z<0.94$ and $0.21<z<0.52$. The time of change seems to be interestingly related to the feature we discuss below.

The shape of the early/late-type 1.6\,$\mu$m LFs seem fairly unchanged with redshift at the highest luminosities (Figure~\ref{c4fig:lfhz}), but at gradually fainter magnitudes they do seem to show a mass build up, resulting in the growth of the hump which, at low redshifts, peaks at absolute magnitudes of $-23<\rm{M}_{1.6}<-22$ and source densities of $\log(\Phi[\rm{Mpc^{-3}\,\Delta{M}}])\sim-3$. It is interesting to notice that the growth of the hump stabilises by $0.21<z<0.52$. By this time, the AGN activity in such bright galaxies has dramatically decreased when compared to higher redshifts, being found in less luminous classes at those low redshifts.

At rest-frame 3.3\,$\mu$m, the LFs of early, late, and starburst spectral-type populations show a shift to fainter absolute luminosities with decreasing redshift (Figure~\ref{c4fig:lf3z}), while at rest-frame 1.6\,$\mu$m, this is only seen in the starburst population at a smaller level. These differences are, of course, driven by different contributions between stellar and hot-dust emissions at these rest-frame wavelengths. While at 1.6\,$\mu$m we see early- and late-type SEDs dominated by stellar light, at 3.3\,$\mu$m (and at 1.6\,$\mu$m for starbursts) there is already a significant hot-dust contribution which varies with time (inducing the observed shift). This extra flux comes from hot-dust emission around TP-AGB stars and AGN. Although we already select a sample of AGN sources, the extra flux (specially at 3.3\,$\mu$m) may originate from lower-luminosity AGN which don't dominate the IR SED enough to be selected by our AGN criterion.

\section{Dust Luminosity Density Functions} \label{c4sec:dlfs}

\begin{figure}
  \begin{center}
    \includegraphics[width=1\columnwidth]{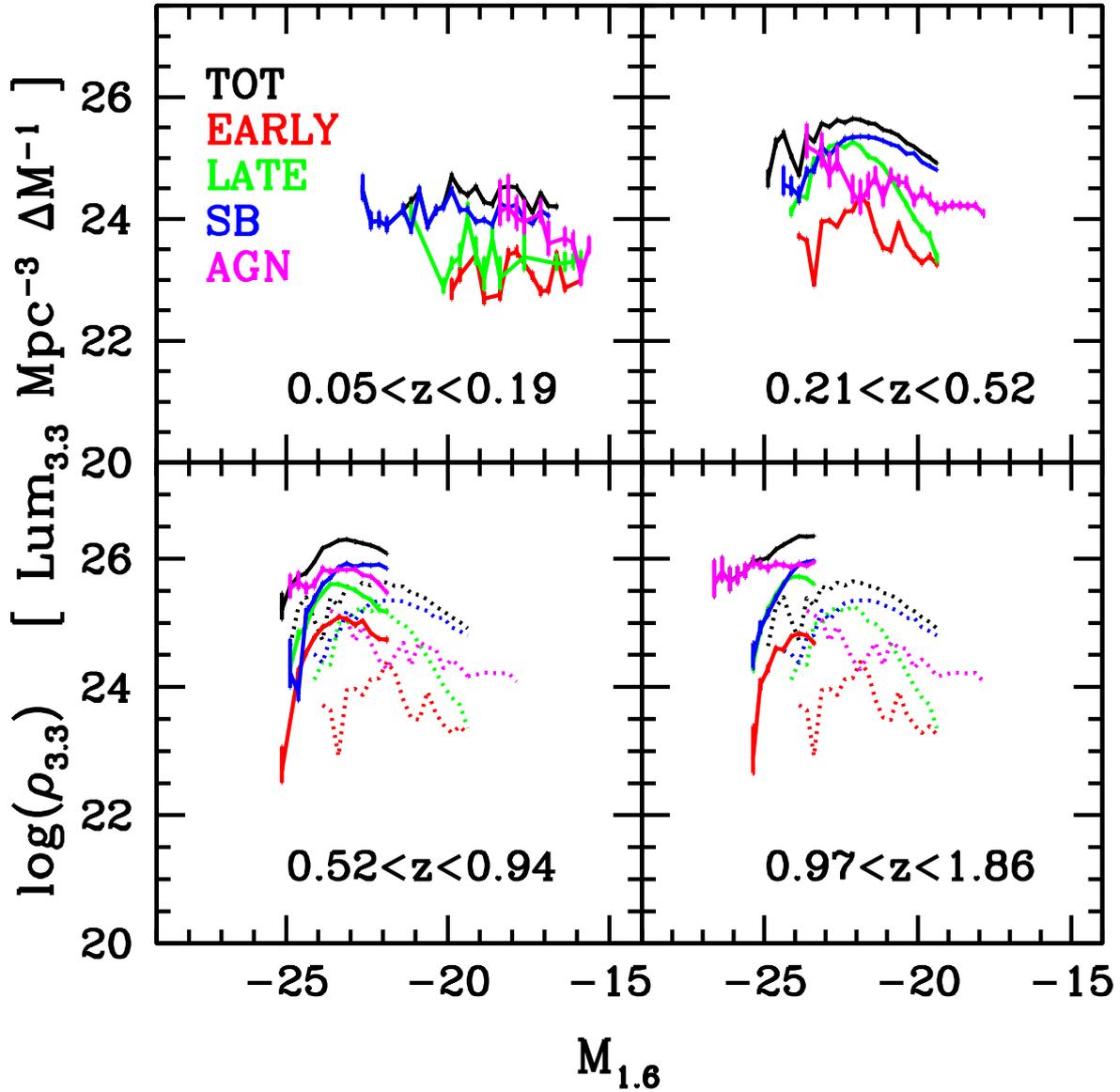}
  \end{center}
  \caption[Rest-frame 3.3$\mu$m dust LDFs]{Rest-frame 3.3$\mu$m dust LDFs depending on distance and galaxy type: Total population (black), Early (red), Late (green), Starburst (blue), and AGN (magenta). Due to the irregular trends in the low redshift panel, the trend shown by each galaxy population in the $0.21<z<0.52$ redshift panel is displayed instead in the subsequent panels for comparison. The dust LDFs were trimmed according to the cuts adopted in Figure~\ref{c4fig:lfh}.}
  \label{c4fig:lf33d}
\end{figure}

\begin{figure}
  \begin{center}
    \includegraphics[width=1\columnwidth]{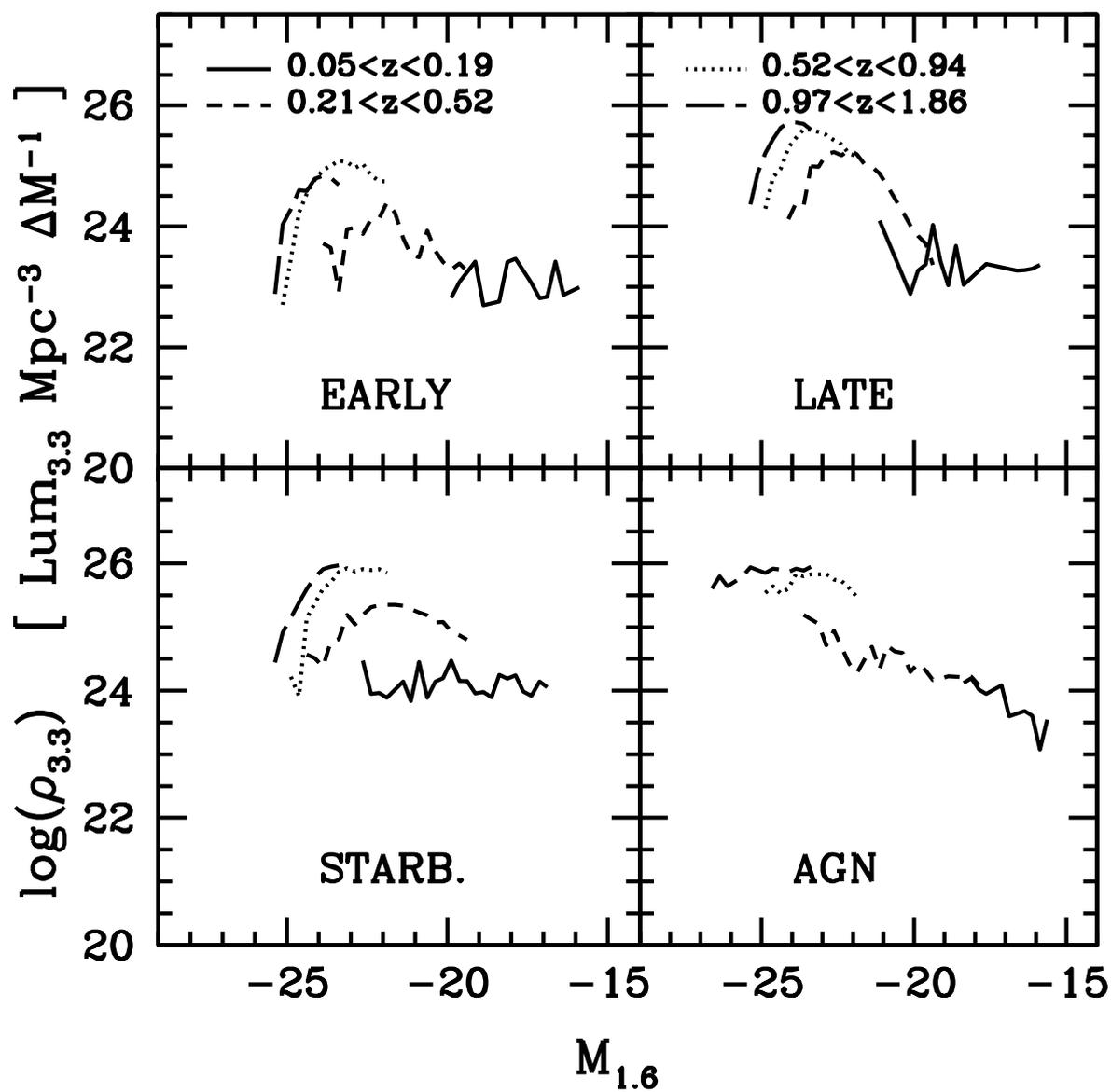}
  \end{center}
  \caption[Rest-frame 3.3$\mu$m dust LDFs evolution with redshift]{Comparing the rest-frame 3.3\,$\mu$m dust LDFs for each galaxy population between redshift bins. Line coding as in Figure~\ref{c4fig:lfhz}. The dust LDFs were trimmed according to the cuts adopted in Figure~\ref{c4fig:lfh}.}
  \label{c4fig:lf33dz}
\end{figure}

\begin{figure}
  \begin{center}
    \includegraphics[width=0.5\columnwidth]{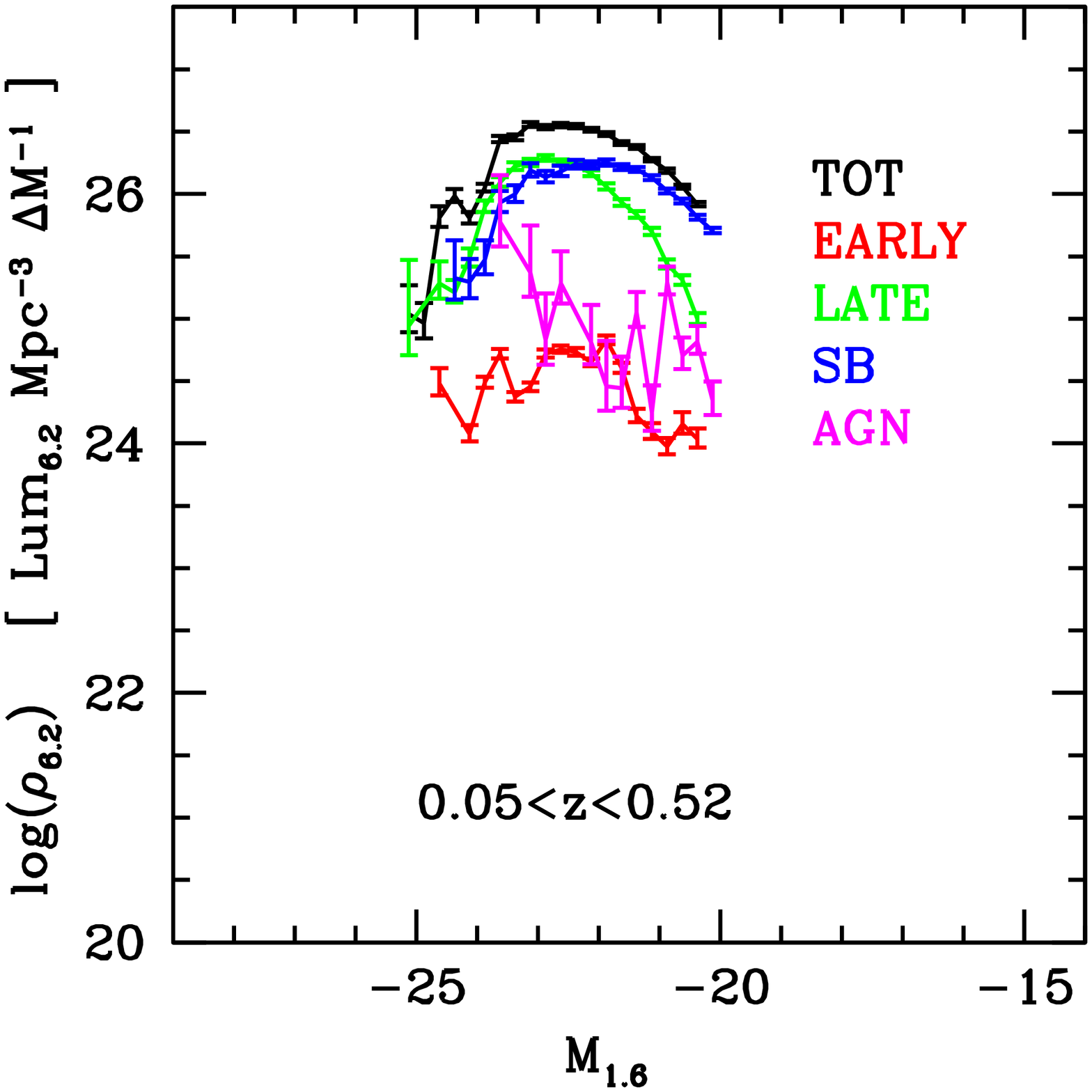}
  \end{center}
  \caption[Rest-frame 6.2$\mu$m dust LDFs]{Local 6.2$\mu$m dust LDFs depending on galaxy type: Total population (black), Early (red), Late (green), Starburst (blue), and AGN (magenta). The dust LDFs were trimmed according to the cuts adopted in Figure~\ref{c4fig:lfh}.}
  \label{c4fig:lf62d}
\end{figure}

Dust luminosity density functions (LDFs) are presented at rest-frame 3.3\,$\mu$m (Figures~\ref{c4fig:lf33d} and \ref{c4fig:lf33dz}) and 6.2\,$\mu$m (Figure~\ref{c4fig:lf62d}). These plots enable us to evaluate how much dust is emitting in the IR in each galaxy population at a given rest-frame 1.6\,$\mu$m luminosity. The choice of the rest-frame 1.6\,$\mu$m absolute magnitude as the x-axis in the figures allows us to know where source count densities, used for the LDFs, are incomplete. Furthermore, this allows to estimate how much dust emission there is in each galaxy luminosity class in Figure~\ref{c4fig:lfh}. Note that each of these luminosity classes can be taken as a proxy to stellar mass classes providing that 1.6\,$\mu$m is dominated by stellar emission.

Although we have shown that at low-$z$ AGN hosts are few (2\% of the overall population), this population contributes significantly to the faint-end of the dust LDF (and maybe even at a comparable level as the much more numerous starbursts at the faintest magnitudes). At high-$z$, the opposite clearly takes place. The AGN population is by far the largest contributor to the bright-end of the 3.3\,$\mu$m dust LDF, as already expected from the LF shown previously (Figure~\ref{c4fig:lf33}). Analysing the rest-frame 6.2\,$\mu$m dust LDFs, however, one realises how strong the contribution of the PAH features and hot dust can be to the overall IR LDF of late-type and starburst galaxies, clearly surpassing that from nearby AGN.

\begin{figure}
  \begin{center}
    \includegraphics[width=0.9\columnwidth]{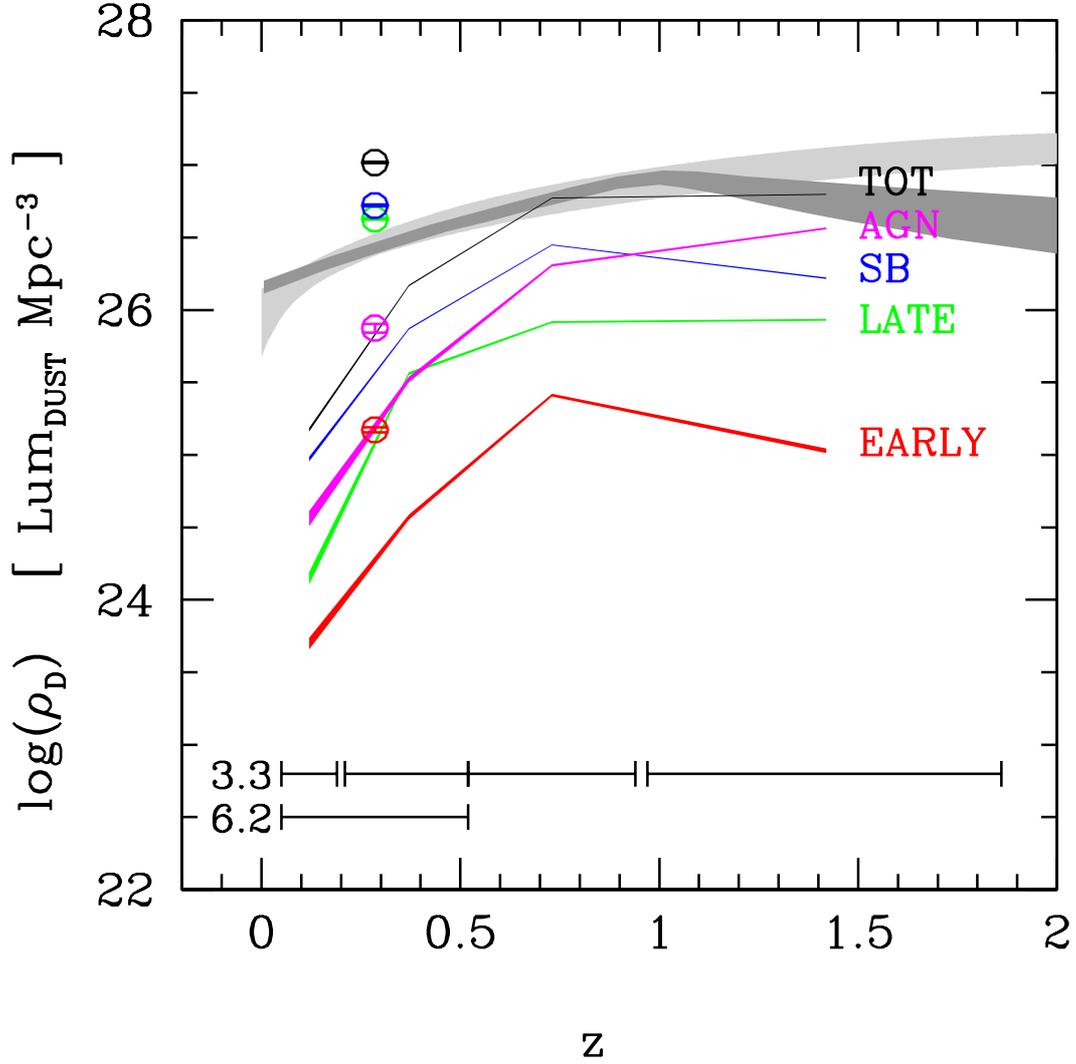}
  \end{center}
  \caption[Dust luminosity densities evolution with redshift]{Rest-frame 3.3 and 6.2$\mu$m dust luminosity densities ($\rho_{\rm D}$) depending on redshift and galaxy type: Total population (black), Early (red), Late (green), Starburst (blue), and AGN (magenta). Rest-frame 3.3\,$\mu$m estimates appear connected as coloured regions (the width represents the statistical uncertainty), while 6.2\,$\mu$m estimates appear as open circles. The redshift intervals corresponding to each data point of each rest-frame wavelength (Table~\ref{c4tab:zrange}) are indicated as error bars at the bottom. The $3\sigma$ trends of the total SF history, as compiled by \citet[][light grey shaded region]{HopkinsBeacom06}, and of the obscured SF history, estimated by \citet[][dark grey shaded region]{CharyPope10}, are shown for comparison. Both trends are scaled to the $\rho_{\rm D}$ value of the total population at $0.52<z<0.94$.}
  \label{c4fig:zlum}
\end{figure}

As a final remark, Figure~\ref{c4fig:zlum} shows how the dust luminosity density has evolved since $z\sim$1--2 (rest-frame 3.3\,$\mu$m estimates appear connected). With a significant drop since $z\sim1$ and a flattening at $z\gtrsim1-2$, it resembles the evolution trend of the SF history of the universe \citep[light grey shaded region,][scaled to the total population luminosity density at $0.52<z<0.94$]{HopkinsBeacom06}. However, comparing the luminosity density at present time to that at $0.52<z<0.94$, the drop in 3.3$\,\mu$m luminosity density is more significant than that of the SF history, by around 1\,dex more. This difference still holds even when comparing to the obscured SF history \citep[dark grey shaded region][]{CharyPope10}. Some interpretations for this may be given: either the reduced star-formation at low redshifts is unable to heat enough quantities of dust for it to dominate at 3.3\,$\mu$m; dust may be farther from UV/optical sources at lower redshifts (blown by stellar winds, as seen, e.g., in Orion nebula); or there is an actual decrease in the dust content in galaxies\footnote{Completeness effects would affect more the higher redshift estimates. A correction for such effect would increase those estimates and, consequently, increase the decay we observe since $0.52<z<0.94$.}. A recent study with \textit{Herschel Space Observatory} data may support the latter \citep{Dunne11}.

The AGN sample appears as the main contributor to the overall galaxy dust luminosity density at $z>1$. At $z<1$, the starburst sample is the largest contributor with AGN hosts still comprising, nonetheless, a significant contribution to the overall dust luminosity density. We stress, however, the completely different source numbers of each of these two populations (Table~\ref{c4tab:numb}), where the starburst population is significantly more numerous, meaning that hot-dust emission in AGN sources in significantly more than in the remainder of the populations. Overlaid in Figure~\ref{c4fig:zlum} are also the data points for rest-frame 6.2$\mu$m (open circles) in the nearby universe. It shows how much more dust is contributing to the galaxy SED at 6.2\,$\mu$m when compared to 3.3\,$\mu$m. For instance, the dust luminosity density at rest-frame 6.2\,$\mu$m at $z<0.52$ is still larger than the dust luminosity density at rest-frame 3.3\,$\mu$m at $z>1$. Table~\ref{c4tab:fracld} details the contributions of each of the galaxy populations to the overall dust luminosity density at rest-frames 3.3 and 6.2\,$\mu$m, depending on the redshift.

\ctable[
   cap     = Luminosity density fractions,
   caption = The contribution of the different galaxy samples to the rest-frame 3.3 and 6.2$\mu$m dust luminosity densities,
   label   = c4tab:fracld,
   pos     = ht
]{ccrrrr}{}{ \FL
Rest-Frame & $z_{\rm BIN}$ & EARLY & LATE & STARB. & AGN \NN
$[\mu\rm{m}]$ &  & $[\%]$ & $[\%]$ & $[\%]$ & $[\%]$ \ML
3.3 & $0.05<z<0.19$ & 3 & 9 & 62 & 26 \NN
& $0.21<z<0.52$ & 3 & 25 & 50 & 22 \NN
& $0.52<z<0.94$ & 4 & 14 & 47 & 35 \NN
& $0.97<z<1.86$ & 2 & 14 & 26 & 58 \ML
6.2 & $0.05<z<0.52$ & 1 & 41 & 50 & 7 \LL
}


\newpage

\section{Conclusions} \label{c4sec:conclusions}

In this chapter, the hottest regime of dust emission was explored. Our new approach considers both stellar and dust emissions separately, as well as the separation of the IR galaxy population into early, late, starburst, and AGN host galaxies. This allows to evaluate the IR luminosity functions depending on galaxy-type and distance, as well as to estimate how much dust is contributing to the IR emission. We have concluded the following:

\begin{itemize}
 \item The upturn seen in the IR LFs at the faint-end is probably linked to AGN activity, but may not be directly due to AGN activity. Instead, AGN activity is believed to be the consequence, and not the cause, of the upturn (e.g., dwarf galaxy disruption). Nevertheless, AGN may help speed up the transition process between star-forming dwarf galaxy into passive dwarf galaxy.

 \item AGN ``downsizing'' between $z\sim1$ and $z\sim0$ is probably a selection effect in our study, as fainter AGN hosts are missed at the highest redshift ranges. But it is clear that the AGN activity in the most luminous objects shuts down between $0.21<z<0.52$ and $0.52<z<0.94$ range. Interestingly, this is the range where the IR LF hump at $-24<\rm{M}<-21$ seems to be finally assembled, hinting for a possible relation between the two episodes.

 \item The observed AGN flux boost at 1.6\,$\mu$m has important implications for any high redshift study on galaxy stellar mass when estimated photometrically (specially at high redshift), as it results in a systematic overestimate of the stellar masses. The shapes of the AGN LFs support the reliability of our results, when comparing with work in the literature.

 \item Although significantly less numerous, AGN comprise a significant contribution to the overall dust emission at rest-frame 3.3\,$\mu$m since $z\sim1-2$.

 \item Evolution with redshift of the hot dust luminosity densities resembles that of SF history of the universe, but it drops more steeply (about 1\,dex more). The reason for such result is, however, still indeterminate. Nonetheless, recent work by \citet{Dunne11} points to an actual decrease of dust content in galaxies since previous times.
\end{itemize}
\chapter{Future prospects}
\label{ch:fut}
\thispagestyle{empty}

With the main focus on the IR spectral regime, this thesis addresses three science key subjects for the understanding of galaxy evolution: extremely red galaxies (ERGs), active galactic nuclei (AGN), and dust.

With a new statistical approach, meaningful results are obtained for Extremely Red Galaxy (ERG) populations (Chapter~\ref{ch:erg}). By separating the sample into pure and common ERGs (respectively, galaxies belonging to only one or to the three ERG groups considered in this study), we show that pure-EROs (pEROs) are mostly passively evolved galaxies, while the common galaxies (mostly IEROs and DRGs) show evidences for a dusty starburst nature. However, a morphology study and the non-existent colour bimodality (there is a continuum in $J-K$ colours from pEROs to DRGs) point to a link among the ERG population, where the more star-forming ERGs will later turn into more passively-evolved ERGs. Hence, the frequently referred difficulty to separate ERGs into distinct populations, either morphologically \citep{Moustakas04} or photometrically \citep{Pierini04,Fontanot10}, is probably a result of this smooth transition during the ERG phase.

The new KI and KIM criteria showed to be more reliable than commonly used IR criteria and are of great use for the study of AGN populations undetected at X-rays or optical wavelengths (Chapter~\ref{ch:agn}). For instance, in Chapter~\ref{ch:lfs} the KI criteria allowed to see that AGN activity is closely related to different features seen in the IR luminosity functions (LFs), and that hot dust in AGN host galaxies may emit significantly at short IR wavelengths probably biasing systematically any stellar mass study at high-$z$ (where the effect seems to be stronger).

The overall results of this thesis reveal the possibility for further applications in a wide variety of science projects, and for improvements to the work itself. In this chapter, future lines of research are considered. Following-up on the developed work, these comprise further testing of the considered techniques (such as the KI and KIM criteria), the need to overcome the present limitations (e.g., small areal coverage), and new proposed projects making use of the experience gained during the course of this thesis (e.g., the study of high redshift passive disc galaxies).

\section{On the application to other surveys}

Before presenting any of the future projects, it should be stressed that extending the work to other fields is not always the answer. It depends on the science goals and survey characteristics. If each of these factors are not properly taken into account, sometimes one will be comparing apples and oranges.

With deep (near-)IR imaging (needed for the study of ERGs), we find, for instance, larger surveys than GOODSs like COSMOS (1.8\,deg$^2$), VISTA Deep Extragalactic Observations (VIDEO 12\,deg$^2$, PI. Matt Jarvis) and the UKIDSS Ultra Deep Survey \citep[UDS, 0.8\,deg$^2$,][]{Warren06}. All these wide-field deep surveys make use of optical to near-IR imaging that reach survey depths comparable to those achieved in the Great Observatories Origins Deep Survey (GOODS) fields, however, except for radio frequencies (with upcoming all-sky surveys reaching 10\,$\mu$Jy $rms$ levels), the remainder spectral regimes tend to be a whole different scenario.

Take the X-ray coverage as an example, it is unlikely that, in the next decade or so, surveys like those referred above will ever reach the 2\,Ms depth achieved in the northern and southern GOODS fields (and even less the recent 4\,Ms on GOODS-South, GOODSs). This holds because of the large areal coverage of these surveys, which $Chandra$ will not be able to cover till such depths in a reasonable time-length. The extended ROentgen Survey with an Imaging Telescope Array \citep[eROSITA, to be launched in 2012/2013,][]{Predehl10} will cover the whole sky in the X-rays (0.5--10\,keV), albeit to depths 2\,dex brighter than those reached by $Chandra$ and with a half energy width (HEW) 25 times larger. The Wide Field X-ray Telescope \citep[WFXT, with a HEW five times larger than $Chandra$,][]{Rosati11} will cover as well the whole sky, and later cover $\sim$100\,deg$^2$ down to the deepest $Chandra$ sensitivity. However, this mission is not yet scheduled for launch.

Also, \textit{Spitzer} Space Telescope (\textit{Spitzer}) is now on `warm mode', meaning that observations at higher wavelengths ($>5\,\mu$m) are no longer possible. Although these higher wavelength filters were the least sensitive even on `cryogenic mode', they were (and are) fundamental for the study of the high redshift Universe by probing rest-frame near-IR (1--3\,$\mu$m at $z>1$), allowing for, e.g., proper stellar mass estimates. The all-sky survey performed by Wide-field Infrared Survey Explorer \citep[WISE,][]{Wright10} reaches magnitude (source confused) limits at $3\,\mu$m brighter than those available in COSMOS at $8\,\mu$m, limiting the studies to the brightest of the sources at high redshifts. Finally, the small $\sim$4\,arcmin$^2$ field of view of \textit{James Webb} Space Telescope makes it more a follow-up science telescope than a survey science one. Hence, the currently available fields resulting from the combination between area and depth of \textit{Spitzer} at $>5\,\mu$m will still be the best in the next years to come.

\section{Extremely red galaxies} \label{c2sec:future}

Although this study is based in one of the deepest data-sets ever assembled, it does not provide a large enough sample to constrain, for instance, the SFR values for the $2\leq{z}\leq3$ non-AGN ERG sample. At this point, we believe that similar methodology to this thesis is only possible in GOODS north (GOODSn), given the similar deep multi-wavelength coverage. This is, in Chapter~\ref{ch:erg}, if we had restricted the AGN classification to the shallower X-ray and IR flux limits available in other surveyed fields (referred above), some of the sources classified as AGN in this thesis would no longer be so, hence being included in the non-AGN stack instead, boosting the stacking flux signal and resulting in a detection. By considering GOODSn, similar flux levels are considered and a proper statistical improvement is possible, even though the overall areal coverage is still small ($\sim$300\,arcmin$^2$ in total for the two fields).

We do believe, however, that larger fields will probably help confirming if the second proposed dip seen in the mass functions (MFs) at $z>1$ (Section~\ref{c2sec:masssec}) is real or a result of a methodology bias. This second dip is found at relatively high masses, hence, shallow IR surveys will still be able to probe it.


\subsection{Dependencies on clustering}

ERGs are known to be found in over-dense regions of the Universe \citep[e.g.,][]{Roche03,Grazian06b,Kong09}, and among ERGs, there are differences between those galaxies showing evolved stellar populations and those known to be dusty starbursts, where the former are found in the densest of the environments \citep[up to twice the clustering amplitude,][]{Daddi02,Roche02,Kong09}. But can one actually see any evolution from $2<z<3$ to $1<z<2$? Also, if each of these two ERG populations indeed track different density environments (although both in dense regions), just by separating them, one can follow the SF history and mass assembly dependency on clustering from high redshifts in ERG populations.

\subsection{Morphology evolution}

Another interesting question to be answered is, if indeed ERGs are all the same population, which of the two populations --- passively evolved or dusty starburst --- grows faster and larger in order to fit the spheroid sizes found in the local Universe \citep{Trujillo06,Trujillo07,Buitrago08}? Or do they turn into the same kind of local spheroids? Will they differ in size nonetheless? In order to assess these morphology-related questions, different observed wave-bands (optical to near-IR) are required to follow the same rest-frame wavelength up to high redshifts. Most of the work done until today on deep and far galaxy samples was based on a comparison of $HST$-ACS with $HST$-NICMOS imaging. The latter unavoidably limited the studies to smaller patches of the sky \citep[like in GOODS NICMOS Survey,][]{Conselice11} due to its smaller field of view and integration efficiency. With the advent of the Wide Field Camera 3 (WFC3) installed on $HST$ in May 2009, nIR imaging is enabled down to unprecedented depths and with improved resolution\footnote{Ground-based telescopes are limited by Earth's atmospheric molecular, ionic, and continuum emission \citep{Mountain09}, and have significantly higher telescope thermal emission comparatively to $HST$ \citep{Mountain09}. Unless aided with a (laser) guiding star and an active and adaptive optic system, the image resolution will always be limited by the atmospheric seeing.}, closer to that achieved with ACS. The Cosmic Assembly Near-infrared Deep Extragalactic Legacy Survey\footnote{http://candels.ucolick.org/} (CANDELS) team was granted 902 $HST$ orbits of WFC3 and ACS observing time (started on October 2010) to cover significant portions of some of the best studied extragalactic fields so far (GOODS north and south, Ultra Deep Survey, Extended Groth Strip, and COSMOS). The first orbits of CANDELS were reserved for GOODSs and some of the data is already available for public use.

As can be seen in Section~\ref{c5sec:pdgs}, the CANDELS \citep[and the WFC3 Early Release Science\footnote{Program 11359 (PI R. W. O'Connell), http://archive.stsci.edu/prepds/wfc3ers/} observations,][]{Windhorst11} data are already being used in one of the most interesting follow-up studies arising from this thesis, the study of Passive Disc Galaxies. Another subject that will take advantage of such data-set is the study of pure-DRGs (Section~\ref{c2sec:pdrgs}). The near-IR imaging from WFC3 (probing the rest-frame optical at $z>2$) will allow to follow the older and colder stellar population present in these galaxies, and, by matching with the observed optical (rest-frame UV) ACS imaging, analyse possible differences in the dynamics between the old and young stellar populations producing, respectively, the characteristic red $J-K$ and monochromatic/blue $i_{775}-K$ colours in each pDRG SED. Grism observations will also allow for a faster spectral coverage of this population, which is still scarce (only about 10\% of the pDRGs have a good quality spectrum). While those are not available for a significant patch of GOODSs, we have applied for 20\,h of observation time with FORS2\footnote{http://www.eso.org/sci/facilities/paranal/instruments/fors/overview.html\#fors2} at the Very Large Telescope, in order to get spectroscopic redshifts for 30 of the brightest pDRGs. The requested time length is that needed to get enough signal to noise to allow for a proper AGN census (relying on line ratios and high ionization emission line detections), a type of activity which is believed to be still in action in pDRGs and is related to a possible recent evolution of these galaxies (Section~\ref{c2sec:pdrgs}).

\subsection{Stacking algorithm}

In parallel, the stacking procedure is being tested in order to search for possible improvements. We focus on both the stacking procedure itself and the pre-analysis of each stamp to be stacked. We aim to constrain the bias towards a given type of source (e.g., flux or apparent extension dependency). Should the science image or a source-removed image be used for stacking? Do we actually understand the statistics behind the stacking analysis? Can we go deeper than the expected $rms$ decrease with $1/\sqrt{N}$ by means of a pre-stacking procedure? These are some of the issues we are exploring. For that purpose, real radio data are being considered and simulations ran depending on both parent galaxy properties and telescope capabilities. This will be of great importance for all-sky surveys done with near-future radio facilities such as ASKAP\footnote{http://www.atnf.csiro.au/SKA/}, MeerKAT\footnote{http://www.ska.ac.za/meerkat/specsci.php}, and MWA\footnote{http://www.mwatelescope.org/}.

\section{High-$z$ passive disc galaxies} \label{c5sec:pdgs}

This has been one of the most interesting outcomes of this thesis, enough to have its own section --- this one --- detailing its implications and what will be done henceforth.

The quest for understanding the formation mechanism of our own galaxy and other local disc galaxies has lead the community to search for their ancestors at high redshift. The Atacama Large Millimetre Array\footnote{http://almascience.eso.org/} (ALMA), in its full mode, will be able to detect Milky-Way-like systems up to $z\sim3$, but the question remains ``do such evolved disc galaxies exist at such high redshifts?''. The presence of disc galaxies at high-$z$ ($z\gtrsim2$) has been first predicted \citep[e.g.,][]{NavarroWhite94,Weil98,SommerLarsen03} and only then confirmed in the beginning of this millennium \citep{Labbe03}. However, the existence of such type of galaxies at such redshifts remains to be fully understood.

Our current knowledge indicates that galaxies with prominent disc profiles (e.g., $n_{\rm Sersic}<2$) are actual frequent and represent a high fraction ($\sim65\%$) of the most massive ($>5\times10^{10}$) high-$z$ galaxies \citep{vanderWel11,Weinzirl11}. Among these, we observe disc galaxies with effective radii comparable to or larger than that of Milky-Way \citep[$r_e\gtrsim5$\,kpc, e.g.,][]{Labbe03,Weinzirl11}, as well as more compact ones \citep[$r_e\lesssim2$\,kpc, e.g.,][]{Buitrago08,vanderWel11}. Overall, these galaxies appear to be clumpy and highly star-forming \citep{Labbe03,Genzel06,Cresci09,Forster09,Cava10}. This goes along with what we expect. It is known that disc instabilities speed up galaxy growth, this is, the clumpier the disc, the quicker and more efficiently its gas content collapses gravitationally and forms stars \citep[e.g.,][and references therein]{LiMacLowKlessen06,Bournaud10}. Also, the disc formation mechanism had to be gradual and gentle in order to have the gas settling onto a disc before converting into stars \citep{Bournaud11a,vanderWel11}. This has been observed by \citet{Genzel06} and simulated by \citet{Dekel09}, where smooth and clumpy cold streams were shown to maintain an unstable gas-rich disc, producing, for several Gyr, giant clumps which would convert into stars and could eventually migrate to the bulge. Hence, the evolution of high-$z$ disc galaxies is expected to be driven by gravitational instability in an inside-out formation scenario, rather than by merger activity \citep{Elmegreen05,Elmegreen09,Dekel09,BournaudElmegreen09,Bournaud11a,Genzel11}. Finally, we expect that Milky-Way-size galaxies at $z\sim2$ evolve to local spheroid systems \citep[either because massive, $\sim10^{11}\,$M$_\odot$, galaxies at $z\sim2$ are found in denser regions or due to the high star-formation and instability in the disc, e.g.,][]{Quadri07p,Quadri07,Scannapieco09,Hartley10,vanDokkum10,Bournaud11a}, while the more compact disc galaxies are the actual candidates for the local spiral population.

However, there has been recent evidence for the existence of compact galaxies showing prominent disc profiles, but already in a passive stage of evolution \citep[][and Section~\ref{c2sec:agncont}]{Stockton08,vanderWel11}. High-$z$ discs are expected to form at $z\sim3-7$ (when the Universe was $\lesssim2$\,Gyr old) and a disc takes $\lesssim$1\,Gyr to form \citep{Eggen62,Scannapieco09,Bournaud11a}. Hence, the rapid evolution seen in this evolved disc population can only be understood through a feedback mechanism which stops cold gas inflows and/or clumps to be formed in the disc. Wind-type feedback (from super-nov\ae or nuclear activity) produces either spheroid or large disc-dominated systems, and cannot explain such compact passive disc galaxies \citep{Okamoto05,Governato07,Scannapieco08,Agertz11}. Morphological quench \citep[MQ,][]{Martig09,Dekel09}, on the other hand, could provide a successful explanation. While the disc evolves through gas instabilities, a bulge is being formed at the same time through the migration of star-forming clumps to the central bulge \citep{Genzel06,Dekel09}. As the bulge grows in mass, its gravitational influence stabilizes the gas distribution in the disc preventing gas clumps to form and produce more stars.

\begin{figure}
  \begin{center}
    \includegraphics[width=0.8\columnwidth]{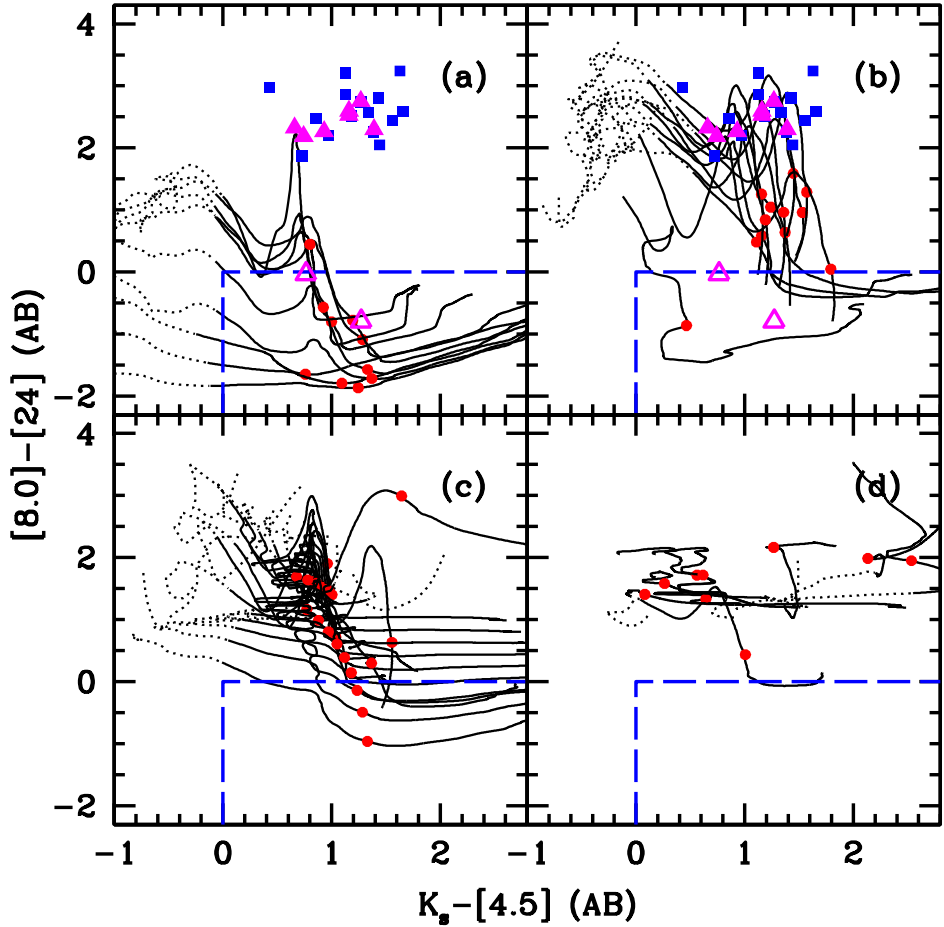}
  \end{center}
  \caption[Selecting high-$z$ passive disc galaxies]{The selection of high-$z$ PDGs. Panel (a) is reserved for early and late-type galaxies, (b) for starburst systems, (c) for hybrid sources, and (d) for pure AGN sources. The colour tracks extend from $z=0$ up to $z=7$. The dotted portion of the tracks indicates the $z<1$ range. The red circle marks $z=3$. The dashed line demarks the selection region where blue dust-free starbursts also fall. These are easily discarded by means of a visual inspection based on their small light profile and/or bright optical detections. For a complete description and discussion of the considered templates, see Chapter~\ref{ch:agn} of this thesis. The mass-selected sample from \citet{Cava10} is shown for comparison in the upper panels as blue squares ($n<1.5$, disc-like) and magenta triangles ($n>1.5$, spheroid-like), revealing an overall non-passive nature as revealed in that work.}
  \label{c5fig:seldiscs}
\end{figure}

\begin{figure}
  \begin{center}
    \includegraphics[width=1\columnwidth]{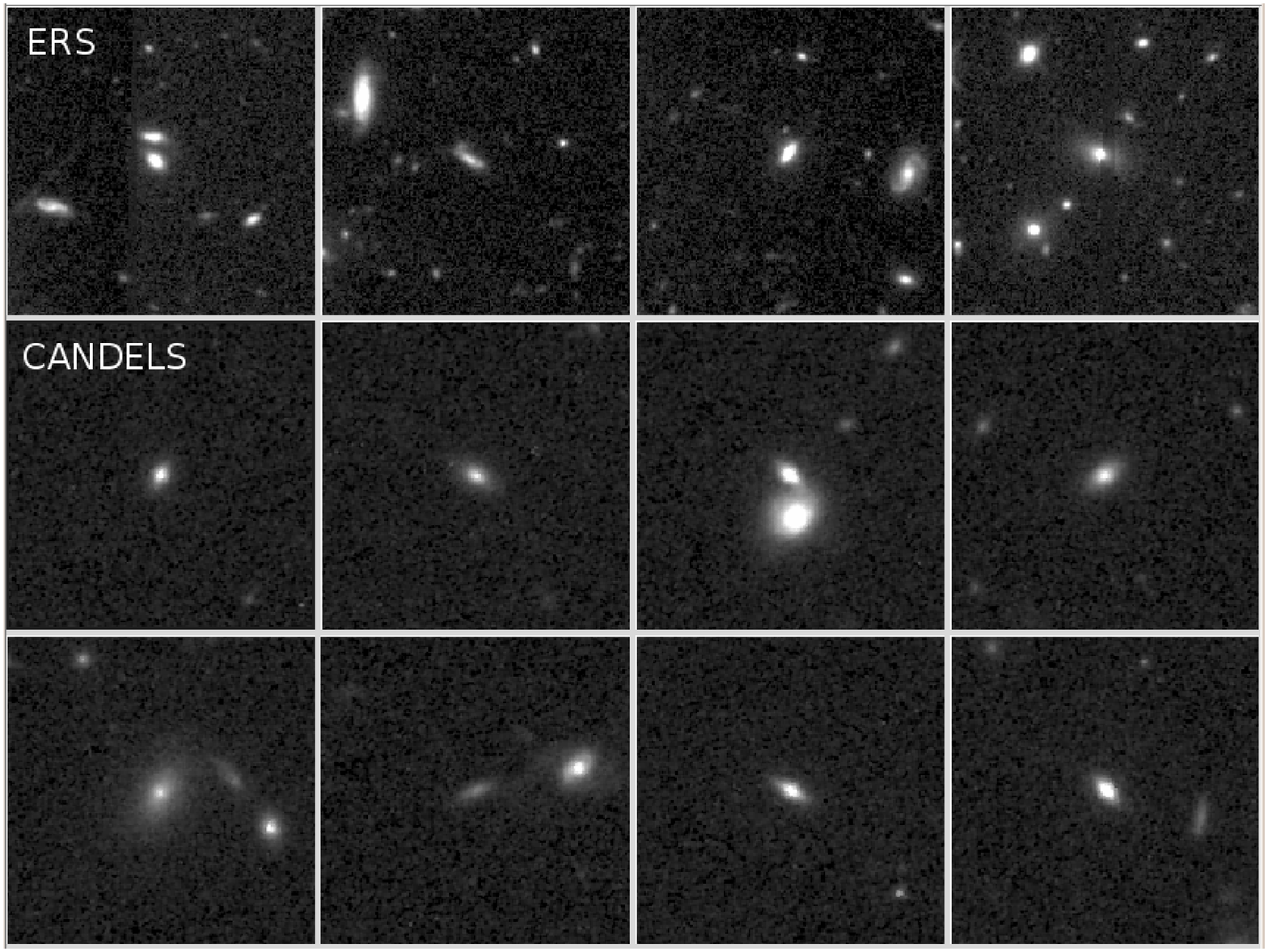}
  \end{center}
  \caption[WFC3-H160 imaging of passive disc galaxies]{Twelve examples of PDGs found in GOODSs are presented. These are 10'' wide WFC3-$H160$ cut-outs from ERS (top row) and CANDELS (two bottom rows) observations. In most, the disc profile is clearly observed.}
  \label{c5fig:discimag}
\end{figure}

The samples of passive disc galaxies (PDGs) are, nevertheless, still scarce. A larger number of such galaxies should be assembled in order to establish their nature and their local counterparts. The search for such PDGs has been based so far on the very generic $J-K$ colour (or comparable colour cuts), or rest-frame colours of mass selected samples. While the first is known to select extremely obscured star-forming systems (including edge-on discs), the latter technique may miss lower-mass passive disc samples, require multi-wavelength photometry, and is very model dependent. Also, many of these candidates are still selected in observed optical bands, where a $z\sim2$ passive system is faint and hard to detect. For instance, the PDG candidate referred in this work (Section~\ref{c2sec:mir25}) is not detected even in the reddest ACS-$HST$ band, $z_{850}$. We have thus developed a new IR-morphology selection of PDGs shown in Figure~\ref{c5fig:seldiscs}. The $K_s-[4.5]$ colour is highly redshift dependent, so different cuts imply different universe epochs. We adopt $K_s-[4.5]>0$ for the selection of $z>1$ sources. The $[8.0]-[24]<0$ cut ensures the selected objects do not have significant obscured star-formation nor AGN emission (i.e., are mostly passive). One can see that practically all spectral energy distributions at $1<z<3$ entering the lower right corner are early/late type systems. The exception being blue (low-metalicity/dust-free) star-forming systems (upper right panel), which can be weed-out by optical inspection (identifying sources which are bright and with no disc component). Deep optical/NIR imaging is then used to select the final candidate sample of PDGs at $1<z<3$. A consistent sample has now been assembled in CDFS using the MUSIC catalogue \citep{Santini09}. Being composed by $\sim$40 objects at $1<z<3$, is more numerous than any of the samples seen in previous works. A few examples of PDGs found in the final sample are shown in Figure~\ref{c5fig:discimag}.

\begin{figure}
  \begin{center}
    \includegraphics[width=1\columnwidth]{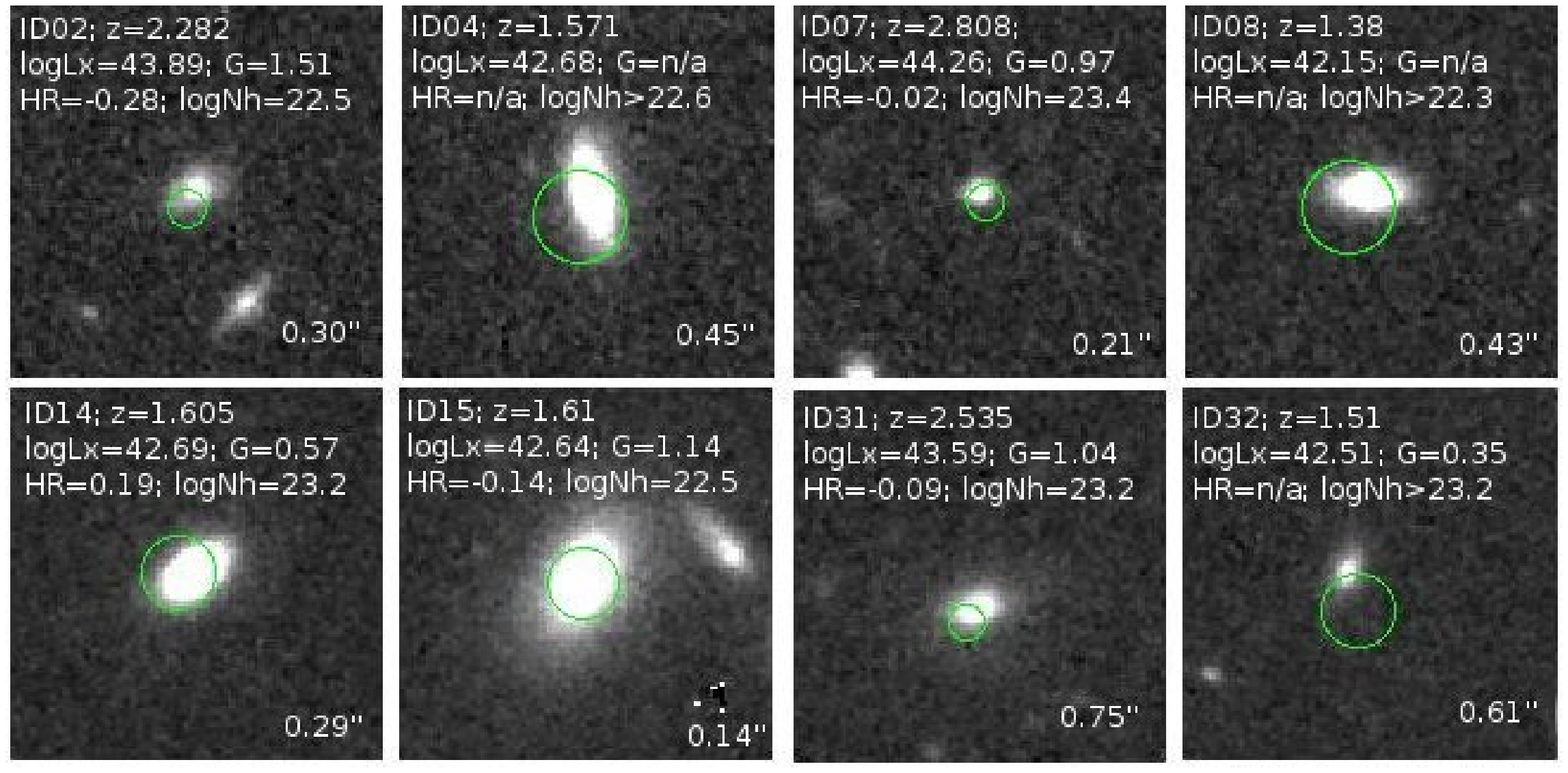}
  \end{center}
  \caption[X-rays detections in passive disc galaxies]{WFC3-$HST$ imaging \citep{Koekemoer11} of the X-rays detected PDGs. Stamp size of $5''\times5''$. Information includes: redshift, X-ray luminosity, observed photon index (G), hardness-ratio, column density ($\rm{N_H}$), and distance to counterpart (bottom right). Green circles show $3\sigma$ confidence astrometry of X-rays detection.}
  \label{c5fig:pdgxr}
\end{figure}

The immediate objective of this project will be the census of $1<z<3$ passive discs and aim to unveil the cause for their existence. Due to their nature, a prominent $1.6\,\mu$m stellar emission bump is present in the SEDs of PDGs. Such feature is probed by the IRAC channels in the $1<z<3$ range, enabling high-quality photometric redshift estimates ($(z_{\rm spec}-z_{\rm spec})/(1+z_{\rm spec})\simeq0.05$ at $1\lesssim{z}\lesssim1.6$\footnote{Until now, only 12 PDGs have available spectroscopy, and they are found in the $1\lesssim{z}\lesssim1.6$ range.}). Furthermore, if MQ is indeed the mechanism to create such systems, a non-negligible fraction of the gas still remains, and is expected to be detected in PDGs. For instance, X-ray emission has been detected for 20\% of our CDFS PDG sample \citep[$<0.8''$ search radius in the 4\,Ms CDFS coverage,][]{Xue11}, and all present high obscuring column-densities ($\rm{\log(N_H[cm^{-2}])>22}$, Figure~\ref{c5fig:pdgxr}). This probably means that a significant fraction of the left-over gas may have migrated to the nuclear region \citep{Bournaud11b}, producing the high obscuration in the X-rays (which is unlikely due to dust, provided the monochromatic/blue $[8.0]-[24]$ colours of PDGs). If such quantity of gas is present, spectroscopy observations in the near-IR (observing H$_\alpha$) and (sub-)millimetre (observing molecular gas) will allow us to study the distribution of gas in PDGs (is it concentrated in the nucleus or inefficiently forming stars in the disc?) and constrain its total amount (are PDG passive because of actual gas-drought or due to MQ?).

With upcoming new observational facilities, the mystery behind PDGs will potentially be unveiled. \textit{James Webb Space Telescope} ($JWST$) will allow the selection of deep and large samples of PDGs, as well as a detailed morphological and near-IR spectroscopy study of these galaxies. ALMA, that has now started, with 16 antennas, observing the science targets proposed for Cycle-0, will also play a major role. The final assembly of this telescope is rapidly reaching its end. At that stage, the study of PDGs will largely benifit from the unique sensitivity and resolution power of ALMA, enabling us to understand such peculiar population which may also explain the origins of our own home.

\section{The search for the most obscured AGN}

It is now clear that the development of the KI and KIM criteria (Chapter~\ref{ch:agn}) has great implications to the science to be made with \textit{James Webb Space Telescope} ($JWST$). One of the prime objectives of these criteria is the search for the most obscured AGN sources in the Universe. In the literature, many groups have attempted to select the so-called compton-thick AGN in deep hard-band X-ray surveys or with extremely red optical-to-MIR colours. While the former is known to miss a significant portion of the obscured AGN sample, the latter is contaminated by extremely obscured non-AGN galaxies, unless stringent constraints are considered. One of this rigorous constraints is high MIR flux cuts ($f_{24\mu\rm{m}}>1$\,mJy), attainable only by the brightest AGN sources. \citet{Fiore08} considered even fainter sources on the assumption that extremely red $R-K$ colours would select a higher fraction of AGN sources. However, as seen in Figure~\ref{c2fig:coragn} (Section~\ref{ch:agn}), although \citet{Fiore08} are right in their assumption, a $J-K$ colour cut is more efficient on achieving that goal, with the advantage that the selected sources will be at higher redshifts with increasingly redder $J-K$ colour cuts (Figure~\ref{c2fig:jkvsz}).

One final class of objects holds a place of interest. In Chapter~\ref{ch:agn}, it was shown the importance of the $K-[4.5]$ colour for the selection of AGN hosts and the characterization of high-$z$ sources. Hence, searching for those sources appearing in deep 4.5\,$\mu$m coverages while remaining undetected in the $K$-band will certainly provide samples of the most extreme sources in the Universe. In combination with higher wavelength bands ($\geq8\,\mu$m), a reliable sample of extremely obscured AGN will be assembled. While GOODSs is already being surveyed for such class of AGN (relying on the MUSIC catalogue, which considers $z_{850}$-, $K_s$-, and 4.5\,$\mu$m-selected sources), COSMOS is likely another deep survey to explore in the search for such class of objects. Also, as soon as SERVS data (deep 3.6 and 4.5\,$\mu$m coverage, PI M. Lacy) is available on VIDEO (a 12\,deg.$^2$ deep near-IR survey, PI M. Jarvis), a large statistical sample of extreme 4.5\,$\mu$m-detected $K$-undetected sources will be assembled for further study, providing useful constraints to both galaxy and AGN evolution models (e.g., number densities, luminosity distributions).

Nonetheless, the IR AGN selection may still require some fine tuning. For instance, all the IR AGN diagnostics have never been tested against the emission from TP-AGB stars \citep[which is known to peak at 2\,$\mu$m,][]{Maraston05}. This is crucial to the high-redshift regime where a larger incidence of systems with enhanced TP-AGB stellar emission is known to reside \citep{Maraston05,Henriques11}. If such effect in the IR regime significantly affects IR AGN selection, than we are forced to use only the most restrictive AGN criteria \citep[like the bright IR excess sources, e.g.,][]{Polletta06,Dey08} or to rely solely on the remainder spectral regimes, which sometimes is not the ideal scenario. However, by peaking at 2\,$\mu$m, the emission from TP-AGB circumnuclear discs is unlikely to produce a red (power-law) SED in the 2--8(--24)\,$\mu$m range probed by the KI and KIM criteria.

\section{Direct comparison of the evolution of hot and cold dust}

Chapter~\ref{ch:lfs} has shown that the evolution of 3.3\,$\mu$m dust emission is declining much more rapidly than the overall SF history \citep[and hence the colder dust emission,][]{CharyPope10} in the Universe. But this study allows to go even further. With the large galaxy numbers available in the COSMOS field, it will be possible to obtain cold-dust luminosity density functions like those for hot-dust in Figure~\ref{c4fig:lf33d}. Hence, a direct comparison of hot and cold dust at each given luminosity bin will be possible. The comparatively shallow FIR/millimetre surveys poses a problem that can be circumvented in this project, as the large galaxy numbers available in each of these bins will allow for a proper stacking analysis, providing average FIR/mm fluxes for each of the luminosity classes. With this we aim to track the interplay between the hot and cold dust regions since high redshifts to the present.

\section{Closing remarks}

There is no doubt that the IR spectral regime has revolutionised our understanding of the Universe we live in. The presence of numerous galaxy populations undetected at optical wavelengths has showed us that there is much more to see beyond the narrow optical spectral window. The journey is still far from finished. Our ability to probe the multi-wavelength Universe to amazing depths is improving by the day, either through instrumental advances or better facilities. Although one should never underestimate archive-based science, the coming observational facilities will provide data-sets that will revolutionise once again our way of thinking, and show that the Universe is as unpredictable as it is fun.

\singlespacing
\addcontentsline{toc}{chapter}{Bibliography}
\bibliography{tesebib}


\end{document}